\documentclass{aa}

\def\HI{\hbox{\rm H\,{\sc i}}}

\def\emph#1{{\sl #1}}
\newcommand{\ltsima} {$\; \buildrel < \over \sim \;$}
\newcommand{\gtsima} {$\; \buildrel > \over \sim \;$}
\newcommand{\lta} {\lower.5ex\hbox{\ltsima}}
\newcommand{\gta} {\lower.5ex\hbox{\gtsima}}

\usepackage{latexsym, graphicx, natbib, longtable}
\usepackage{longtable}
\usepackage{lscape}
\usepackage{txfonts}
\begin{document}
   \title{The WSRT Virgo {\HI} filament survey I}

   \subtitle{Total Power Data}

   \author{A. Popping
          \inst{1} \inst{2} \inst{3}
          \and
          R. Braun\inst{3}
          }

   \offprints{A. Popping \email{attila.popping@oamp.fr} }

   \institute{Laboratoire d'Astrophysique de Marseille, 38 Rue Fr\'{e}d\'{e}rique Joliot-Curie, 13388 Marseille Cedex 13, France
    \and 
    Kapteyn Astronomical Institute, P.O. Box 800, 9700 AV Groningen, the Netherlands
    \and
    CSIRO -- Astronomy and Space Science, P.O. Box 76, Epping, NSW 1710, Australia}

   \date{}

% \abstract{}{}{}{}{} 
% 5 {} token are mandatory
 
  \abstract
   {Observations of neutral hydrogen can
  provide a wealth of information about the kinematics of galaxies. To
  learn more about the large scale structures and accretion processes,
  the extended environment of galaxies have to be observed. Numerical
  simulations predict a cosmic web of extended structures and gaseous
  filaments.}
   {To observe the direct vicinity of galaxies, column
  densities have to be achieved that probe the regime of Lyman limit
  systems. Typically {\HI} observations are limited to a brightness
  sensitivity of $N_{HI} \sim 10^{19}$ cm$^{-2}$ but this has to be
  improved by $\sim2$ orders of magnitude.}
   {With the Westerbork
  Synthesis Radio Telescope (WSRT) we map the galaxy filament
  connecting the Virgo Cluster with the Local Group. About 1500 square
  degrees on the sky is surveyed, with Nyquist sampled pointings. By
  using the WSRT antennas as single dish telescopes instead of the
  more conventional interferometer we are very sensitive to extended
  emission.}
   {The survey consists of a total of 22,000 pointings and
  each pointing has been observed for 2 minutes with 14 antennas. We
  reach a flux sensitivity of 16 mJy beam$^{-1}$ over 16 km s$^{-1}$,
  corresponding to a brightness sensitivity of $N_{HI} \sim 3.5\times10^{16}$ cm$^{-2}$ 
       for sources that fill the beam. At a typical distance of 10 Mpc 
       probed by this survey, the beam extent corresponds to 
       about 145 kpc in linear scale. Although the processed data cubes are affected
  by confusion due to the very large beam size, we can identify most
  of the galaxies that have been observed in HIPASS. Furthermore we
  made 20 new candidate detections of neutral hydrogen. Several of the
  candidate detections can be linked to an optical counterpart. The
  majority of the features however do not show any signs of stellar
  emission. Their origin is investigated further with
  accompanying {\HI} surveys which will be published in follow up papers.}
   {}

   \keywords{galaxies:formation -- 
                galaxies: intergalactic medium  }

   \maketitle
%
%________________________________________________________________

\section{Introduction}

Unbiased, wide-field sky surveys are very important in improving 
understanding of our extended extragalactic environment. They provide
information about the clustering of objects and the resulting large scale
structures. Furthermore, they are essential in providing
a complete sample of galaxies, their mass function and physical
properties. Several outstanding examples are the SDSS (Sloan Digital
Sky Survey) \citep{2000AJ....120.1579Y} at optical wavelengths, and
HIPASS ({\HI} Parkes All Sky Survey) \citep{2001MNRAS.322..486B} and
ALFALFA (The Arecibo Legacy Fast ALFA Survey)
\citep{2005AJ....130.2598G} in the 21cm line of neutral hydrogen. All
these surveys have been important milestones, that significantly
improved our understanding of the distribution of galaxies in the
universe. But despite the impressive results, these surveys can only
reveal the densest structures in the Universe like galaxies, groups and
clusters.

In the low redshift Universe, the number of detected baryons is
significantly below expectations, indicating that not all the baryons
are in galaxies. According to cosmological measurements the baryon
fraction is about 4$\%$ at $z \sim 2$ (\cite{2003ApJS..148....1B};
\cite{2003ApJS..148..175S}). This is consistent with actual numbers of
baryons detected at $z > 2$ (\cite{1997ApJ...490..564W};
\cite{1998ARA&A..36..267R}). In the current epoch however, at $z \sim
0$ about half of this matter has not been directly observed
(\cite{1998ApJ...503..518F}; \cite{1999ApJ...514....1C};
\cite{2000ApJ...534L...1T}; \cite{2002ApJ...564..631S};
\cite{2004ApJS..152...29P}).

Recent hydrodynamical simulations give a possible solution for the
``Missing Baryon'' problem (\cite{1999ApJ...514....1C};
\cite{2001ApJ...552..473D}; \cite{2002ApJ...564..604F}). Not all the
baryons are in galaxies, that are just the densest concentrations in the
Universe. Underlying them is a far more tenuous Cosmic Web, connecting the
massive galaxies with gaseous filaments. The simulations predict that
at $z$~=~0 cosmic baryons are almost equally distributed amongst three
phases (1) the diffuse IGM, (2) the warm hot intergalactic medium
(WHIM), and (3) the condensed phase. The diffuse phase is associated
with warm, low-density photo-ionized gas. The WHIM consists of gas
with a moderate density, that has been heated by shocks during
structure formation. The WHIM has a very broad temperature range from
$10^5$ to $10^7$ K. The condensed phase is associated with cool
galactic concentrations and their halos. These three components are
each coupled to a decreasing range of baryonic over-density:
log$(\rho_H/\bar{\rho_H})$ $<$ 1, 1--3.5, and $>$ 3.5 and are probed by QSO
absorption lines with specific ranges of neutral column density:
log$(N_{HI})$ $<$ 14, 14--18 and $>$ 18 \citep{2005ASPC..331..121B}.

\subsection{Cosmic Web}
The Warm Hot Intergalactic Medium is thought to be formed during
structure formation. Low density gas is heated by shocks 
during its infall onto the filaments that define the large scale
structure of the Universe. Most of these baryons are still
concentrated in unvirialized filamentary structures of highly
ionized gas.

The WHIM has been observationally detected in QSO absorption line
spectra using lines of NeVIII (Savage et al. 2005), OVI (e.g. Tripp et
al. 2008), broad Ly$\alpha$ (Lehner et al. 2007) and X-ray absorption
\citep{2005AdSpR..36..721N}. Of course, absorption studies alone do
not give us complete information on the spatial distribution of the
WHIM. Emission from the Cosmic Web would give entirely new information
about the distribution and kinematics of the intergalactic gas.

Direct detection of the WHIM is very difficult in the EUV and X-ray
bands \citep{1999ApJ...514....1C}. The gas is ionized to such a
degree, that it becomes ``invisible'' in infrared, optical or UV light,
but should be visible in the FUV and X-ray bands
\citep{2005AdSpR..36..721N}. Given the very low density, extremely
high sensitivity and a large field of view is needed to image the
filaments. Capable detectors are not yet available for the X-ray or
FUV (\cite{2003PASJ...55..879Y}; \cite{2005AdSpR..36..721N}).

Due to the moderately high temperature in the intergalactic medium
(above $10^4$ Kelvin), most of the gas in the Cosmic Web is highly
ionised. To detect the trace neutral fraction in the photoionized
Ly$\alpha$ forest using the 21-cm line of neutral hydrogen, a column
density sensitivity of $N_{HI}\sim10^{17-18}$ cm$^{-2}$ is
required. At the current epoch we can confidently predict that in
going down from {\HI} column densities of $10^{19}$ cm$^{-2}$ (which
define the current ''edges'' of well studied nearby galaxies in {\HI}
emission) to $10^{17}$ cm $^{-2}$ the surface area will significantly
increase, as demonstrated in \cite{2002ApJ...567..712C},
\cite{2004A&A...417..421B} and \cite{2009A&A...504...15P}.

The critical observational challenge is crossing the ``{\HI} desert'',
the range of log($N_{HI}$) from about 19.5 down to 18 over which
photo-ionization by the intergalactic radiation field produces an
exponential decline in the neutral fraction from essentially unity
down to a few percent (eg. \cite{1994ApJ...423..196D}). Nature is
kinder again to the {\HI} observer below log(N$_{HI}$) = 18, where the
neutral fraction decreases only very slowly with log($N_{HI}$). The
neutral fraction of hydrogen is thought to decrease with decreasing
column density from about 100$\%$ for $\log(N_{HI})$ $\>=$ 19.5 to
about 1$\%$ at $\log(N_{HI})=17$ \citep{1994ApJ...423..196D}. The
baryonic mass traced by this gas is expected to be comparable to that
within the galaxies, as noted above.

To detect the peaks of the Cosmic Web in {\HI}, a blind survey is
required that covers a significant part of the sky, of the
order of at least 1000 square degrees. Furthermore a brightness sensitivity
is required that is about an order of magnitude more sensitive than
HIPASS. 

The Westerbork Synthesis Radio Telescope (WSRT) has been used to
undertake a deep fully sampled survey mapping $\sim 1300$ square
  degrees of sky. The survey covers a slab perpendicular to the plane
  of the local supercluster, centred on the galaxy filament
  connecting the Local Group with the Virgo Cluster.  Due to our
  observing strategy with declinations between $-$1 and 10 degrees and a
  limited velocity range, the survey does not encompass the complete
  Virgo cluster. In an unbiased search for diffuse and extended {\HI}
gas, both the auto-correlation and cross-correlation data are reduced
and analysed. In this paper we will only discuss the total-power
product, as this product is most sensitive to faint and extended
emission. The resulting detections will be further analysed and
compared with the cross-correlation data products and other data in
subsequent papers.

We have achieved an {\sc RMS} sensitivity of about 16~mJy
Beam$^{-1}$ at a velocity resolution of 16~km~s$^{-1}$ over $\sim
1300$ deg$^2$ and between $400 < V_{Hel} < 1600$ km s$^{-1}$. The
corresponding {\sc RMS} column density for emission filling the
$2983\times2956$ arcsec effective beam area is $\sim 3.5 \times
10^{16}$ cm$^{-2}$ over 16~km~s$^{-1}$. Although the flux sensitivity
is similar to HIPASS, that has typically achieved 13.5
mJy Beam$^{-1}$ at a velocity resolution of 18 km s$^{-1}$, the column
density sensitivity is far superior. With the 14 arcmin intrinsic beam
size of the Parkes telescope, the {\sc RMS} column density
sensitivity in HIPASS is $\sim 4 \times 10^{17}$ cm$^{-2}$ over 18 km
s$^{-1}$, which is more than an order of magnitude less sensitive.

In the Westerbork Virgo Filament Survey we detect 129 sources that are
listed in the HIPASS catalogue. We have made 20 new {\HI} detections,
of which many do not have a clear optical counterpart. The outline of
this paper is as follows: in Sect.~\ref{sec:obs} we describe the
survey observations and strategy, directly followed by the reduction
procedures of the auto-correlation data. In Sect.~\ref{sec:results} we
present the results of {\HI} detections of known galaxies and the new
detections. We end with a short discussion and conclusion in
Sect.~\ref{sec:conclusion}. The results of the cross-correlation data
of the Westerbork Virgo Filament Survey and the detailed analysis and
data comparison will be presented in two subsequent papers.

\begin{figure}
  \includegraphics[width=0.5\textwidth]{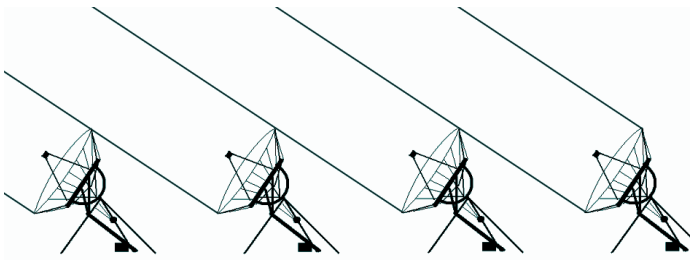}
  \caption{Observing mode of the WSRT dishes; a filled aperture of
    300~m is simulated by placing 12 of the 14 telescopes at regular
    intervals and observing only at extreme hour angles.}
  \label{config4}
\end{figure}

\section{Observations}
\label{sec:obs}
To obtain the highest possible brightness sensitivity in
cross-correlations, the WSRT was configured to simulate a large filled
aperture in projection. Twelve of the 14 WSRT 25~m telescopes were
positioned at regular intervals of 144~m. When observing at very low
declinations and extreme hour angles, a filled aperture is formed (as
can be seen in Fig.~\ref{config4}), which is 300 $\times$ 25~m in
projection. In this peculiar observing mode the excellent spectral
baseline and PSF properties of the interferometer are still obtained
while achieving excellent brightness sensitivity. A deep fully-sampled
survey of the galaxy filament joining the Local Group to the Virgo
Cluster has been undertaken, extending from 8 to 17 hours in RA and
from $-$1 to +10 degrees in declination and covering 40~MHz of
bandwidth with 8~km~s$^{-1}$ resolution.\\

Simultaneously with the cross-correlation data, auto-correlation data
was acquired. These auto-correlation data pertain to the same set of
positions on the sky. Data were acquired in a semi-drift-scan mode,
whereby the 25~m telescopes of the WSRT array tracked a sequence of
positions for a 60~s integration that were separated by one minute of
right ascension (about 15 arcmin) yielding Nyquist-sampling in the
scan direction of the telescope beam. Data was acquired in two 20~MHz
IF bands centered at 1416 and 1398 MHz. The beamwidth of each
telescope is $38\times 37$ arcmin FWHM at an observing frequency of
1416~MHz.\citep{2008A&A...479..903P}. Each drift-scan sequence,
lasting about 9 hours, was separated by 15 arcmin in declination to
give Nyquist sampling. Typically, an observing sequence consisted of a
standard observation of a primary calibration source (3C48 or 3C286) a
drift-scan observation and an additional primary calibration
source. Each session provided a strip of data of $135 \times 0.25$
true degrees. In total 45 of these strips provided the full survey
coverage of 11 degrees in declination. Each of the total of 24,300
pointings was observed two times, once when the sources were rising
and once when they were setting. The total of 90 sessions were
distributed over a period of more than two years, between December 2004
and March 2006.\\

Although the observations cover a large bandwidth in each of two bands,
  we only use the radial velocity range from 400 to 1600 km s$^{-1}$
  in the first band. For lower radial velocities, the emission is too
  confused with Galactic emission and combined with the very
  large beam size, useful analysis was deemed impractical. The
  second IF band with a lower central frequency samples larger
  distances, where the central frequency corresponds to a Hubble-flow
  distance of about 65
  Mpc. The physical beam size at this distance is about 850 kpc.
  Detecting emission which fills such a large beam would be
  very unlikely, while the problem of confused detections
  is more serious.

To minimize solar interference, an effort was made to measure the data
only after local sunset and before local sunrise. Unfortunately this
was not successful for the whole survey and a few runs show the
effects of solar interference.\\

\section{Data reduction}
Auto-correlation and Cross-correlation data were acquired
simultaneously, and were separated before importing them into Classic
AIPS \citep{1981NRAON...3....3F}. We will now only describe the steps
that have been undertaken to reduce the auto-correlation or
total-power data. The reduction method for the cross-correlation data
is significantly different and will be
described in another publication.\\

Every baseline of the drift-scan data of each survey run was inspected
and flagged in Classic AIPS, using the
SPFLG utility. Suspicious features appearing in the frequency or time
display of each auto-correlation baseline were critically
inspected. This was accomplished by comparing the 28 independent
spectral estimates resulting from 14 telescopes, each with two
polarizations. Features which could not be reproduced in the
simultaneous spectra were flagged.\\

Absolute flux calibration of the data was provided by the observed
mean cross-correlation coefficient measured for the standard
calibration sources (3C48 or 3C286) of known flux density. The
measured ratio of flux density to correlation coefficient averaged
over all 14 telescopes and 2 polarizations was $340 \pm 10$
Jy/Beam.\\

Two different methods were employed to generate data-cubes of the
auto-correlation data. The main difficulty with total power data, is
obtaining a good band-pass calibration. The first method employed
taking a robust average of a 30~min sliding window, to estimate the
band-pass as a function of time and an 850~km~s$^{-1}$ sliding window
to estimate the continuum level as a function of frequency. Only the
inner three quartiles of the values were included in these averages,
making them moderately robust to outliers, including {\HI} emission
features, in the data. The big advantage of this method is that it
could be applied blindly in a relative fast way, and it produces
uniform noise characteristics in the resulting cube. In this way, it
is very suitable for detecting faint and diffuse sources. However the
disadvantage is that bright sources with a moderately high level of
{\HI} emission that are extended in either the spatial or velocity
direction produce a local negative artifact. Under these
circumstances, better results are obtained with a more complicated and
time consuming method, described below.\\

\begin{figure*}[t!]
  \includegraphics[width=0.5\textwidth]{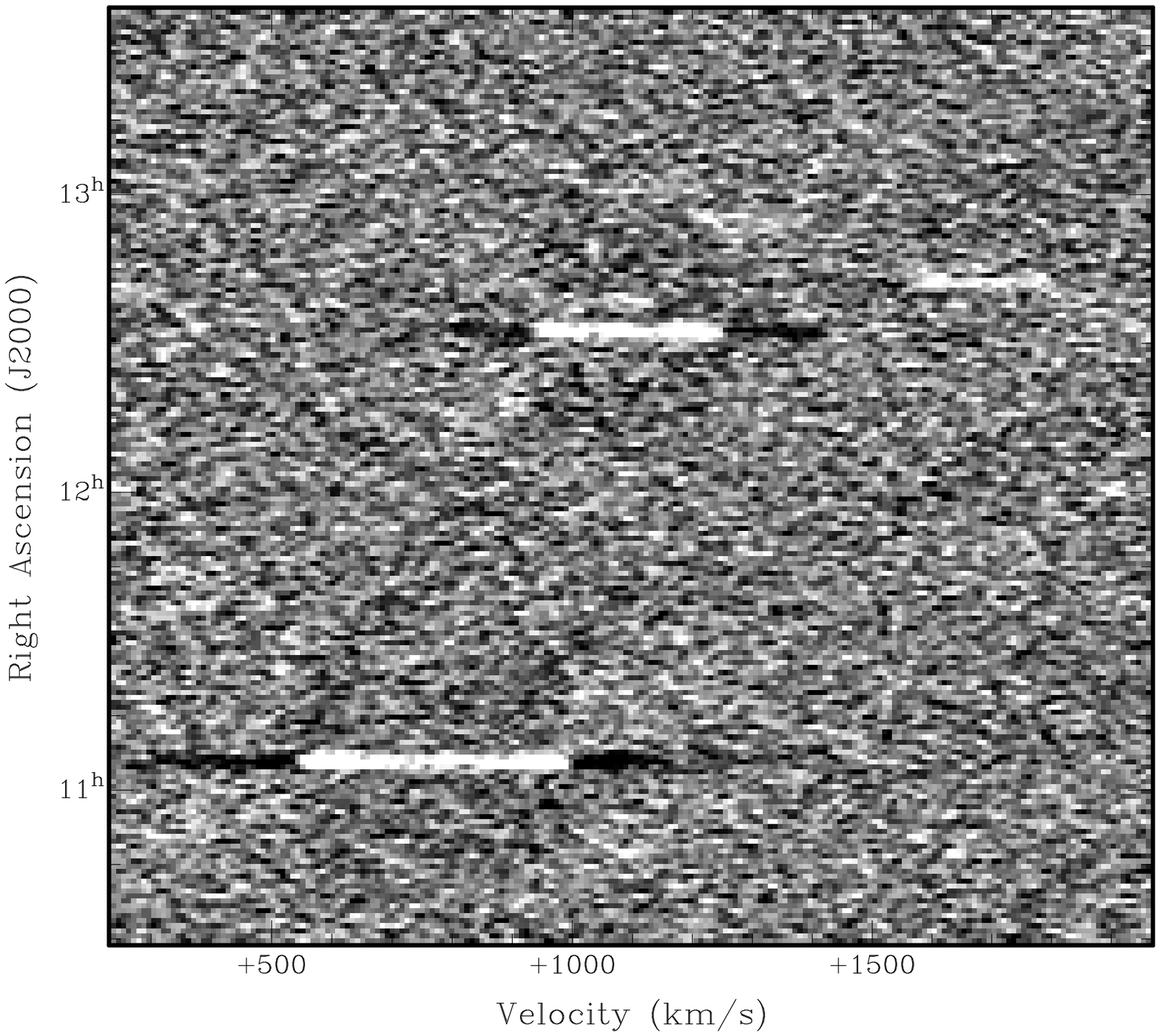}
  \includegraphics[width=0.5\textwidth]{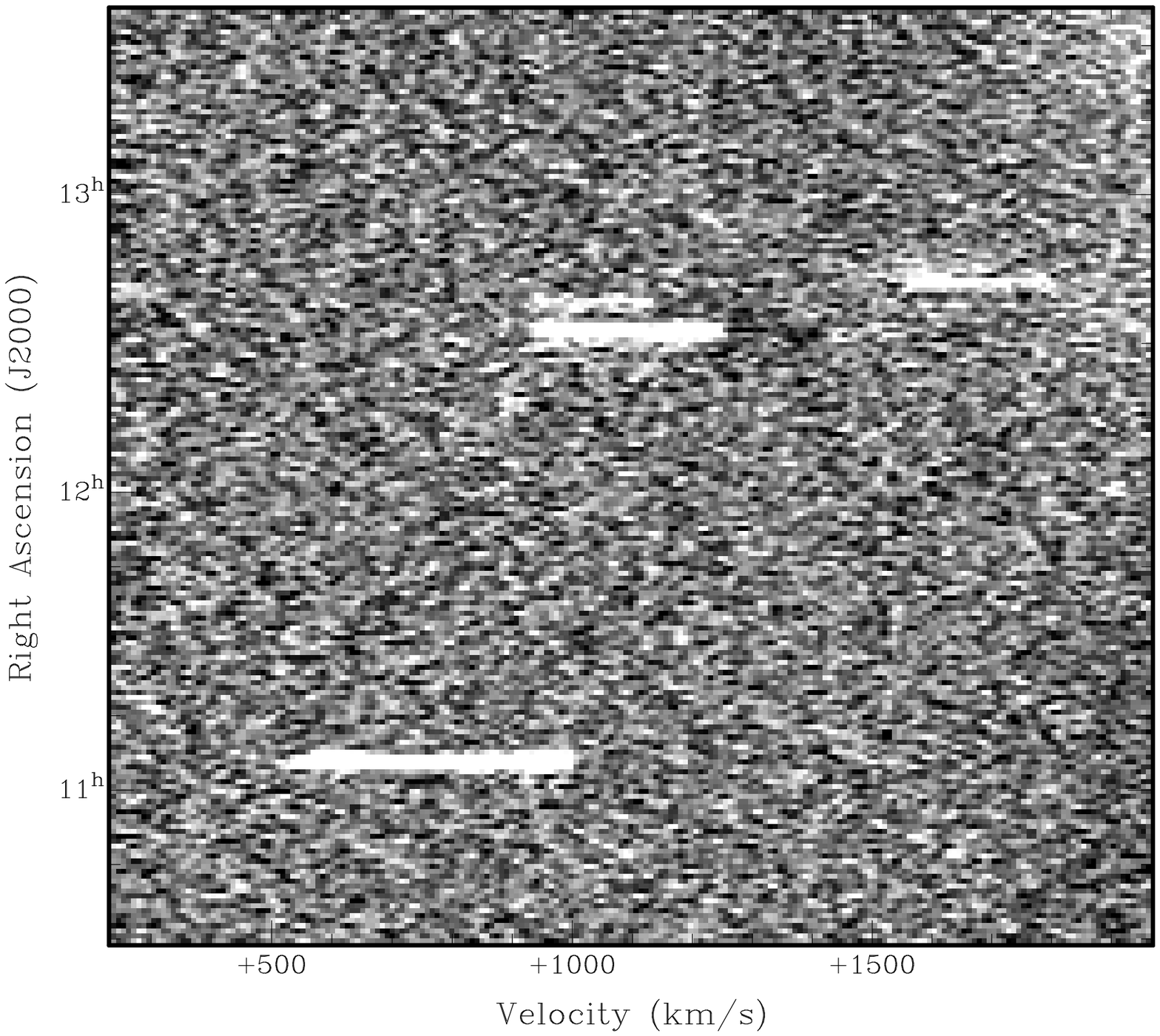}
  \caption{Illustration of the bandpass correction method. In the left
    panel a robust average over a sliding window in both frequency and
    position is used to identify the brightest sources of emission. In
    the right panel the bright sources have been individually masked
    before carrying out a polynomial fit. Both panels show the same
    region (declination is zero) with the same intensity scale.}
  \label{bandpass}
\end{figure*}

The result of the first bandpass-removal method has been used to
create a mask. For each declination the clearly recognisable bright
sources that correspond to galaxies were included by hand. In the
mask, the location of the galaxies was set to zero and the rest of the
declination scan was set to unity. The mask was applied to the raw
data, so only the noise, diffuse sources and the bandpass
characteristics remain. A second order polynomial was then fit in the
frequency direction and the masked data is divided by this polynomial
result. In the next step a zeroth order polynomial is fit in the time
domain and the masked data is divided by this product. Finally a third
order polynomial is applied again in the frequency domain, to remove
small oscillations or artifacts.  Within each declination strip a
correction has been applied to correct for the Doppler shift at the
time of the observation before combining the declinations and creating
a three dimensional cube. The improvement in using the second method
for the bandpass correction is shown in Fig.~\ref{bandpass}. In the
left panel bright sources can be easily identified, however there are
large negative spectral artifacts at the source location. By masking the regions
of bright emission, a much better bandpass estimate could be achieved
that does not suffer from artifacts as can be seen in the right panel
of Fig.~\ref{bandpass}.

\subsection{Doppler Correction}
The drift-scan data were resampled in frequency to convert from the
fixed geocentric frequencies of each observing date to a heliocentric
radial velocity at each observed position. The offsets in velocity
have been determined using the reference coordinate utilities within
aips++\footnote{The AIPS++ (Astronomical Information Processing System) is a
product of the AIPS++ Consortium. AIPS++ is freely available for use
under the Gnu Public License. Further information may be obtained from
http://aips2.nrao.edu.}. This correction depends on the
earth's velocity vector relative to the pointing direction at the time
of an observation and varies between about $-$30 and +30 km s$^{-1}$
during the course of a year. Since the observations have been
undertaken over a time span of several years, this effect has to be
taken into account.

\subsection{Calibration}

\begin{figure}[t!]
\begin{center}
  \includegraphics[width=0.5\textwidth]{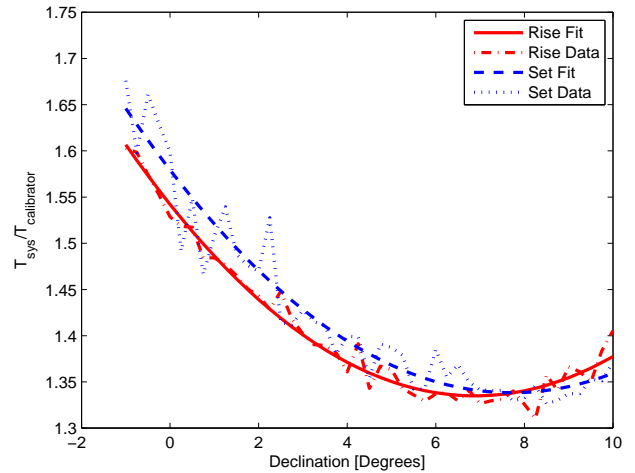}
   \caption{Due to the extreme hour angles at which the observations
     were taken, there is an increased system temperature with respect
     to the calibrators. This correction is dependent on the
     declination. The dash-dotted line represents the calibration
     factors for the {\it rise} data with the best second order
     polynomial fit shown as a solid line. The dotted line corresponds
     to the {\it set} data, with the fit shown as a short-dashed
     line.}
  \label{calibration}
\end{center}
\end{figure}

Due to the extreme hour angles and low declinations of the
observations, there is a larger intervening airmass (between 1.35
  and 1.7) and increased ground pick-up effecting the observed
  emission than in a typical observation. While the attenuation of the
  astronomical signal is minimal (less than 2\%) in view of the low
  zenith opacity at the observing frequency, the system temperature
  increases significantly. This increase is measured directly by
  comparison with a periodically injected noise signal of known
  temperature and can be understood in terms of a combination of
  atmospheric emission and the extended far-sidelobe pattern of the
  telescope response convolved with the telescope environment.  As a
result, the system temperature $(T_{sys})$ of the survey scans was
higher than for the calibrator sources. This effect has to be taken
into account when doing the gain-calibration to get correct flux
values. In Fig.~\ref{calibration} this correction factor is plotted as
function of declination, based on the ratio of system temperatures
seen in the survey scans relative to the associated calibration
scans. The correction that has to be applied is strongly correlated
with declination (since this is directly coupled to elevation); at the
lowest declination of $-$1 degrees, the gains have to be multiplied by
a factor $\sim 1.6$ to get correct flux values. The minimum correction
is near 7.5 degrees. The slight increase in the ratio at higher
declinations may be due to increasing ground pick-up in the spill-over
lobe of the telescope illumination pattern. The scans that observe the
setting of the sources have a slightly higher correction
factor. Antenna 1 (locally known as RT0) suffered from severe blockage
by the trees to the west of the array at these extreme hour angles and
therefore it has not been used. The gain corrections can be fit using
a second order polynomial. These corrections have been applied
independently to both the rise and set data.

\subsection{Data Cubes}
The 45 drift-scans of both the setting and rising data were combined
into two separate data cubes and exported to the MIRIAD software
package \citep{1995ASPC...77..433S}. A combined cube was obtained by
taking the {\sc RMS}-weighted average of the two independent cubes
containing all the data. This cube combines two fully independent
surveys of the same region. A spatial convolution was applied to all
three cubes with a 2000 arcsec FWHM Gaussian with PA=0 to introduce
the desired degree of spatial correlation in the result. A hanning
smoothing was applied with a width of three pixels to smooth the cubes
in the velocity domain, resulting in a velocity resolution of
16~km~s$^{-1}$.

\subsection{Sensitivity}
\begin{figure}[t!]
  \begin{center}
  \includegraphics[width=0.5\textwidth]{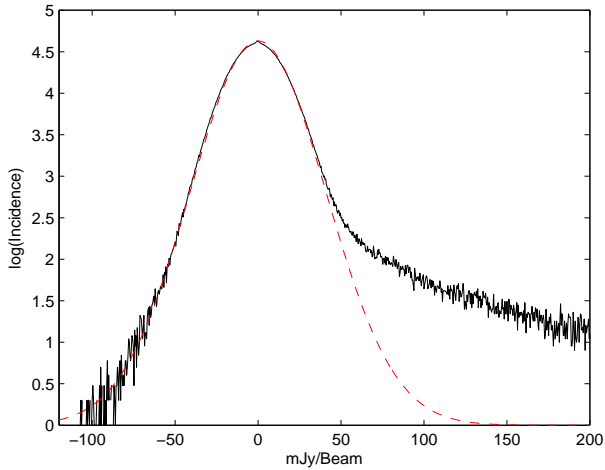}
   \caption{Histogram of the occurrence of brightnesses in the combined
   data cube on a logarithmic scale. The high brightnesses are dominated by
   significant emission, but the noise at low brightnesses can be fitted
   with a Gaussian function with a dispersion that closely agrees with
   the {\sc RMS} value in emission-free regions.}
  \label{noiseH}
\end{center}
\end{figure}

\begin{figure}[t!]
\begin{center}
  \includegraphics[width=0.5\textwidth]{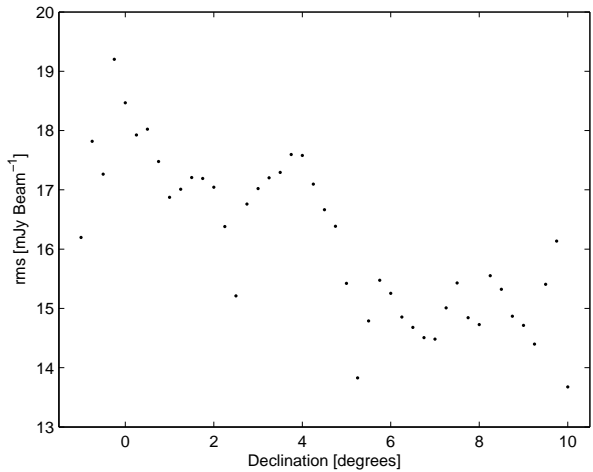}
   \caption{Differences in {\sc RMS} noise as function of
     declination. There is some scatter due to different conditions,
     since each declination is observed on a different date. In
     general low declinations have a slightly elevated noise value,
     due to an increased system temperature at the lowest
     declinations.}
  \label{dec_rms}
\end{center}
\end{figure}

After creating cubes of the combined and individual rise and set data,
sub-cubes were created, excluding Galactic emission and excluding the
edge of the bandpass. The noise in the rise-data is 22 mJy beam$^{-1}$
over 16 km s$^{-1}$, while the noise in the set-data is slightly
worse, 23 mJy beam$^{-1}$ over 16 km s$^{-1}$. The noise in the
combined data cubes is 16 mJy beam$^{-1}$ over 16 km s$^{-1}$, which
is in agreement with what would be expected, as the noise improves
with exactly a factor $\sqrt{2}$. In Fig.~\ref{noiseH} a histogram is
plotted of the flux values in the combined data cube. On the positive
side the flux values are dominated by real emission, however a
Gaussian can be fitted to the noise at negative fluxes. The noise
appears to be approximately Gaussian with a dispersion of 16 mJy
beam$^{-1}$. There is however some dependance of the {\sc RMS} values
on declination as shown in Fig.~\ref{dec_rms}. When observing a
specific declination strip, there is not much difference in the noise
at different right ascensions or in the frequency domain, as all data
points have been obtained under similar circumstances. Since the
declinations strips have been observed on different days, some real
fluctuation in the noise is more likely. We can see a scatter in the
noise for different declinations of 5 to 10 percent. Furthermore,
there is a general trend that the lowest declinations have the highest
noise values, which is expected due to a higher system temperature at
these lowest declinations (as demonstrated in Fig.~\ref{calibration}).

The flux sensitivity can be
converted to a brightness temperature using the equation:
\begin{equation}
T_b = \frac{\lambda^2S}{2k\Omega}
\end{equation}
where $\lambda$ is the observed wavelength, $S$ is the flux density,
$k$ the Boltzmann constant and $\Omega$ is the beam solid angle of the
telescope. When using the 21 cm line of \hbox{\rm H\,{\sc i}}, this
equation can be written as:
\begin{equation}
T_b = \frac{606}{b_{min}b_{maj}}S
\end{equation}
where $b_{min}$ and $b_{maj}$ are the beam minor and
major axis respectively in arcsec and $S$ is the flux in units of mJy/Beam. The
total flux can be converted into an \hbox{\rm
H\,{\sc i}} column density assuming negligible self-opacity using:
\begin{equation}
N_{HI} = 1.823 \cdot 10^{18} \int T_b dv
\end{equation}
with $[N_{HI}]$ = cm$^{-2}$, $[T_b]$ = K and $[dv]$ = km s$^{-1}$,
resulting in a column density sensitivity of $3.5\cdot10^{16}$
cm$^{-2}$ over 16 km s$^{-1}$.

We emphasise that the stated column density limit assumes emission
completely filling the beam. This can only be achieved, if the
emitting structure is larger than the beam. Observations described in
this paper can only resolve very extended structures and have reduced
sensitivity to compact features like dwarf galaxies or the inner parts
of large galaxies. Emission from compact structures will be diluted to
the full size of the beam and a better angular resolution is required
to distinguish compact from extended emission.

\section{Results}
\label{sec:results}
Due to the very large beam of the observations it is impossible to
determine the detailed kinematics of detected objects. Small and dense
objects cannot be distinguished from diffuse and extended structures
as the emission of compact sources will be spatially diluted to the
large beam size. Nevertheless, the total power product of the survey
is still a very important one, as it provides the best {\HI}
brightness sensitivity over such a large region for intrinsically
diffuse structures. There are other surveys with a comparable flux
sensitivity, but with a much smaller beam. These observations would
need to be dramatically smoothed in the spatial domain to get a
similar column density sensitivity as our survey. The diffuse emission
we seek is hidden in the noise at the native resolution and can easily
be affected by bandpass corrections or other steps in the reduction
process. In general, an {\HI} observation is most sensitive to
structures with a size that fill the primary beam of a single dish
observations or the synthesized beam of interferometric data.

We detect many galaxies in the filament connecting the Virgo Cluster
with the Local Group. Detailed analysis of known galaxies is not very
interesting at this stage, as there are other {\HI} surveys like HIPASS
and ALFALFA that have observed the same region with much higher
resolution. These surveys, or deep observations of individual galaxies
are much more suitable to analyse the physical parameters of these
objects. In the Total Power product of the WVFS we are interested in
emission that can not or has not been detected by previous
observations, because it is below their brightness sensitivity limit.

\begin{figure*}[t!]
  \includegraphics[angle=270,width=0.5\textwidth]{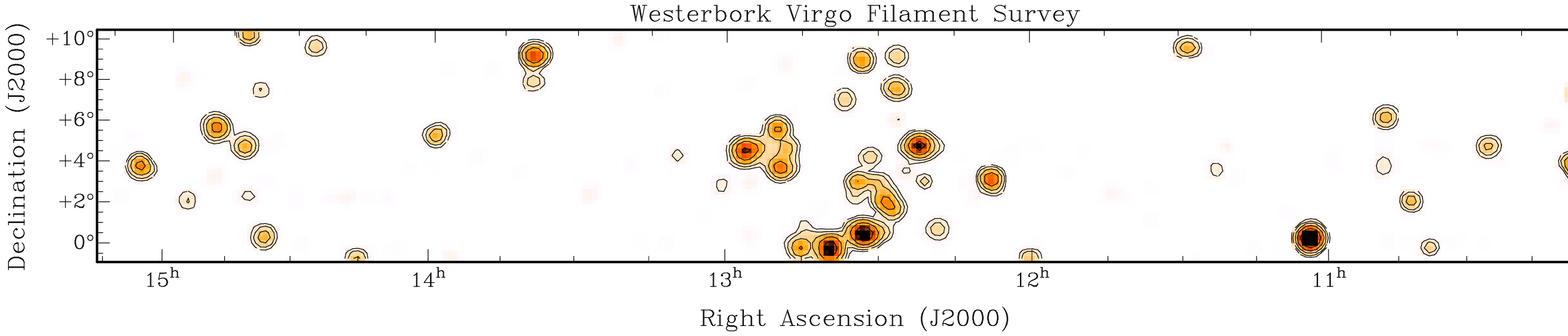}\\
  \includegraphics[angle=270,width=0.97\textwidth]{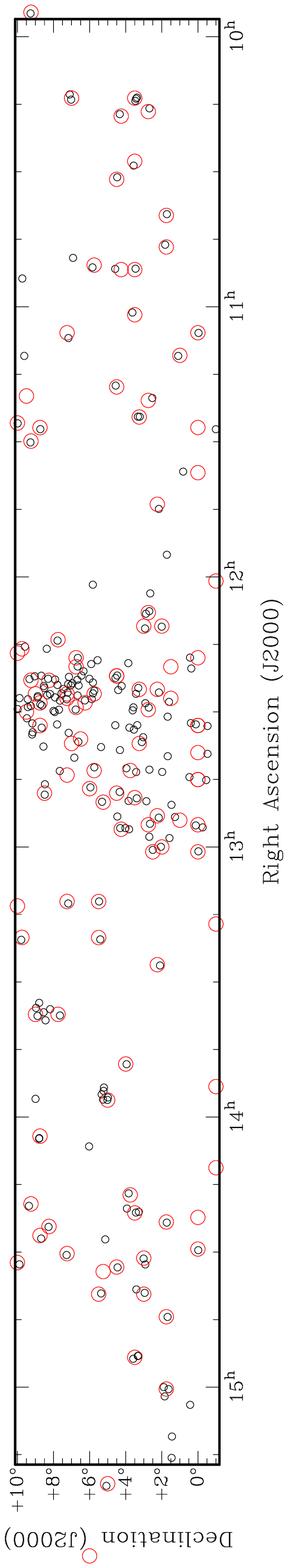}    
   \caption{ Illustration of the central 110 degrees of the WVFS
       region and detections in the velocity interval $400 < V_{Hel} <
       1600$ km s$^{-1}$. The top panel shows the integrated
       brightness levels, with contour levels drawn at 5, 10, 20, 40,
       80 and 160 Jy Beam$^{-1}$ km s$^{-1}$. Note that contour levels
       are chosen very conservatively and do not include faint
       emission near the noise floor. The second panel shows
       the position of all known {\HI}-detected galaxies (small black circles) within the
       redshift range of the WVFS data with the WVFS detections overlaid (large red
       circles).}
  \label{mom0}
\end{figure*}

An overview of the central 110 degrees in Right Ascension of the
survey sky coverage is given in Fig.~\ref{mom0} together with contours
of the brightest emission. The image shows the zeroth moment map
integrating the velocity interval $400< V_{Hel} < 1600$ km
s$^{-1}$. Contour levels are drawn at 5, 10, 20, 40, 80 and 160 Jy
Beam$^{-1}$ km~s$^{-1}$. The second panel shows the location of
galaxies for which {\HI} has been detected previously within the same
redshift interval as the WVFS total-power data (small black circles),
all WVFS detections are indicated by large red circles. The known
galaxies where selected from the HyperLeda \citep{1989A&AS...80..299P}
database, by looking for galaxies with a known {\HI} component within
the spatial and spectral range of WVFS.  While we do detect most known
galaxies, the survey suffers from confusion, especially in the densely
populated central part of the survey region. When multiple galaxies
with overlapping velocity structures are within one beam, these result
in only one detection. A couple of galaxies for which {\HI} has been
detected before are not found in our data, when carefully looking into
the data cubes for some cases a tentative signal can be observed,
however this does not reach a three $\sigma$ level as the {\HI} flux
is too much diluted by the large beam.

An attempt was made to detect sources using the source finding
algorithm {\it Duchamp} \citep{2008glv..book..343W} and by applying
masking algorithms within the MIRIAD \citep{1995ASPC...77..433S} and
GIPSY \citep{1992ASPC...25..131V} software packages. None of these
automatic methods appeared to be practical due to the very large
intrinsic beam size of the data. All sources are unresolved and there
is a lot of confusion between sources at a similar radial velocity
where the angular separation is smaller than the beam-width.

A list of candidate sources was determined from visual inspection of
subsequent channel maps, using the KVIEW task in the KARMA package
\citep{1996ASPC..101...80G}. The combined cube containing both the
{\it rise} and {\it set} data, as well as the individual {\it rise}
and {\it set} cubes were each inspected. Features were accepted if
local peaks exceeded the 3$\sigma$ limit in at least two subsequent
channels in the combined data cube and if they exceeded the 2$\sigma$
limit in the individual rise and set data products. This cutoff level
is very low, however the {\it rise} and {\it set} data represent two
completely independent observations undertaken at different times,
giving extra confidence in the resulting candidates. Furthermore we
are looking for diffuse extended structures, which are expected to
occur at those low flux levels. Using a high clipping level will
significantly reduce the chances for including such diffuse emission
features in an initial candidate list.

In total we found 188 candidate sources of which the properties are
estimated in detail. The integrated line strengths have been determined
for each candidate by extracting the single spectrum with the highest
flux density from both the {\it rise} and {\it set} cube. As there
were artifacts in the bandpass, a second order polynomial has been
fitted to the bandpass and was subtracted from the spectra. The
average of the two integrated line strengths was determined to get the
best solution. We assume here that all detections are unresolved when
using an effective FWHM beamsize of $2982 \times 2956$ arcsec.

Subsequently all candidate detections have been compared with catalogued
detections in the {\HI} Parkes All Sky Survey. The HIPASS database
completely covers our survey region and currently has the best column
density sensitivity.

The list of candidate detections is split into two parts. Detections
with an HIPASS counterpart at a similar position and velocity can be
confirmed and are reliable detections. In total, 129 of our candidates
could be identified in the HIPASS catalogue. When taking into account
the expected overlap of HIPASS objects in our larger spatial beam, we
confirm 146 of the 149 HIPASS detections in this region. The remaining
58 WVFS candidates have not been catalogued in HIPASS.

The corresponding error in flux density was determined over a velocity
interval of $1.5\times W_{20}$, where $W_{20}$ is the velocity width
of the emission profile at 20\% of the peak
intensity and is given by:

\begin{equation}
\sigma = \sqrt{\frac{1.5 \cdot W_{20}}{v_{res}}} \cdot \delta v \cdot rms
\end{equation}

\cite{2002ApJ...567..247R} have shown that in surveys of this type, an
asymptotic completeness of about 90\% is reached at a signal-to-noise
ratio of 8, when considering the integrated flux. Comparison with the
noise histogram shown in Fig.~\ref{noiseH} demonstrates that no
negative peaks occur which exceed this level, suggesting that the
incidence of false positives should also be minimal.  When we adopt
this limit, only 20 detections, with an integrated flux density
exceeding 8 times the associated error remain from the 58 candidates.

We will mention the candidate detections here and give their general
properties, however we leave further analysis to a subsequent paper, when we
incorporate the cross-correlation data and an improved version of the
HIPASS product for comparison. We emphasise here that although the
detections seem obvious in the total-power data at the 8$\sigma$ level,
they are considered as candidate detections. They have to be analysed
and compared using other data-sets, to be able to confirm the
detections and make strong statements.

\subsection{Source Properties of Known Detections}
The properties of all previously known {\HI} detections are summarised
in table~\ref{conf_det}.  The first column gives the names of the
source as given in the Westerbork Virgo Filament Survey. The name
consists of the characters ``WVFS'' followed by the right ascension of
the object in [hh:mm] and the declination in [d:mm]. The second column
gives the more common name of objects for which we have identified the
{\HI} counterpart. In the third and forth column the RA and Dec
positions are given, followed by the estimated heliocentric recession
velocity in the fifth column. In the last two columns we give the
integrated flux in [Jy-km~s$^{-1}$] and the $W_{20}$ line width in [km
  s$^{-1}$]. Spectra of all the confirmed {\HI} detections are shown
in the appendix of this paper.

Several of the detections are at the edge of the frequency coverage of
the cube and are indicated with an asterisk in the table in the column
with the $W_{20}$ values.  The observed spectrum for these sources is
not complete, which results in only a lower limit to the integrated
flux. We will not consider these sources in our further analysis.

\onecolumn
\begin{landscape}
\begin{center}
\begin{longtable}{llccccc}

\caption{Physical properties of confirmed detections in the Westerbork Virgo Filament Survey total-power data.} 
\label{conf_det}\\

%This is the header for the first page of the table...

\hline
\hline
Name           &  Optical ID.              &     RA [hh:mm:ss] & Dec [dd:mm] & $V_{Hel}$ [km s$^{-1}$] & $S$ [Jy km s$^{-1}$] & $W_{20}$  [km s$^{-1}$]\\
\hline   
\endfirsthead

%This is the header for the remaining page(s) of the table...
\hline
\hline
Name           &  Optical ID.              &     RA [hh:mm:ss] & Dec [dd:mm] & $V_{Hel}$ [km s$^{-1}$] & $S$ [Jy km s$^{-1}$] & $W_{20}$  [km s$^{-1}$]\\
\hline
\endhead

%This is the footer for all pages except the last page of the table...
\hline
  \multicolumn{7}{c}{{Continued on Next Page\ldots}} \\
\endfoot

%This is the footer for the last page of the table...
\hline \hline
\endlastfoot

WVFS 0906+0615  & UGC 4781        	      &  09:06:27   &     6:15 &	1419    &     15.0   &   234  \\
WVFS 0908+0515  & SDSS J090836.54+051726.8    &	09:08:27    &    5:15  &	597	&      1.2   &   50  \\
WVFS 0908+0600  & UGC 4797	              &  09:08:27   &     6:00 &	1285	&      4.2   &   120 \\
WVFS 0910+0700	& NGC 2775         	      &  09:10:27   &     7:00 &	1491	&      9.7   &   160 \\
                & NGC 2777		      &		    &	       &		&	     &       \\
WVFS 0943-0045	& UGC 5205	              &  09:43:33   &    -0:45 &	1501	&      8.1   &   115 \\
WVFS 0943+0945	& IC0559	              &  09:43:33   &     9:45 &	522	&      6.2   &   150 \\
WVFS 0944-0045	& SDSS J094446.23-004118.2    &	09:44:32    &   -0:45  &	1194	&      4.2   &   150 \\
WVFS 0951+0745	& UGC 5288	              &  09:51:34   &     7:45 &	539	&     25.9   &   120 \\
WVFS 0953+0130	& NGC3044	              &  09:53:34   &     1:30 &	1300	&     35.6   &   330 \\
WVFS 0954+0915	& NGC 3049	              &  09:54:35   &     9:15 &	1469	&     13.5   &   230 \\
WVFS 1013+0330	& NGC 3169	              &  10:13:38   &	3:30   &	1200	&    110.7   &   510 \\
WVFS 1013+0700	& UGC 5522	              &  10:13:38   &	7:00   &	1194	&     40.4   &   235 \\
WVFS 1016+0245	& UGC 5539	              &  10:16:38   &	2:45   &	1251	&      9.1   &   210 \\
WVFS 1017+0415	& UGC 5551	              &  10:17:38   &	4:15   &	1302	&      5.5   &   120 \\
WVFS 1027+0330	& UGC 5677	              &  10:27:40   &	3:30   &	1169	&      6.1   &   130 \\
WVFS 1031+0430	& UGC 5708	              &  10:31:41   &	4:30   &	1144	&     30.0   &   210 \\
WVFS 1039+0145	& UGC 5797	              &  10:39:42   &	1:45   &	671	&      4.4   &   110 \\
WVFS 1046+0145	& NGC 3365	              &  10:46:43   &	1:45   &	945	&     42.5   &   265 \\
WVFS 1050+0545	& NGC 3423	              &  10:50:44   &	5:45   &	988	&     34.7   &   185 \\
WVFS 1051+0330	& PGC 2807138	              &  10:51:44   &	3:30   &	1053	&     13.1   &   105 \\
WVFS 1051+0415	& UGC 5974	              &  10:51:44   &	4:15   &	1030	&     11.6   &   180 \\
WVFS 1101+0330	& NGC 3495	              &  11:01:46   &	3:30   &	1028	&     27.5   &   330 \\
WVFS 1105+0000	& NGC 3521	              &  11:05:46   &	0:00   &	704	&    275.8   &   480 \\
WVFS 1105+0715	& NGC 3526	              &  11:05:46   &	7:15   &	1418	&      6.0   &   205 \\
WVFS 1110+0100	& CGCG 011-018	              &  11:10:47   &	1:00   &	969	&      4.3   &   75  \\
WVFS 1117+0430	& NGC 3604	              &  11:17:48   &	4:30   &	1527	&      3.2   &   120 \\
WVFS 1119+0930	& SDSS J111928.10+093544.2    &	11:19:49    &	9:30   &	961	&      1.5   &   40  \\
WVFS 1120+0245	& UGC 6345	              &  11:20:48   &	2:45   &	1568	&      9.6   &   100 \\
WVFS 1124+0315	& NGC 3664	              &  11:24:29   &	3:15   &	1380	&     19.0   &   160 \\
WVFS 1125+1000	& IC 0692	              &  11:25:49   &	10:00  &	1127	&      2.8   &   80  \\
WVFS 1126-0045	& UGC 6457	              &  11:26:49   &	-0:45  &	937	&      4.6   &   90  \\
WVFS 1126+0845	& IC 2828	              &  11:26:50   &	8:45   &	1011	&      3.9   &   90  \\
WVFS 1129+0915	& NGC3705	              &  11:29:50   &	9:15   &	1019	&     51.5   &   360 \\
WVFS 1136+0045	& UGC 6578	              &  11:36:51   &	0:45   &	1022	&      5.4   &   115 \\
WVFS 1143+0215	& PGC 036594	              &  11:43:52   &	2:15   &	976	&      5.6   &   55  \\
WVFS 1200-0100	& NGC 4030	              &  12:00:55   &    -01:00&	1418	&     39.5   &   360 \\
WVFS 1207+0245	& NGC 4116 	              &  12:07:56   &	2:45   &	1285	&     89.5   &   230 \\
                & NGC 4123		      &		    &	       &		&	     &       \\
WVFS 1210+0200	& UGC 7178	              &  12:10:56   &	2:00   &	1302	&     10.9   &   100 \\
WVFS 1210+0300	& UGC 7185	              &  12:10:57   &	3:00   &	1269	&     13.6   &   150 \\
WVFS 1213+0745	& UGC 7239	              &  12:13:57   &	7:45   &	1194	&      7.6   &   140 \\
WVFS 1215+0945	& NGC 4207	              &  12:15:58   &	9:45   &	599	&      5.1   &   180 \\
WVFS 1216+1000	& UGC 7307	              &  12:16:57   &	10:00  &	1152	&      2.7   &   65  \\
WVFS 1217+0030	& UGC 7332	              &  12:17:58   &	0:30   &	911	&     19.1   &   85  \\
WVFS 1217+0645	& NGC 4241	              &  12:17:58   &	6:45   &	704	&      8.5   &   140 \\
WVFS 1219+0645	& VCC 0381	              &  12:19:58   &	6:45   &	456	&      1.4   &   40$^*$  \\
WVFS 1219+0130	& UGC 7394	              &  12:19:58   &	1:30   &	1552	&      3.4   &   125 \\
WVFS 1221+0430	& NGC 4301	              &  12:21:59   &	4:30   &	1252	&     20.2   &   135 \\
WVFS 12222+0915	& NGC 4316	              &  12:22:58   &	9:15   &	1244	&      7.1   &   365 \\
WVFS 1222+0430	& M 61	                      &  12:22:00   &	4:30   &	1535	&     95.8   &   185 \\
WVFS 1222+0815	& NGC 4318	              &  12:22:59   &	8:15   &	1402	&      2.8   &   90  \\
WVFS 1223+0215	& UGC 7512	              &  12:24:59   &	2:15   &	1477	&      4.1   &   95  \\
WVFS 1224+0315	& pgc 040411	              &  12:24:59   &	3:15   &	900	&     10.1   &   85  \\
WVFS 1225+0545	& VCC 0848                    &	12:25:59    &	5:45   &	1110	&     13.9   &   175 \\
                & NGC 4376		      &		    &	       &		&	     &       \\
                & NGC 4423		      &		    &	       &		&	     &       \\
WVFS 1225+0715	& IC 3322A	              &  12:25:59   &	7:15   &	1078	&      8.7   &   115 \\ 
WVFS 1225+0900	& NGC 4411         	      &  12:25:59   &	9:00   &	1236	&     20.9   &   110 \\
                & NGC 4411 b		      &		    &	       &		&	     &       \\
WVFS 1226+0130	& pgc135803	              &  12:26:59   &	1:30   &	1265	&     43.3   &   110 \\
WVFS 1226+0715	& UGC 7557	              &  12:26:59   &	7:15   &	920	&     31.9   &   175 \\
WVFS 1227+0615	& NGC 4430	              &  12:27:59   &	6:15   &	1402	&      2.7   &   120 \\
WVFS 1227+0845	& UGC 7590	              &  12:27:59   &	8:45   &	1053	&      4.6   &   95  \\
WVFS 1228+0645	& IC 3414	              &  12:28:59   &	6:45   &	497	&      4.8   &   130$^*$ \\
WVFS 1229+0245	& UGC 7612         	      &  12:29:30   &	2:45   &	1595	&     16.6   &   170 \\
                & UGC 7642		      &		    &	       &		&	     &       \\
WVFS 1230+0930	& HIPASS J1230+09	      &  12:30:00   &	9:30   &	473	&      5.6   &   120$^*$ \\
WVFS 1233+0000	& NGC 4517	              &  12:33:01   &	0:00   &	1078	&    124.1   &   325 \\
WVFS 1233+0030	& NGC 4517A	              &  12:33:01   &	0:30   &	1510	&     31.7   &   175 \\
WVFS 1233+0845	& NGC 4519	              &  12:33:01   &	8:45   &	1186	&     51.8   &   220 \\
WVFS 1236+0630	& IC 3576	              &  12:36:01   &	6:30   &	1045	&     15.2   &   70  \\
WVFS 1237+0315	& UGC 07780	              &  12:37:01   &	3:15   &	1410	&      3.0   &   130 \\
WVFS 1237+0700	& IC 3591	              &  12:37:01   &	7:00   &	1593	&     10.4   &   120$^*$ \\
WVFS 1239-0030	& NGC 4592	              &  12:39:02   &   -00:30 &	1061	&    127.5   &   220 \\
WVFS 1243+0345	& NGC 4630	              &  12:43:01   &	3:45   &	696	&      6.8   &   160 \\
WVFS 1243+0545	& VCC 1918	              &  12:43:02   &	5:45   &	961	&      1.8   &   90  \\
WVFS 1244+0715	& VCC 1952	              &  12:44:02   &	7:15   &	1277	&      1.6   &   70  \\
WVFS 1245-0030	& NGC 4666	              &  12:45:02   &    -00:30&	1527	&     22.0   &   380 \\
WVFS 1245+0030	& UGC 7911	              &  12:45:02   &	0:30   &	1144	&     12.4   &   120 \\
WVFS 1247+0600	& UGC 7943	              &  12:47:03   &	6:00   &	812	&     11.5   &   145 \\
WVFS 1248+0430	& NGC 4688	              &  12:48:03   &	4:30   &	961	&     28.2   &   70  \\
WVFS 1248+0830	& NGC 4698	              &  12:48:03   &	8:30   &	1000	&     26.9   &   130 \\
WVFS 1249+0330	& NGC 4701         	      &  12:49:03   &	3:30   &	704	&     65.5   &   180 \\
                & UGC 7983		      &		    &	       &		&	     &       \\
WVFS 1250+0515	& NGC 4713	              &  12:50:03   &	5:15   &	621	&     51.5   &   195 \\
WVFS 1253+0215	& NGC 4772	              &  12:53:04   &	2:15   &	1044	&     12.5   &   480 \\
WVFS 1254+0100	& NGC 4771	              &  12:54:04   &	1:00   &	986	&      2.1   &   290 \\
WVFS 1255+0015	& UGC 8041	              &  12:55:04   &	0:15   &	1310	&     14.3   &   200 \\
WVFS 1255+0245	& ARP 277	              &  12:55:04   &	2:45   &	889	&     16.7   &   220 \\
WVFS 1256+0415	& NGC 4808                    &	12:56:04    &	4:15   &	721	&    105.4   &   295 \\
                & NGC 4765		      &		    &	       &		&	     &       \\
                & UGC 8053		      &		    &	       &		&	     &       \\
WVFS 1300+0200	& UGC 08105	              &  13:00:00   &	2:00   &	895	&     10.8   &   155 \\
WVFS 1301+0000	& NGC 4904	              &  13:01:05   &	0:00   &	1152	&     10.9   &   195 \\
WVFS 1301+0230	& NGC 4900        	      &  13:01:05   &	2:30   &	937	&     13.0   &   145 \\
                & UGC 8074		      &		    &	       &		&	     &       \\
WVFS 1312+0530	& UGC 8276	              &  13:12:07   &	5:30   &	870	&      3.5   &   75  \\
WVFS 1312+0715	& UGC 8285	              &  13:12:07   &	7:15   &	887	&      5.1   &   150 \\
WVFS 1313+1000	& UGC 8298	              &  13:13:07   &	10:00  &	1127	&      8.0   &   100 \\
WVFS 1317-0100	& UM 559	              &  13:17:07   &   -01:00 &	1227	&      4.0   &   130 \\
WVFS 1320+0530	& UGC 8382	              &  13:20:08   &	5:30   &	953	&      3.0   &   115 \\
WVFS 1320+0945	& UGC 8385	              &  13:20:06   &	9:45   &	1127	&     13.3   &   150 \\
WVFS 1326+0215	& NGC 5147                    &	13:26:09    &	2:15   &	1069	&     10.9   &   150 \\
                & HIPASS J1328+02	      &		    &	       &		&	     &       \\
WVFS 1337+0745	& UGC 8614	              &  13:37:11   &	7:45   &	1011	&     18.6   &   190 \\
WVFS 1337+0900	& NGC 5248                    &	13:37:11    &	9:00   &	1119	&     87.2   &   290 \\
                & UGC 8575		      &		    &	       &		&	     &       \\
                & UGC 8629		      &		    &	       &		&	     &       \\
WVFS 1348+0400	& NGC 5300	              &  13:48:13   &	4:00   &	1153	&     11.0   &   210 \\
WVFS 1353-0100	& NGC 5334	              &  13:53:14   &    -01:00&	1360	&     16.1   &   220 \\
WVFS 1356+0500	& NGC 5364         	      &  13:56:14   &	5:00   &	1202	&     51.5   &   320 \\
                & NGC 5348		      &		    &	       &		&	     &       \\
WVFS 1404+0845	& UGC 8995	              &  14:04:15   &	8:45   &	1218	&     10.9   &   190 \\
WVFS 1411-0100	& NGC 5496	              &  14:11:16   &    -01:00&	1535	&     34.9   &   270 \\
WVFS 1417+0345	& PGC 140287	              &  14:17:18   &	3:45   &	1370	&     12.6   &   180 \\
WVFS 1419+0915	& UGC 9169                    &	14:19:18    &	9:15   &	1250    &     22.7   &   160 \\
                & SDSS J142044.53+083735.8    &		    &	       &		&	     &       \\

WVFS 1421+0330	& NGC 5577	              &  14:21:18   &	3:30   &	1468	&      9.8   &   225 \\
WVFS 1422-0015	& UGC 5584	              &  14:22:18   &    -00:15&	1635	&     14.0   &   165$^*$ \\
WVFS 1423+0145	& UGC 9215	              &  14:23:19   &	1:45   &	1368	&     19.8   &   255 \\
WVFS 1424+0815	& UGC 9225	              &  14:24:19   &	8:15   &	1244	&      6.4   &   160 \\
WVFS 1426+0845	& UGC 9249	              &  14:26:19   &	8:45   &	1335	&      6.4   &   155 \\
WVFS 1429+0000	& UGC 9299	              &  14:29:20   &	0:00   &	1518	&     45.2   &   220 \\
WVFS 1430+0715	& NGC5645	              &  14:30:20   &	7:15   &	1335	&     18.4   &   200 \\
WVFS 1431+0300	& IC 1024	              &  14:31:20   &	3:00   &	1435	&      9.0   &   240 \\
WVFS 1432+1000	& NGC 5669	              &  14:32:20   &	10:00  &	1343	&     36.7   &   210 \\
WVFS 1433+0430	& NGC 5668	              &  14:33:20   &	4:30   &	1535	&     30.8   &   120 \\
WVFS 1434+0515	& UGC 9385	              &  14:34:20   &	5:15   &	1601	&      9.4   &   130$^*$ \\
WVFS 1439+0300	& UGC 9432	              &  14:39:21   &	3:00   &	1560	&      8.4   &   110 \\
WVFS 1439+0530	& NGC 5701	              &  14:39:21   &	5:30   &	1468	&     57.7   &   150 \\
WVFS 1444+0145	& NGC 5740	              &  14:44:22   &	1:45   &	1577	&     23.5   &   300$^*$ \\
WVFS 1453+0330	& NGC 5774                    &	14:53:23    &	3:30   &	1535	&     63.9   &   205 \\
                & HIPASS J1452+03	      &		    &	       &		&	     &       \\
WVFS 1500+0145	& NGC 5806	              &  15:00:25   &	1:45   &	1236	&      5.4   &   245 \\
WVFS 1521+0500	& NGC 5921	              &  15:21:28   &	5:00   &	1435	&     28.8   &   210 \\
WVFS 1537+0600	& NGC 5964	              &  15:37:30   &	6:00   &	1418	&     37.6   &   215 \\
WVFS 1546+0645	& UGC 10023	              &  15:46:32   &	6:45   &	1402	&      3.7   &   100 \\
WVFS 1606+0830	& CGCG 079-046	              &  16:06:35   &	8:30   &	1310	&      3.7   &   90  \\
WVFS 1607+0730	& IC 1197	              &  16:07:35   &	7:30   &	1335	&     18.1   &   280 \\
WVFS 1609+0000	& UGC 10229	              &  16:09:36   &	0:00   &	1477	&      4.4   &   95  \\
WVFS 1618+0145	& CGCG 024-001	              &  16:18:37   &	1:45   &	1526	&      6.4   &   150 \\
WVFS 1618+0730	& NGC 6106	              &  16:18:37   &	7:30   &	1401	&     22.3   &   270 \\
WVFS 1655+0800	& HIPASS J1656+08	      &  16:55:43   &	8:00   &	1435	&      2.1   &   80  \\

\end{longtable}
\end{center}

\end{landscape}
\twocolumn

\subsection{Confused Sources}
Source confusion is a significant problem in the determination of
{\HI} fluxes for some of the detections. Due to the large intrinsic
beam size of the WVFS, many sources are spatially overlapping and
cannot be distinguished individually. This also complicates the
comparison with HIPASS and fluxes from the HyperLeda database
\citep{1989A&AS...80..299P}. When we suspect that a WVFS detection
contains several sources which are individually listed in the HIPASS
catalogue, this is indicated in table~\ref{conf_det}. In our
comparison with other catalogues we will take this into account, by
integrating the LEDA or HIPASS fluxes of the relevant galaxies in the
case of a confused detection.

A general consequence of source confusion is that only a portion of
the combined flux is tabulated, in comparison to the HIPASS data. This
is because the group of confused galaxies listed as one WVFS object
are often significantly larger than the intrinsic beam size, while
only the spectrum containing the brightest emission peak is
integrated, in keeping with the assumption that all detected objects
are unresolved.

\subsection{Optical ID's}

The NASA/IPAC Extragalactic Database (NED)\footnote{The NASA/IPAC
  Extragalactic Database (NED) is operated by the Jet Propulsion
  Laboratory, California Institute of Technology, under contract with
  the National Aeronautics and Space Administration.} has been used to
look for catalogued optical counterparts of the {\HI}
detections. Counterparts were sought within a 30 arcmin radius, since
this radius corresponds to the radius of the first null in the primary
beam of the WSRT telescopes. Only objects within this radius can have
a significant contribution to the measured {\HI} fluxes.

Furthermore, all new {\HI} detections are compared with optical images in
the red band from the second generation DSS. Only 2 of the 20 new {\HI}
detections have a clear optical counterpart and belong to objects for
which the {\HI} component has not previously been detected.

\subsection{New Detections}
\begin{figure*}
  \begin{center}

  \includegraphics[width=0.3\textwidth]{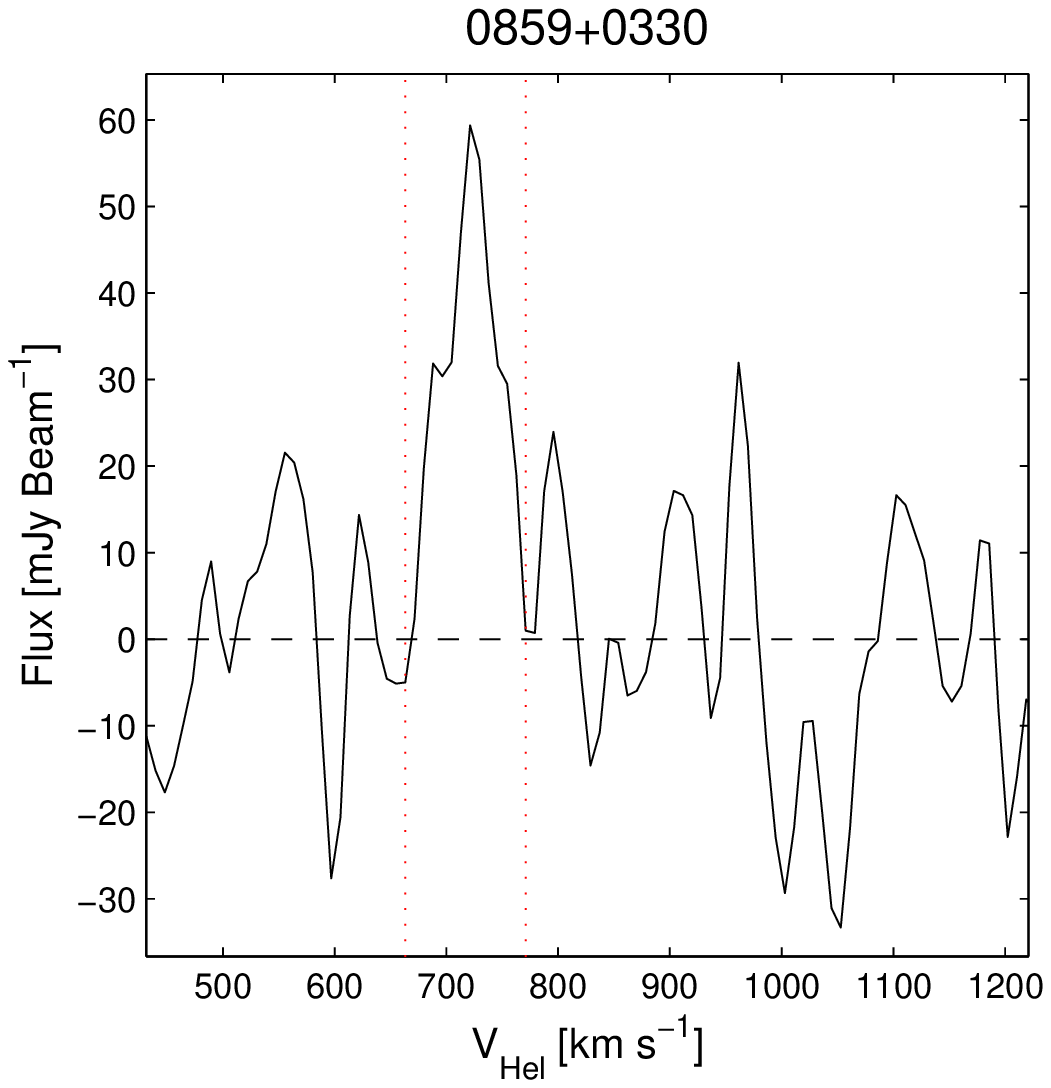}
  \includegraphics[width=0.3\textwidth]{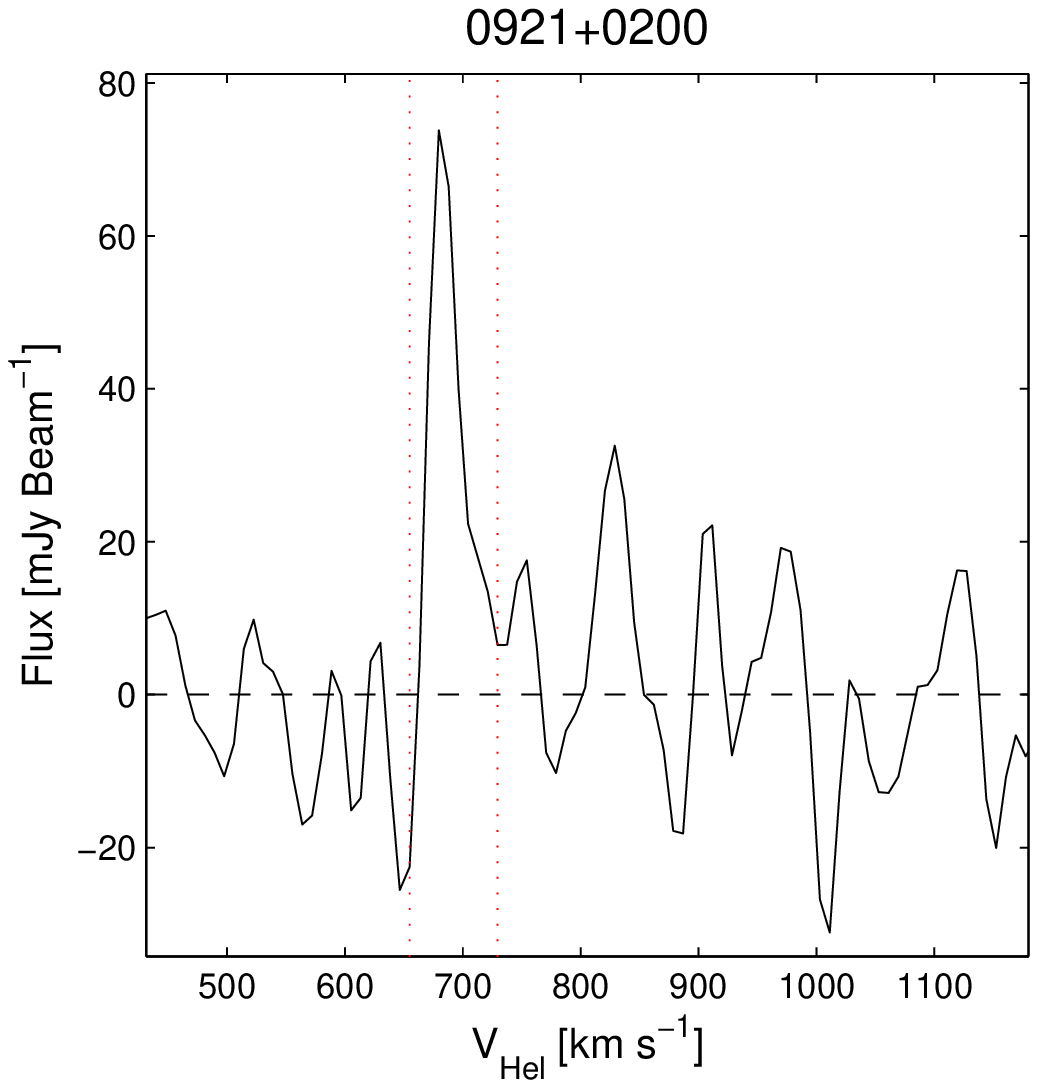}
  \includegraphics[width=0.3\textwidth]{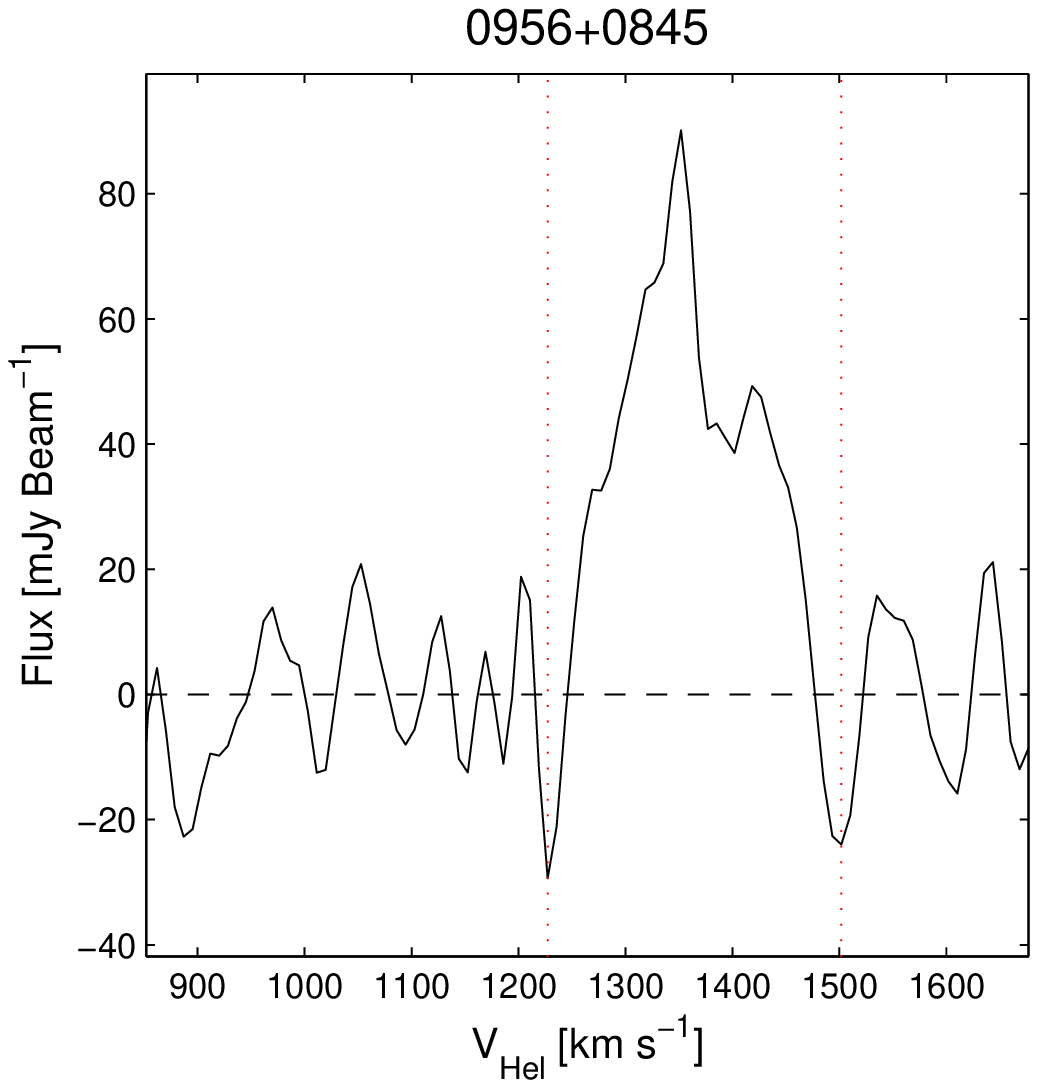}
 
 \includegraphics[width=0.3\textwidth]{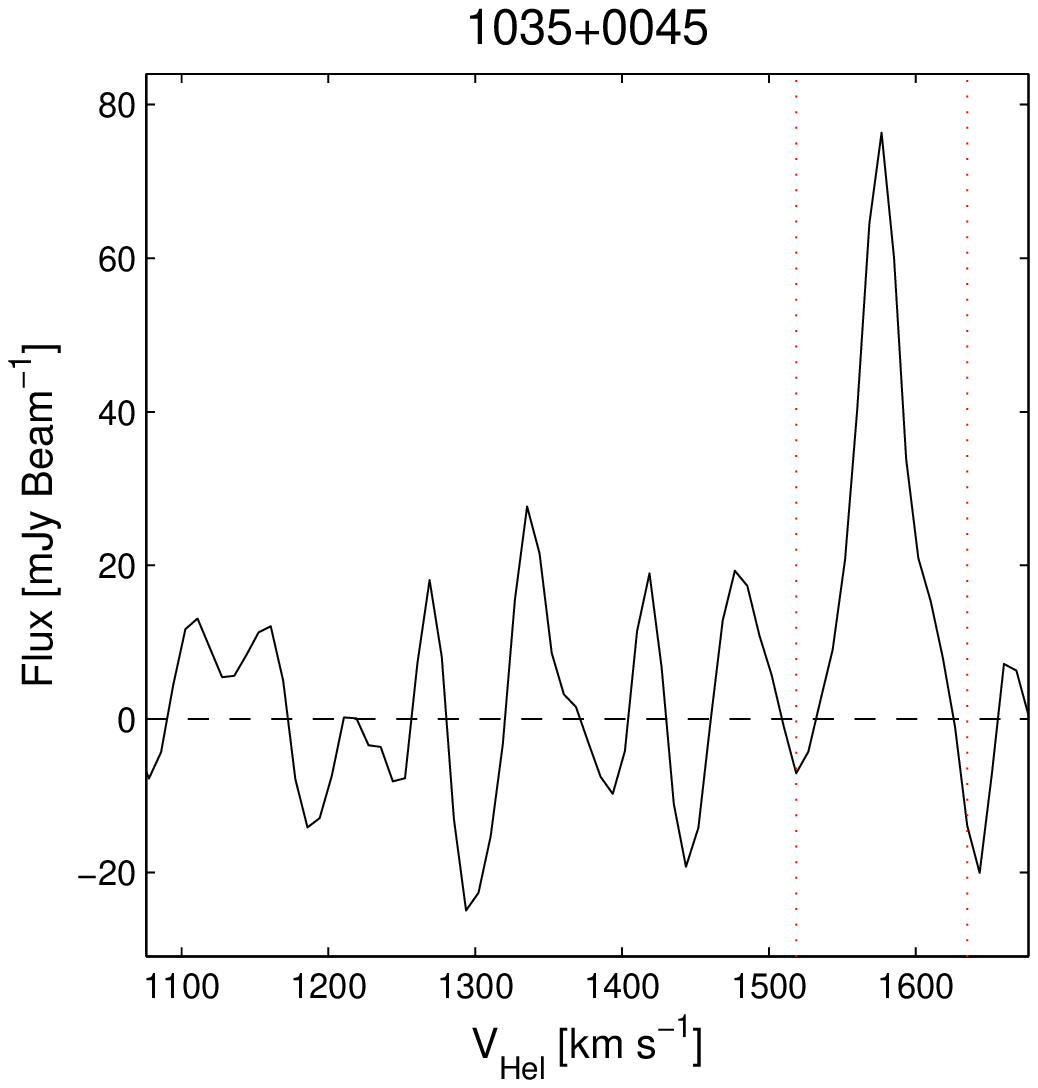}
 \includegraphics[width=0.3\textwidth]{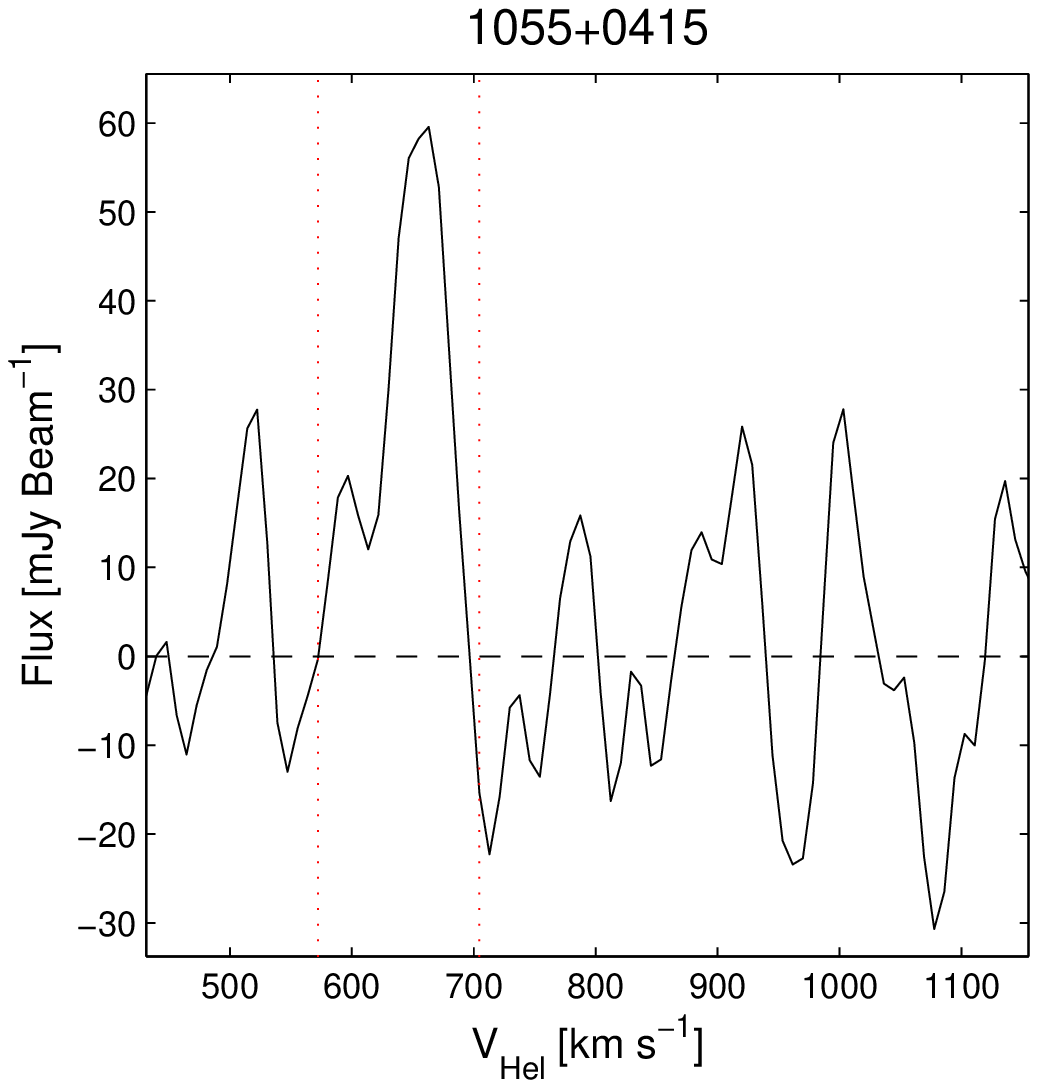}
 \includegraphics[width=0.3\textwidth]{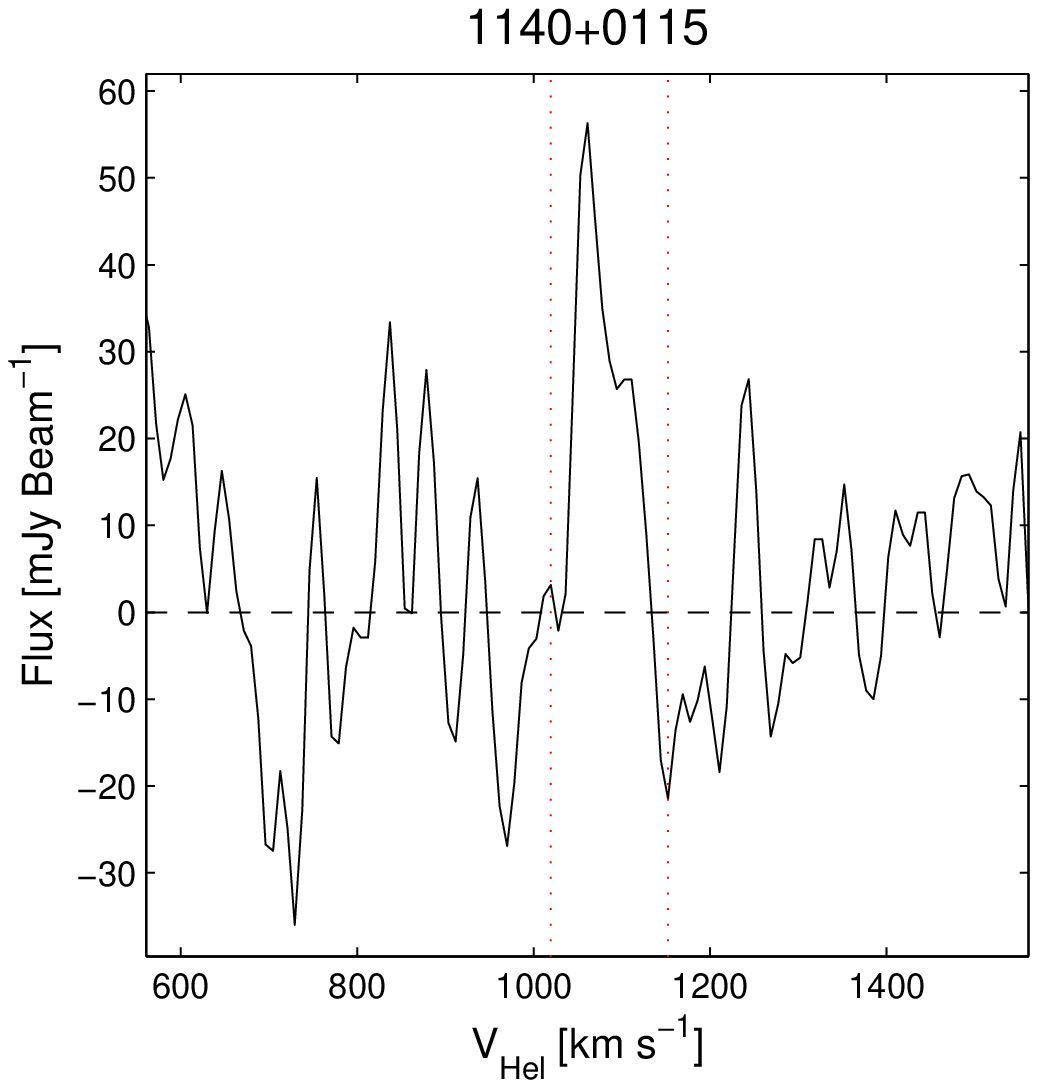}

 \includegraphics[width=0.3\textwidth]{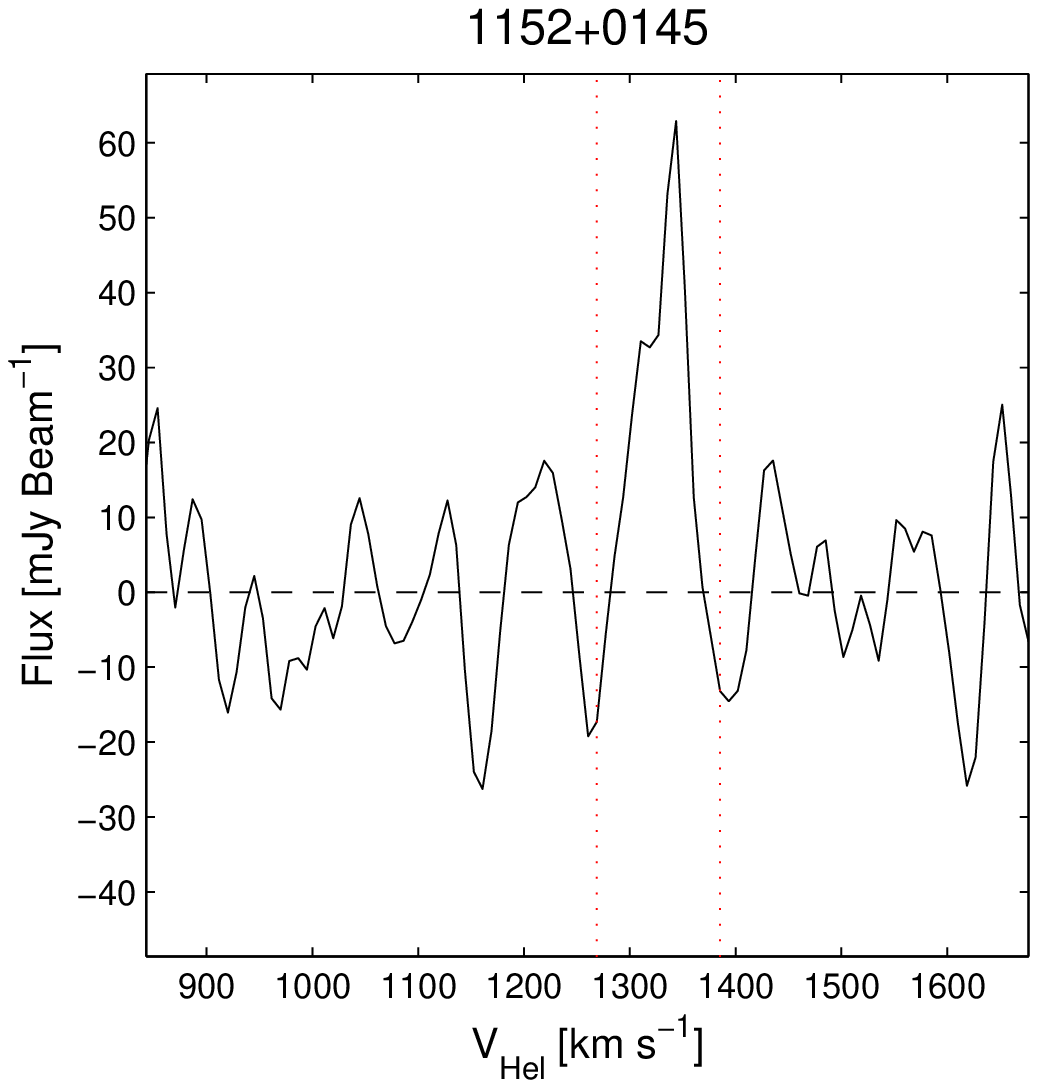}
 \includegraphics[width=0.3\textwidth]{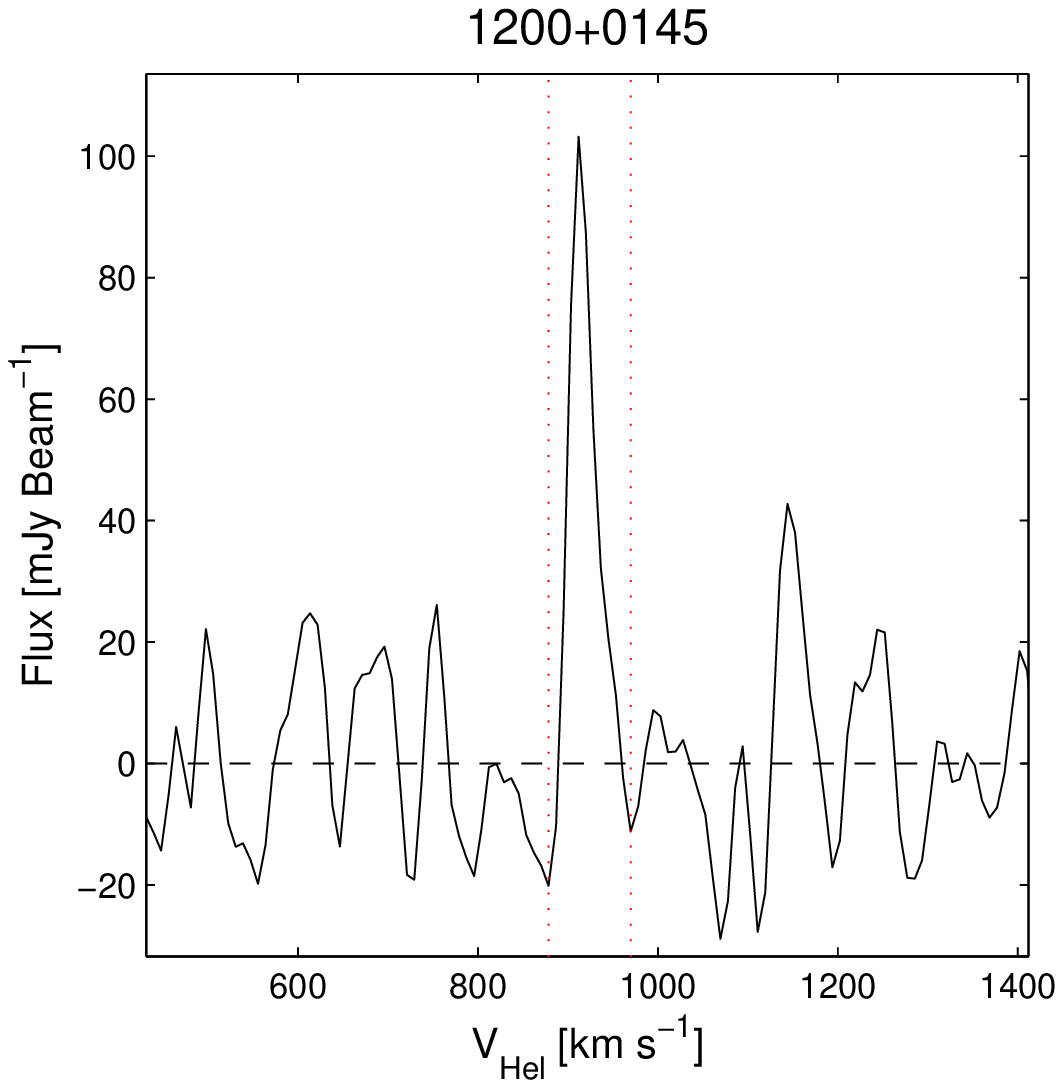}
 \includegraphics[width=0.3\textwidth]{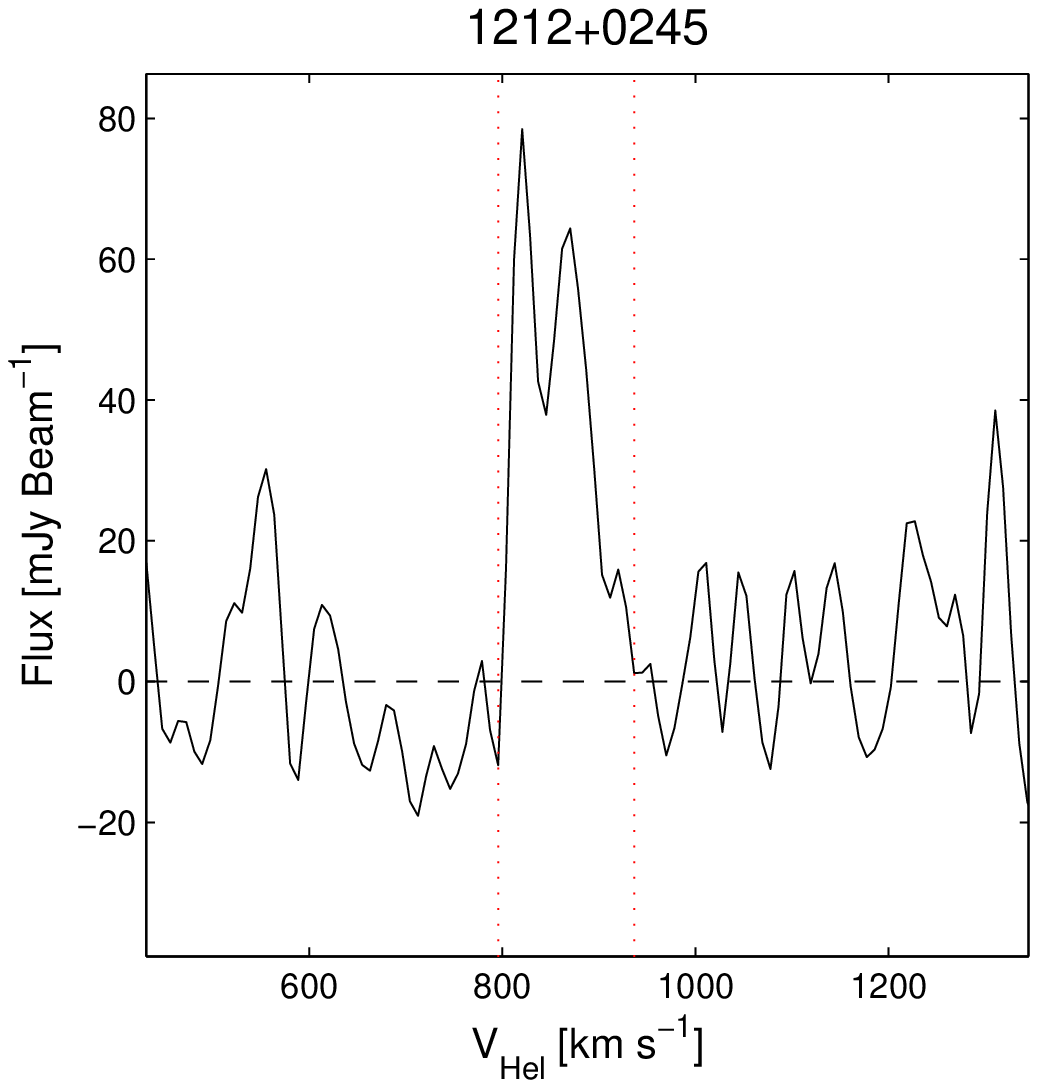}

 \includegraphics[width=0.3\textwidth]{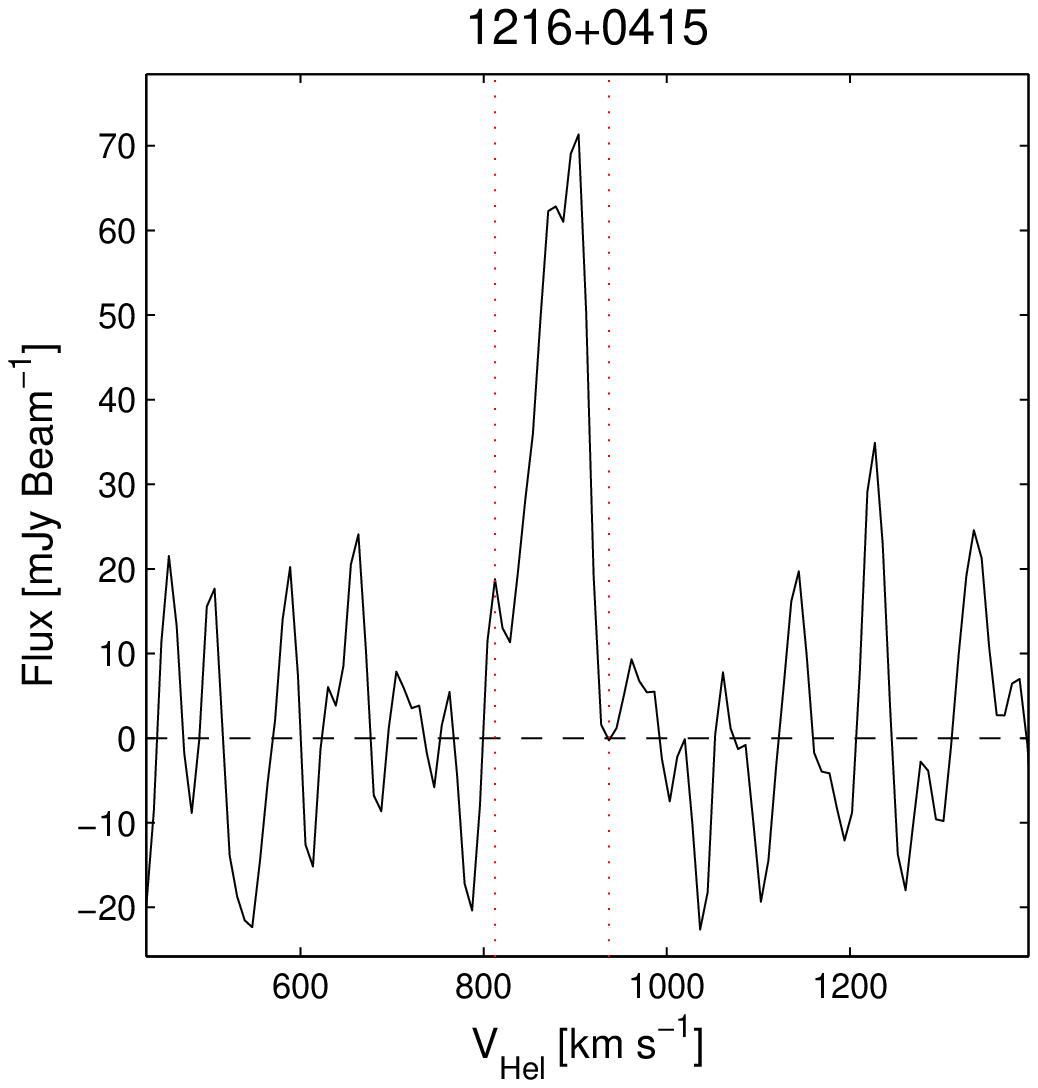}
 \includegraphics[width=0.3\textwidth]{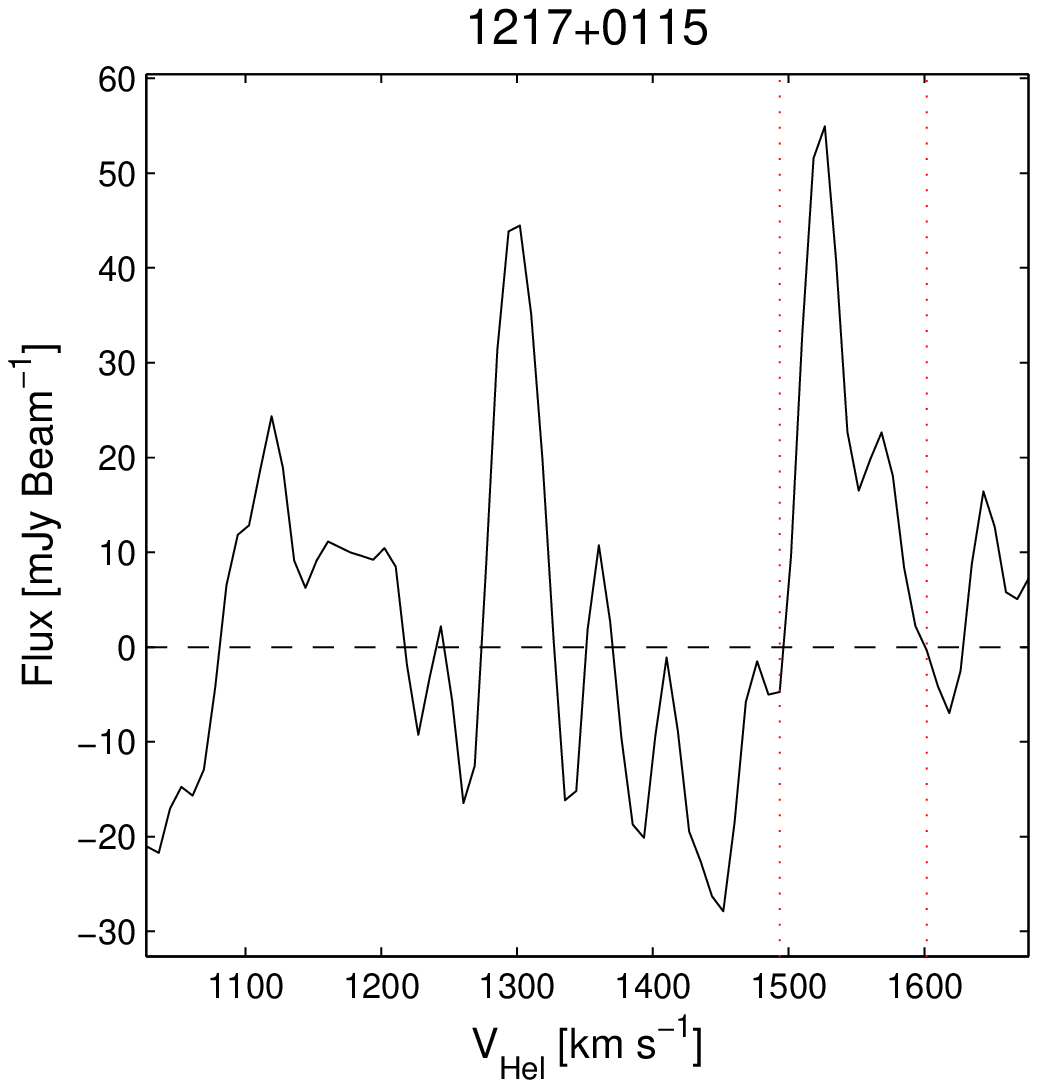}
 \includegraphics[width=0.3\textwidth]{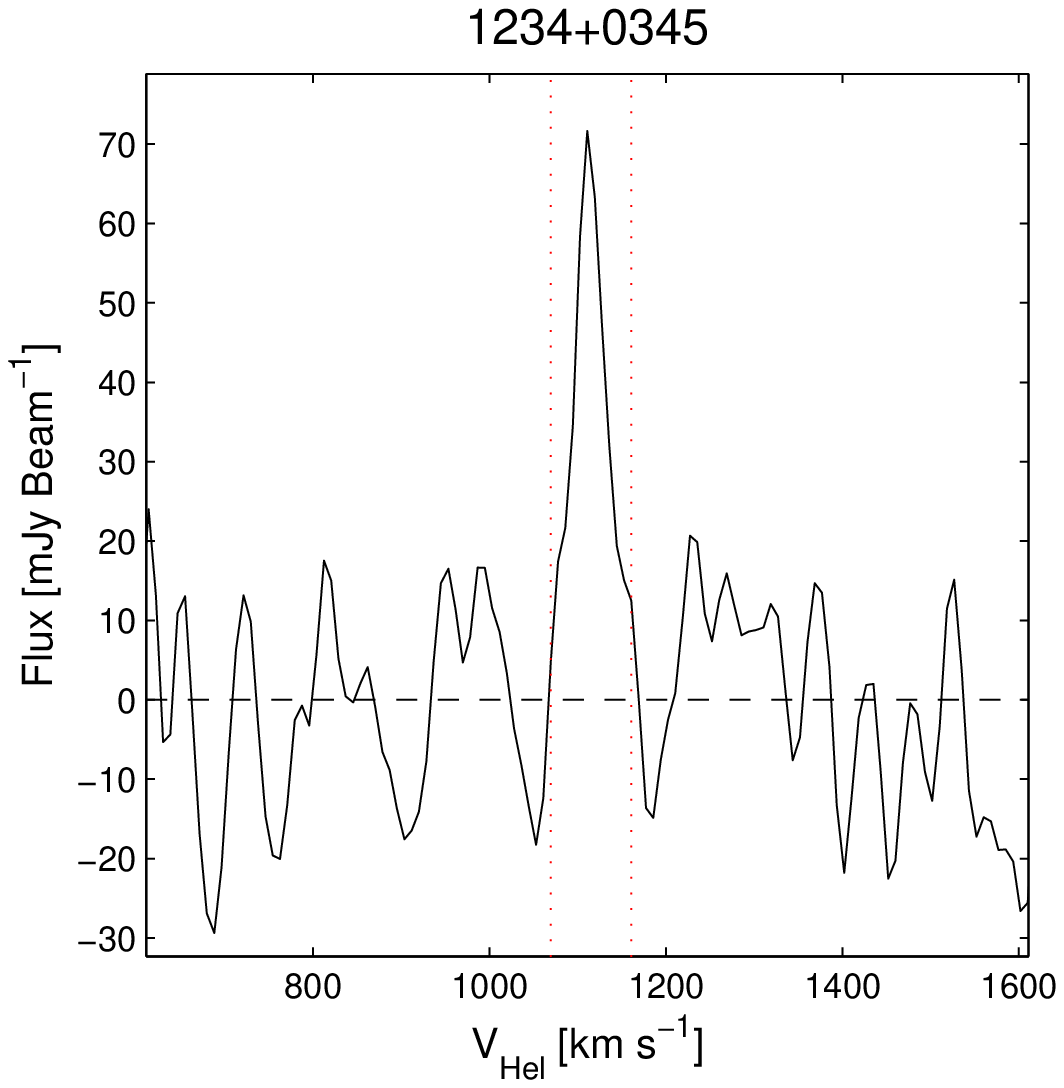}

 \end{center}
  \caption{{\HI} spectra of the new detections in the Westerbork Virgo
  Filament Survey at the position of the highest peak flux. The
  velocity interval over which the integrated line strength has been
  determined is indicated by the two vertical dashed lines.}
  \label{spectra}
\end{figure*}

The spectrum that has been derived for each new {\HI} detection is
plotted in Fig.~\ref{spectra}. The two dashed vertical lines indicate
the velocity range over which the spectrum has been integrated to
determine the total line strengths of the detections. All
physical properties of the new detections are listed in
Table~\ref{new_det}. The first column gives the WVFS name, which is
constructed as for the previously confirmed
detections. The second and third columns give the position of the
detections as accurately as possible followed by the heliocentric
recession velocity.

The spatial resolution of the WVFS data is very coarse due to the
intrinsic beam size of 30'. The centroid positions of all new
detections is determined as accurately as possible from a Gaussian or
parabolic fit to the peak of integrated {\HI} line strength over the full
line width of a new detection. The accuracy of the centroid position
is based on the intrinsic beam size and the signal-to-noise ratio as
$\textrm{HWHM}/(s/n)$. For a signal-to-noise ratio of eight, which is
the lower limit of our detections, this corresponds to a position
accuracy of $\sim4$ arcmin in both $\alpha$ and $\delta$.\\

Column 5 and 6 in Table~\ref{new_det} give the integrated flux and the
velocity width at 20\% of the peak flux of each detection. Based
  on these two values the rms noise level $(\sigma)$ and the
  signal-to-noise ratio are calculated in the last two columns.

We tabulate all basic properties of these sources, but will leave
further detailed analysis to a later paper where we will incorporate
the cross-correlation data for comparison.  Some features of each
object are noted below. We note again that when column densities
  are mentioned, these values assume emission completely filling the
  beam. Since the beam is very large, the detections are often not
  resolved spatially and it is possible that higher column densities
  do occur at smaller scales.\\

\begin{table*}
\begin{center}
  \begin{tabular}{cccccccc}
\hline
\hline
\small Name              &      \small RA         & \small DEC         & \small $V_{Hel}$     & \small $S$             & \small $W_{20}$        & \small $\sigma$  & \small $S/N$ \\
\small                        &      \small [hh:mm:ss] & \small [dd:mm:ss]  & \small [km s$^{-1}$] & \small [Jy km s$^{-1}$]  & \small  [km s$^{-1}$] & \small [Jy km s$^{-1}$]  & \small   \\
\hline
\small WVFS 0859+0330	& 	\small 08:59:22   &	\small 3:28:57  &	\small 721   &  \small 3.9  & \small 	90      & \small 0.37 & \small 10.5\\
\small WVFS 0921+0200	&	\small 09:21:20   &	\small 2:00:09  &	\small 680   &	\small 2.6  & \small   55      & \small 0.29 & \small 9.0\\
\small WVFS 0956+0845	&	\small 09:56:34   &	\small 8:45:05  &	\small 1343  &	\small 11.1 & \small   215    & \small 0.57& \small 19.4\\
\small WVFS 1035+0045	&	\small 10:36:48   &	\small 0:37:56  &	\small 1576  &	\small 3.1  & \small   65       & \small 0.32 & \small 9.7\\
\small WVFS 1055+0415	&	\small 10:55:50   &	\small 4:03:17  &	\small 655   &	\small 4.2  & \small   110     & \small 0.41& \small 10.2\\
\small WVFS 1140+0115	&	\small 11:41:10   &	\small 1.28:44  &	\small 1079  &	\small 3.0  & \small   85       & \small 0.36 & \small 8.3 \\
\small WVFS 1152+0145	&	\small 11:52:54   &	\small 1:53:42  &	\small 1335  &	\small 2.6  & \small   70       & \small 0.33 &  \small 8.0 \\
\small WVFS 1200+0145	&	\small 12:00:45   &	\small 1:46:14  &	\small 912   &	\small 3.5  & \small   50       & \small 0.28 & \small 12.5\\
\small WVFS 1212+0245	&	\small 12:12:09   &	\small 2:50:26  &	\small 845   &	\small 6.3  & \small   100     & \small 0.39 &  \small 16.2\\
\small WVFS 1216+0415	&	\small 12:17:07   &	\small 4:19:03  &	\small 895   &	\small 5.6  & \small   90       & \small 0.37&  \small 15.1\\
\small WVFS 1217+0115	&	\small 12:19:22   &	\small 1:29:49  &	\small 1527  &	\small 2.8  & \small   80        & \small 0.35 & \small 8.0\\
\small WVFS 1234+0345	&	\small 12:34:18   &	\small 3:33:52  &	\small 1111  &	\small 3.9  & \small   80        & \small 0.35 & \small 11.1\\
\small WVFS 1253+0145	&	\small 12:52:18   &	\small 1:49:38  &	\small 837   &	\small 2.5  & \small   50        & \small 0.28 & \small 8.9\\
\small WVFS 1324+0700	&	\small 13:23:46   &	\small 6:59:14  &	\small 531   &	\small 3.0  & \small   70        & \small 0.33 & \small 9.1\\
\small WVFS 1424+0200	&	\small 14:24:24   &	\small 1:58:57  &	\small 539   &	\small 3.9  & \small   70        & \small 0.33 & \small 11.8\\
\small WVFS 1500+0815	&	\small 15:00:46   &	\small 8:16:53  &	\small 1426  &	\small 3.3  & \small   105      & \small 0.40 & \small 8.3\\
\small WVFS 1524+0430	&	\small 15:24:17   &	\small 4:32:33  &	\small 1086  &	\small 2.5  & \small   55        & \small 0.29 & \small 8.6\\
\small WVFS 1529+0045	&	\small 15:29:30   &	\small 0:41:37  &	\small 679   &	\small 3.5  & \small   50        & \small 0.28 & \small 12.5\\
\small WVFS 1547+0645	&	\small 15:47:54   &	\small 6:43:07  &	\small 613   &	\small 2.3  & \small   55        & \small 0.29 & \small 8.0\\
\small WVFS 1637+0730	&	\small 16:37:17   &	\small 7:29:26  &	\small 1343  &	\small 2.9  & \small 	60        & \small 0.30 & \small 9.7\\
\hline
\hline

  \end{tabular}
\end{center}

\caption{Source properties of  candidate {\HI} detections in the Westerbork Virgo
Filament Survey.}
\label{new_det}
\end{table*}

{\bf WVFS 0859+0330}: This detection does not seem to have an optical
counterpart and is not in the vicinity of another galaxy. The velocity
width is about 90 km s$^{-1}$, and the highest measured column
  density at this resolution is $N_{HI}\sim 4.7\times 10^{17}$
cm$^{-2}$.\\

{\bf WVFS 0921+0200}: Detection with no visible optical counterpart in
the DSS image, and no known galaxy within four degrees. This object
has a narrow line width of only 55 km s$^{-1}$ and an integrated
column density of $N_{HI} \sim 3.5\times 10^{17}$ cm$^{-2}$, 
  assuming the emission fills the beam.\\

{\bf WVFS 0956+0845}: {\HI} detection in the immediate neighbourhood
of NGC 3049 at a projected distance of only $\sim 0.7$ degrees,
although the central velocity is offset by about 150 km s$^{-1}$. This
detection has a relatively weak, but very broad profile of $\sim 200$
km s$^{-1}$, it could be related to NGC 3049. The total flux of this
detection is 11 Jy km s$^{-1}$, corresponding to a column density of
$N_{HI} \sim 1.4\times 10^{18}$ cm$^{-2}$, integrated over the full
line width.\\

{\bf WVFS 1035+0045}: Isolated {\HI} detection with no nearby galaxy
at a similar radial velocity. At angular distances of 2 and 4 degrees,
there are strong indications for other {\HI} detections with a similar
profile at exactly the same radial velocity. These detections did not
pass the $8\sigma$ detection limit and therefore are not listed in the
table of detections. WVFS 1035+0045 could be the brightness component
of a much more extended underlying filament, the velocity width is 65
km s$^{-1}$, with an integrated column density of $N_{HI} \sim 4.1\times
10^{17}$ cm$^{-2}$.\\

{\bf WVFS 1055+0415}: A relatively strong {\HI} detection in the direct
vicinity of NGC 3521, at an offset of 2.5 degrees. The radial velocity
is comparable, although 100 km~s$^{-1}$ offset from the systematic velocity
of NGC 3521. Note, however, the more than 500 km~s$^{-1}$ linewidth of
this galaxy. When assuming a distance to this galaxy of 7.7 Mpc, the
projected separation of WVFS 1055+0415 is $\sim 350$ kpc. It has a 110
km~s$^{-1}$ line width and an integrated column density of $N_{HI}
\sim 5.4\times 10^{17}$ cm$^{-2}$.\\

{\bf WVFS 1140+0115}: There seems to be a bridge connecting this
source with UGC 6578, which is a relatively small galaxy. The angular
offset to UGC 6578 is about 1.1 degree, which corresponds to 300 kpc
at a distance of 15.3 Mpc. WVFS 1140+0115 has a line width of 85 km
s$^{-1}$ and a column density of $N_{HI} \sim 3.8\times 10^{17}$
cm$^{-2}$.\\

{\bf WVFS 1152+0145}: This detection is about 3.5 degrees separated
from two massive galaxies, NGC 4116 and NGC 4123. These two galaxies
are confused in our data cubes and appear as one source. The radial
velocity of WVFS 1152+0145 is similar to the two galaxies, and when
using a distance of 25.4 Mpc to NGC 4116, the projected separation of
the filament is 1.5 Mpc. An interesting fact is that the spectral
profile of NGC 4116/4123 shows an enhancement at exactly the velocity
of WVFS 1152+0145, indicating that there is extra {\HI} at this
velocity. WVFS 1152+0145 has a line width of 70 km s$^{-1}$ and an integrated
column density of $N_{HI} \sim 3.3\times 10^{17}$ cm$^{-2}$.\\

{\bf WVFS 1200+0145}: An {\HI} detection at exactly the same radial velocity
as UGC 7332 at a separation of 4.4 degrees. UGC 7332 has a likely
distance of 7 Mpc, which means that the projected distance between the
galaxy and WVFS 1200+0145 is about 500 kpc. We note that there
are several other galaxies at a very similar radial velocity, but
slightly more separated from WVFS 1200+0145. This new {\HI} detection has
a line width of only 50 km s$^{-1}$ and a column density of
$N_{HI} \sim 4.4\times 10^{17}$ cm$^{-2}$.\\

{\bf WVFS 1212+0245}: This detection is most likely related to PGC
135791, as both position and velocity of the {\HI} detection agree very
well. It is the first time that an {\HI} component has been detected for
this dwarf galaxy at a distance of 5.3 Mpc. The {\HI} detection is quite
strong, with a total estimated flux of 6.3 Jy km s$^{-1}$ when
integrating over the full line width of 100 km s$^{-1}$, which
corresponds to a column density of $N_{HI} \sim 8.2\times 10^{17}$
cm$^{-2}$.\\

{\bf WVFS 1216+0415}: This is a relatively bright new detection, with
a total flux of 5.6 Jy km s$^{-1}$ integrated over the 90 km s$^{-1}$
line width,  which corresponds to a column density of $N_{HI} \sim
7.2\times 10^{17}$ cm$^{-2}$. There are several galaxies in the
projected vicinity of WVFS 1216+0415 for which the redshift and
distance are unknown. Most apparent is SDSS J121643.27+041537.7, a
diffuse dwarf galaxy, listed in the Sloan Digital Sky Survey (SDSS)
archive. Although the centroid in the WVFS data is imprecise due to
the low resolution, SDSS J121643.27+041537.7 is within the 90\%
contour of the peak flux. Higher resolution {\HI} data could provide a
better indication whether the detected {\HI} is related to this object.
Separated by 2.2 degrees (corresponding to 500 kpc at assuming
distance of 13.1 Mpc) from WVFS 1216+0415 is PGC 040411. This {\HI}
detection could be related to the spiral galaxy PGC 040411, because of
the relatively small projected distance and the matched radial
velocity.\\

{\bf WVFS 1217+0115}: This detection is in the vicinity of several
galaxies, at different distances, therefore it is difficult to say
whether there is a relation between WVFS 1217+0115 and any of these
galaxies. The most nearby galaxy is UGC 7394, separated by 0.8
degrees, which corresponds to 370 kpc, at a distance of 27 Mpc. At a
distance of 13.1 Mpc are three galaxies: M61, UGC 7612 and UGC 7642, all
separated by $\sim 3$ degrees from WVFS 1217+0115, or 700 kpc. There
is reasonable correspondence in velocity with all of the
aforementioned galaxies. Because of the large over-density it is most
likely that WVFS 1217+0115 belongs to the group containing M61. The
line width of this detection is 80 km s$^{-1}$, with an integrated
column density of $N_{HI} \sim 3.5\times 10^{17}$ cm$^{-2}$.\\

{\bf WVFS 1234+0345}: This object is the {\HI} counterpart of UGC 7715, at
the same position and velocity. This galaxy is not listed in the
HIPASS catalogue, however is not a completely new detection as the
LEDA database gives a flux of 1.7 Jy km s$^{-1}$. We detect an almost
two times larger flux of 3.9 Jy km s$^{-1}$ and a line width of 80
s$^{-1}$, which corresponds to an {\HI} column density of $N_{HI}
\sim 4.9\times 10^{17}$ cm$^{-2}$.\\
  
{\bf WVFS 1253+0145}: The line width of this detection is only 50 km
s$^{-1}$, with an integrated column density of $N_{HI} \sim 3.1\times
10^{17}$ cm$^{-2}$. This {\HI} detection is possibly the counterpart of
SDSS J125249.40+014404.3, a dwarf Elliptical listed in the SDSS
archive with a radial velocity of 883 km $^{-1}$ and a distance of 5.8
Mpc. Another possibility is a relation with NGC 4772, this galaxy is
at a larger distance of 13.0 Mpc. There is a connecting bridge
of only half a degree and the radial velocity matches the peak of this
object. The peak of this companion is slightly brighter than the
galaxy itself, which is a little bit suspicious. \\

{\bf WVFS 1324+0700}: This detection is very isolated, and there does
not seem to be any relationship to a nearby galaxy out to
a few degrees. WVFS 1324+0700 has a line width of 70 km
s$^{-1}$ and a column density of of $N_{HI} \sim 3.9\times
10^{17}$ cm$^{-2}$.\\

{\bf WVFS 1424+0200}: This detection appears to be very isolated,
without a recognisable connection to a galaxy. The DSS image shows an
optical galaxy, this is UGC 9215 at a radial velocity of 1397 km
s$^{-1}$, this is about 850 km s$^{-1}$ different from WVFS 1424+0200,
so any relation is very unlikely. The line width of WVFS 1424+0200 is
70 km s$^{-1}$ and it has an integrated column density of $N_{HI} \sim
4.7\times 10^{17}$ cm$^{-2}$.\\

{\bf WVFS 1500+0815}: There are several massive galaxies with a
systemic velocity within 100 km s$^{-1}$ of the velocity
of WVFS 1500+0815 (NGC 5964, NGC 5921, NGC 5701, NGC 5669 and NGC
5194). All these galaxies are at a distance of about 24 Mpc. At this
distance the projected separation to WVFS 1500+0815 would be between 2.5
and 5 Mpc. A direct connection to any of the galaxies is not obvious,
unless there is a very large diffuse envelope between them,
which is perhaps not unreasonable, as the radial velocities of the
galaxies are all very similar. The highest measured column density of WVFS
1500+0815 is $\sim 4.1\times 10^{17}$ cm$^{-2}$, when integrated over
the full line width of 105 km $^{-1}$. As for all the new detections,
there are many optical detections in the projected vicinity of the {\HI}
detection, but without redshift information. Worth special mention is
SDSS J150103.32+081936.5, a dwarf galaxy that based on visual
assessment could be at the relevant distance.\\

{\bf WVFS 1524+0430}: There are no known galaxies with a comparable
radial velocity in the vicinity or WVFS 1524+0430. In the DSS images
we find two dwarf galaxies that could be related to the {\HI} detection:
SDSS J152444.50+043302.3 and SDSS J152445.97+043532.5. Higher
resolution {\HI} data would be needed to resolve the {\HI} and provide more
information about the exact position. Based on visual inspection both
SDSS sources could be at a relevant distance, as the optical
appearance is similar to dwarf galaxies with a known radial
velocity. The line width is only 55 km s$^{-1}$ and it has a column
density of $N_{HI} \sim 3.1\times 10^{17}$ cm$^{-2}$.\\

{\bf WVFS 1529+0045}: This is also an isolated {\HI} detection without a
clear optical counterpart. With a line width of only 50 km
$s^{-1}$ and an integrated flux of 3.5 Jy km s$^{-1}$ it is a
relatively narrow, but strong detection compared to the other isolated
detections. The peak column density of WVFS 1529+0045 is $N_{HI} \sim
4.4\times 10^{17}$ cm$^{-2}$.\\

{\bf WVFS 1547+0645}: Another isolated detection without any nearby known
galaxy or optical counterpart. With a velocity width of 50 km s$^{-1}$
and a total flux of only 2.2 Jy km s$^{-1}$ this is the weakest
detection that passed the detection threshold. The column density of
WVFS 1547+0645 is only $N_{HI} \sim 2.9\times 10^{17}$
cm$^{-2}$.\\

{\bf WVFS 1637+0730}: The last new detection in the survey, NGC 6106 is at an
offset of 4.75 degrees to WVFS 1637+0730, corresponding to a
projected distance of 2 Mpc, at an assumed 
distance of 23.8 Mpc. The radial velocity of this galaxy is 1448 km
s$^{-1}$ which is about 100 km s$^{-1}$ offset from WVFS
1637+0730. Because of the relatively large projected distance and the
significant offset in velocity, a direct relation between WVFS
1637+0730 and NGC 6106 is not very obvious, and WVFS 1637+0730 is more
likely another isolated detection. In the DSS image an optical galaxy
can be identified, this is UGC 10475, a background galaxy with a
radial velocity of 9585 km s$^{-1}$. The velocity width of WVFS
1637+0730 is 60 km s$^{-1}$ and it has a column density of
$N_{HI} \sim 3.6\times 10^{17}$ cm$^{-2}$.\\

Only very few of the new {\HI} detections have a clear optical
counterpart and can be assigned to a known galaxy. There are several
isolated detections, but most of the detections could potentially be
related to a known, usually massive, galaxy. These {\HI} detections have
a radial velocity that is very comparable to the systemic velocity of
the major galaxy. The projected separation of these detection ranges
from 300~kpc up to 2~Mpc. Smaller offsets from galaxies can not be
identified, as the primary beam size of the survey already spans
150~kpc at a distance of 10~Mpc. Any object within 300~kpc of a galaxy
would very likely be confused and could not be identified as an
individual object.

All new {\HI} detections have a line width between $\sim50$ and
$\sim100$ km s$^{-1}$, with the exception of WVFS 0956+0845. The
column densities  at the resolution of the primary beam and
  integrated over the velocity width, vary between $N_{HI} \sim
2.9\times 10^{17}$ and $\sim8\times 10^{17}$ cm$^{-2}$.\\

\begin{figure*}
 \begin{center}
  
 \includegraphics[width=0.3\textwidth]{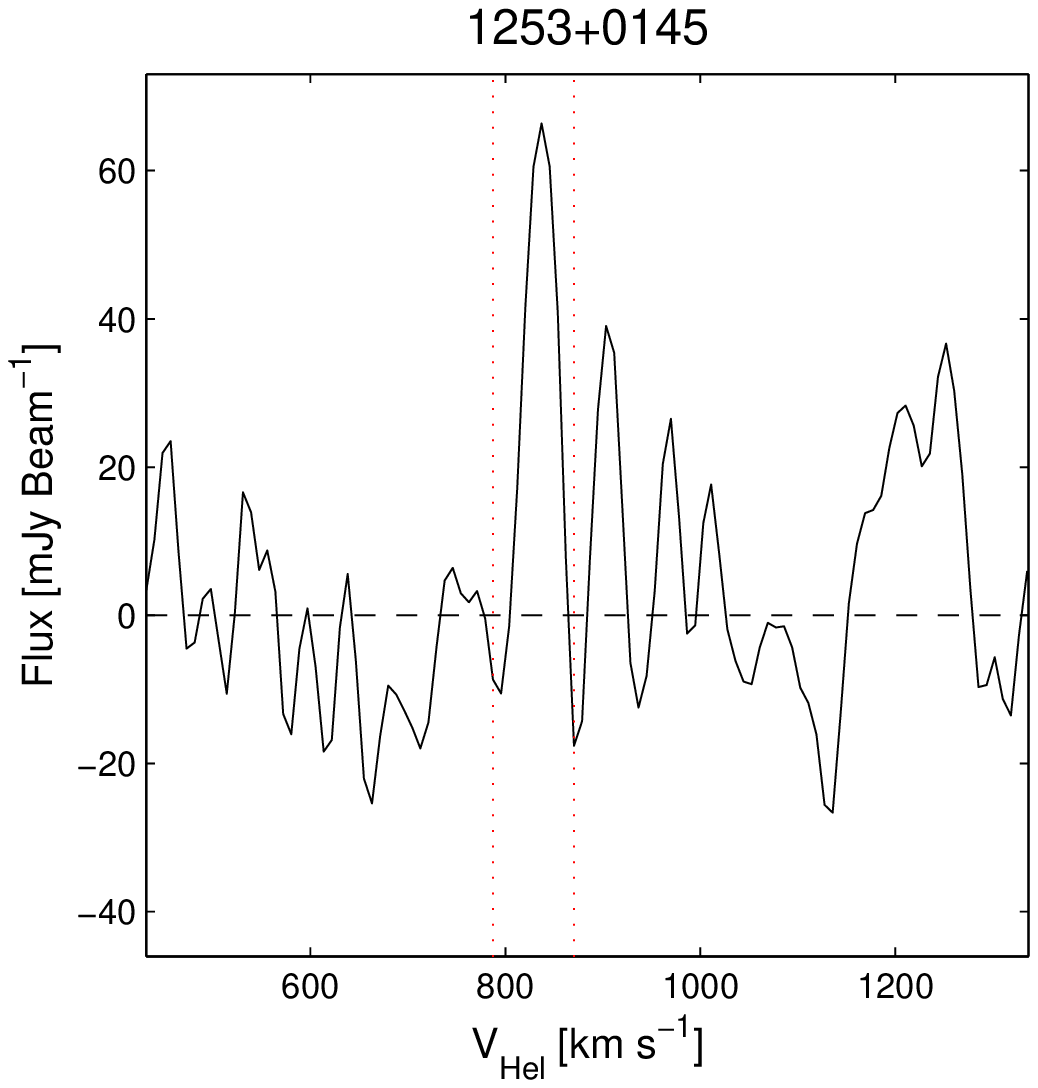}
 \includegraphics[width=0.3\textwidth]{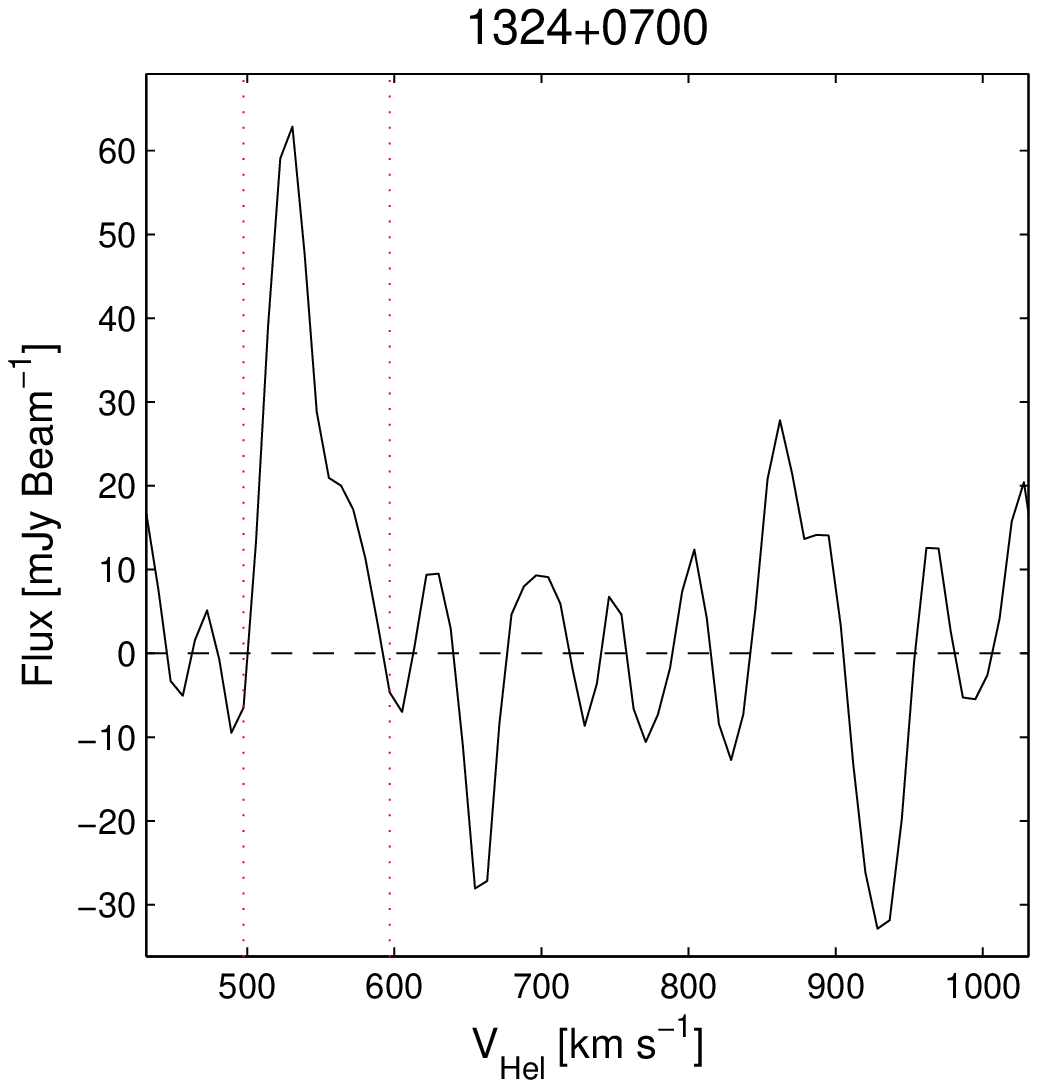}
 \includegraphics[width=0.3\textwidth]{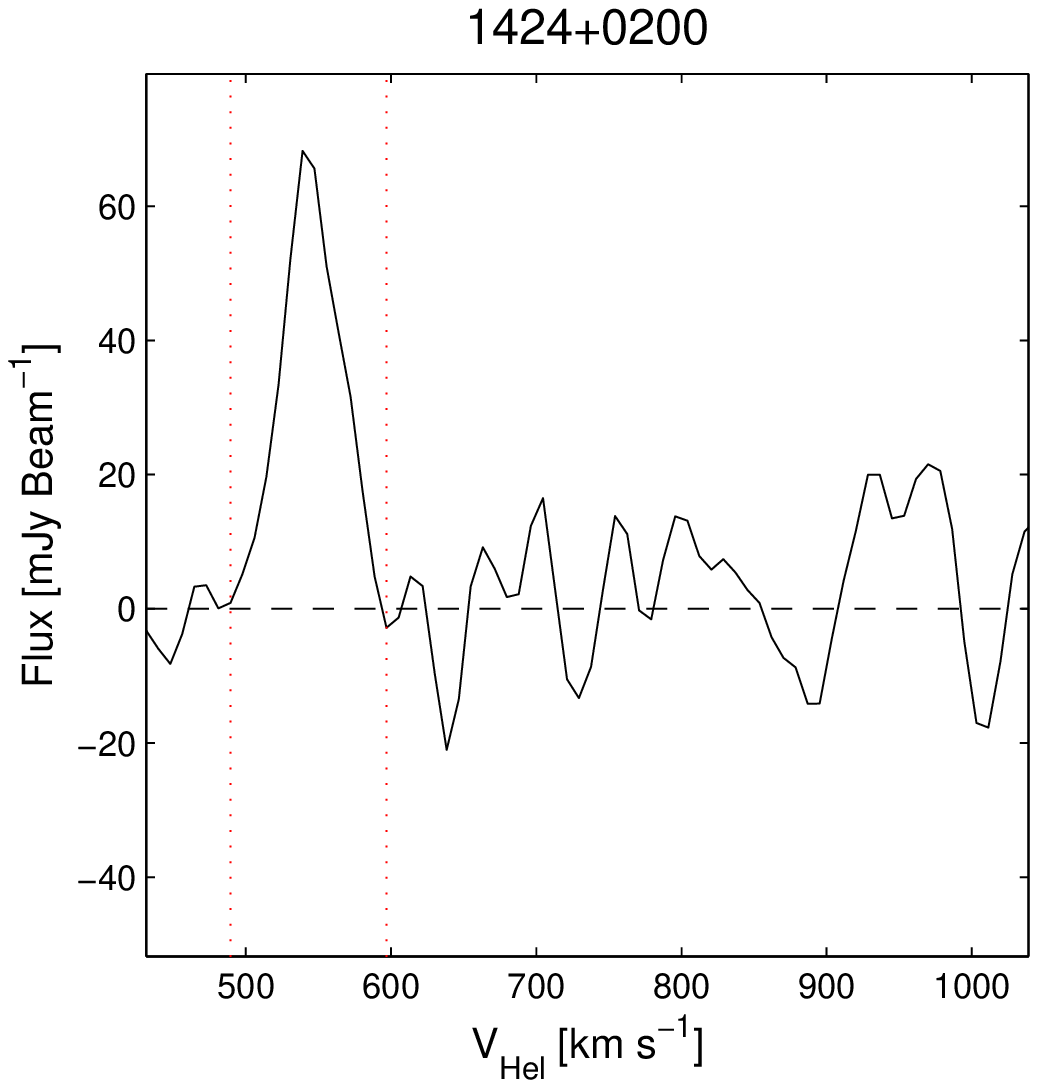}

 \includegraphics[width=0.3\textwidth]{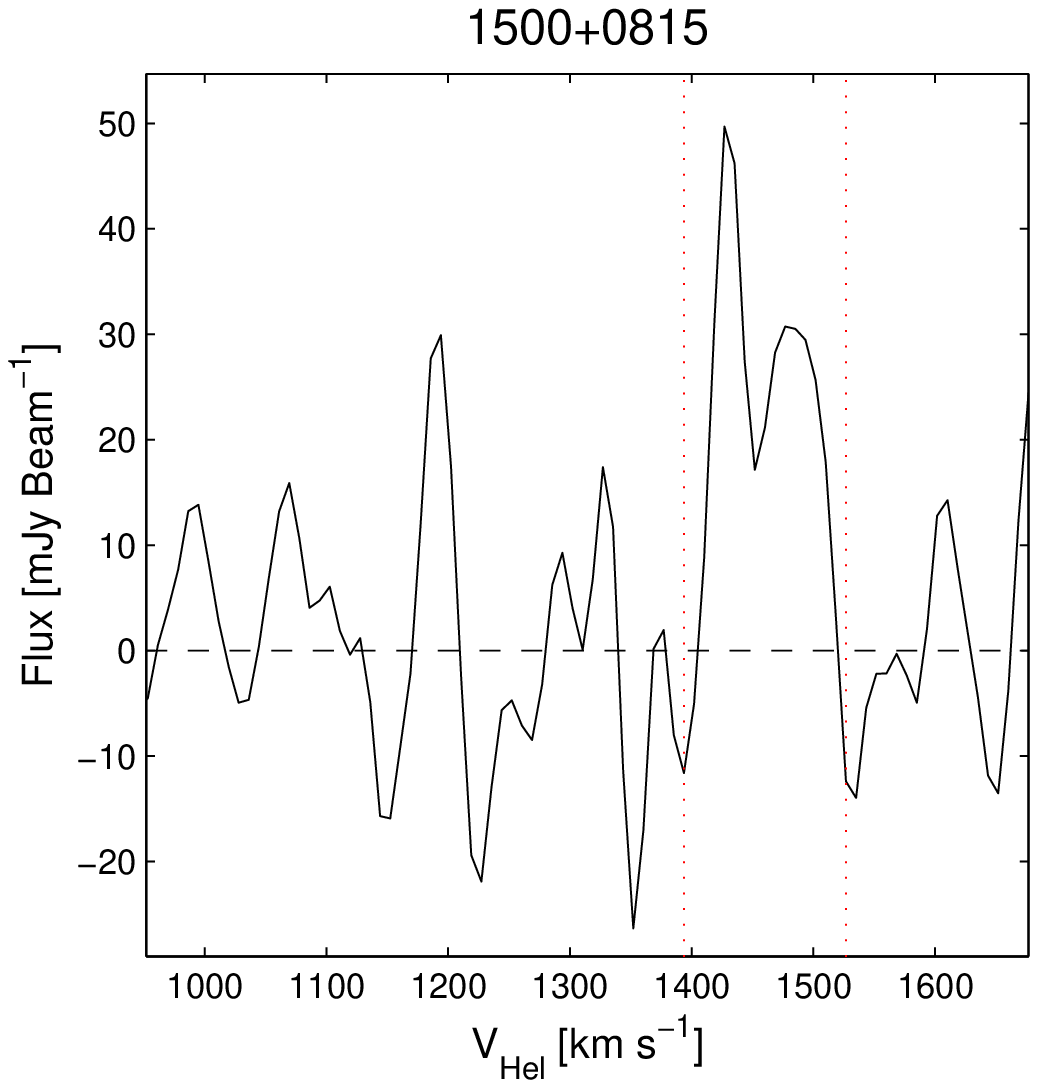}
 \includegraphics[width=0.3\textwidth]{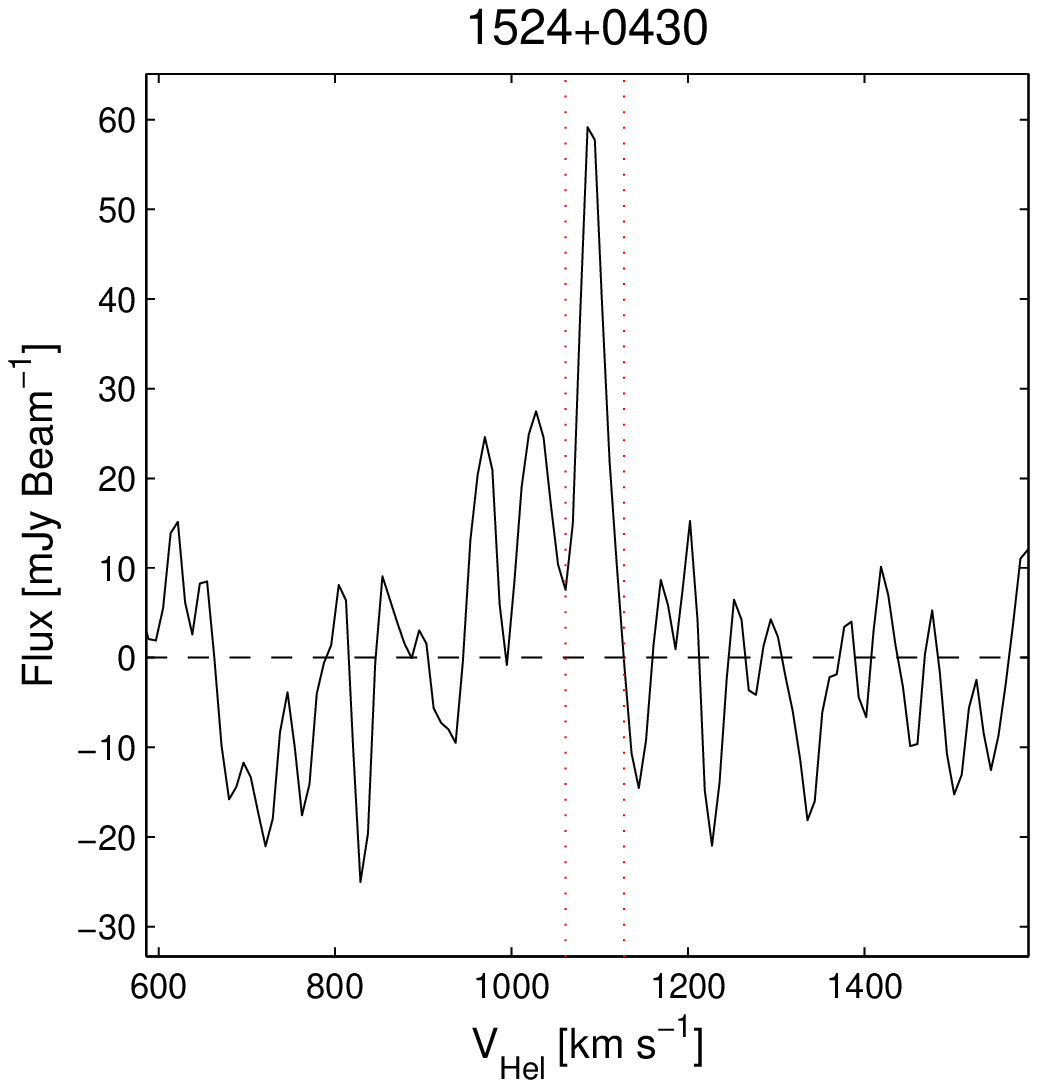}
 \includegraphics[width=0.3\textwidth]{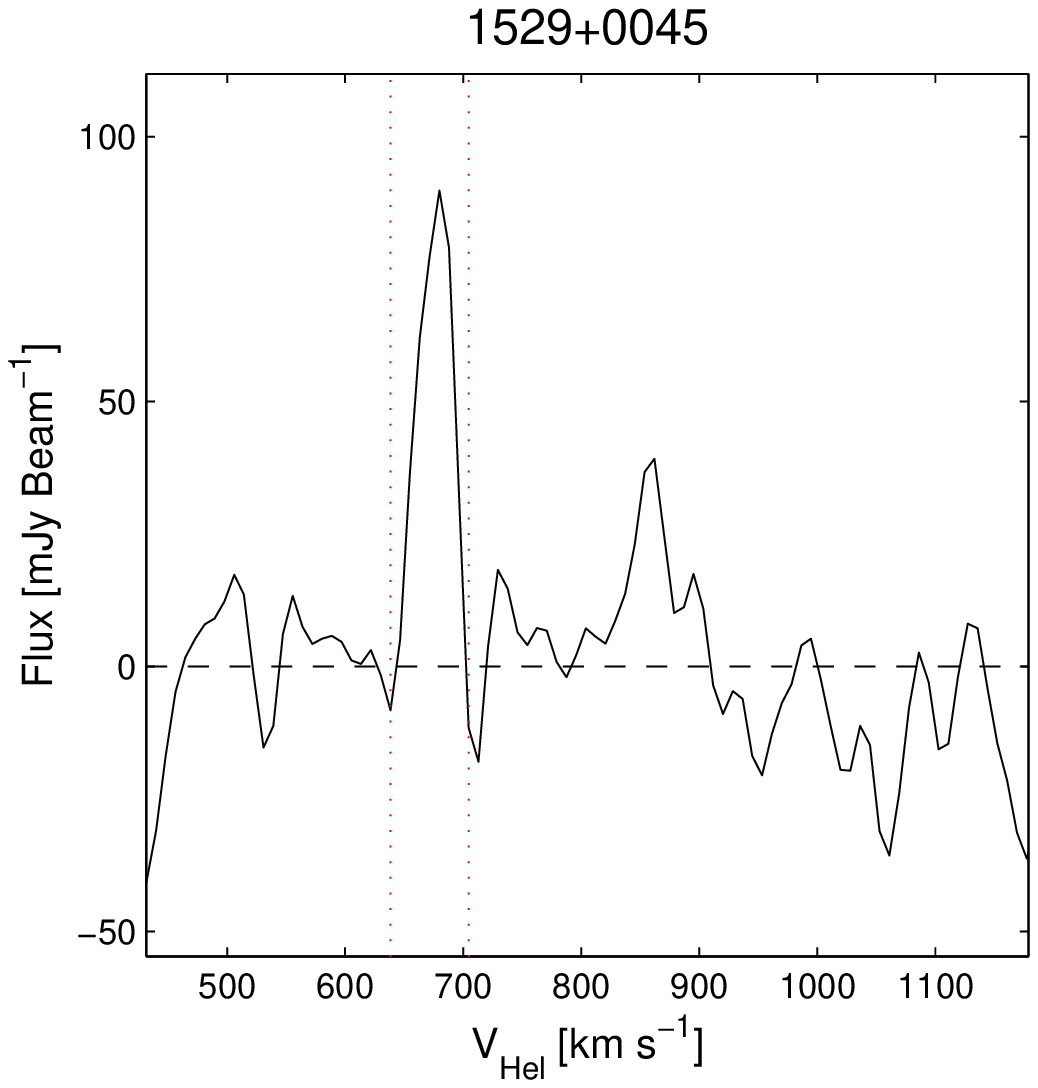}

 \includegraphics[width=0.3\textwidth]{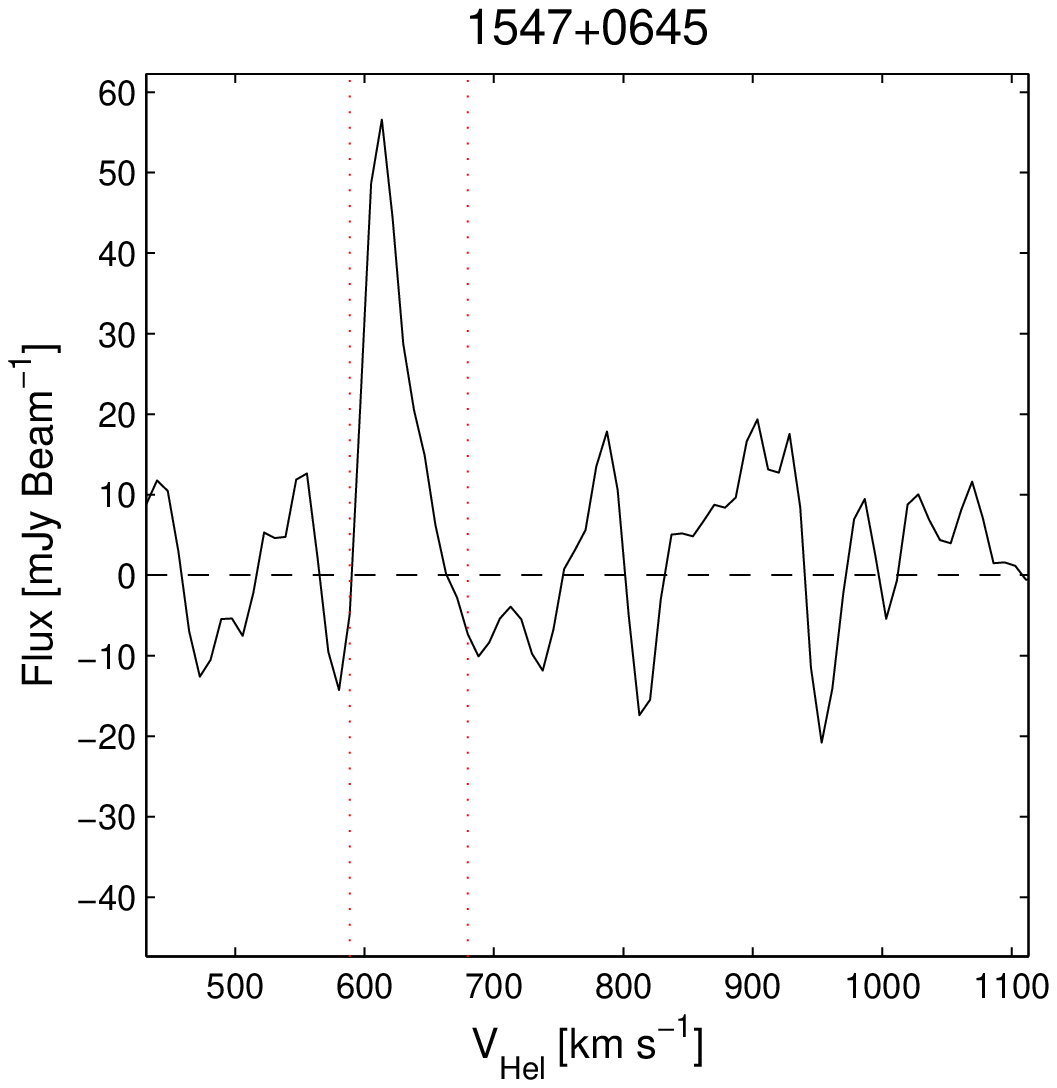}
 \includegraphics[width=0.3\textwidth]{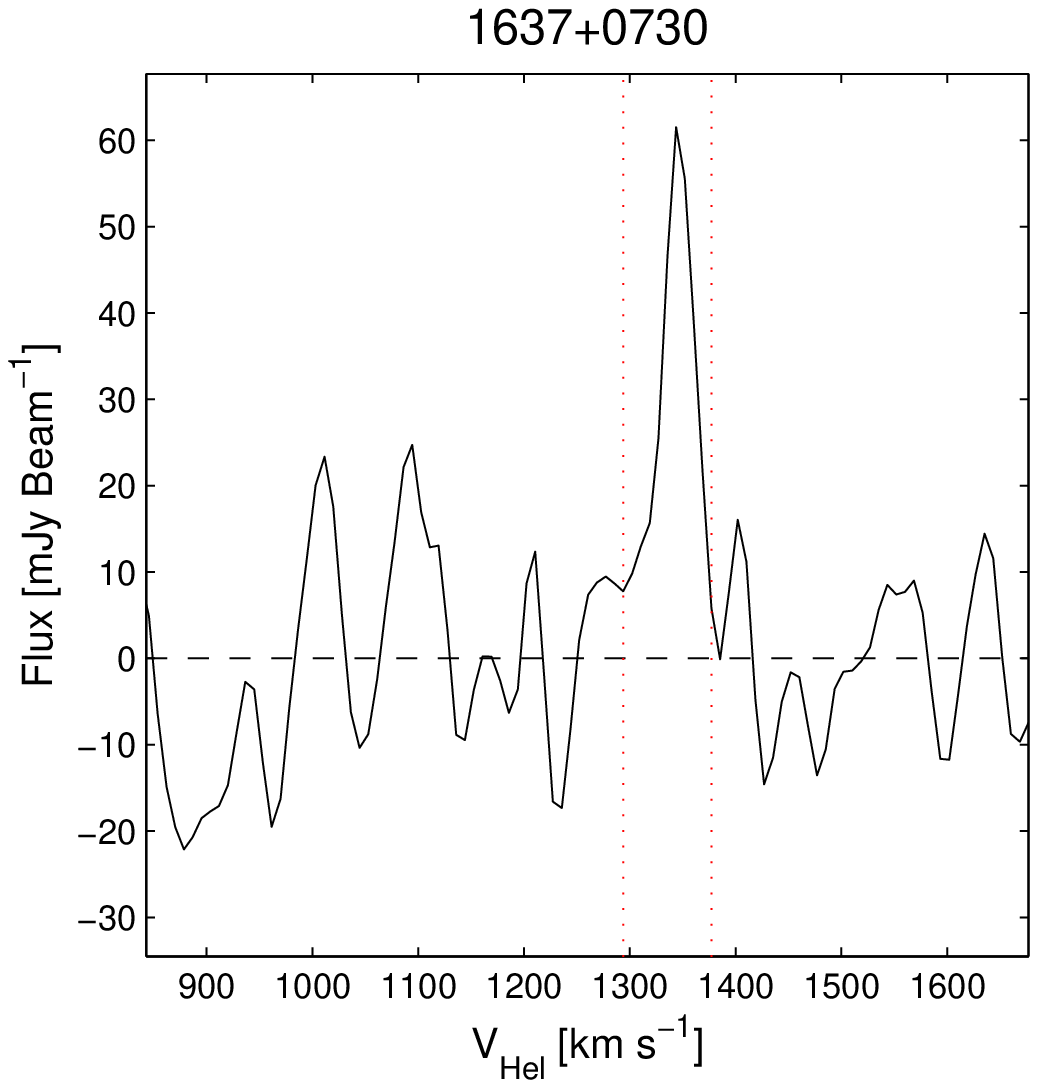}
\end{center}

{\bf Fig~\ref{spectra}.} (continued)					 
\end{figure*}

\begin{figure*}[h]
 
  \includegraphics[width=0.5\textwidth]{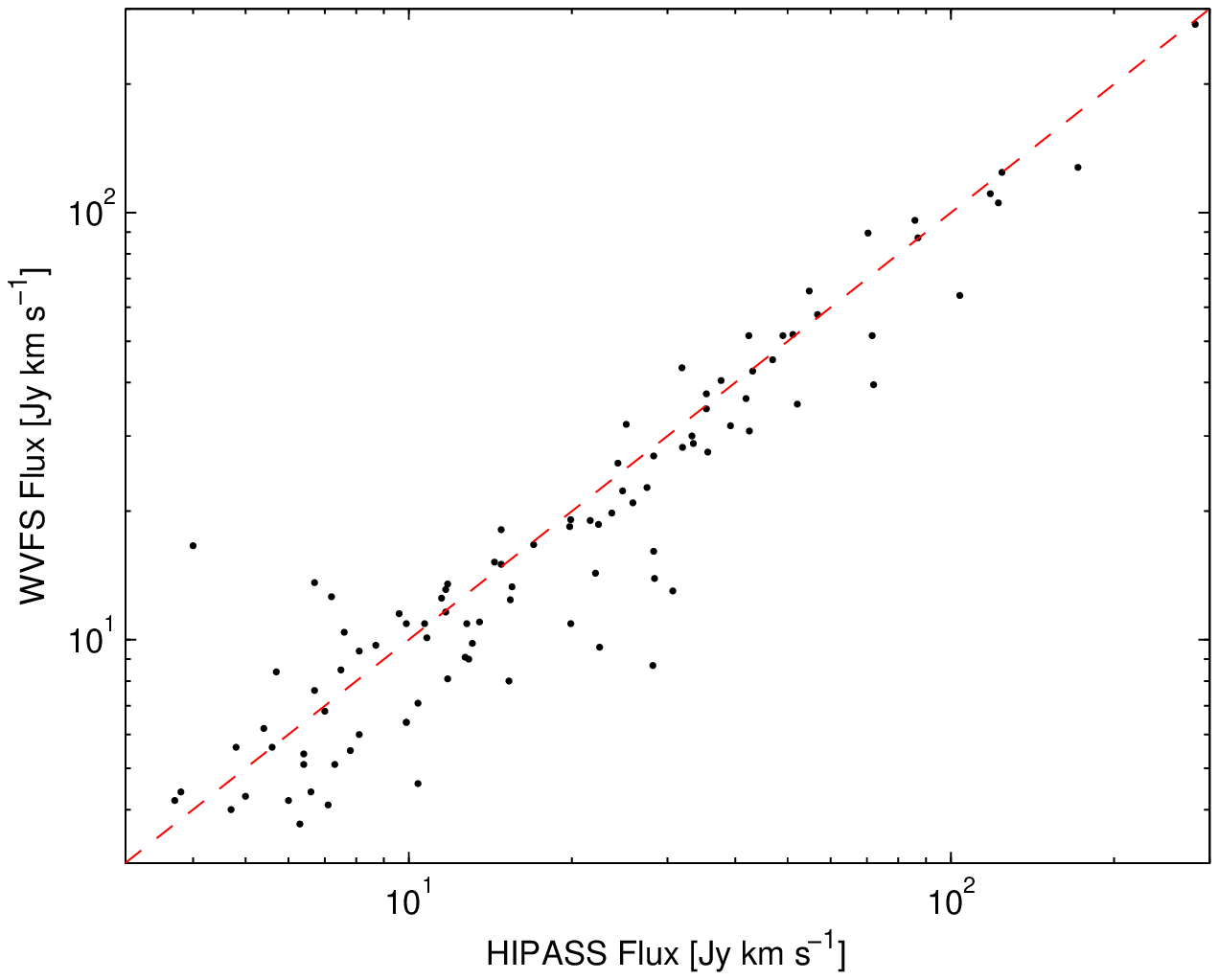}
  \includegraphics[width=0.5\textwidth]{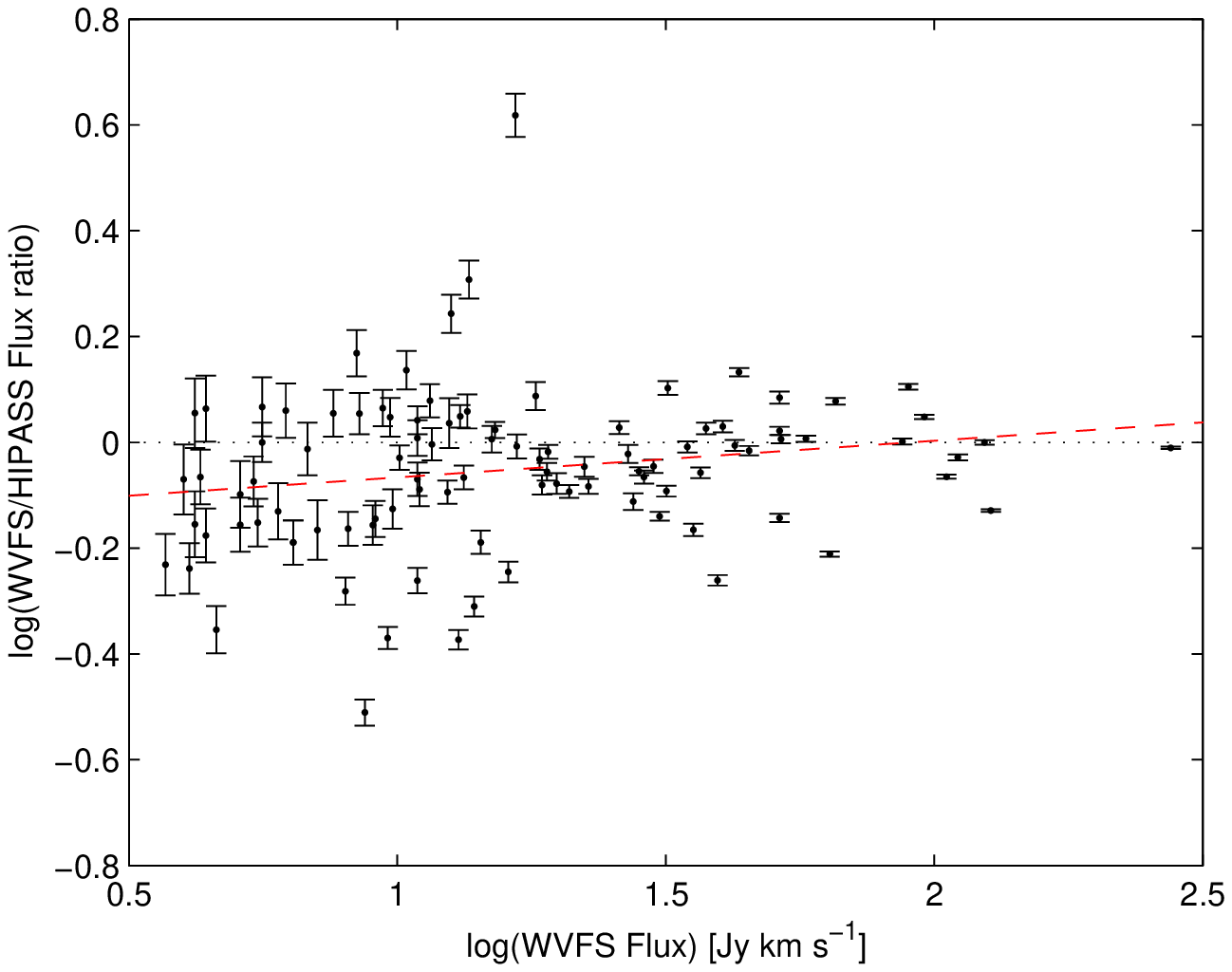}
  
  \includegraphics[width=0.5\textwidth]{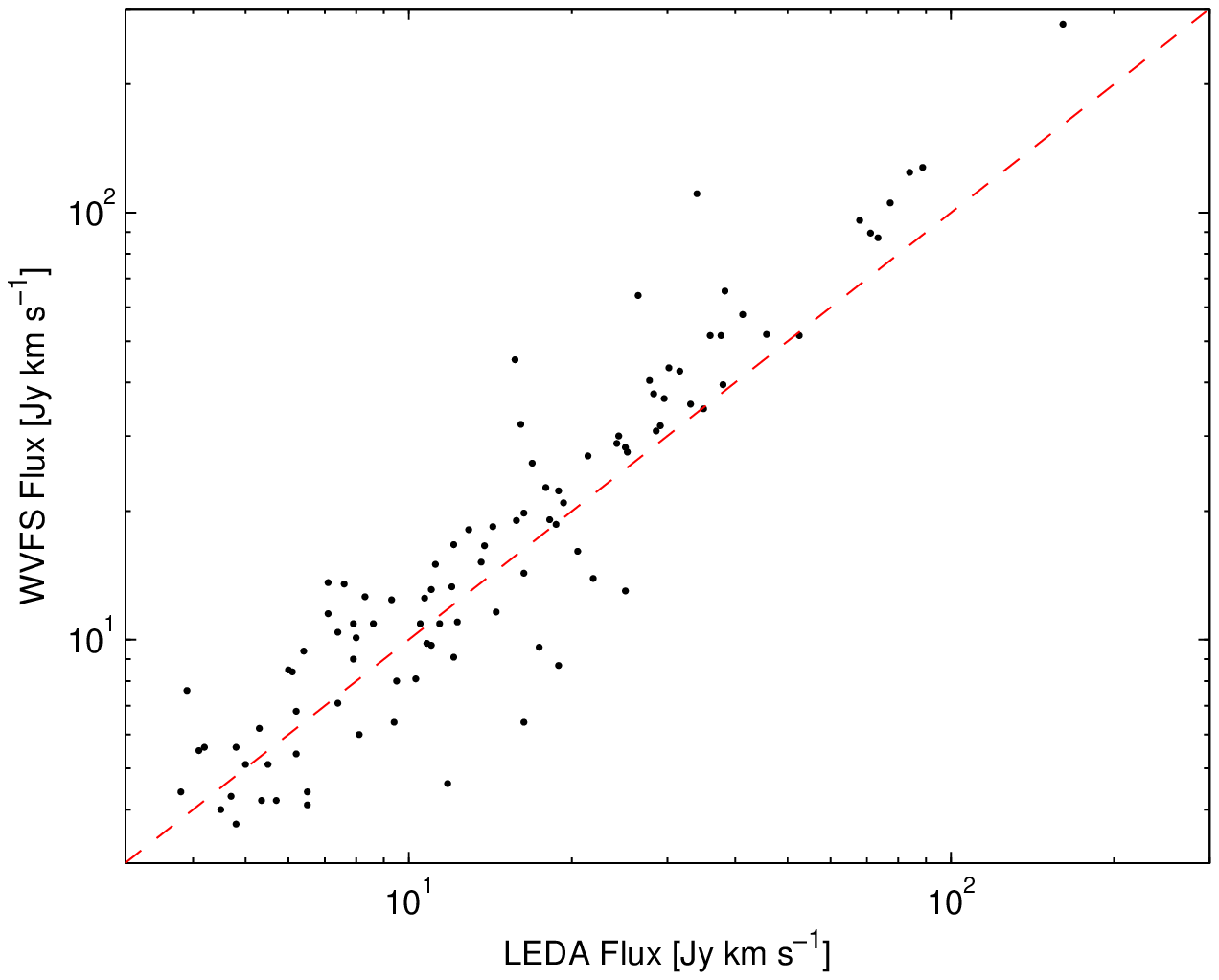}
  \includegraphics[width=0.5\textwidth]{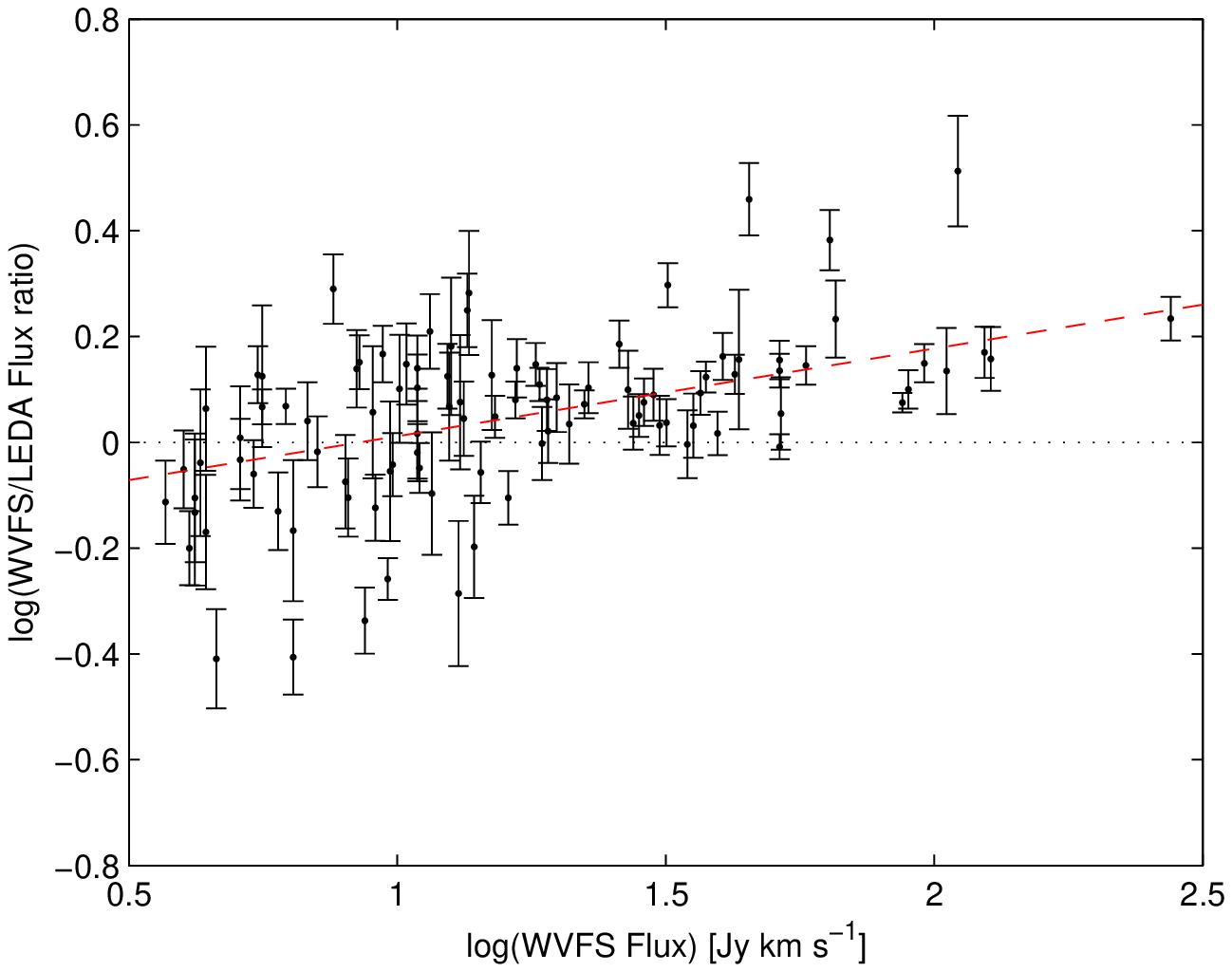}

  \includegraphics[width=0.5\textwidth]{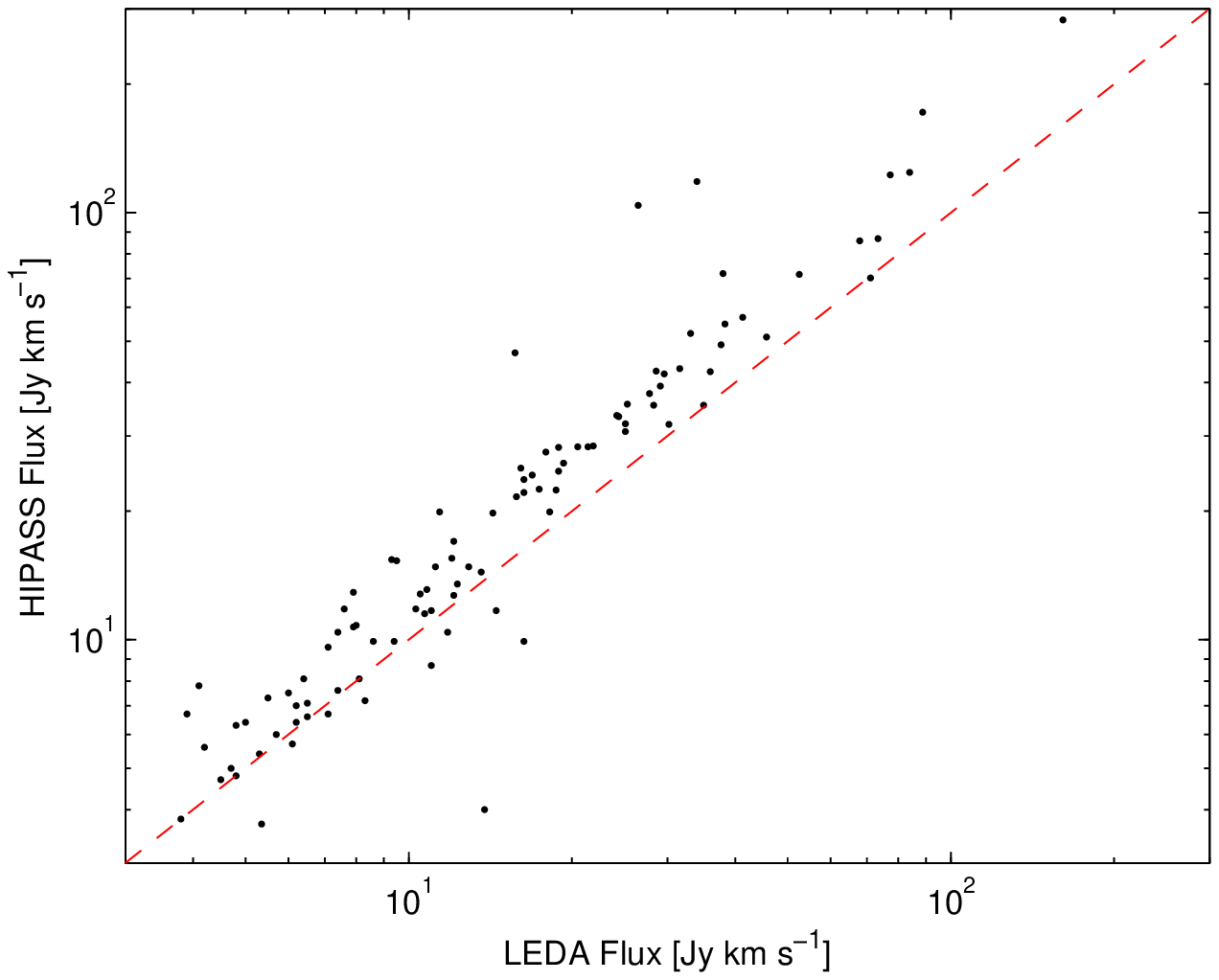}
  \includegraphics[width=0.5\textwidth]{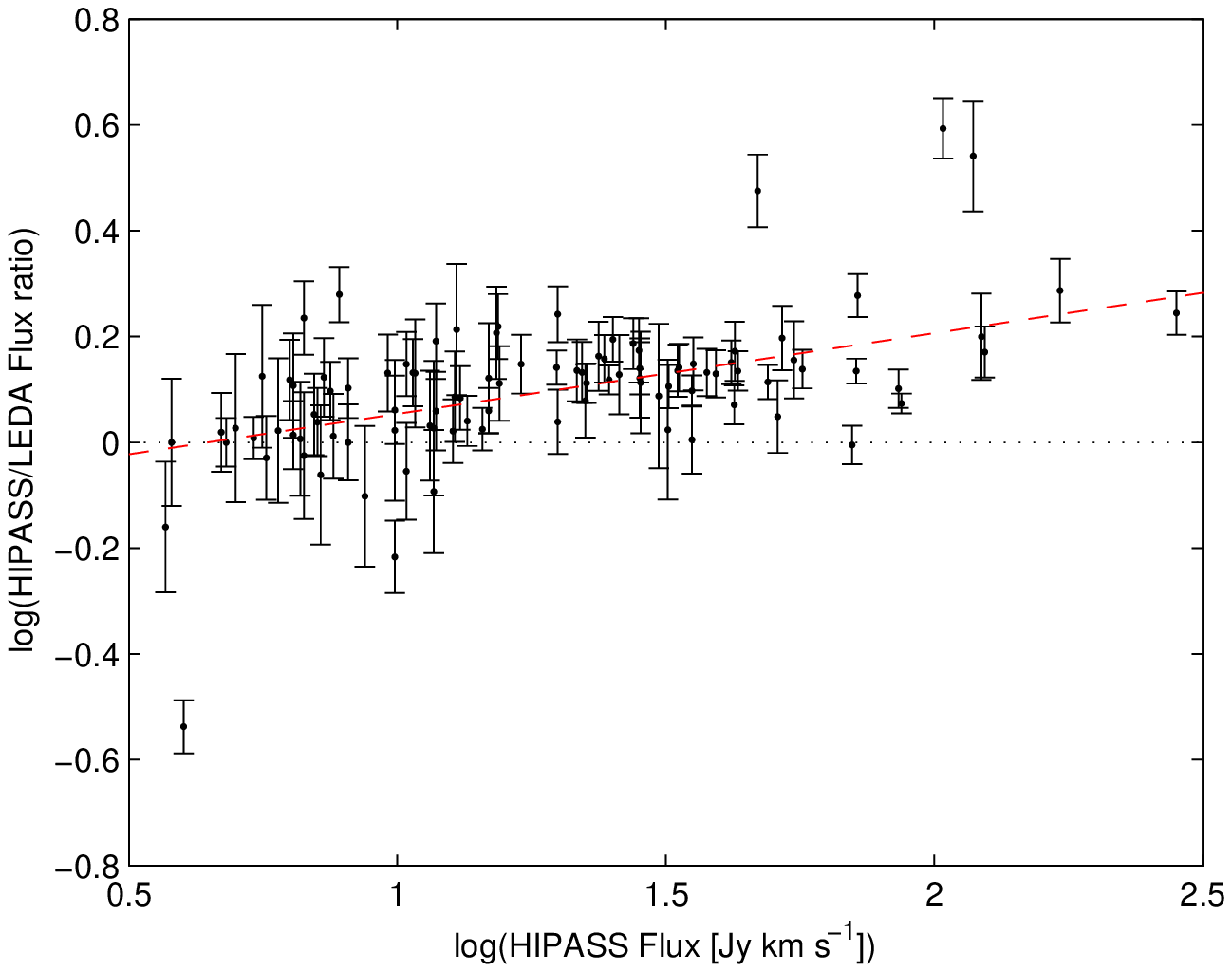}

 \caption{Comparison of determined {\HI} fluxes with values obtained from
 HIPASS and LEDA. The left panels show the direct relation between the
 different catalogues, with the red line indicating the points where
 fluxes are equivalent. The right panels show the ratio between two
 catalogues as function of flux, both on a logarithmic scale, the red
 line indicated here the best power law fit through the data
 points. The first row shows the comparison between WVFS and HIPASS
 fluxes, while the second row shows the comparison between WVFS and
 LEDA. As a reference, the comparison between HIPASS and LEDA fluxes
 is plotted in the bottom row.}
  \label{ratios}
\end{figure*}

\subsection{{\HI} in the extended galaxy environment}
We compare our measured galaxy fluxes with the fluxes measured by the
{\HI} Parkes all sky survey (HIPASS) and fluxes tabulated in LEDA. Only
those sources are considered for which the integrated signal-to-noise
ratio is larger than 8 in both the WVFS and HIPASS
surveys. Furthermore, galaxies have been excluded which occur at the
edge of the WVFS band, as no complete spectrum can be derived for
these sources, resulting in an integrated flux value that is known to
be only a lower limit.

It is interesting to look for any systematic differences in total flux
between the several catalogues. Flux values derived from both WVFS and
HIPASS have undergone a uniform calibration procedure that was similar for all
sources. Both surveys are single dish surveys with a relative large
primary beam sizes of 15' for HIPASS and 49' for WVFS after spatial
smoothing. At a distance of 10~Mpc, these beam sizes correspond to 40
and 140~kpc respectively, comparable to or larger than the typical {\HI}
diameter of a galaxy.

The LEDA fluxes are compiled from measurements made with very different
telescopes, yielding much greater variety in calibration
procedures. Because the fluxes are obtained from different telescopes,
it is not possible to relate the fluxes to one specific beam size.

In the left panels of Fig.~\ref{ratios} the integrated flux values of
the three catalogues are compared, with WVFS
vs. HIPASS in the top panel, WVFS vs. LEDA in the middle panel and
HIPASS vs. LEDA in the bottom panel. The dashed line goes through the
origin of the diagram, with a slope of one, indicating identical
fluxes. The best correspondence is between the HIPASS and WVFS data as
the points are scattered around the red line. When looking at the
WVFS-LEDA comparison, there is agreement for fluxes below $\sim 20$ Jy
km s$^{-1}$, but for larger fluxes all WVFS fluxes seem to be
systematically higher. The same effect is apparent in the HIPASS-LEDA
comparison.

To have a better understanding of the differences, the flux ratios of
the different catalogue pairs are plotted in the right panel. Fluxes
and flux ratios are both plotted on a logarithmic scale, equivalent
flux values are indicated by the black dotted line at zero. The data
points in each plot are fitted with a power law function, indicated by
the dashed line.

The scatter in the WVFS-HIPASS comparison is almost perfectly centered
around zero. The fitted power law has a slope of $a=0.069\pm 0.07$ and
a scaling factor of $b=-0.14\pm 0.09$. There is one source which is
significantly stronger in the WVFS data, which is WVFS 1210+0300 or
UGC 7185. The reason for this large discrepancy is not clear. There
are quite a few sources for which the measured flux in HIPASS is
significantly higher. This can be partially ascribed to confusion
effects, as has been described earlier.

The flux ratios between WVFS and LEDA show substantial deviations
especially for larger flux values. The power law fit has a relatively
steep slope of $a=0.15\pm0.05$ and a scaling factor of
$b=-0.1\pm0.08$. Above a 20 Jy km s$^{-1}$ flux limit, the WVFS values
are brighter than the LEDA values without any exception.

Because this is quite a dramatic result, the same comparison has been
done between the HIPASS and LEDA fluxes in the bottom right panel of
Fig.~\ref{ratios}. Although the power law fit has a very similar slope
compared to the WVFS data of $a=0.16\pm0.05$, the scaling factor of
$b=-0.15\pm0.08$ is marginally larger.

The general conclusion is that both WVFS and HIPASS find significantly
more {\HI} in galaxies than LEDA. This effect is strongest for objects
with an {\HI} flux above 20 Jy km s$^{-1}$. Above this level the excess
in {\HI} flux for both these single dish surveys is $\sim 40$\%.

A possible explanation is that both HIPASS and WVFS are more sensitive
to diffuse emission, due to the large intrinsic beam sizes. The flux
values listed by LEDA, are based on a combination of fluxes obtained
in different measurements. Although we cannot access these individual
values, a large number of the flux values were likely obtained with
smaller intrinsic beam sizes, e.g. interferometric data. In general, a
smaller intrinsic beam is much less sensitive to diffuse emission than
a large beam, and therefore will miss diffuse emission
preferentially. However, the differences between WVFS and LEDA are
remarkably large and systematic which is a point of concern. For some
individual targets we have compared the flux values of LEDA with all
available flux values given by the NASA Extragalactic Database
(NED). Here we find a similar trend: flux values listed in NED are
generally much higher than the values given by LEDA. To derive {\HI}
fluxes, the LEDA team do not merely calculate a weighted average of
available flux values from the literature. Several corrections are
applied in an attempt to get more uniformity among the fluxes, and the
result is then scaled to fluxes obtained with the Nancay telescope.

We have confidence in the calibration of the WVFS data and the derived
fluxes of our detections and see excellent correspondence with the
HIPASS catalog. We have serious reservations regarding the accuracy of
the LEDA-tabulated {\HI} fluxes.

\subsection{Line width and Gas accretion modes}
\begin{figure}[h]
  \includegraphics[width=0.5\textwidth]{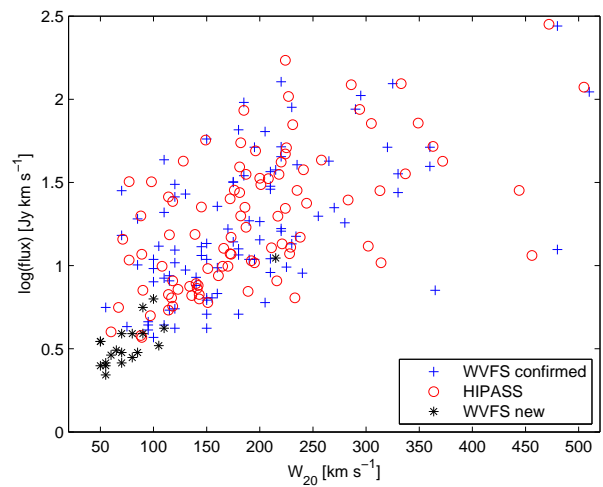}
 \caption{ Flux as function of $W_{20}$ for the WVFS detections
     and the same detection in HIPASS. The behaviour of HIPASS and
     WVFS detections agree very well, in general objects with a larger
     flux have a broader line-width.}
  \label{flux_w20}
\end{figure}

In Fig.~\ref{flux_w20} the flux of each detection is plotted as
  function of $W_{20}$, the line-width at 20\% of the peak. Known and
  confirmed detections are shown with blue plus signs, while the new
  detections are plotted as black stars, the known detections are
  compared with the same objects from the HIPASS database, shown as
  red circles. The same basic trend is apparent in both the HIPASS and
  WVFS tabulations, with brighter detections generally accompanied by
  a larger line-width. The new WVFS detections simply extend this
  trend to low brightnesses and the lowest line-widths.\\

By measuring the line-widths of the detections, an estimate can
  be given of the upper limit of the kinetic temperature, using the
  equation:
\begin{equation}
T_{kin} \leq \frac{m_H \Delta v^2}{8 k_B \textrm{ln}2 }
\end{equation}
where $m_H$ is the mass of an hydrogen atom, $k_B$ is the Boltzmann
constant and $\Delta v$ is the FWHM {\HI} line-width. Apart from
one detection with a velocity width of 215 km s$^{-1}$, the velocity
widths of all the detections are between 50 and 110 km s$^{-1}$. When
assuming that the lines are not broadened by internal turbulence or
rotation, the maximum temperatures range between $\sim 5 \cdot 10^4$
and $\sim 3 \cdot 10^5$ K. If the new detections without an optical
counterpart are indeed related to the cosmic web, then this gas could
be examples of the cold accretion mode as described in
\cite{2005MNRAS.363....2K}, where gas is directly accreting from the
intergalactic medium onto the galaxies at temperatures of $\sim 10^5$
K, without being shock-heated to very high temperatures. We note
that the neutral fraction of gas is expected to drop
very rapidly for temperatures above $10^5$ K and hence it is very
unlikely that high {\HI} column densities would be associated with
thermally dominated linewidths greatly exceeding 100 km s$^{-1}$.

\subsection{Non Detections}
The {\HI} Parkes All Sky Survey completely covers the region observed in
the WVFS. In this region a total of 147 objects are listed in the
HIPASS catalogue within the velocity coverage of WVFS. Most of these
sources could be detected and confirmed by the WVFS, although some of
them could not be identified individually, due to confusion. For three
sources listed in HIPASS we could not determine {\HI} emission in the
WVFS.

NGC 4457 has a flux of 7.2~Jy~km~s$^{-1}$ in HIPASS and
4.4~Jy~km~s$^{-1}$ is listed in the LEDA database. Although there is
substantial discrepancy between those numbers, the source has
significant flux and should be easily detected in the WVFS. There is a
tentative detection in the WVFS data at the expected position and
velocity, however it does not pass our detection limit.

HIPASS J1233-00 has a flux of 3.0 Jy km s$^{-1}$ in HIPASS and 2.9 Jy
km~s$^{-1}$ in LEDA. Although these numbers are consistent, it is a weak
detection, especially when taking into account the $W_{20}$ value of
112 km s$^{-1}$ listed in the HIPASS catalogue. The source does not
appear in WVFS, but it would be near the detection limit. 

HIPASS J1515+05 has a flux of 2.5 km s$^{-1}$ in HIPASS with a
$W_{20}$ value of 121 km s$^{-1}$, making this a very weak
detection. The integrated line strength has a signal-to-noise value of
only 4, when taking into account the sensitivity of HIPASS. There is
no indication for {\HI} in the WVFS, but also none in the LEDA and NED
databases.
  
Since we only expect a source completeness level of about 90\% at our
8$\sigma$ significance threshold (\cite{2002ApJ...567..712C}) it is
not surprising that several faint cataloged sources are not redetected
independently in the WVFS.

\section{Discussion and Conclusion} 
\label{sec:conclusion}

We have carried out an unbiased wide-field {\HI} survey of $\sim1300$
deg$^{2}$ of sky, mapping the galaxy filament connecting the Local
Group with the Virgo cluster. In the total power data we are
especially sensitive to very diffuse and extended emission, due to the
large intrinsic beam size of the observation. Apart from three
sources, we can confirm all detections that have been obtained with
the {\HI} Parkes All Sky Survey in this region, when taking into
account confusion effects. Apart from previously known sources, we
identify 20 new candidate detections with an integrated {\HI}
flux exceeding 8$\sigma$. These candidates have a typical 
  integrated column density of only $\sim 3\times 10^{17}$ cm$^{-2}$,
when assuming that the emission is filling the beam. The velocity
width at 20\% of the peak ranges between $\sim50$
and $\sim100$ km s$^{-1}$ with the exception of one object with a
significantly broader line width of 215 km~s$^{-1}$.

If these candidates are intrinsically diffuse structures, then they
could not have been detected in HIPASS or any other currently available
wide-field {\HI} survey, as the WVFS column density sensitivity is
about an order of magnitude better. The objects would be at
the one sigma level in the full resolution HIPASS data, which makes
identification extremely difficult, even assuming that spatial
smoothing were applied after-the-fact.

For most of our new candidates we can not find a clear optical
counterpart, making direct confirmation difficult. As our data is so
sensitive, we are exploring a new realm in detecting very diffuse
and extended {\HI} and there is not much data available in the literature
to compare with. The detection limits have been set fairly
conservatively in that the integrated flux has to exceed a
signal-to-noise of 8. In addition, we only accept candidates that are
individually apparent in both the {\it rise} and {\it set} data, which
are two independent observations.

The new  candidate detections have properties similar to the {\HI} filament
connecting M31 and M33, as described in
\cite{2004A&A...417..421B}. This filament has a very comparable column
density to the WVFS detections of $3\times 10^{17}$ cm$^{-2}$ without
a clearly identified optical counterpart.

Follow up observations, at higher resolution but with similar
brightness sensitivity are critical. This can not only confirm the
detections, but also put more constraints on the actual peak column
densities. A possible scenario might be that our candidate
detections are actually collections of discrete bright clumps, the
flux of which is diluted in our large beam. This is unlikely, as in
that case the clumps should have been detected individually by HIPASS,
which achieves a slightly better point source sensitivity than the
WVFS.

If these candidates and their low intrinsic column densities can be
confirmed, we can for the first time identify a whole class of objects
related to filaments of the Cosmic Web; very extended gas clouds with
extremely low neutral column densities in the intergalactic medium.\\

The original HIPASS data and the WVFS cross-correlation data will
serve as follow-up observations for the sample presented
here. Although the brightness sensitivity of both these surveys is not
as good as for the WVFS total power data, gas clumps with slightly
higher column densities can be easily identified. As mentioned
previously, the comparison with these surveys and detailed analysis
will be explained in forthcoming papers. With all three survey
coverages in hand, the data can be interpreted more effectively. We
hope to confirm several of the intergalactic {\HI} detections and put
more light on the intergalactic reservoir of gas in the vicinity of
galaxies. By looking at the kinematics and line widths of the
detections, we hope to learn more about galaxy and AGN feedback and
whether galaxies are fueled preferentially through hot-mode or
cold-mode accretion processes.\\

\begin{acknowledgements}
 The Westerbork Synthesis Radio Telescope is operated by the ASTRON
 (Netherlands Foundation for Research in Astronomy) with support from
 the Netherlands Foundation for Scientiﬁc Research (NWO)

\end{acknowledgements}

\bibliographystyle{aa}
\bibliography{names,thesisbibliography}

\newpage

\onecolumn
\begin{appendix}
\section{Spectra of confirmed {\HI} detections in the WVFS total power data.}

\begin{figure*}[!h]
  \begin{center}

 \includegraphics[width=0.22\textwidth]{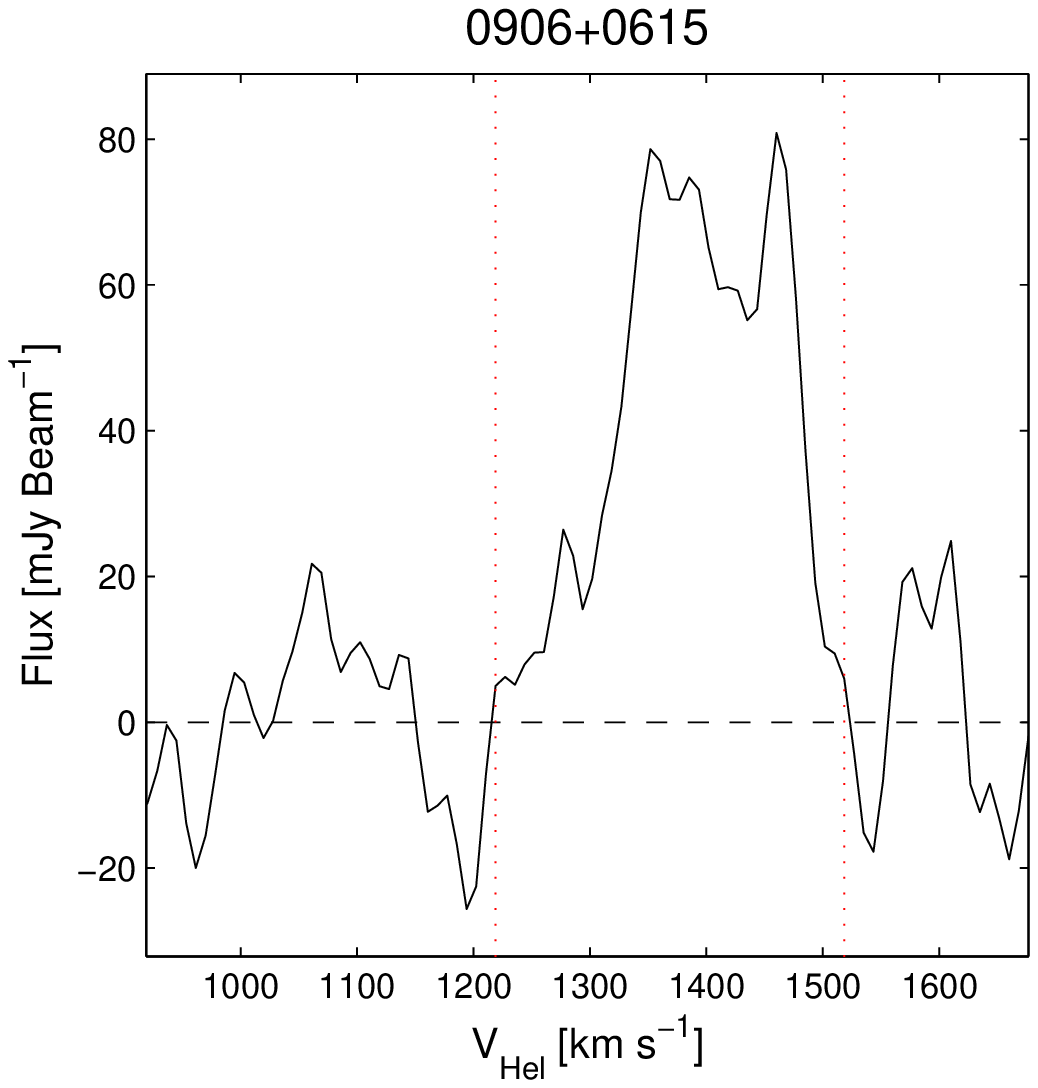}
 \includegraphics[width=0.22\textwidth]{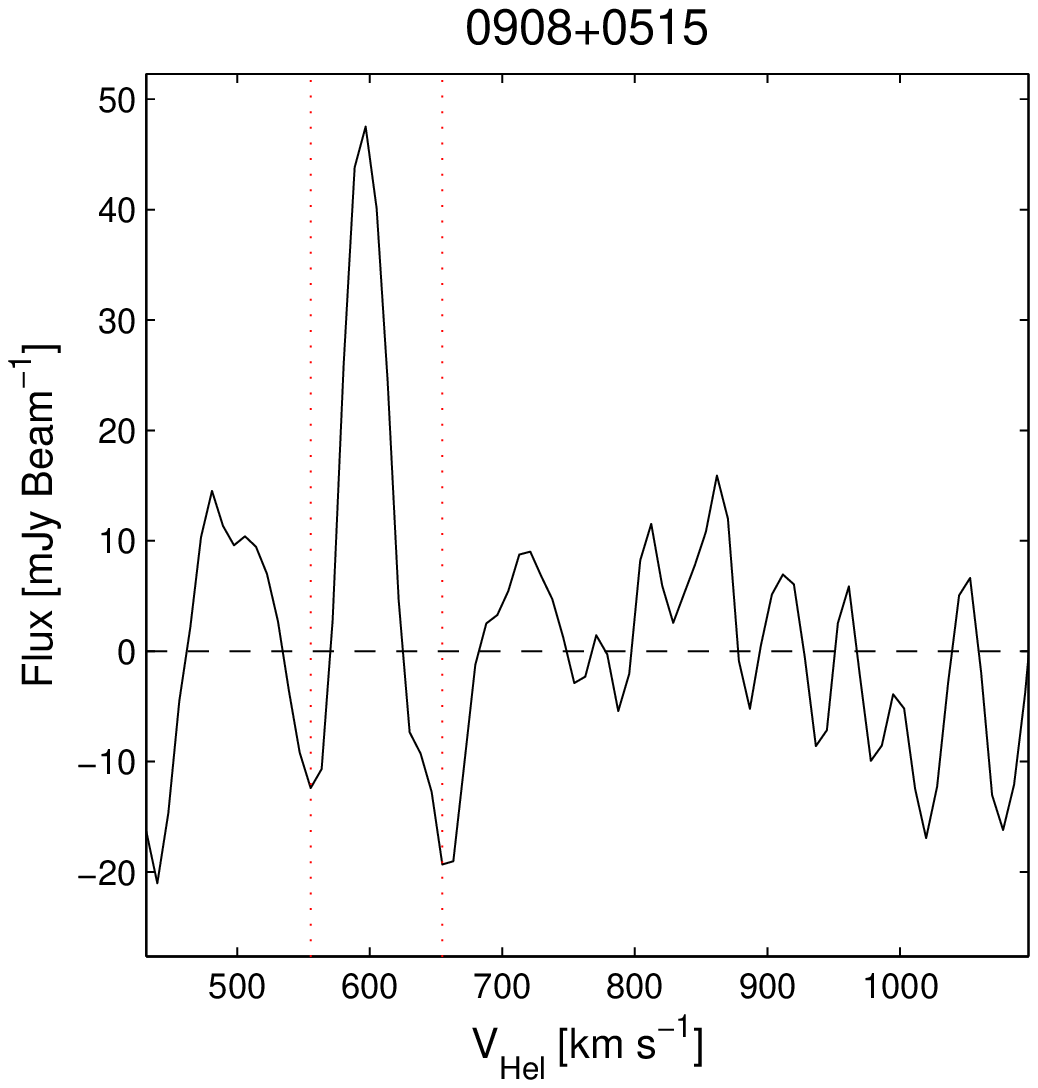}
 \includegraphics[width=0.22\textwidth]{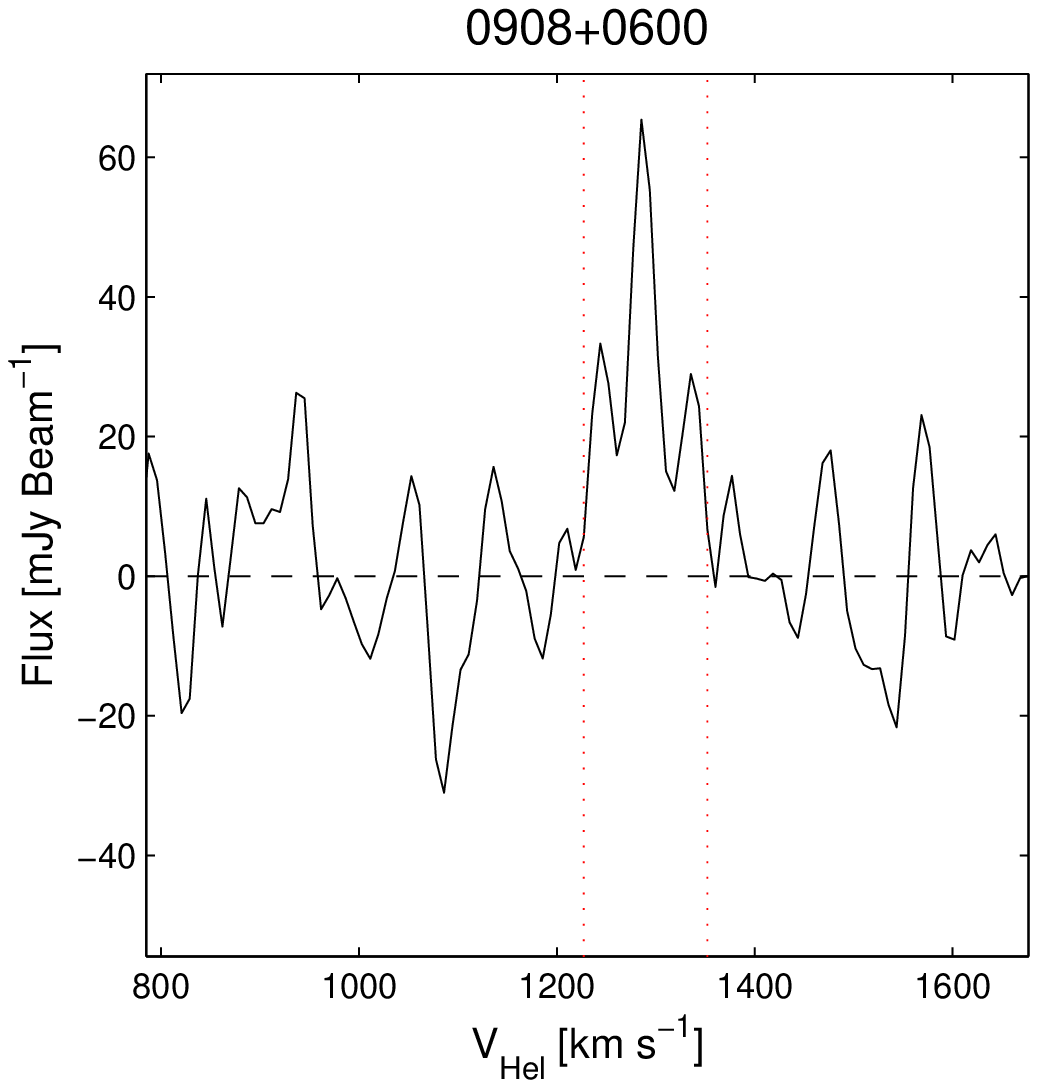}
 \includegraphics[width=0.22\textwidth]{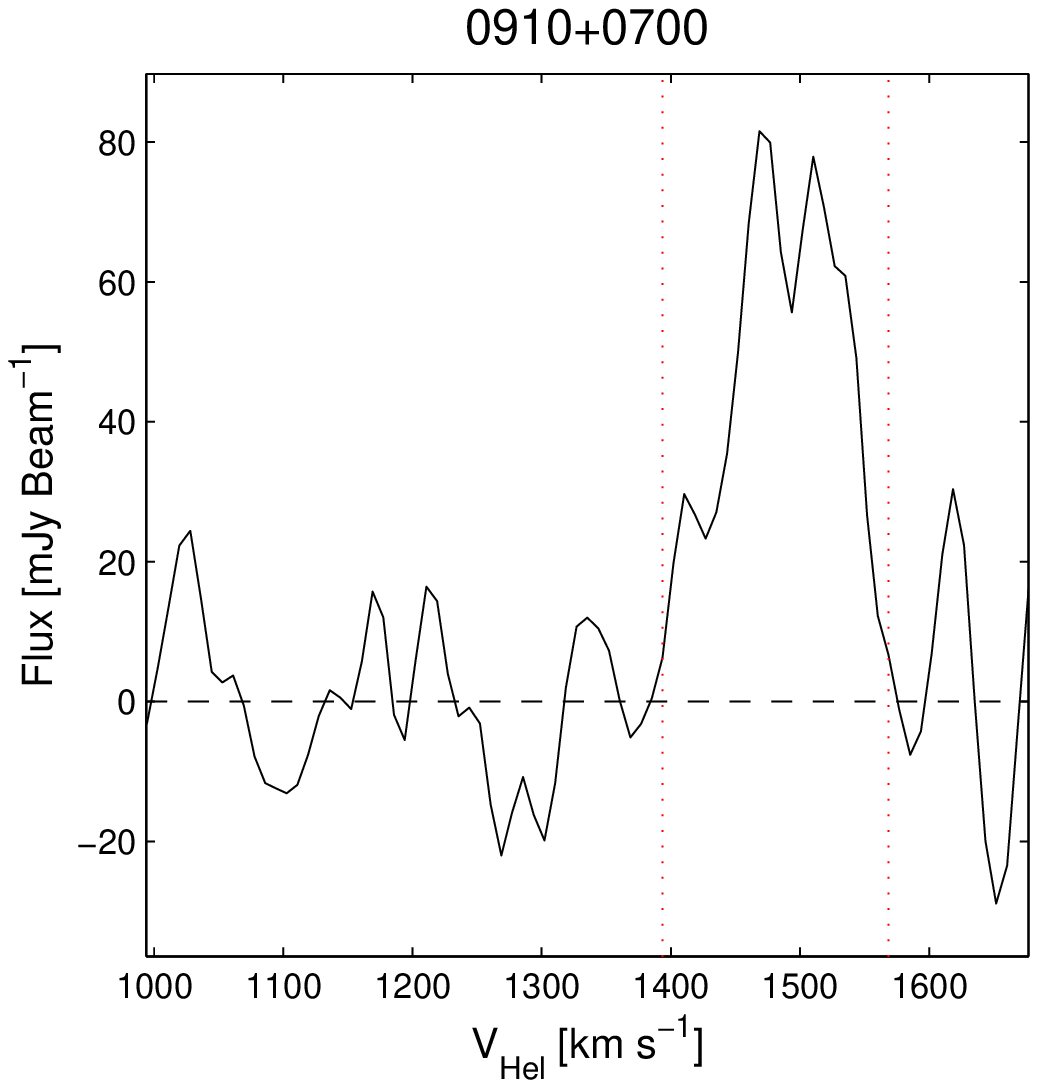}
 \includegraphics[width=0.22\textwidth]{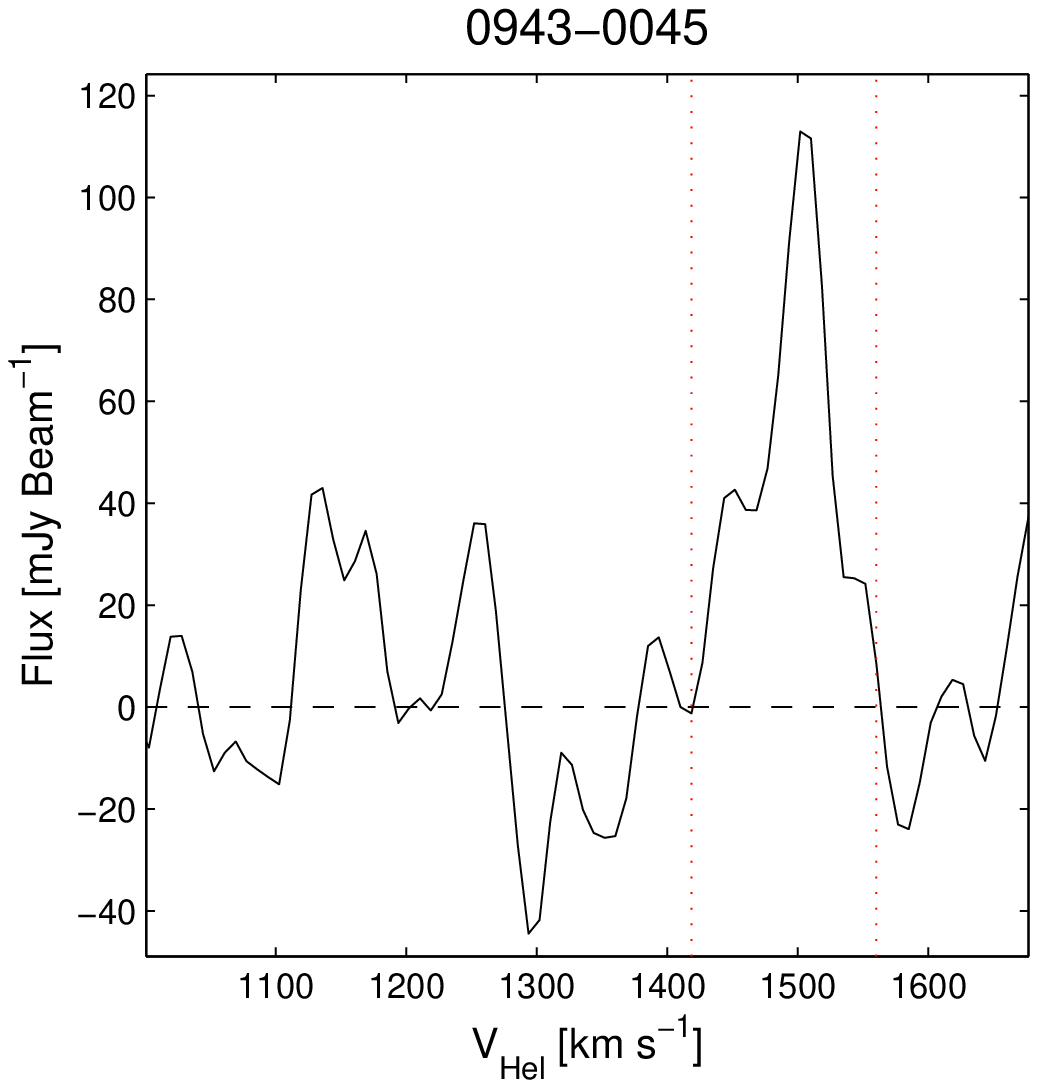}
 \includegraphics[width=0.22\textwidth]{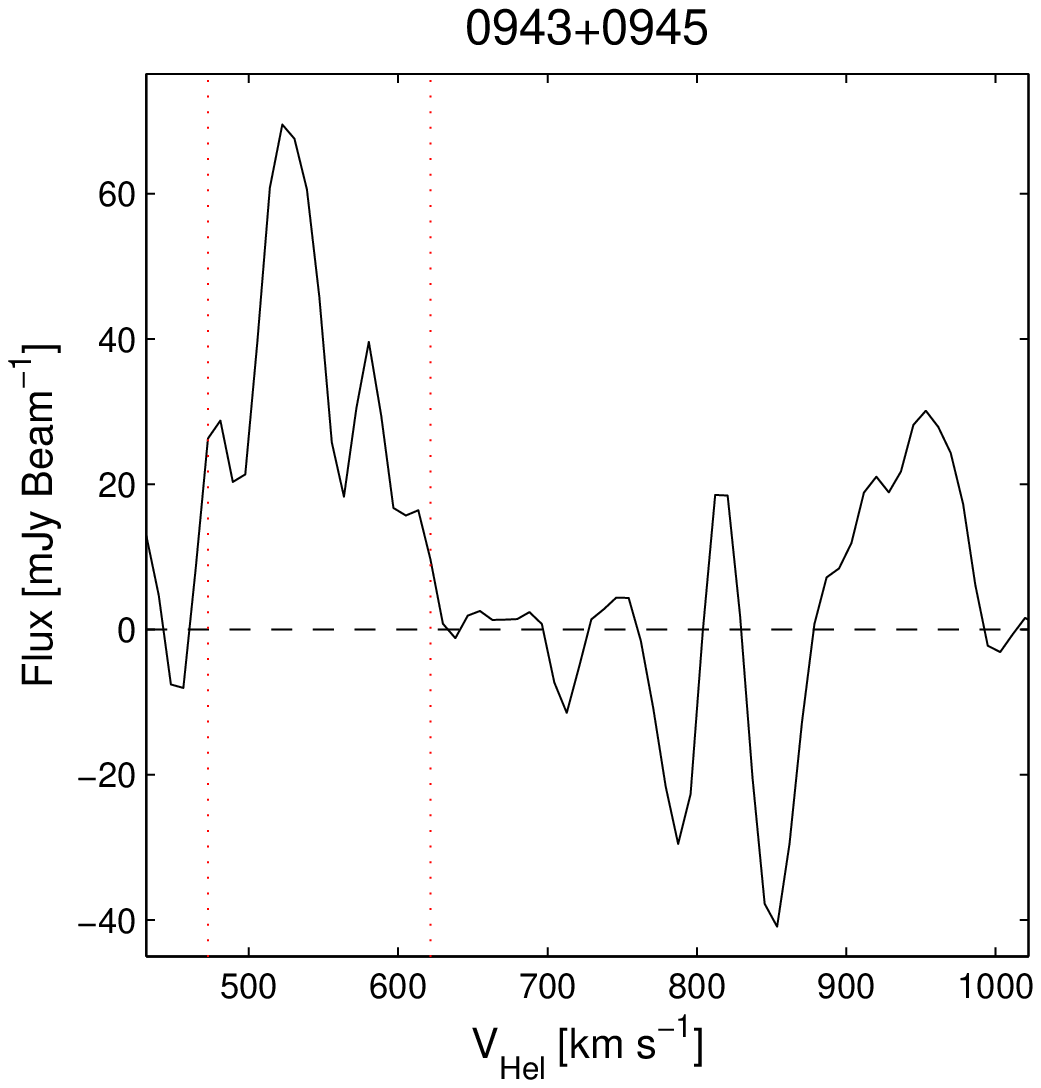}
 \includegraphics[width=0.22\textwidth]{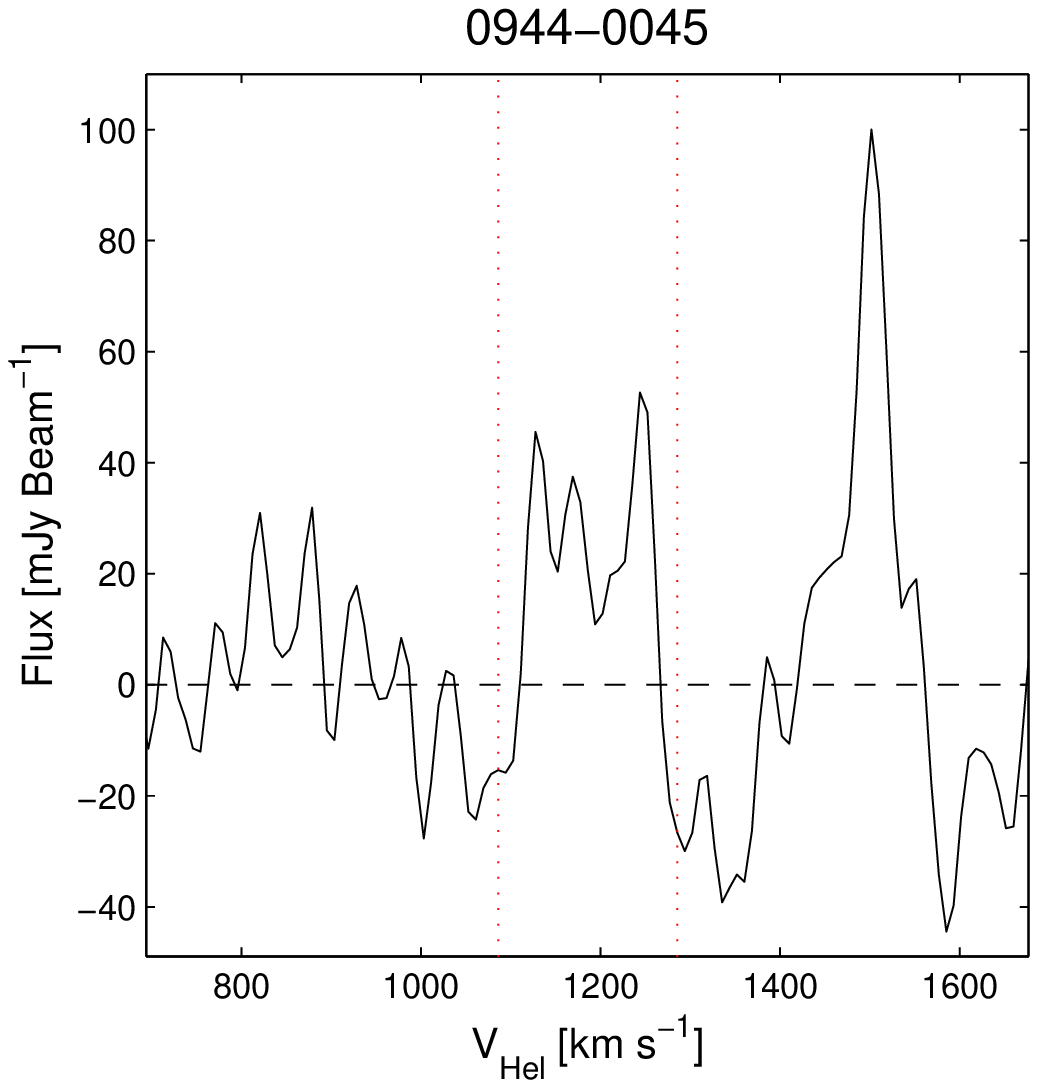}
 \includegraphics[width=0.22\textwidth]{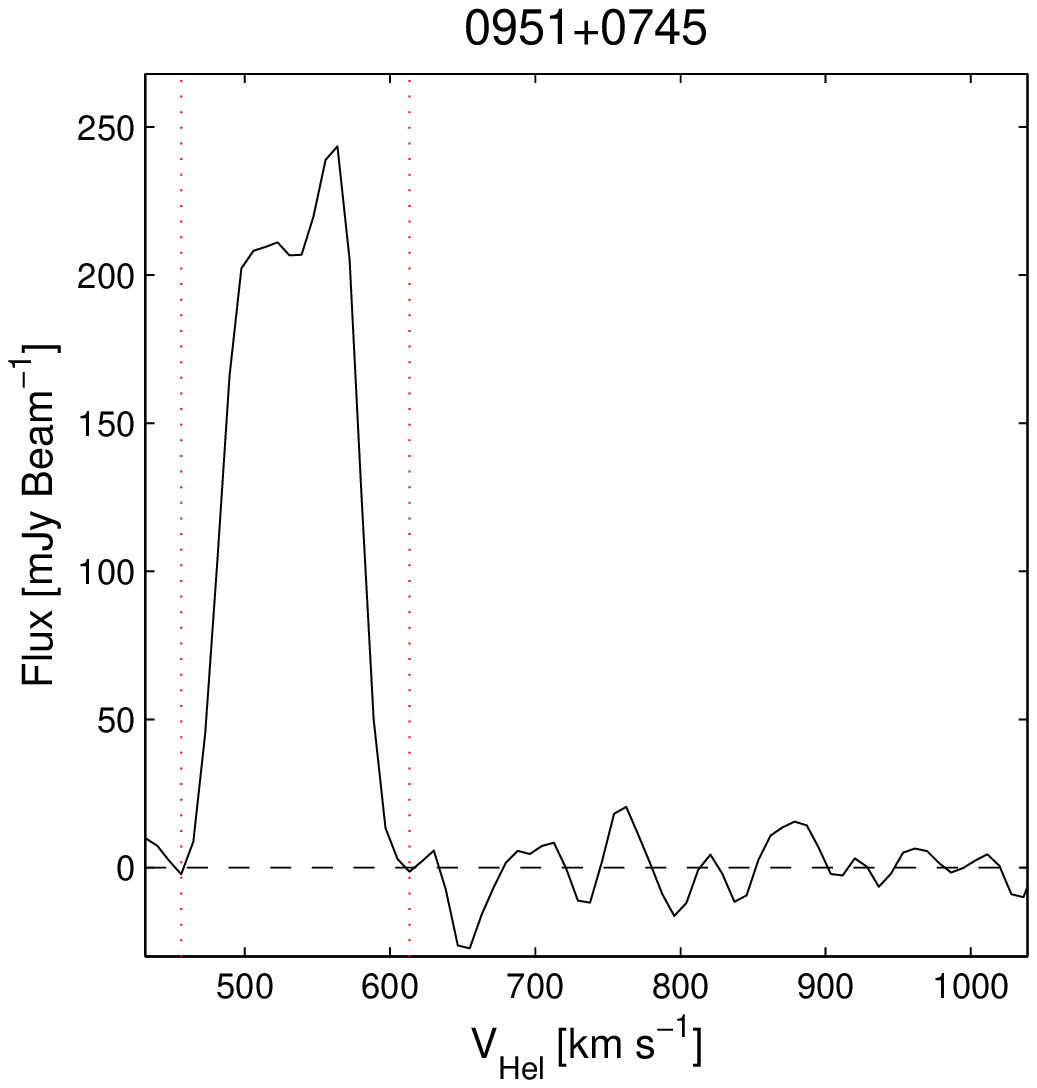}
 \includegraphics[width=0.22\textwidth]{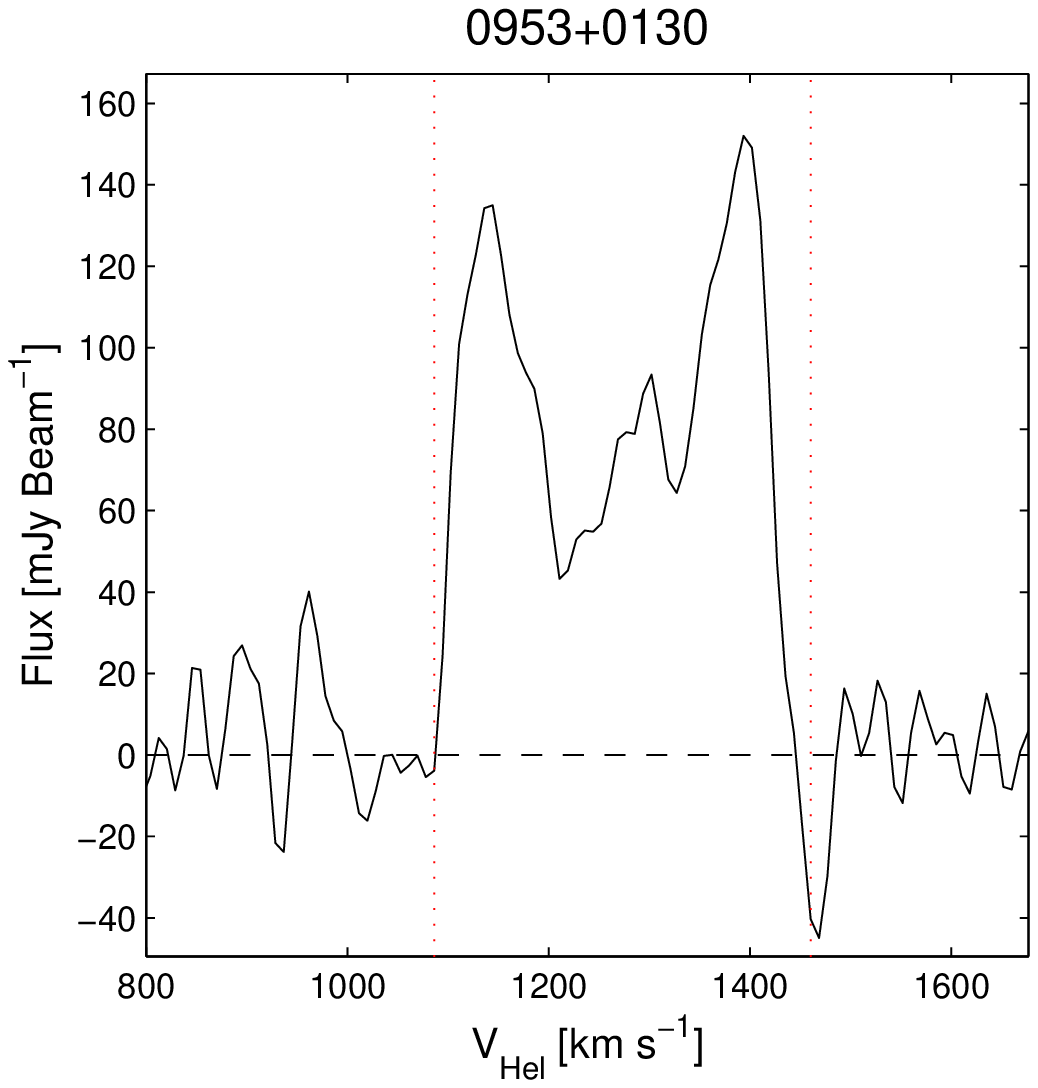}
 \includegraphics[width=0.22\textwidth]{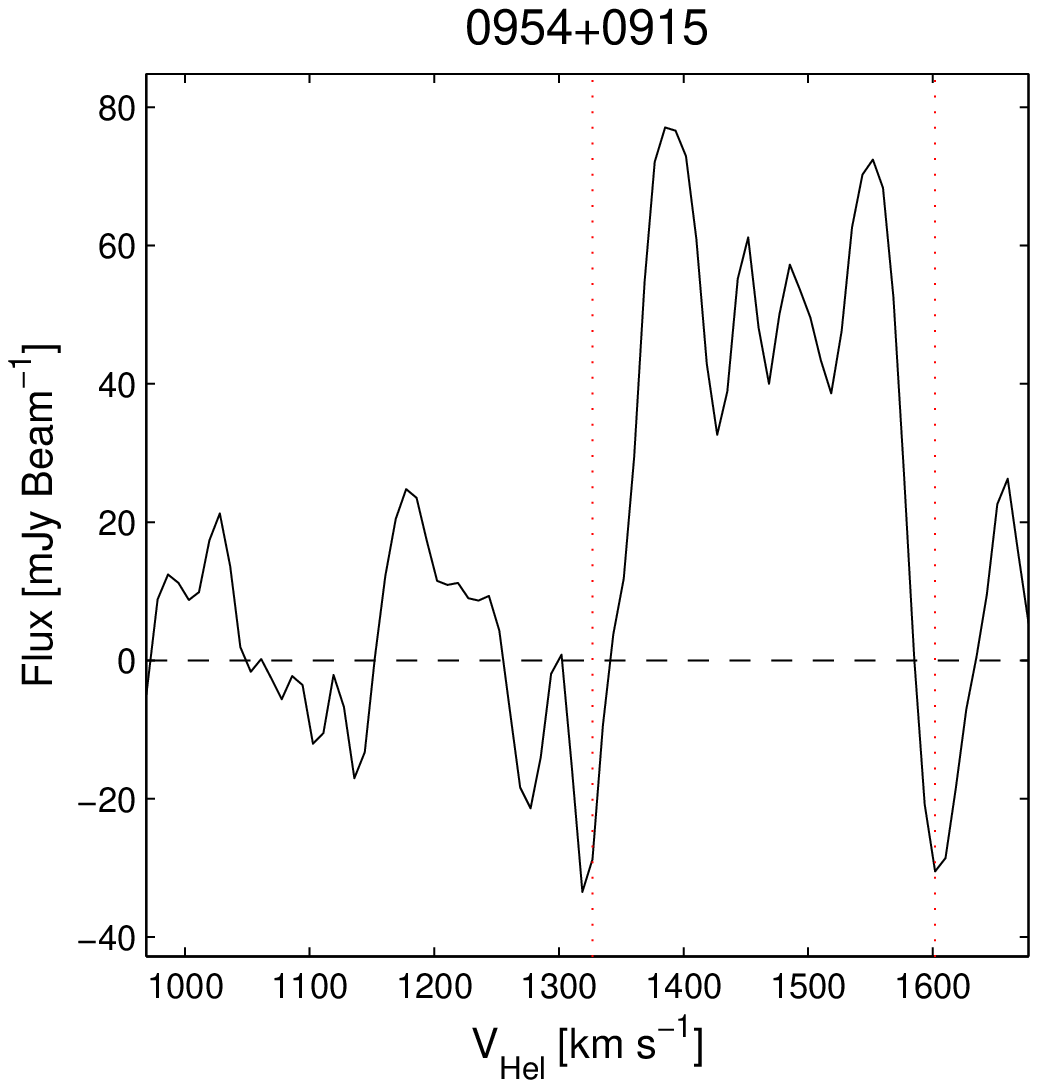}
 \includegraphics[width=0.22\textwidth]{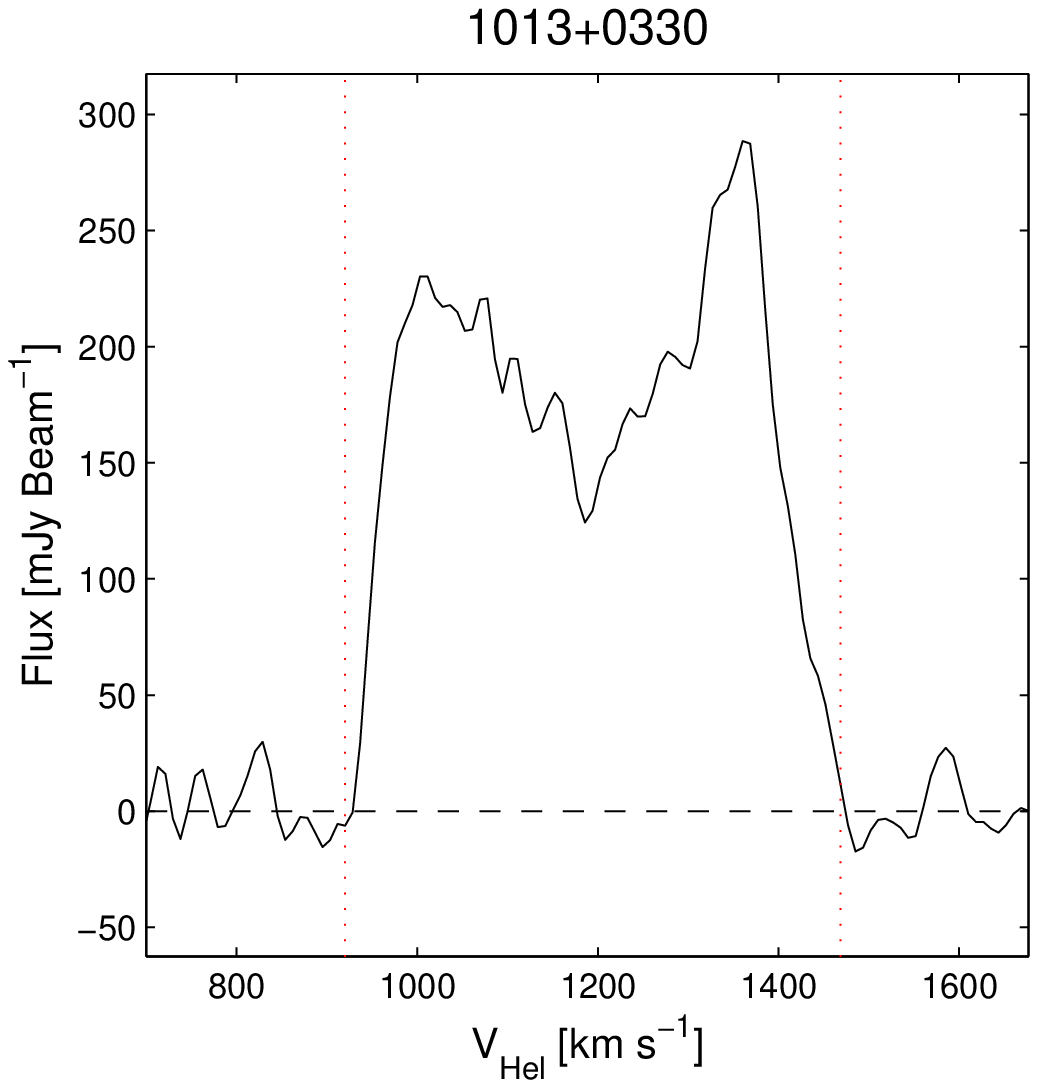}
 \includegraphics[width=0.22\textwidth]{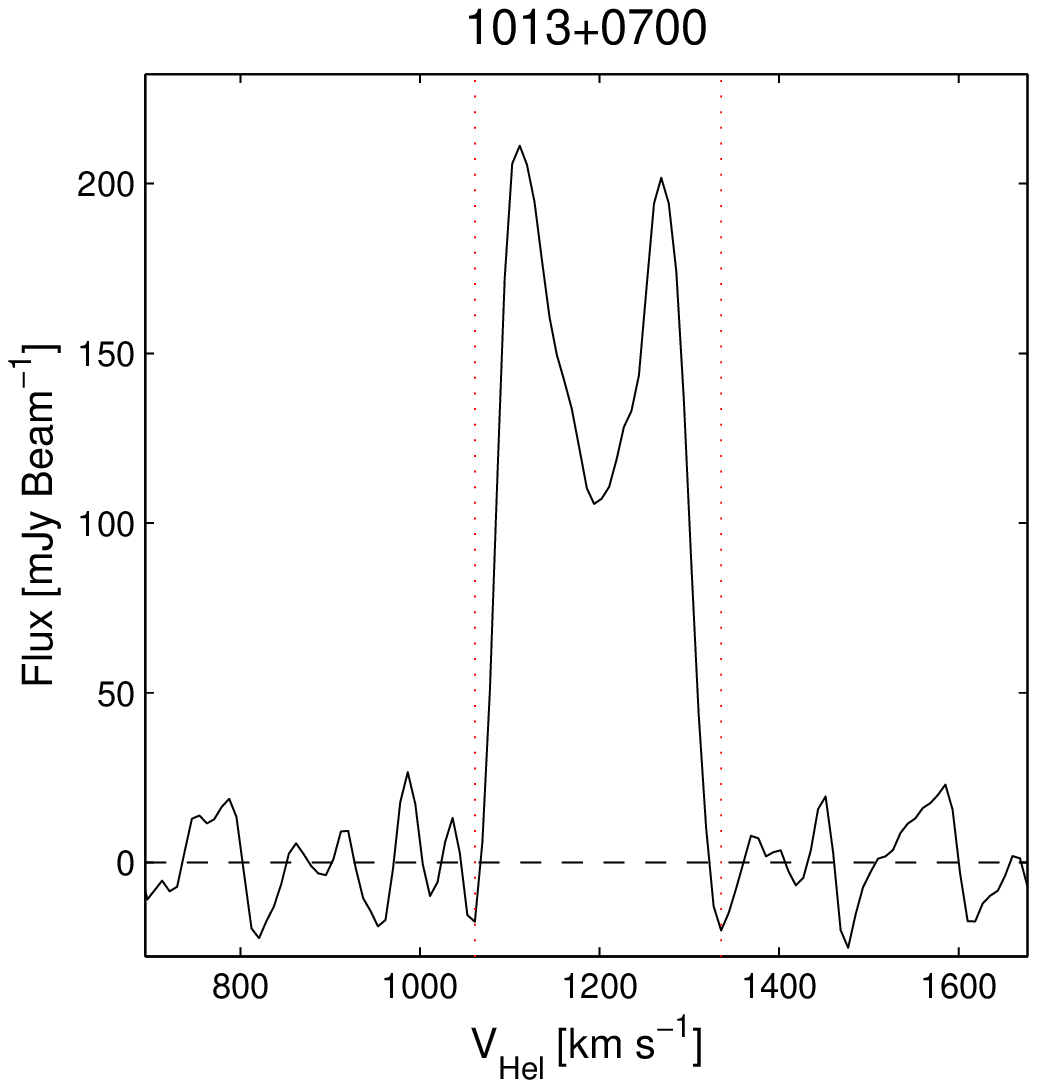}
 \includegraphics[width=0.22\textwidth]{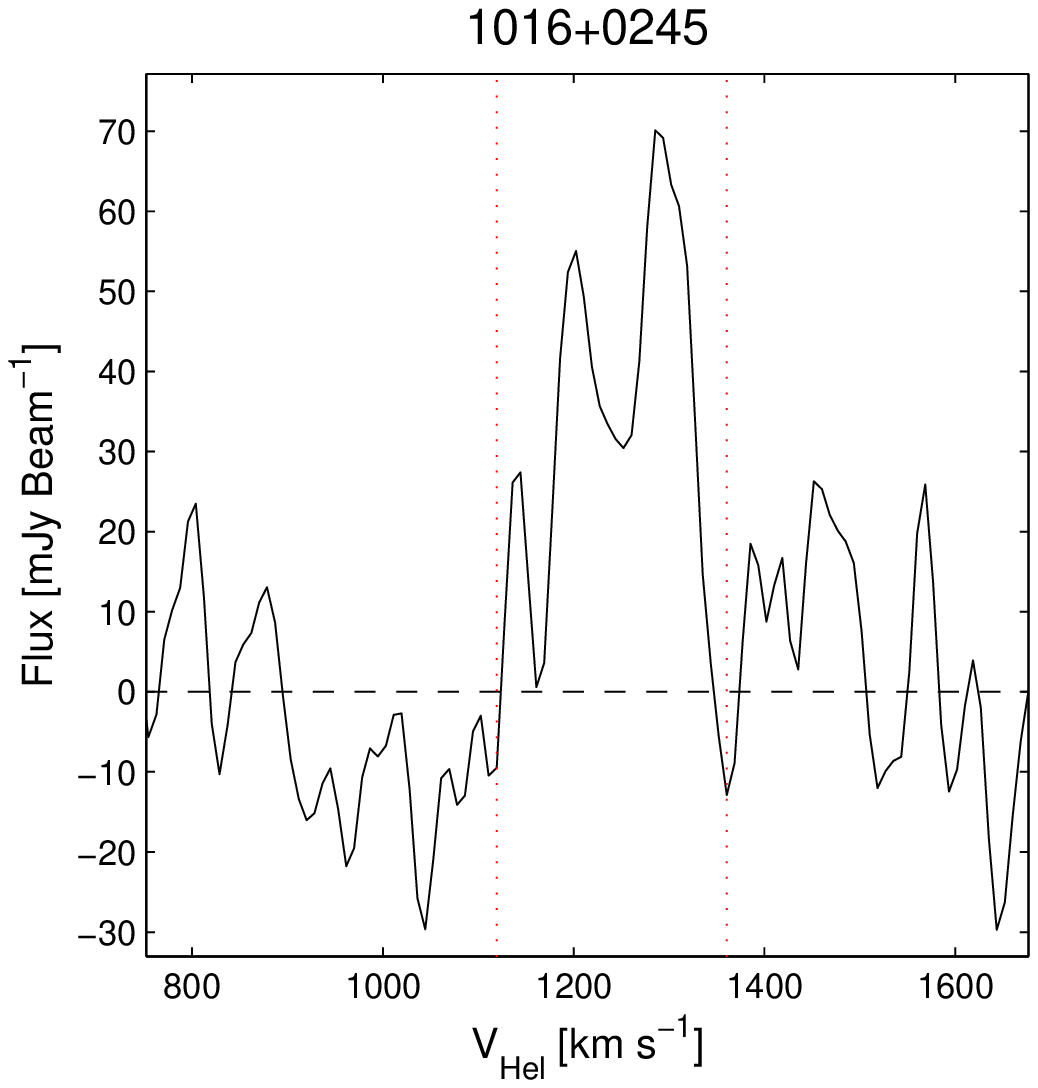}
 \includegraphics[width=0.22\textwidth]{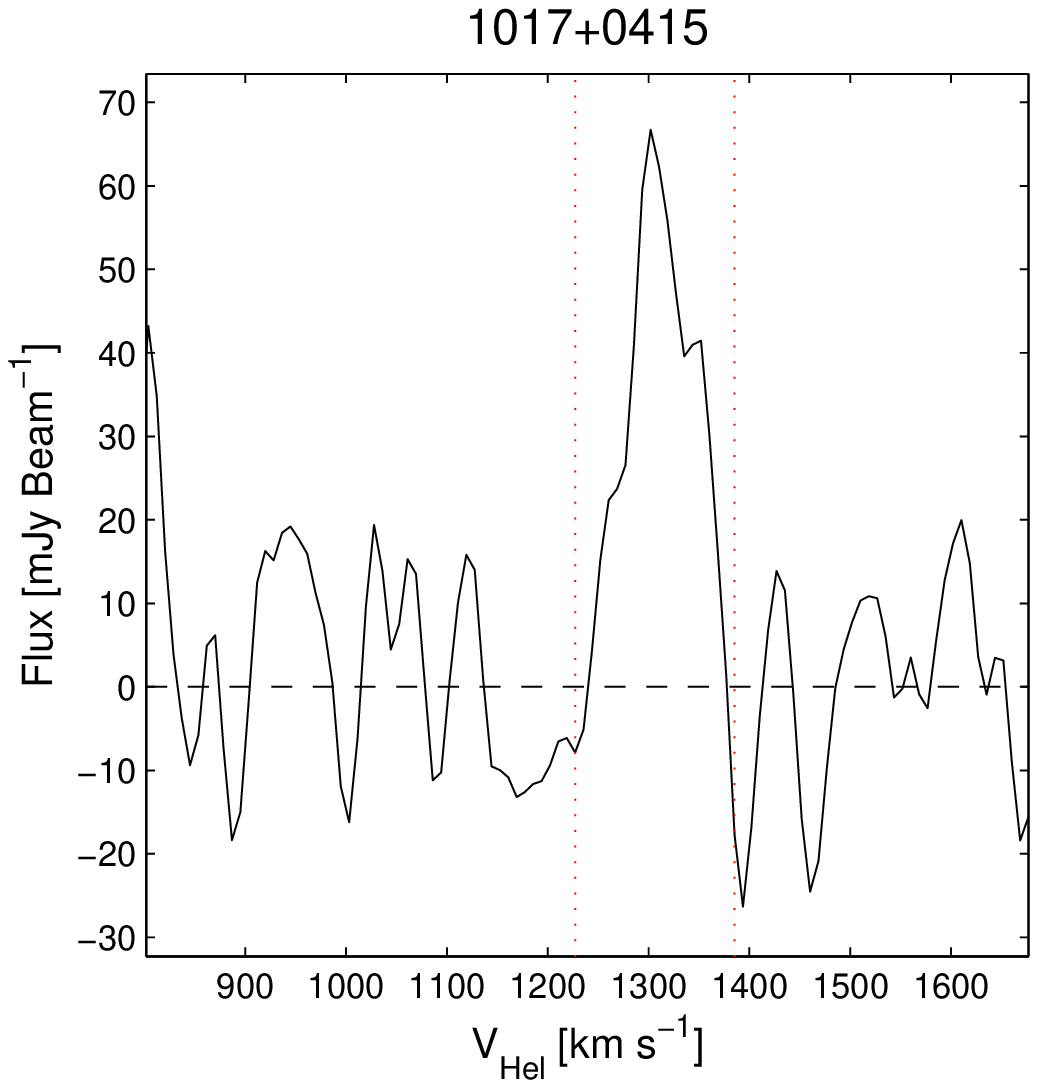}
 \includegraphics[width=0.22\textwidth]{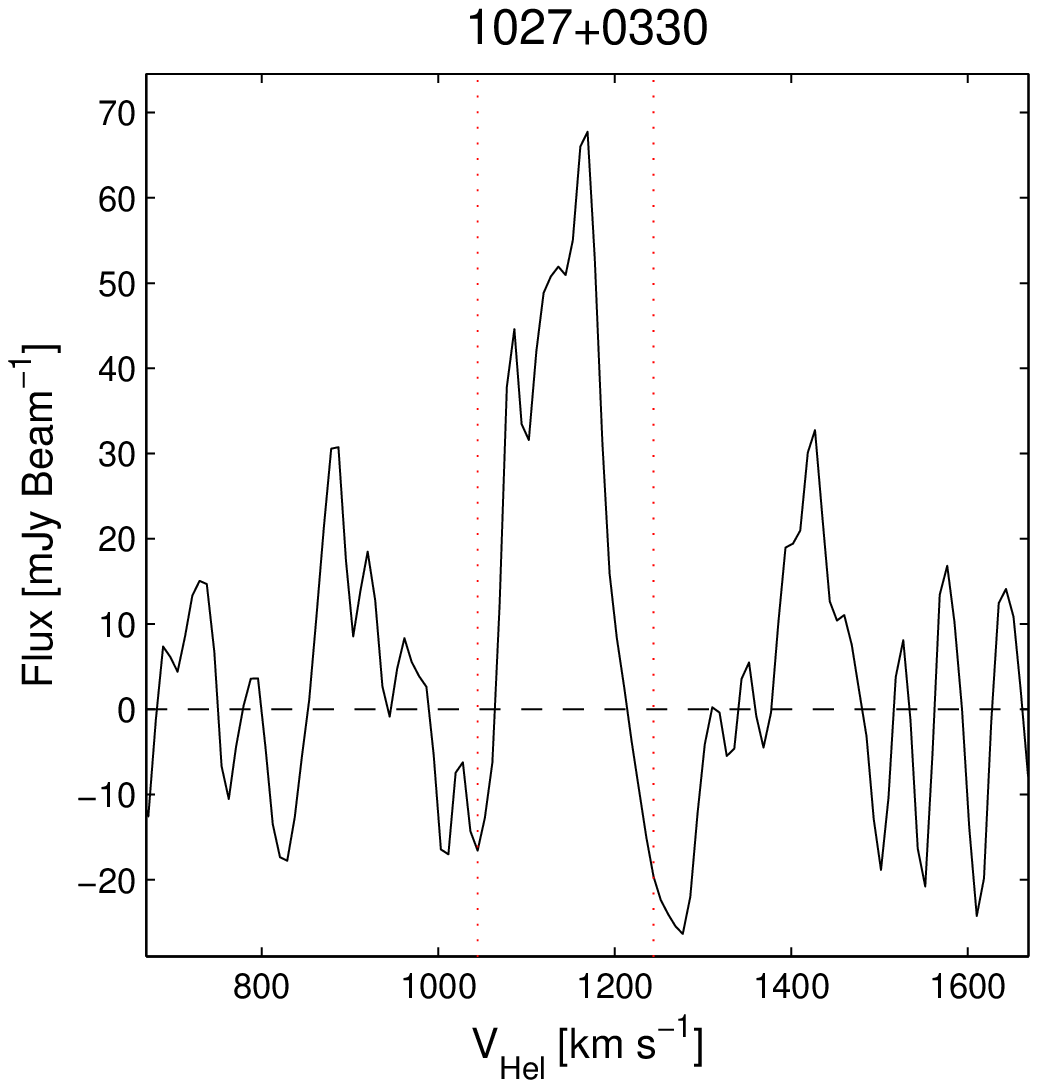}
 \includegraphics[width=0.22\textwidth]{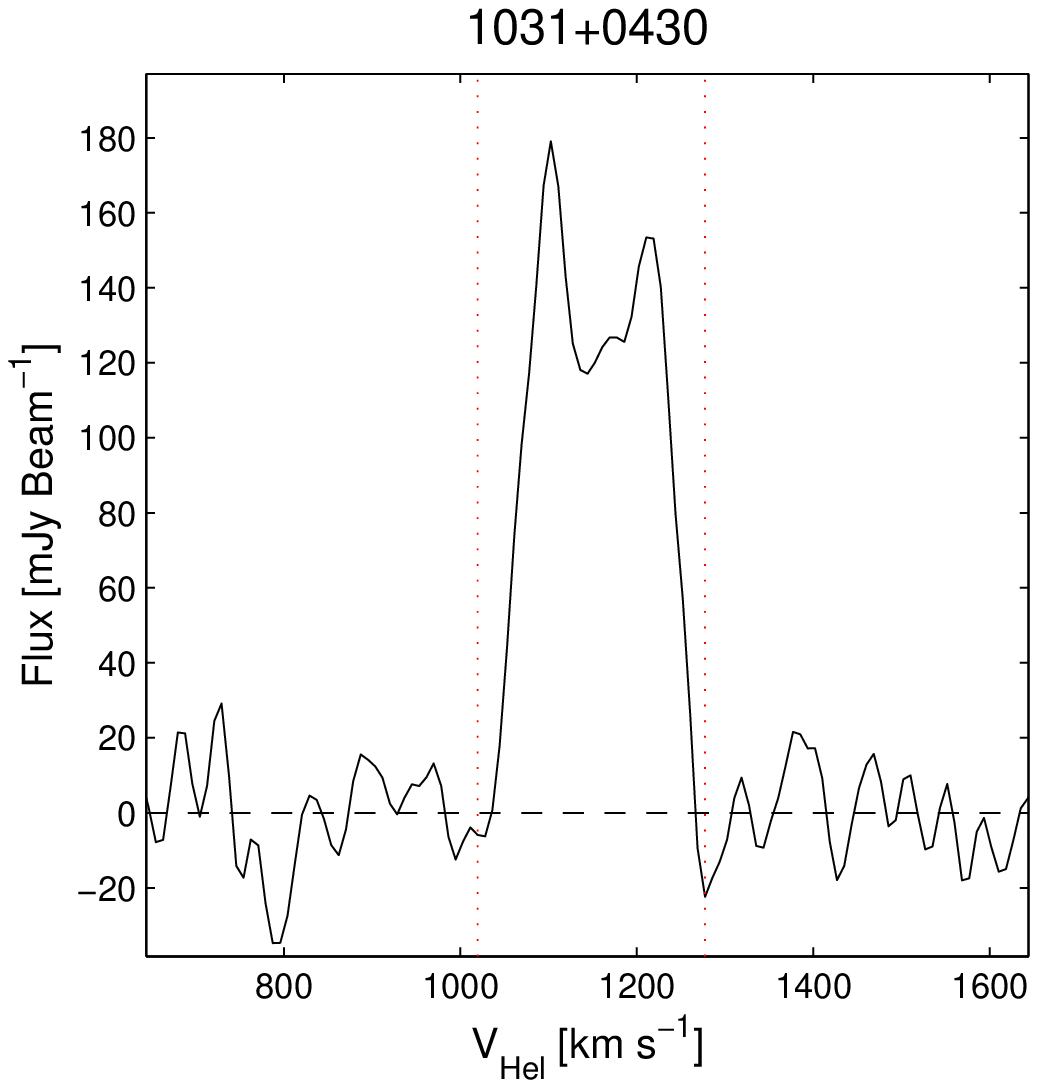}
 \includegraphics[width=0.22\textwidth]{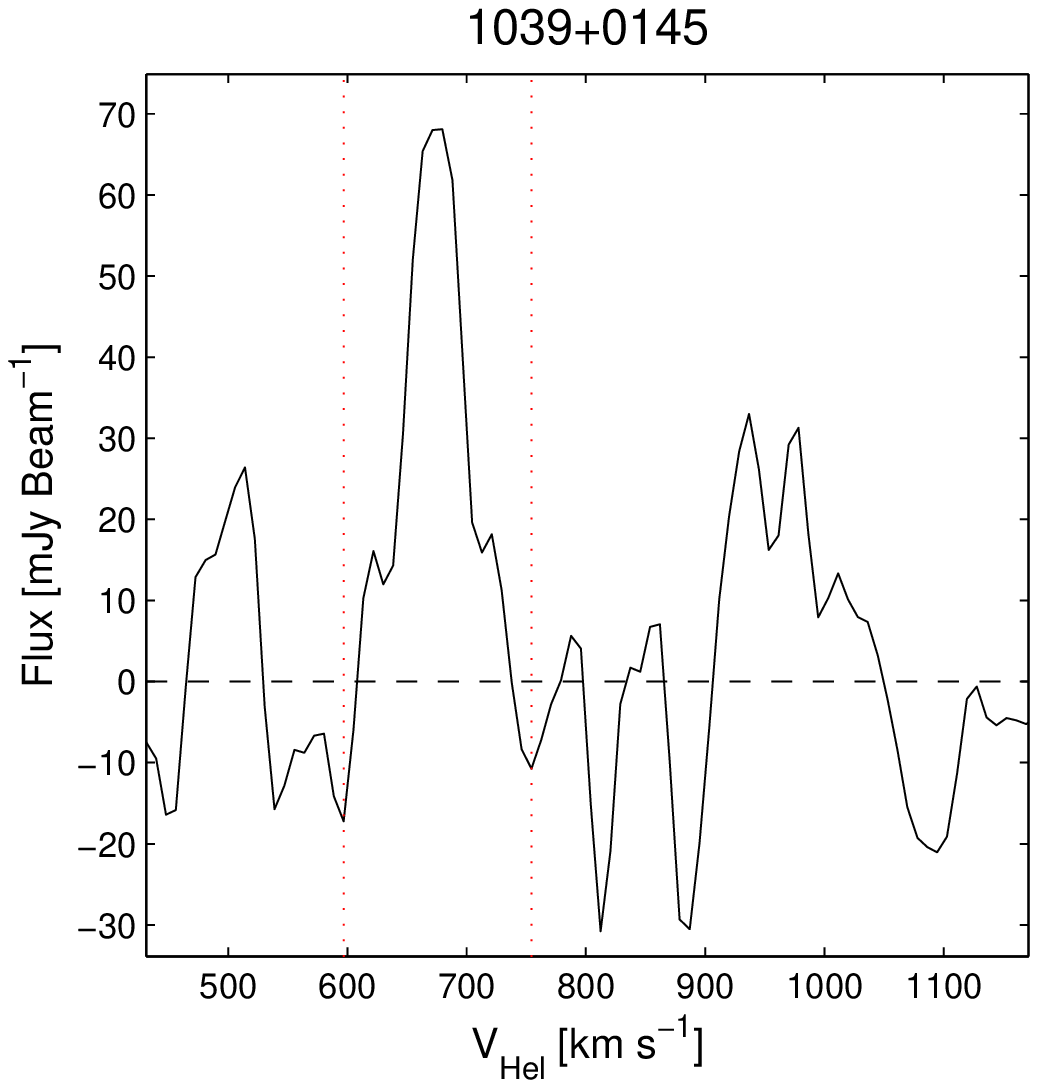}
 \includegraphics[width=0.22\textwidth]{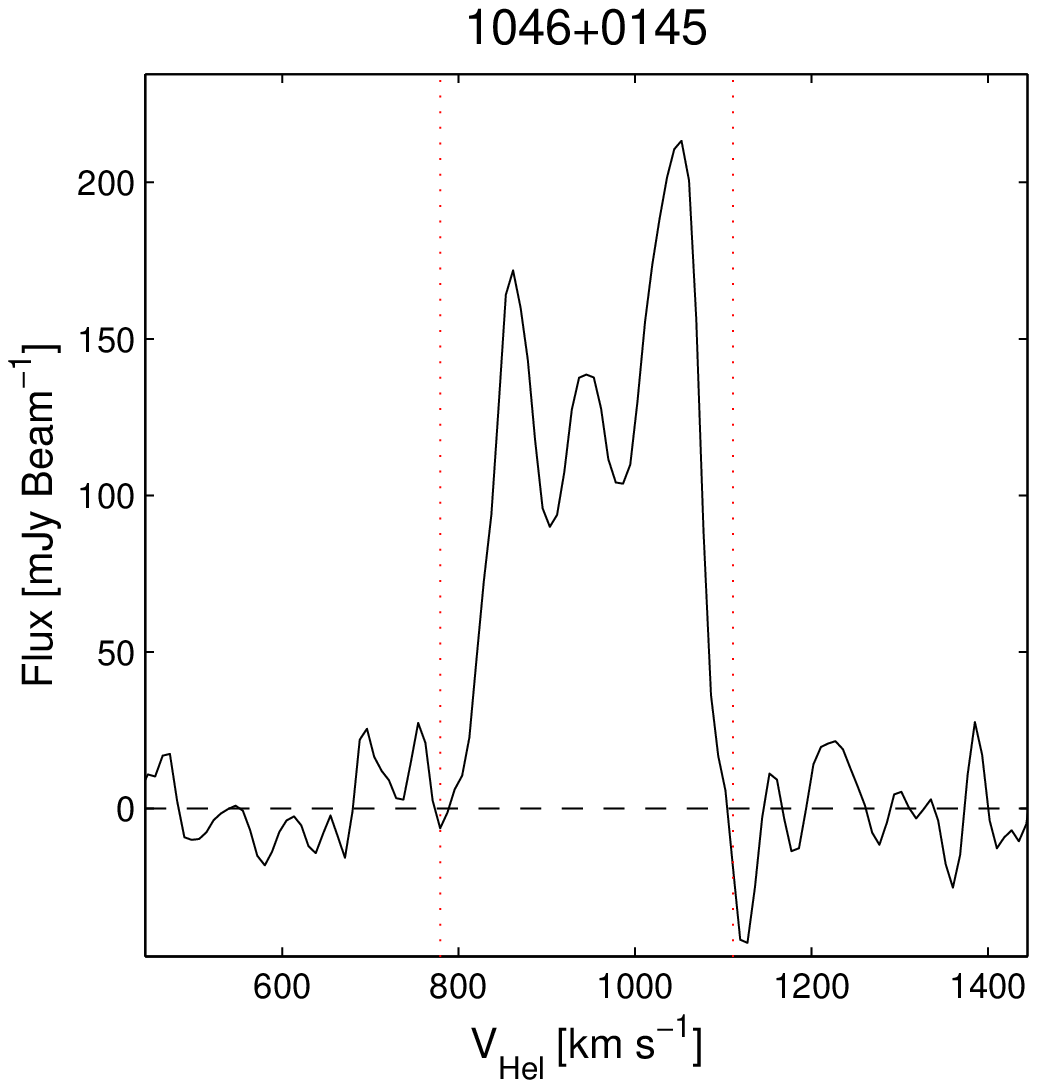}
 \includegraphics[width=0.22\textwidth]{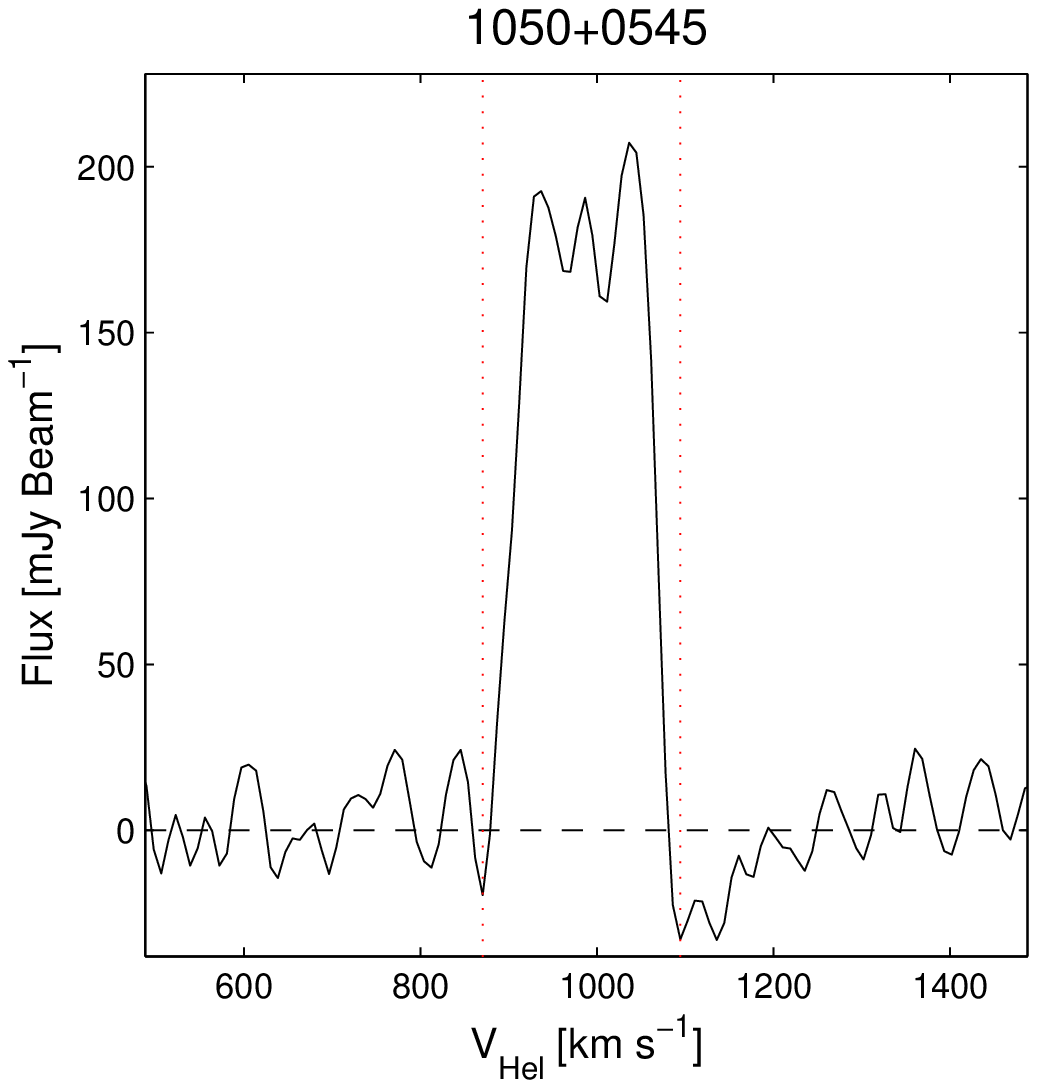}
 \includegraphics[width=0.22\textwidth]{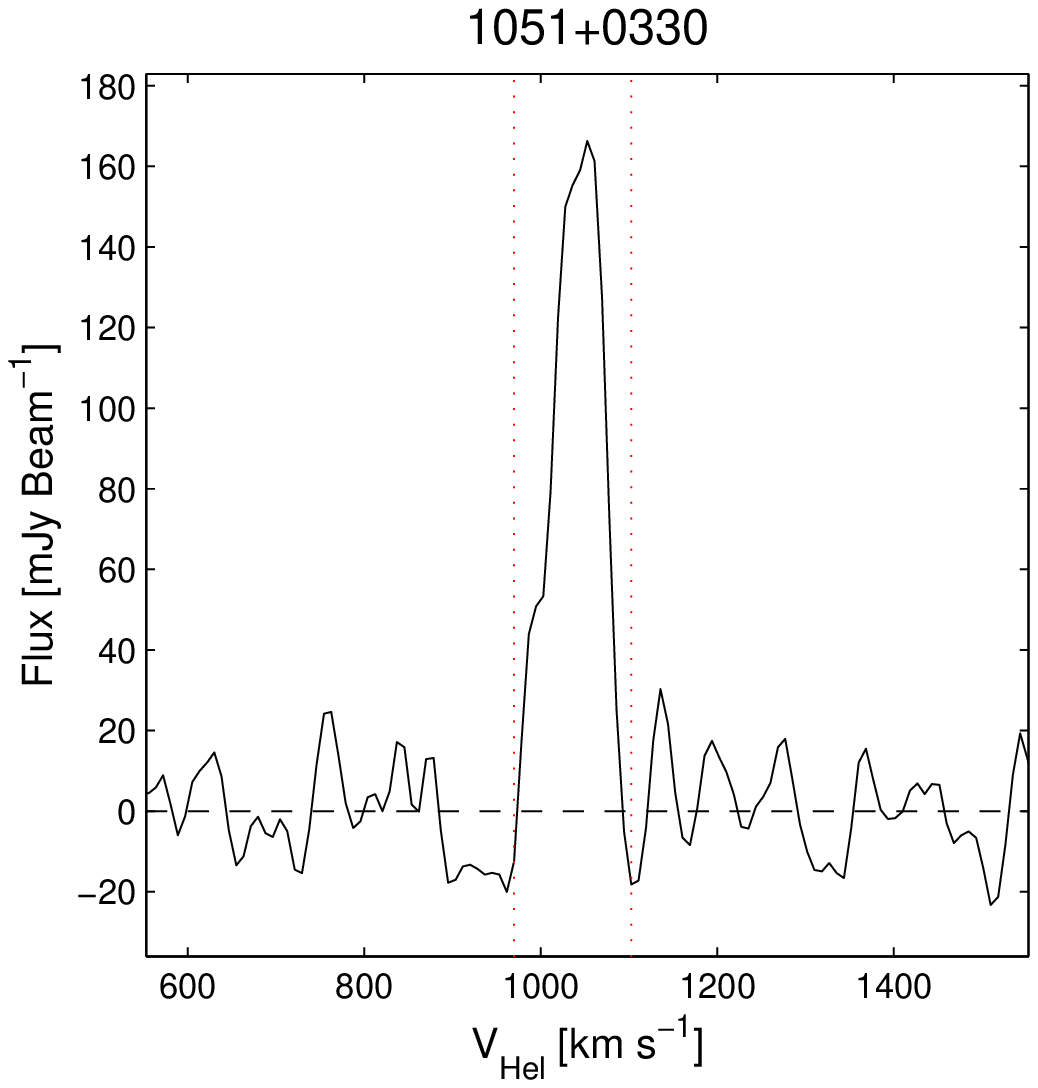}

 \end{center}                                            
  \caption{Spectra of all detections of neutral hydrogen in the WVFS
    total power data. The velocity width of each object is indicated
    by the two vertical dotted lines.}
  \label{all_spectra}                                    
\end{figure*}

\begin{figure*}
  \begin{center}

 \includegraphics[width=0.22\textwidth]{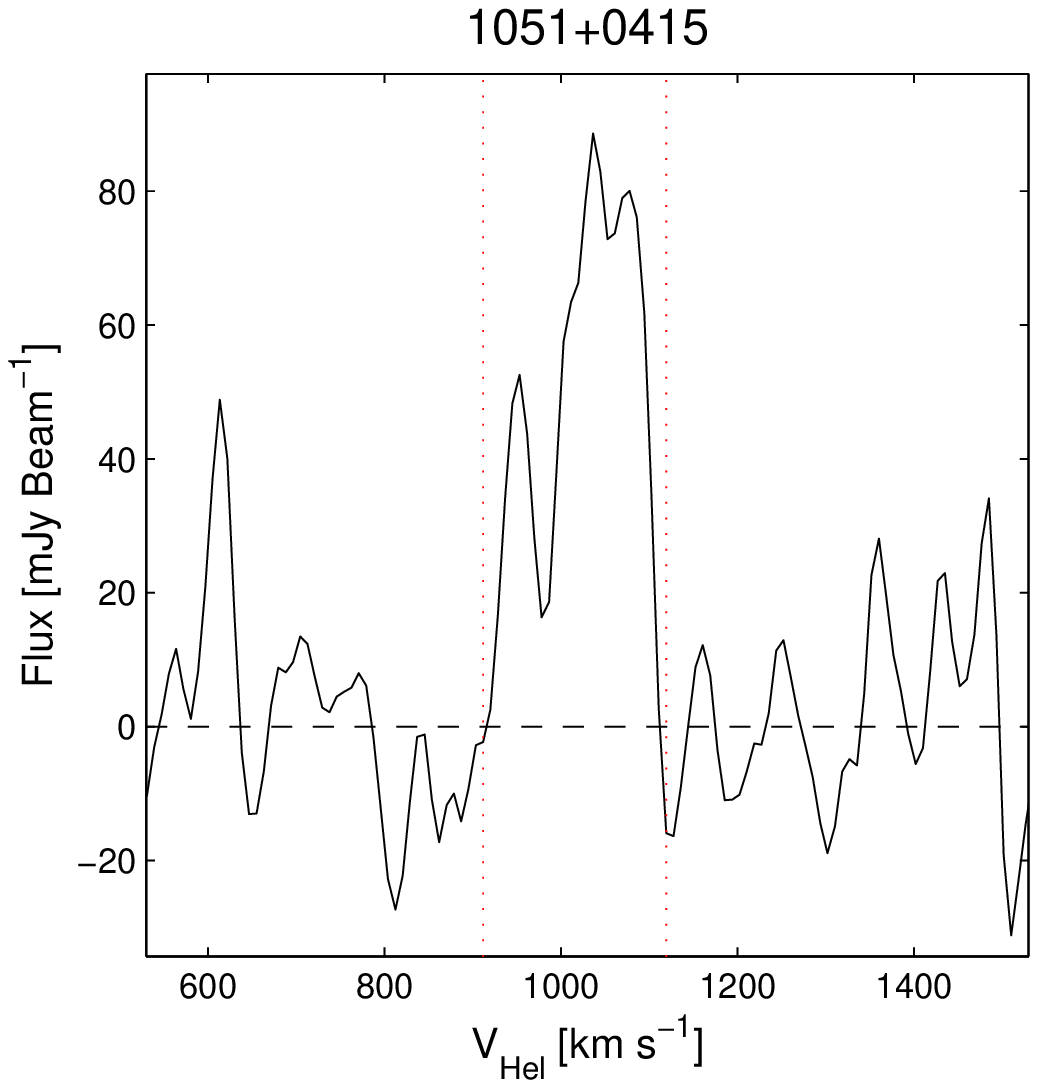}
 \includegraphics[width=0.22\textwidth]{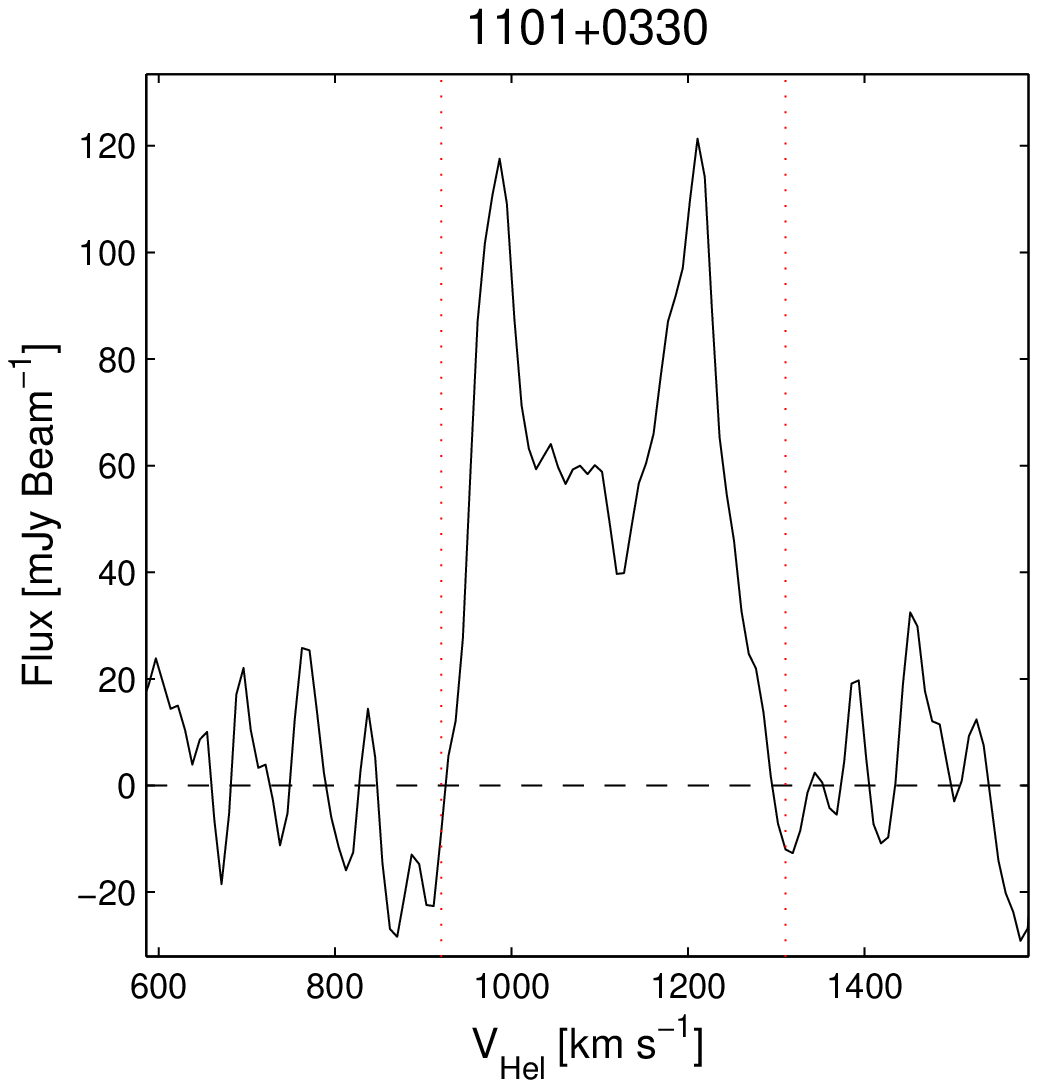}
 \includegraphics[width=0.22\textwidth]{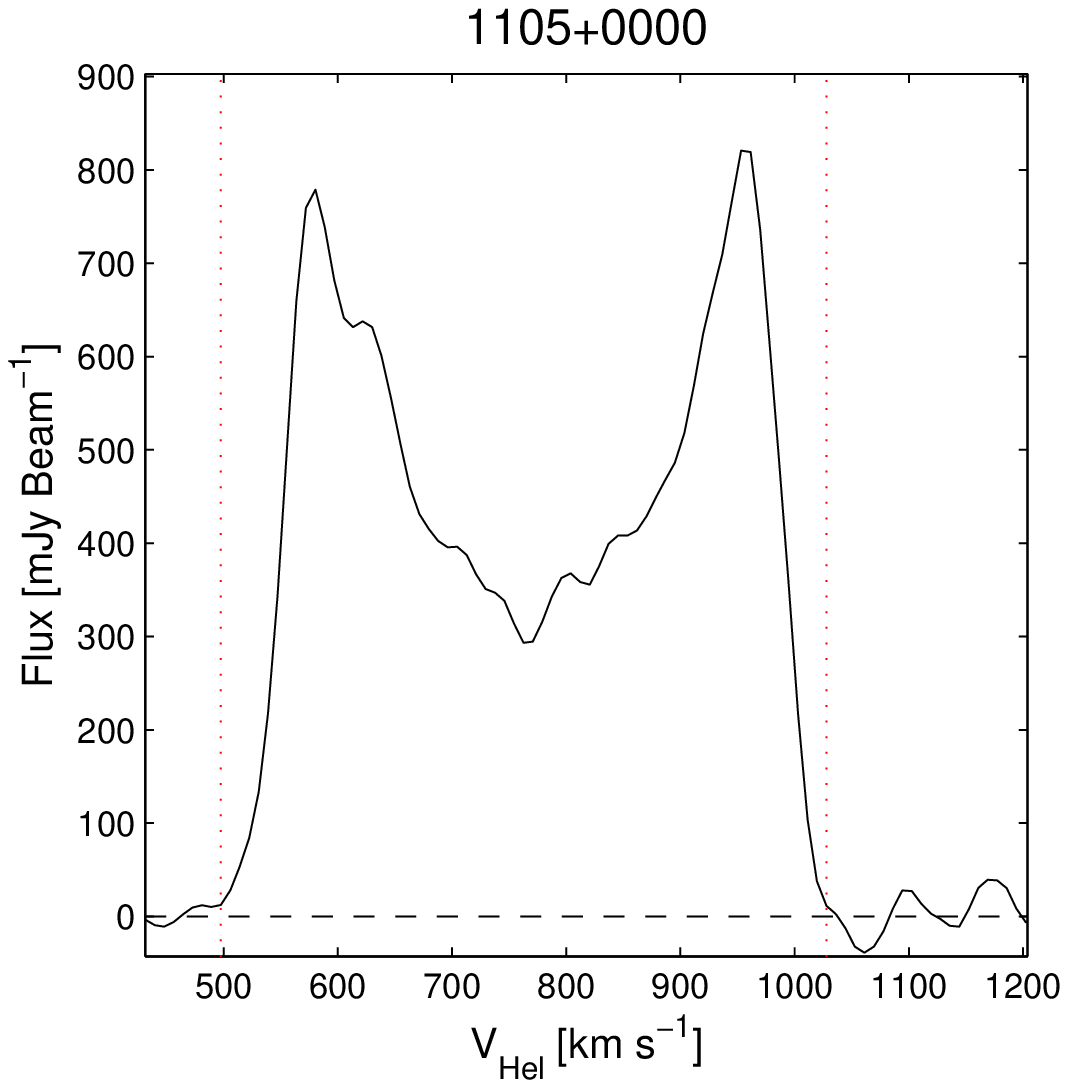}
 \includegraphics[width=0.22\textwidth]{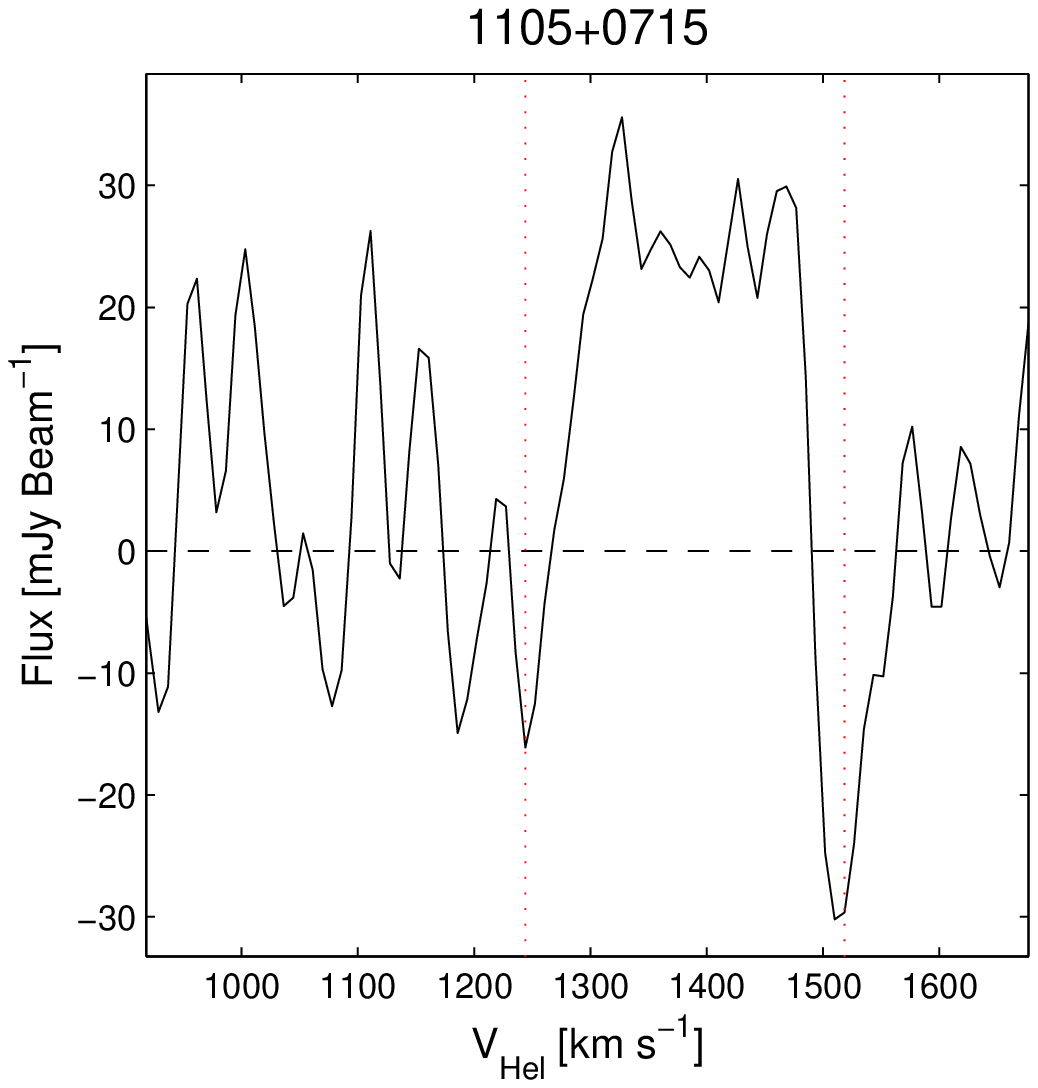}
 \includegraphics[width=0.22\textwidth]{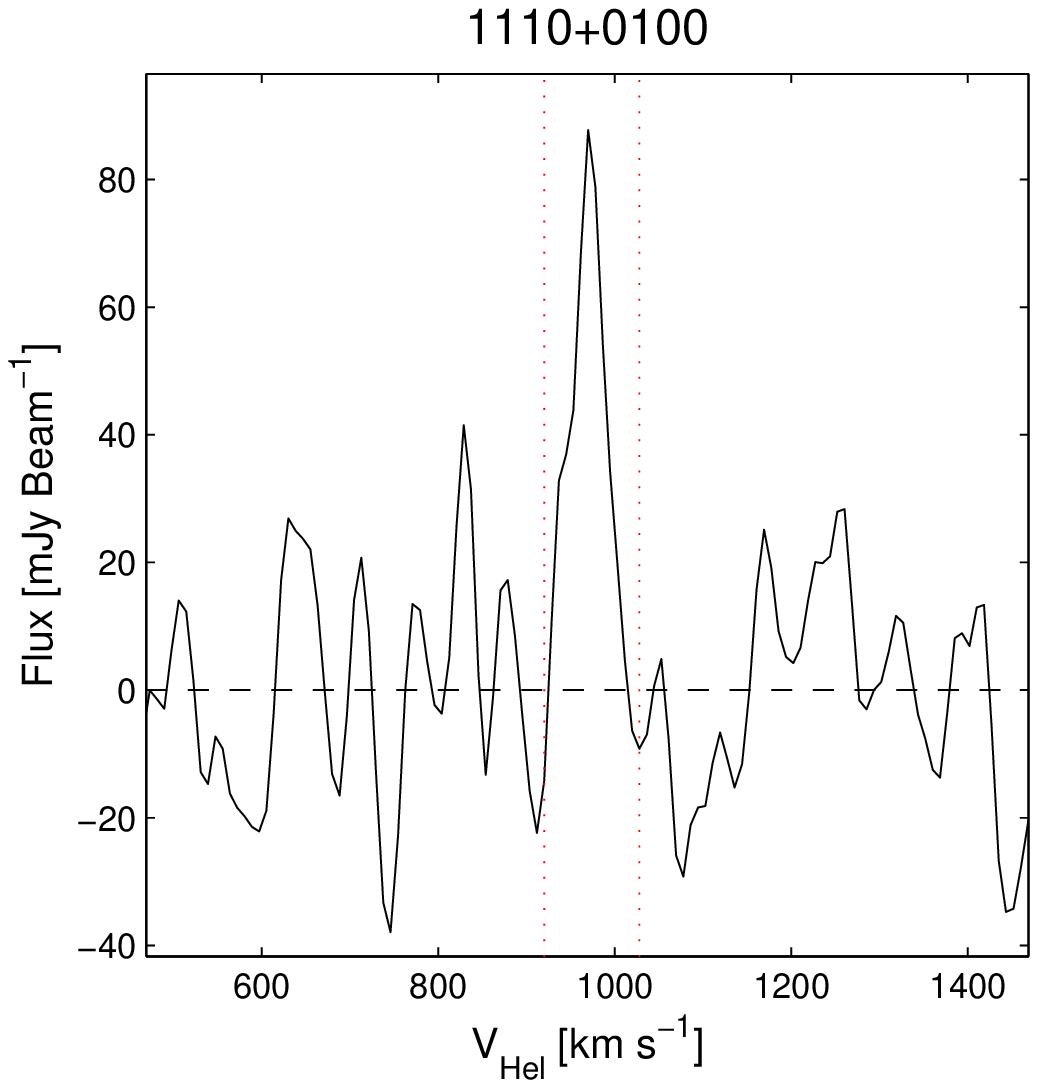}
 \includegraphics[width=0.22\textwidth]{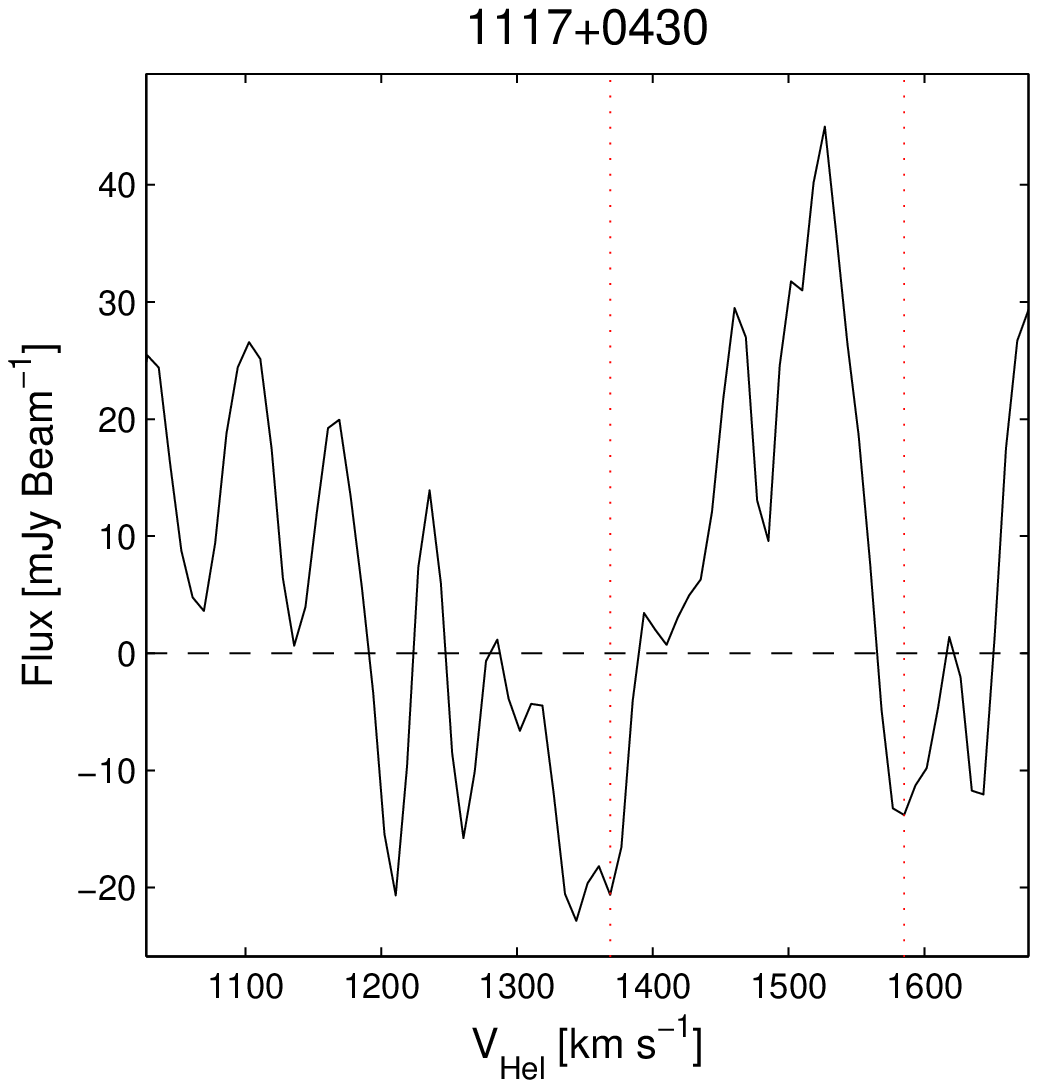}
 \includegraphics[width=0.22\textwidth]{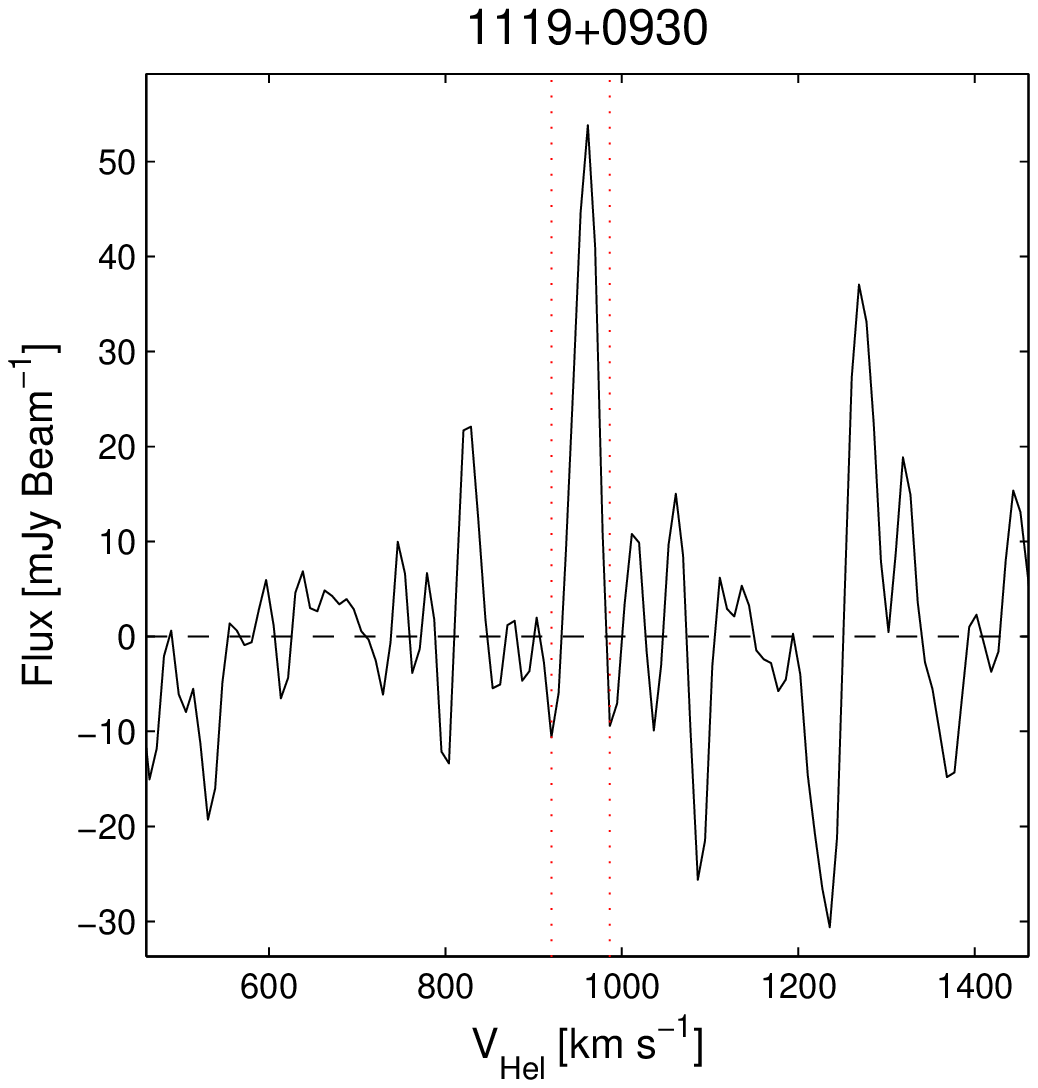}
 \includegraphics[width=0.22\textwidth]{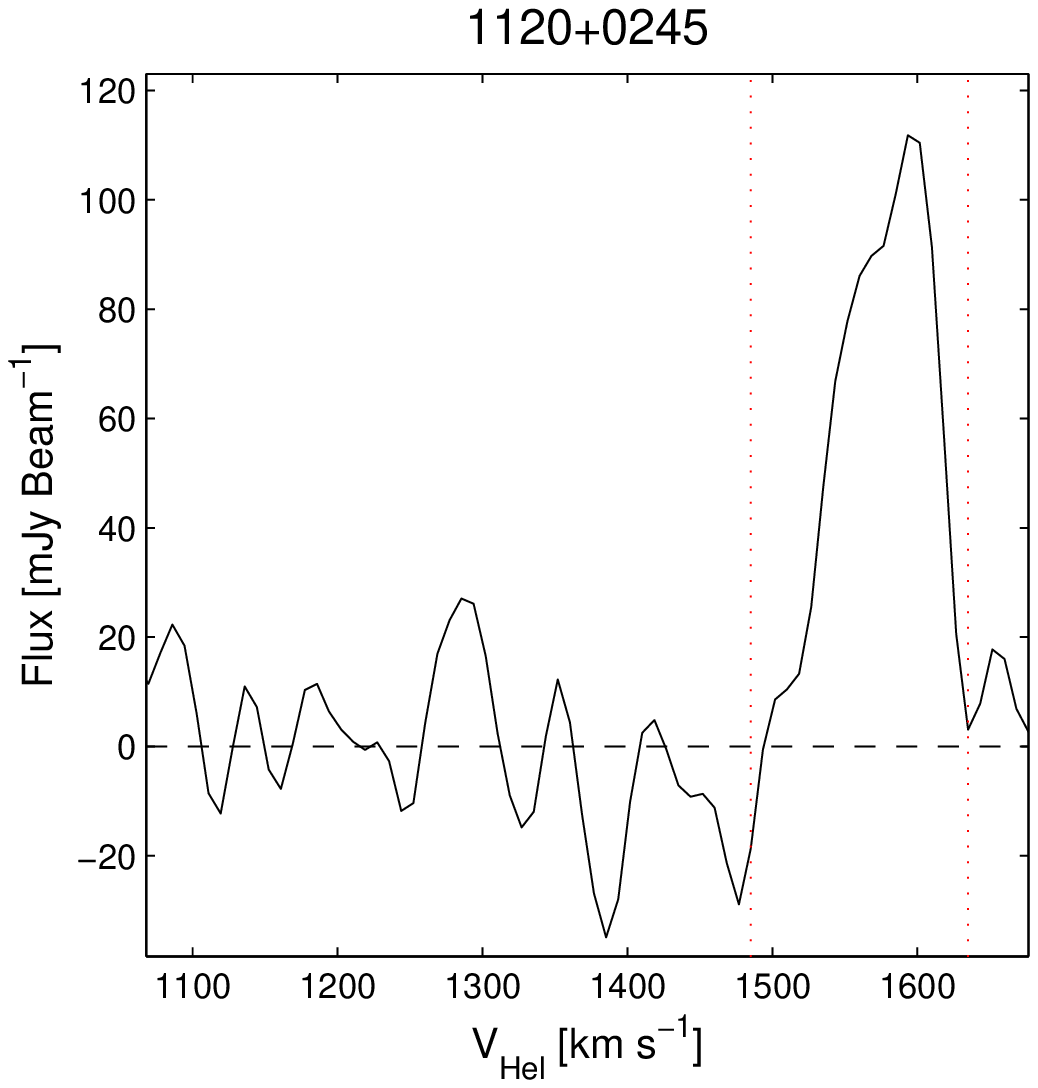}
 \includegraphics[width=0.22\textwidth]{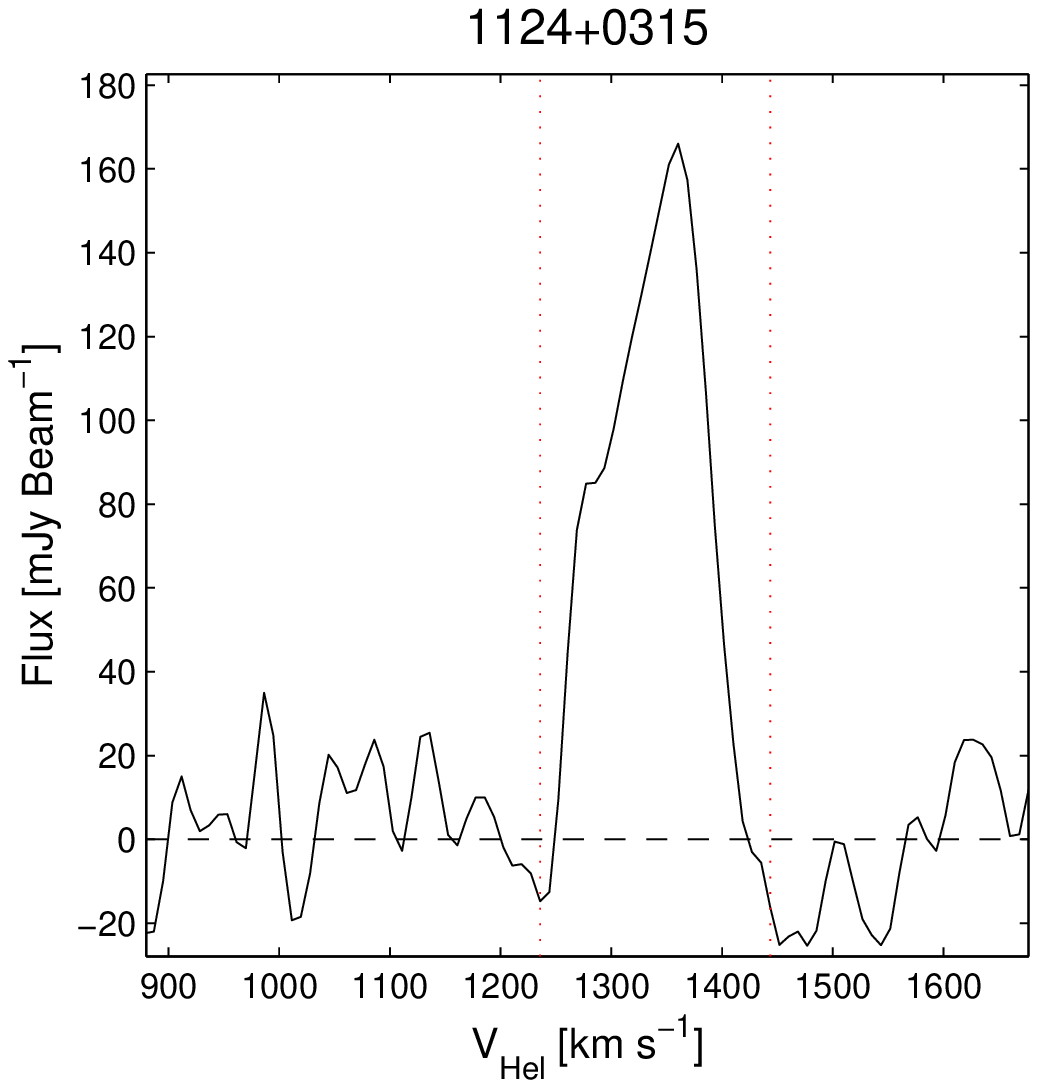}
 \includegraphics[width=0.22\textwidth]{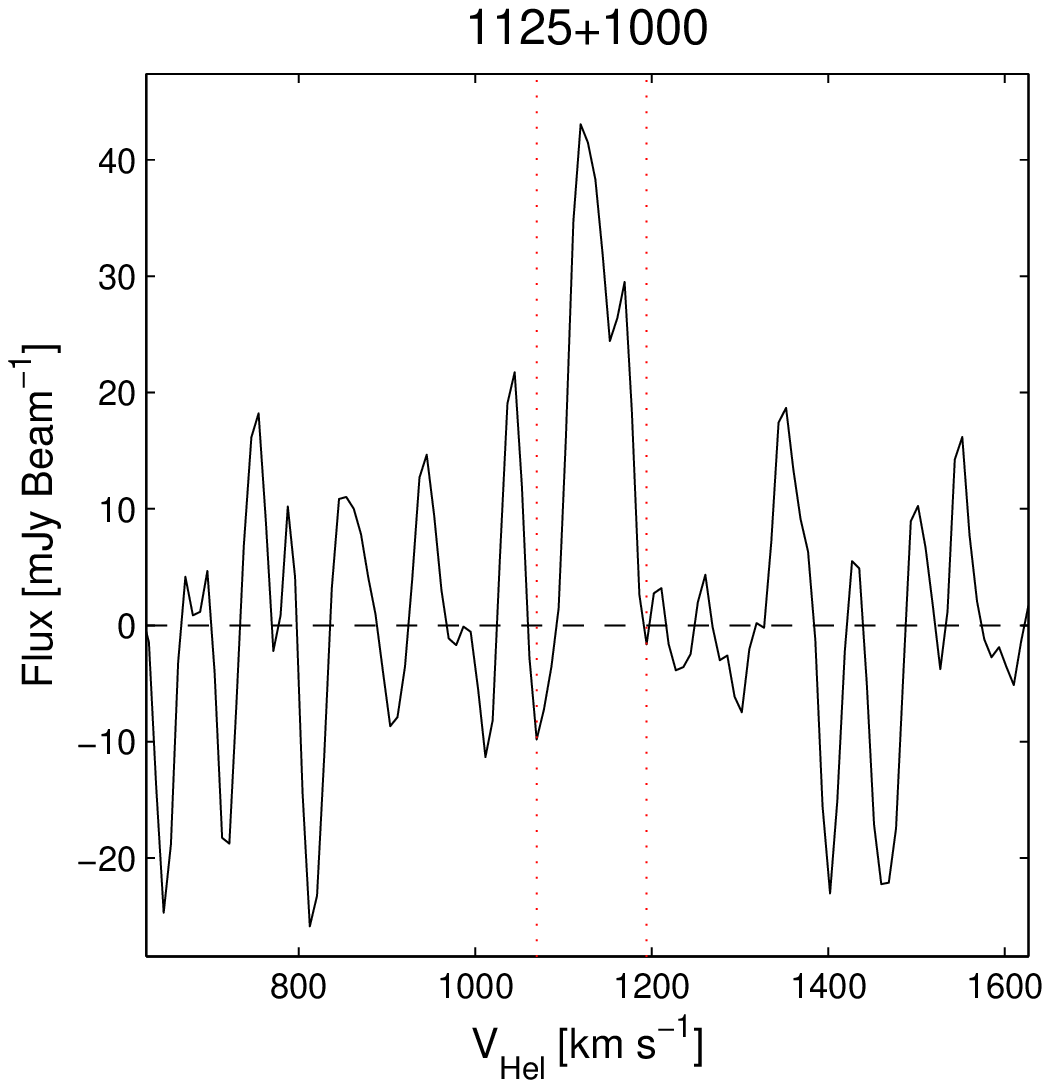}
 \includegraphics[width=0.22\textwidth]{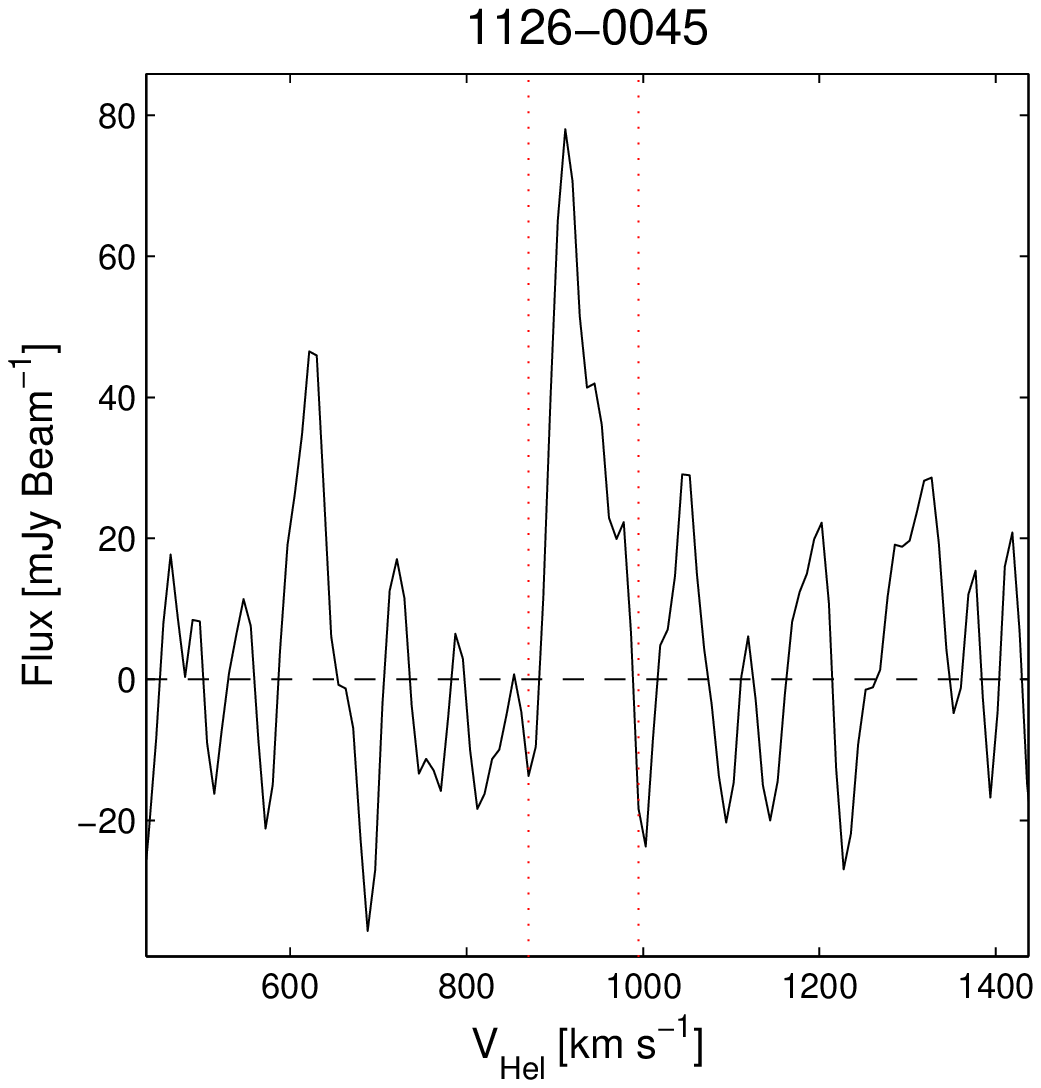}
 \includegraphics[width=0.22\textwidth]{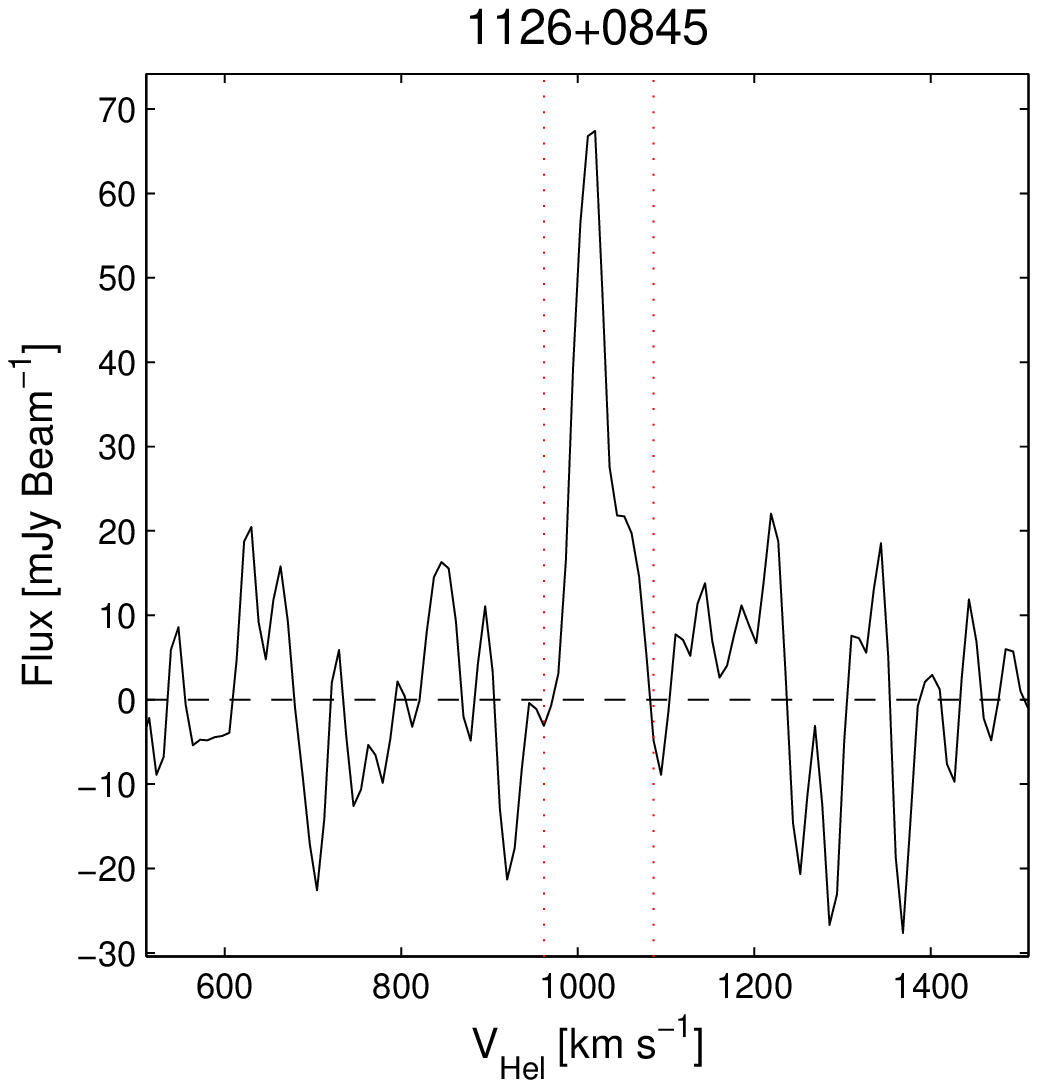}
 \includegraphics[width=0.22\textwidth]{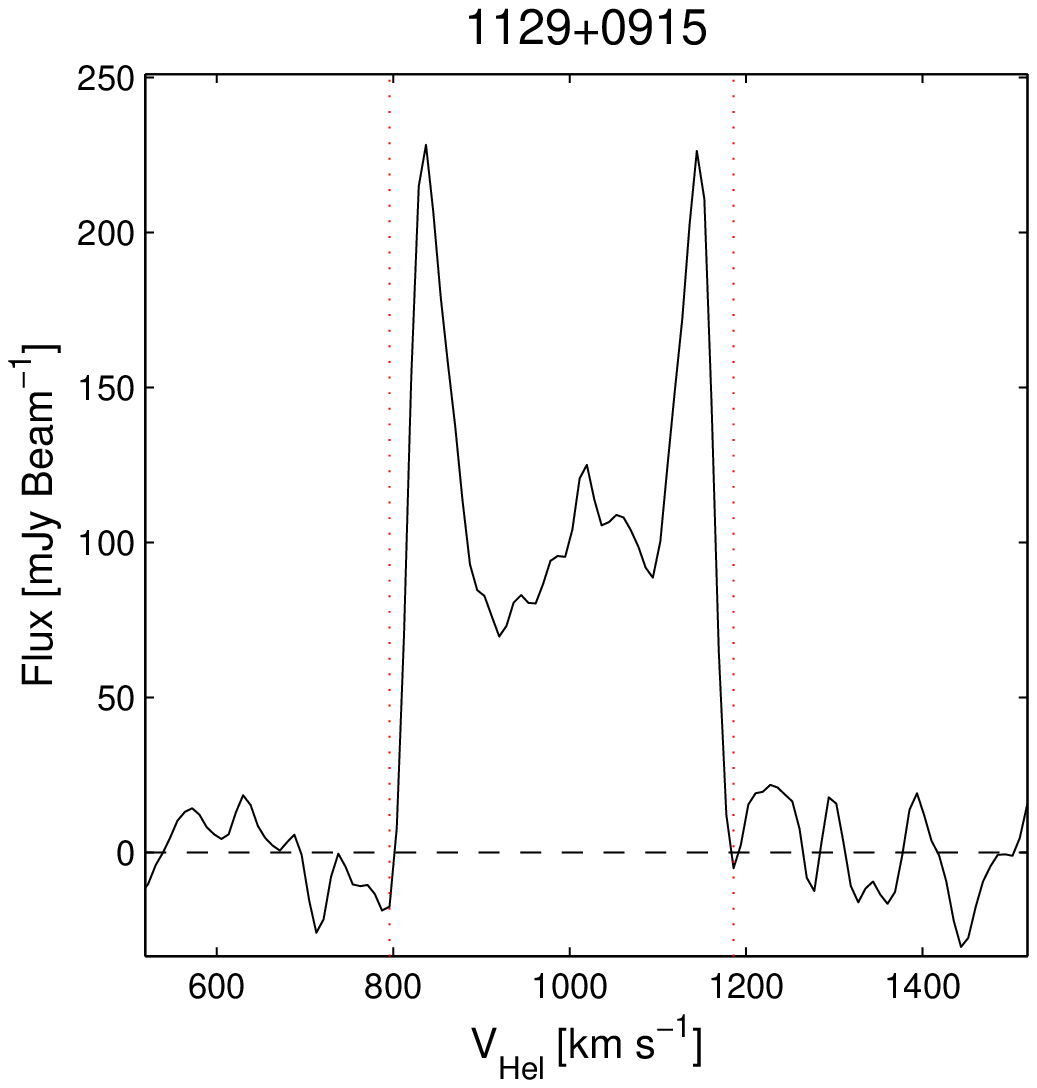}
 \includegraphics[width=0.22\textwidth]{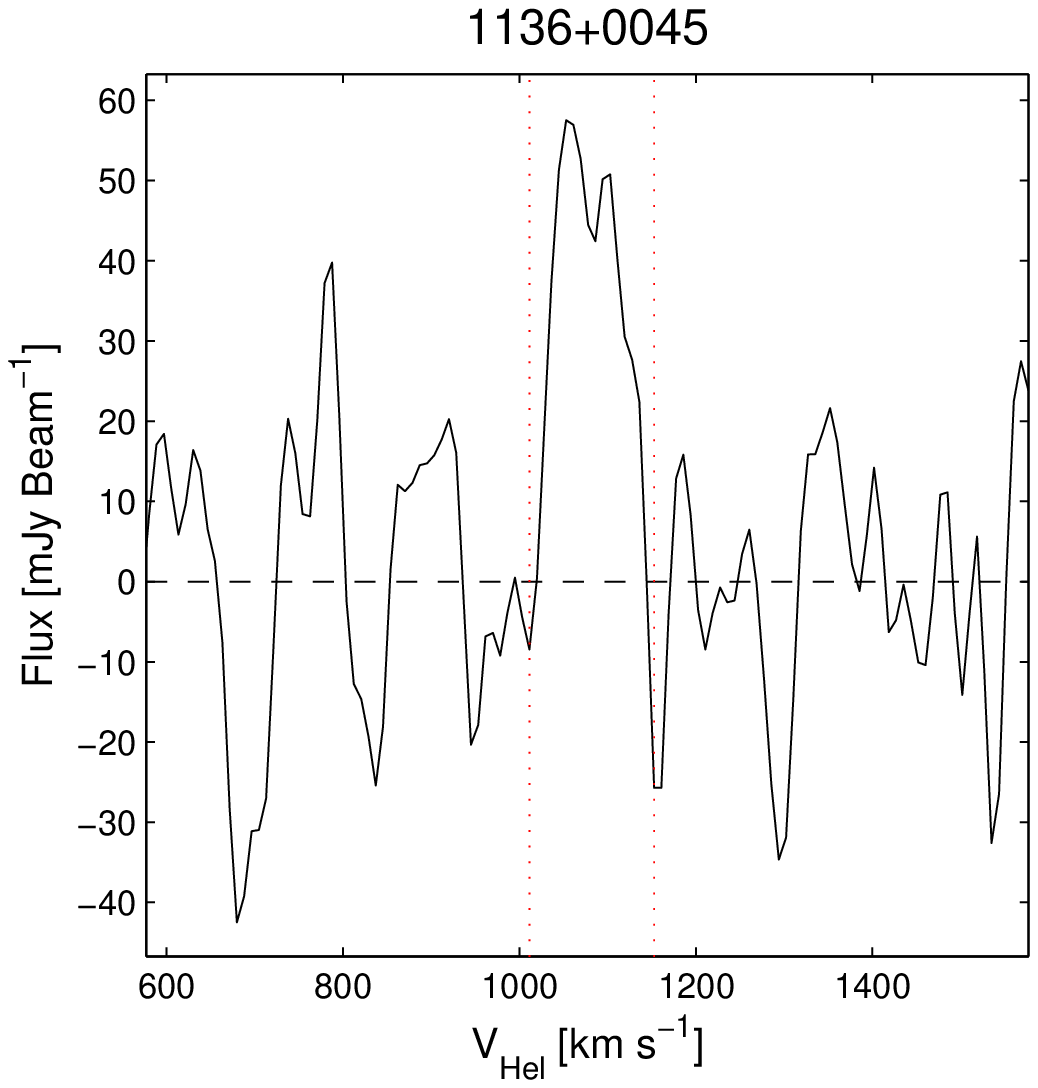}
 \includegraphics[width=0.22\textwidth]{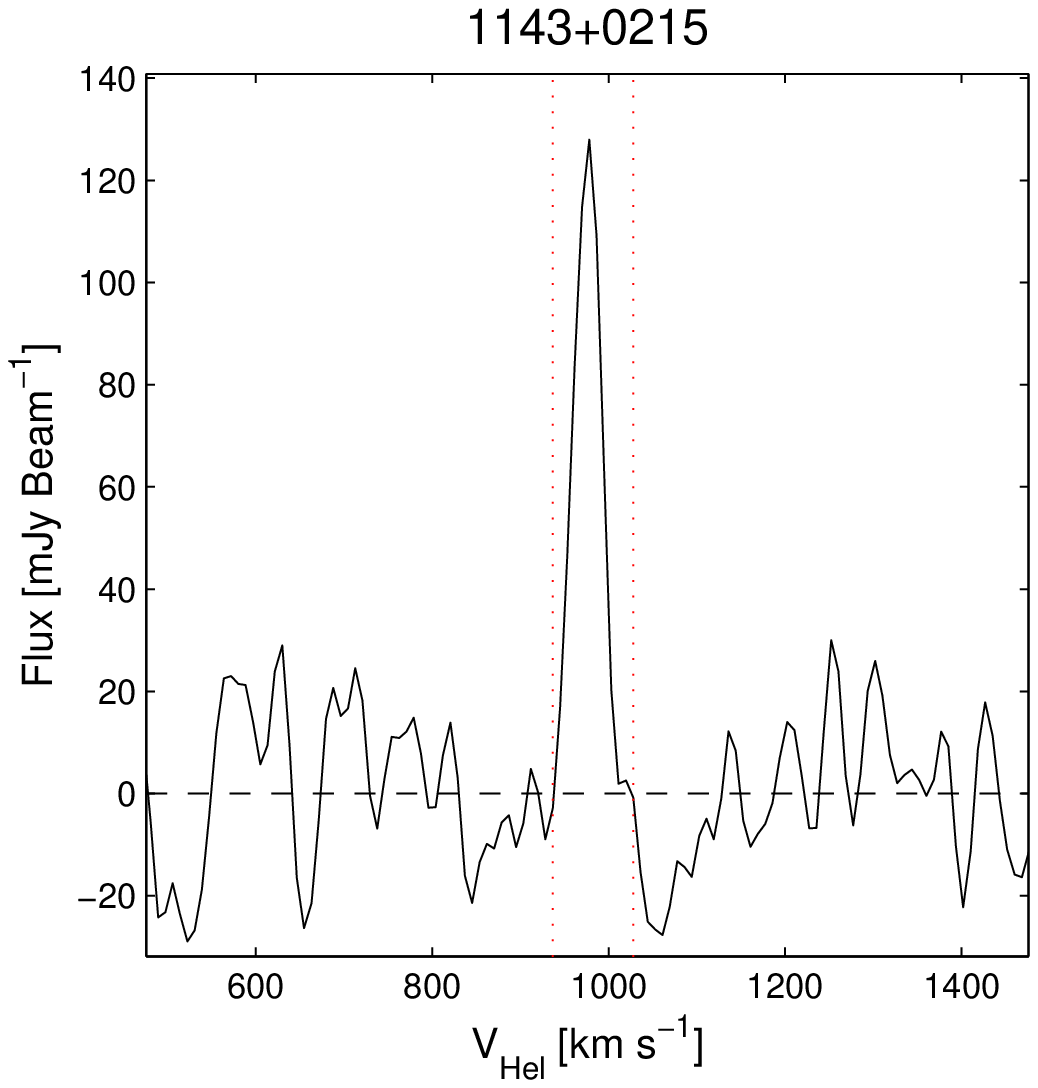}
 \includegraphics[width=0.22\textwidth]{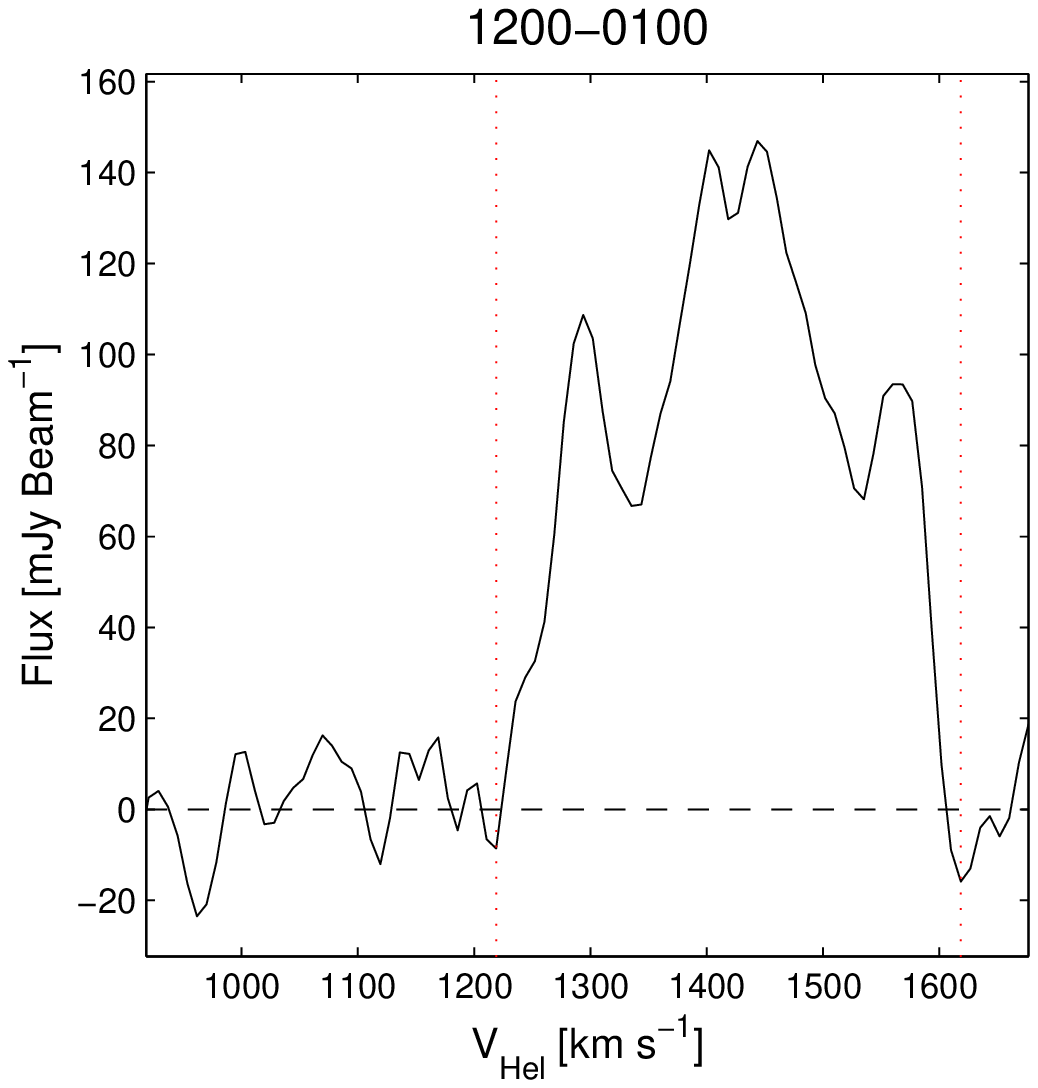}
 \includegraphics[width=0.22\textwidth]{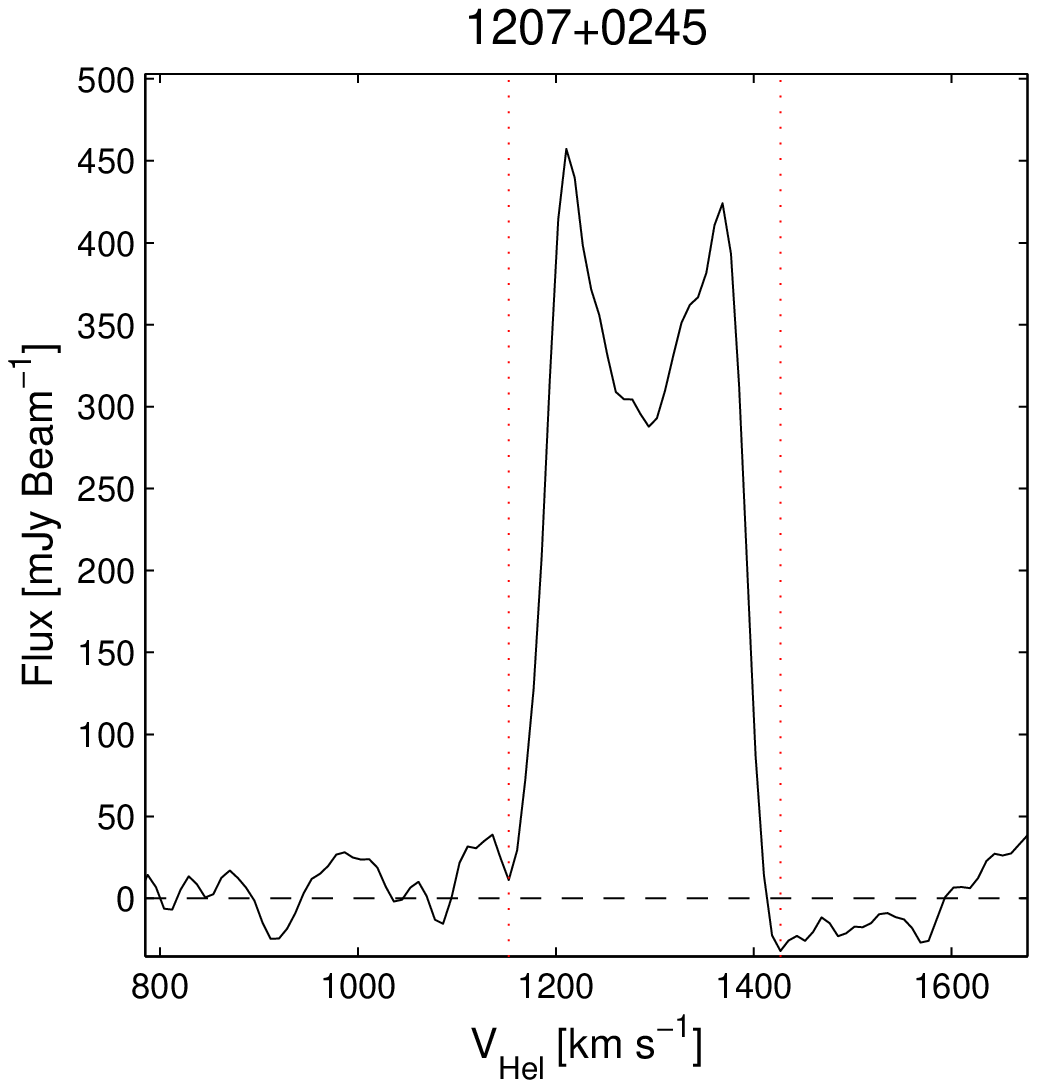}
 \includegraphics[width=0.22\textwidth]{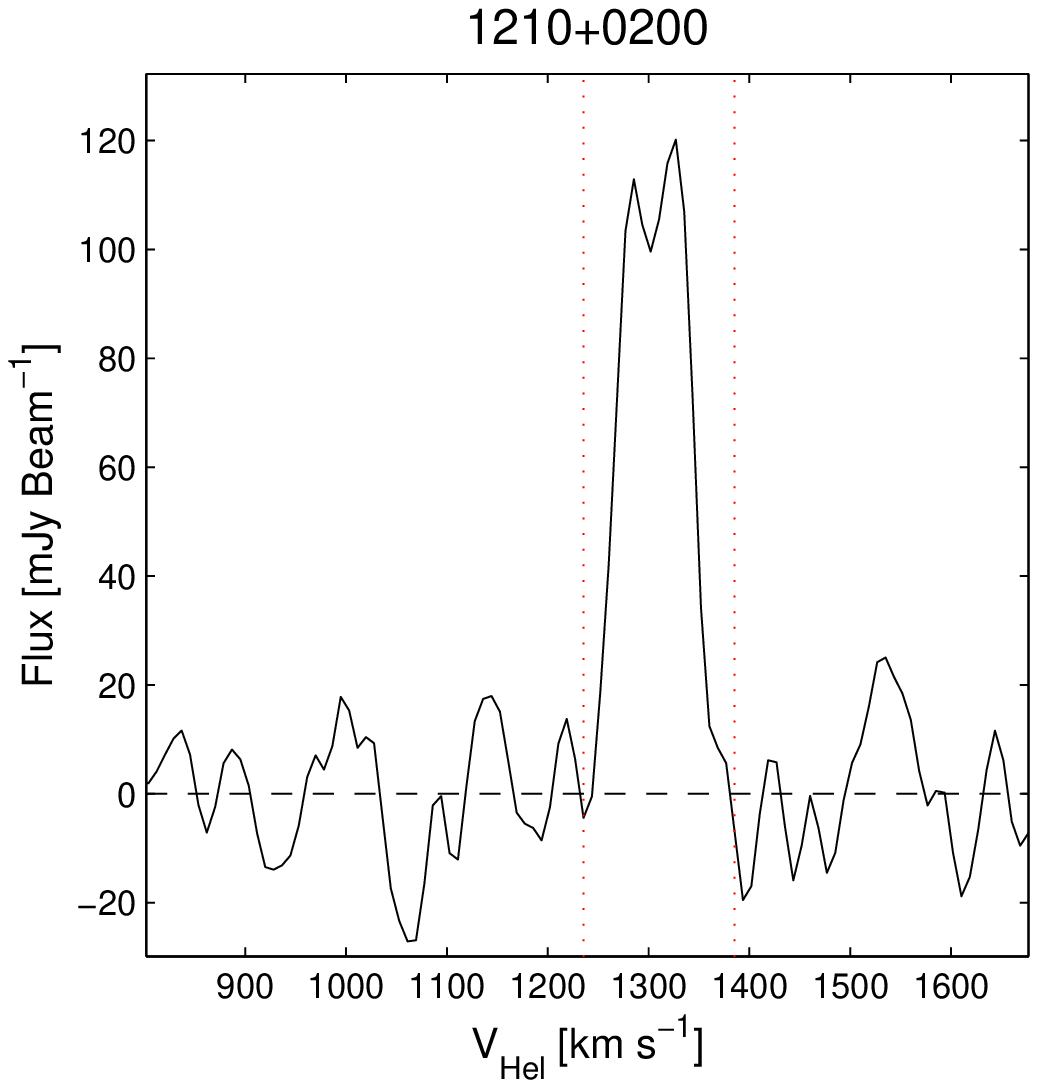}
 \includegraphics[width=0.22\textwidth]{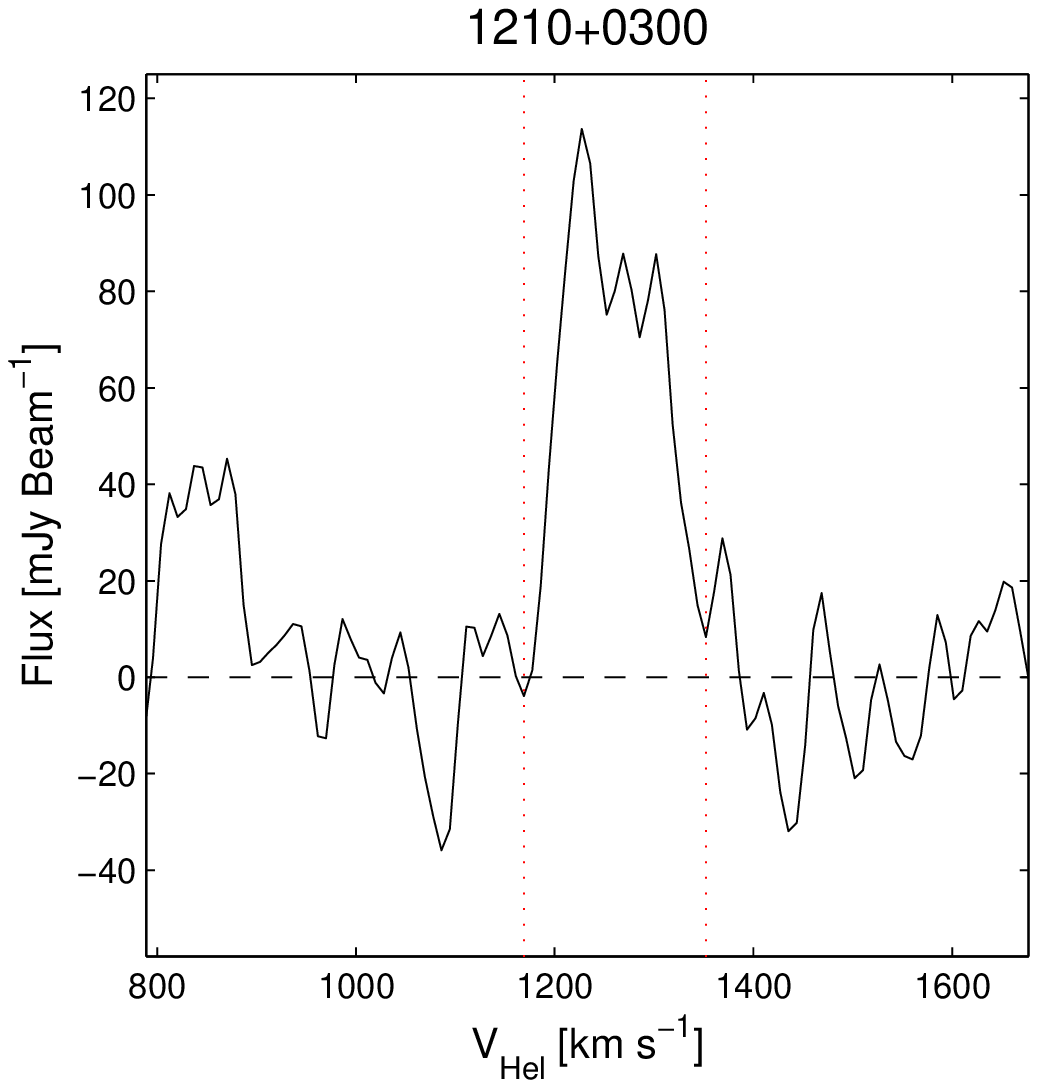}
 \includegraphics[width=0.22\textwidth]{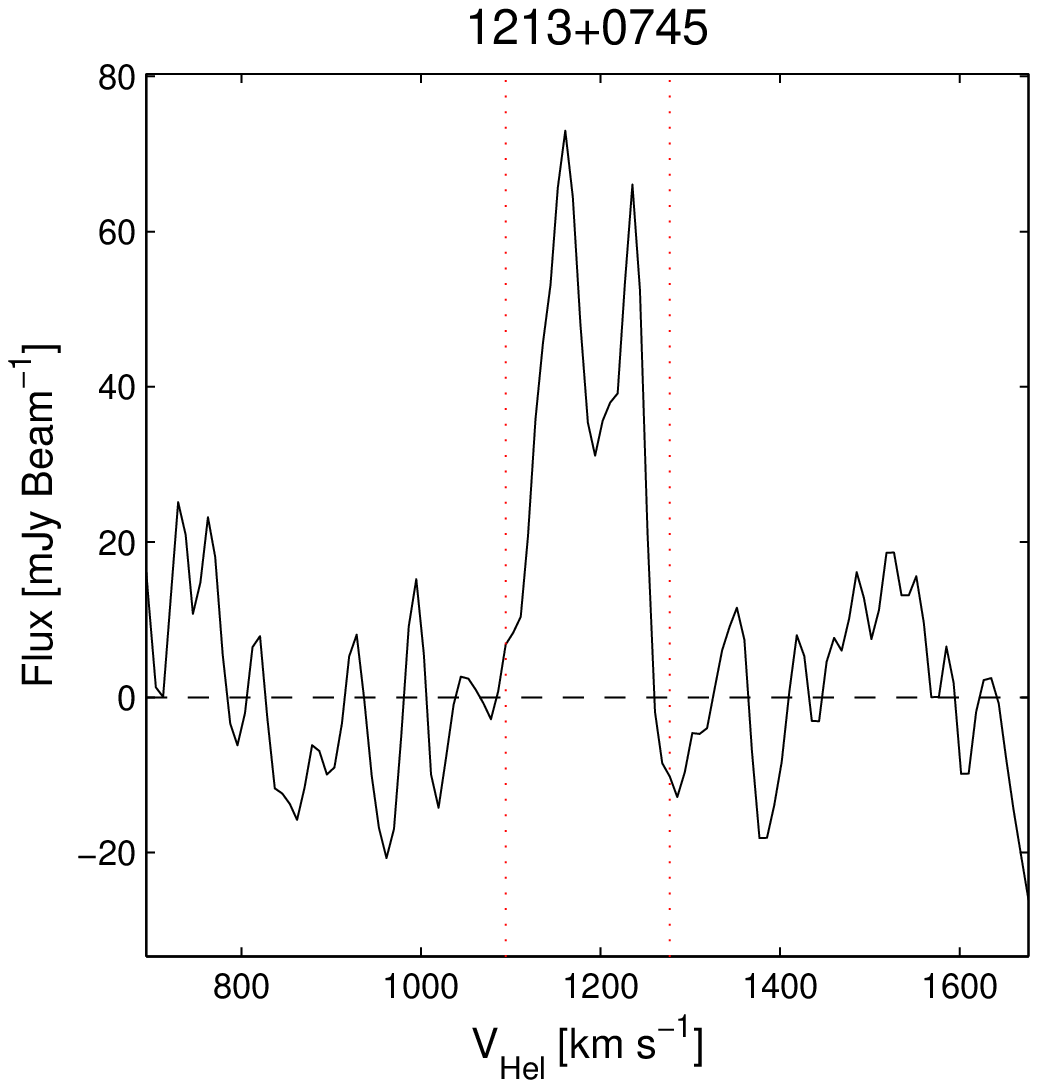}

 \end{center}                                            
{\bf Fig~\ref{all_spectra}.} (continued)                                        
 
\end{figure*}

\begin{figure*}
  \begin{center}

 \includegraphics[width=0.22\textwidth]{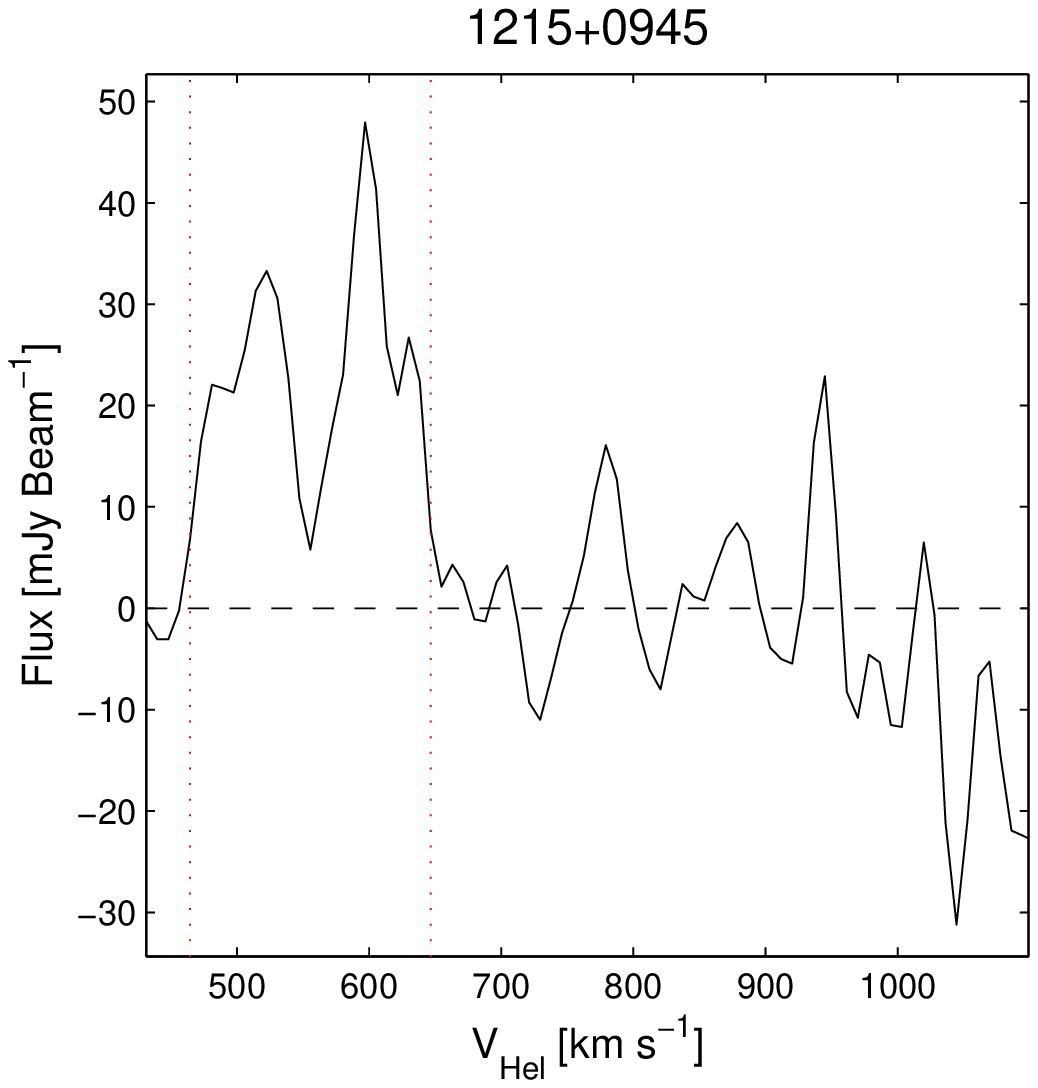}
 \includegraphics[width=0.22\textwidth]{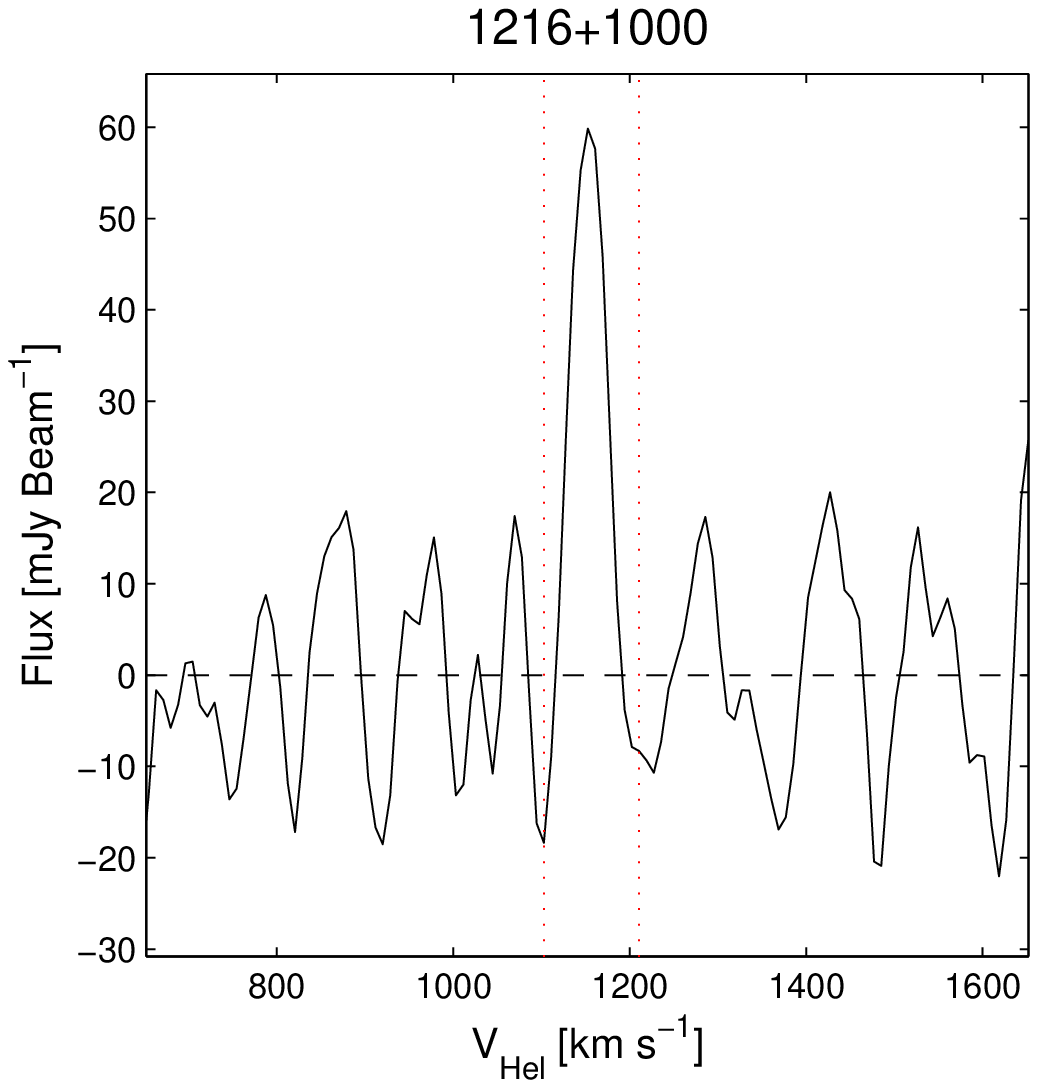}
 \includegraphics[width=0.22\textwidth]{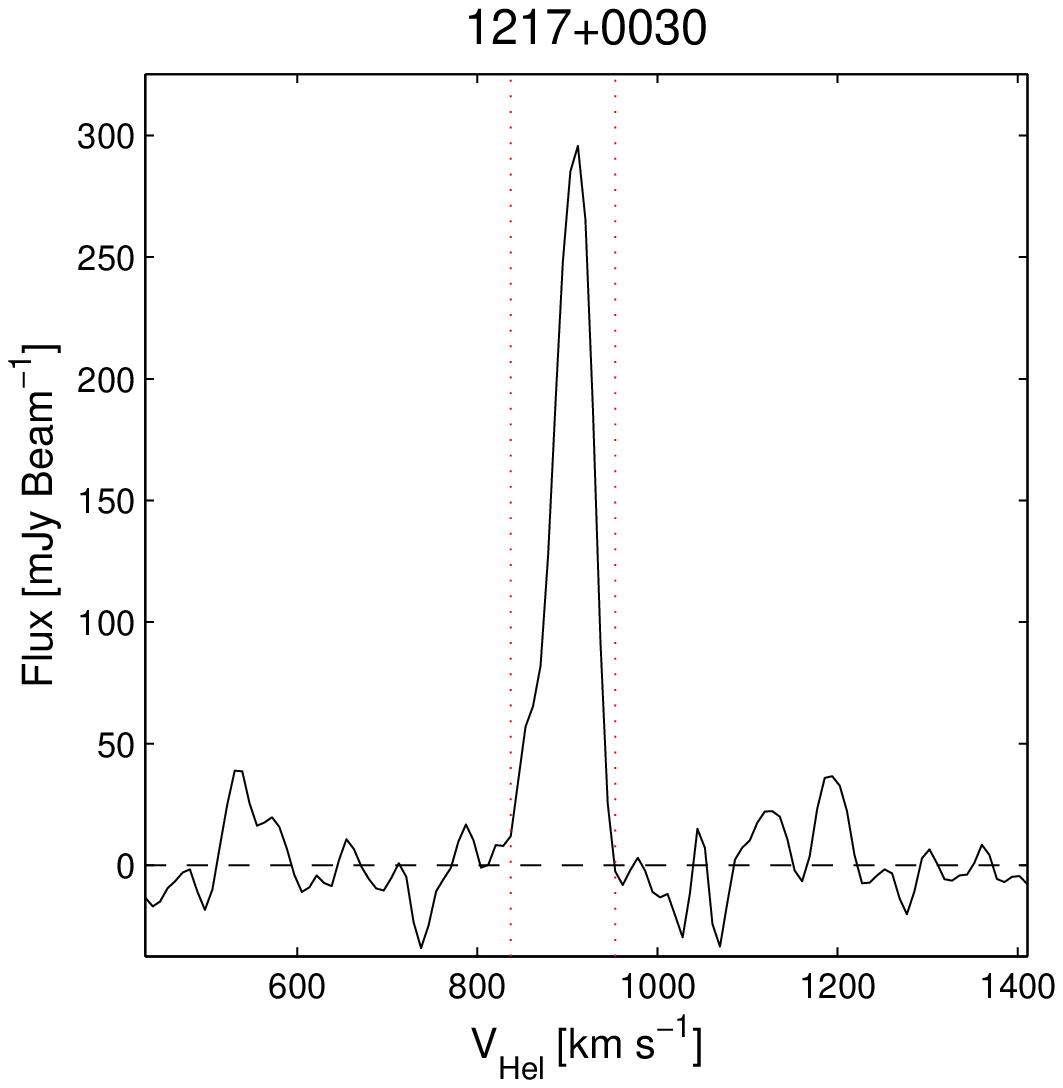}
 \includegraphics[width=0.22\textwidth]{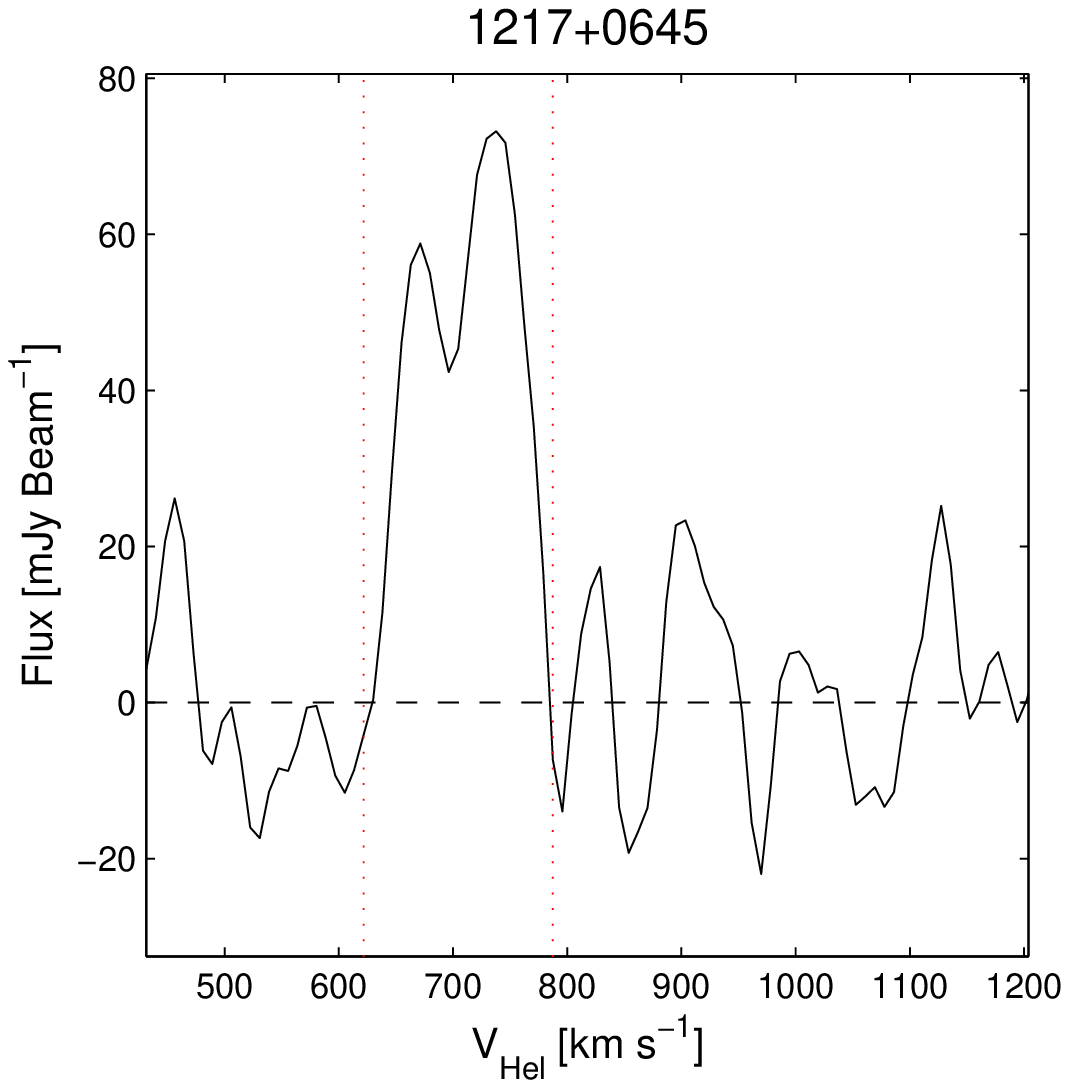}
 \includegraphics[width=0.22\textwidth]{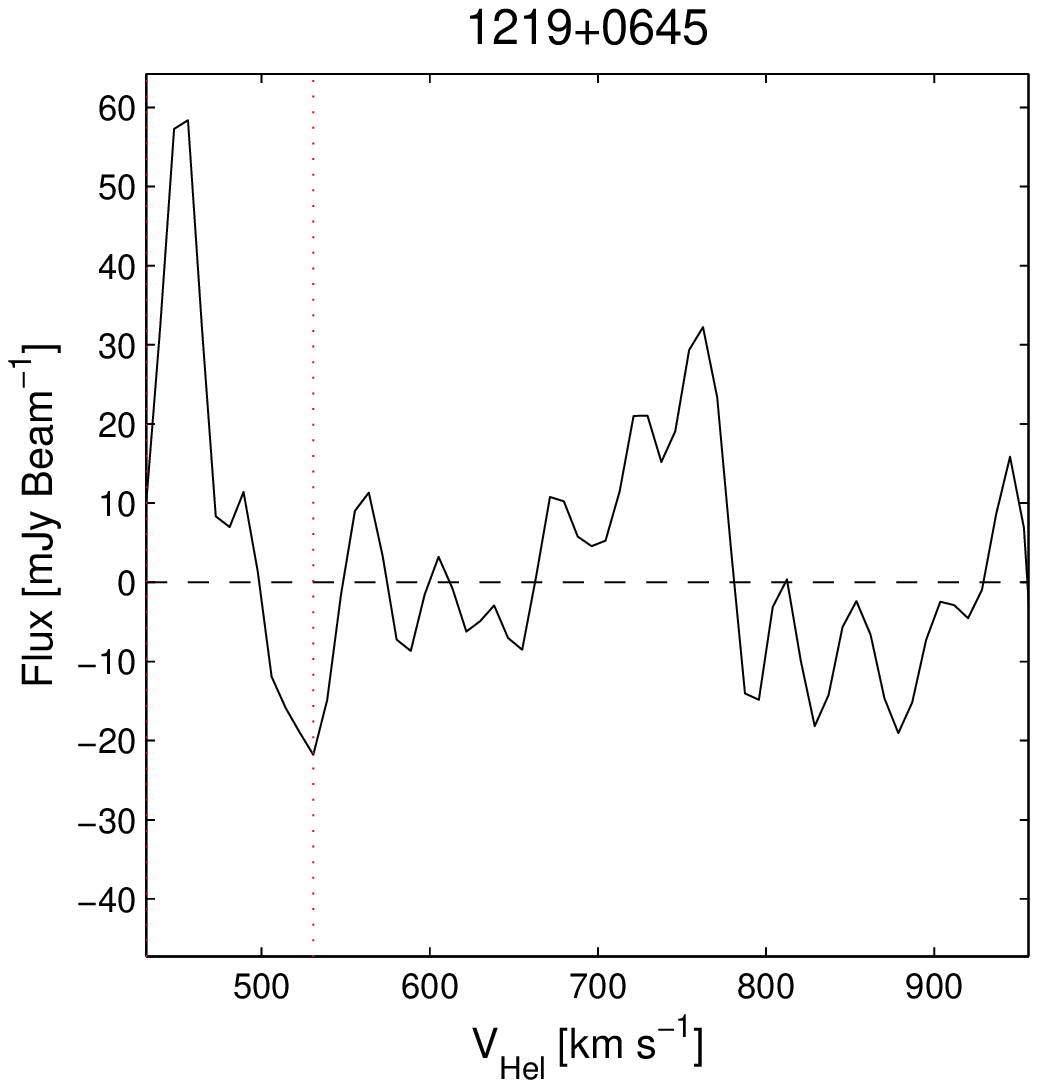}
 \includegraphics[width=0.22\textwidth]{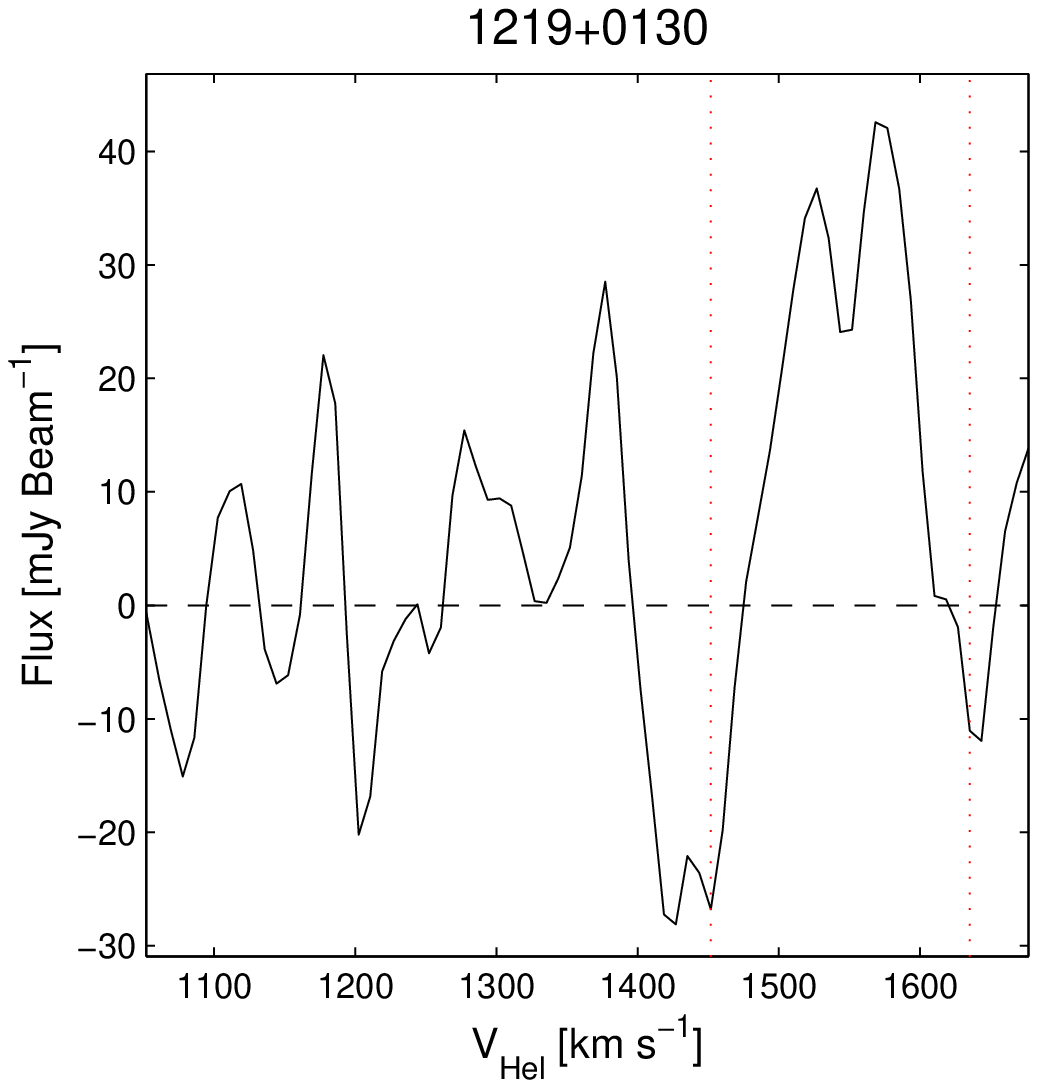}
 \includegraphics[width=0.22\textwidth]{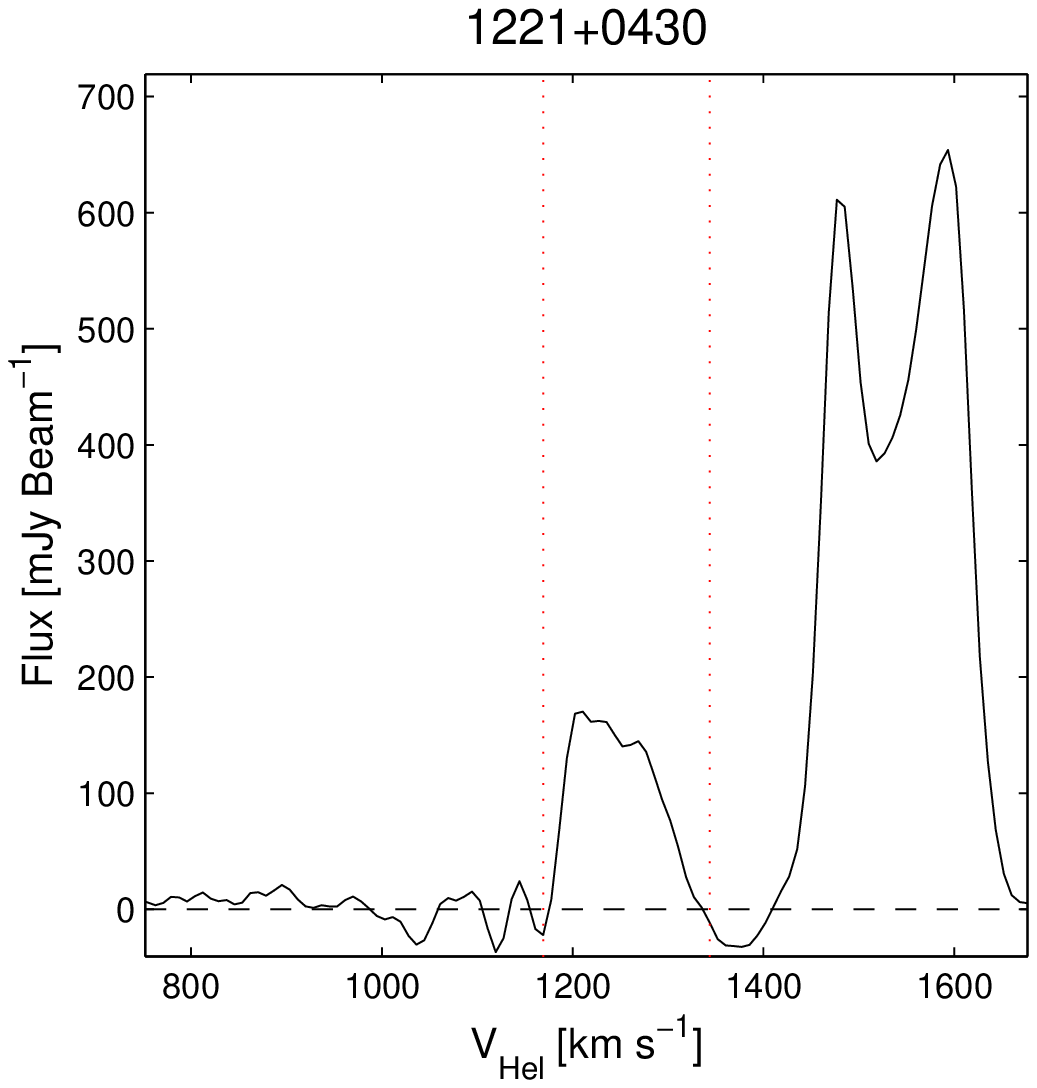}
 \includegraphics[width=0.22\textwidth]{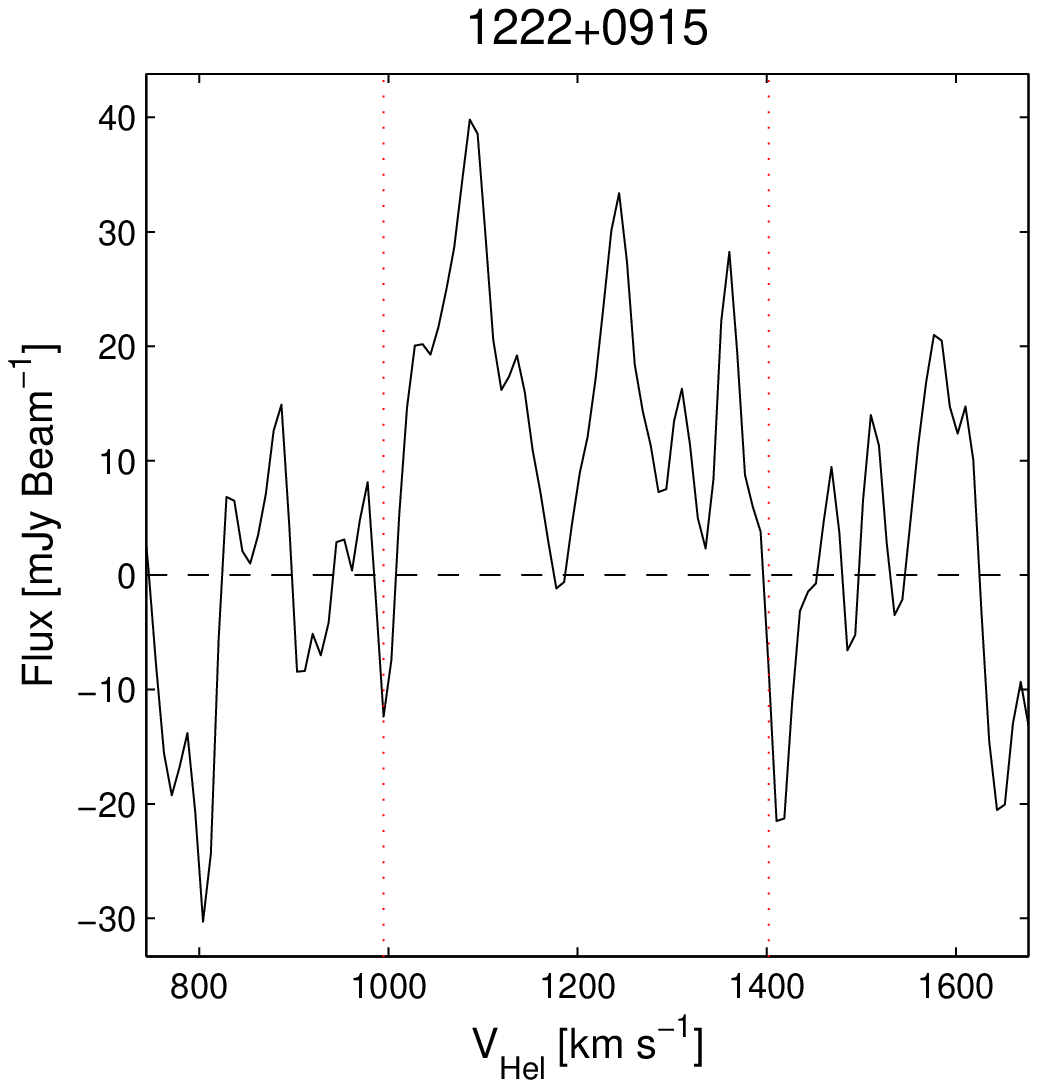}
 \includegraphics[width=0.22\textwidth]{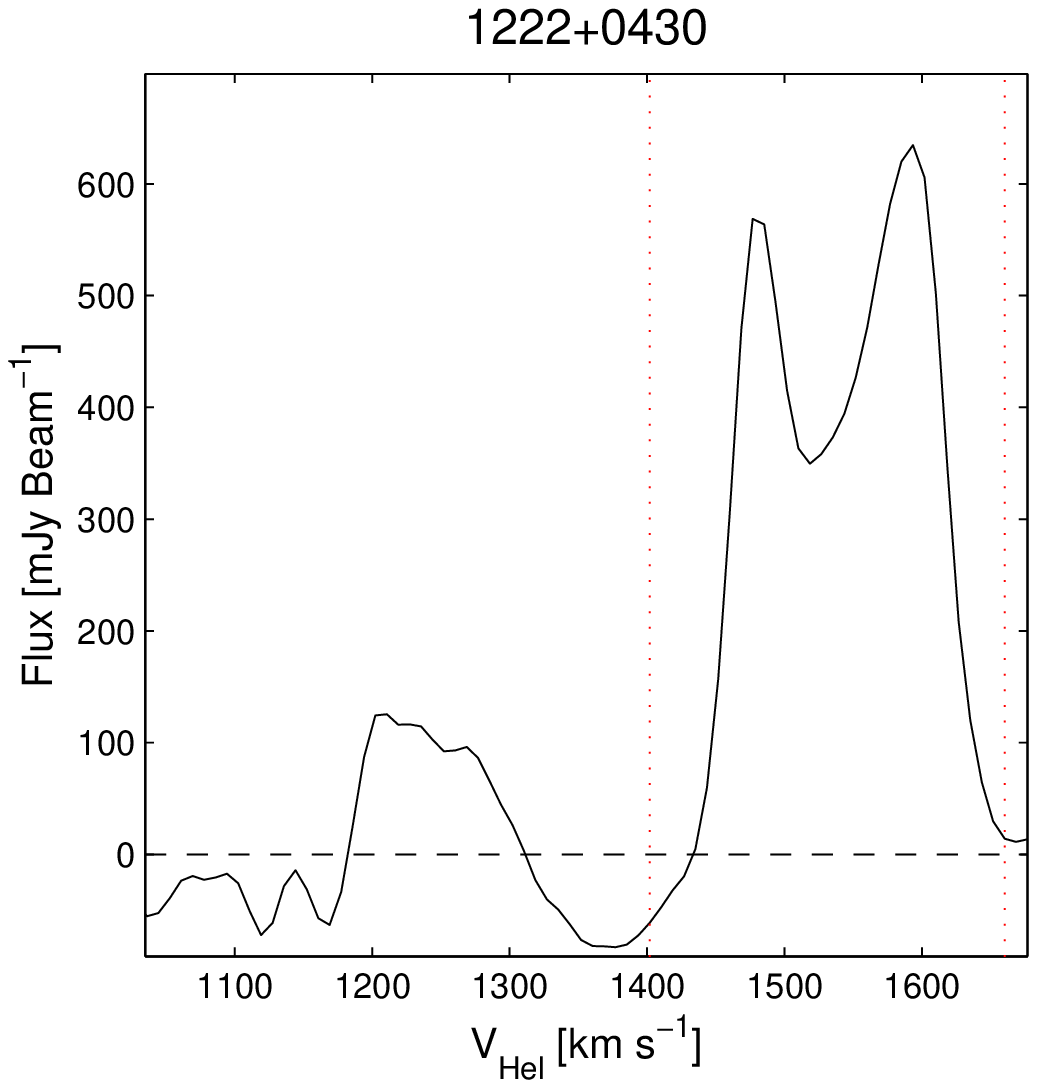}
 \includegraphics[width=0.22\textwidth]{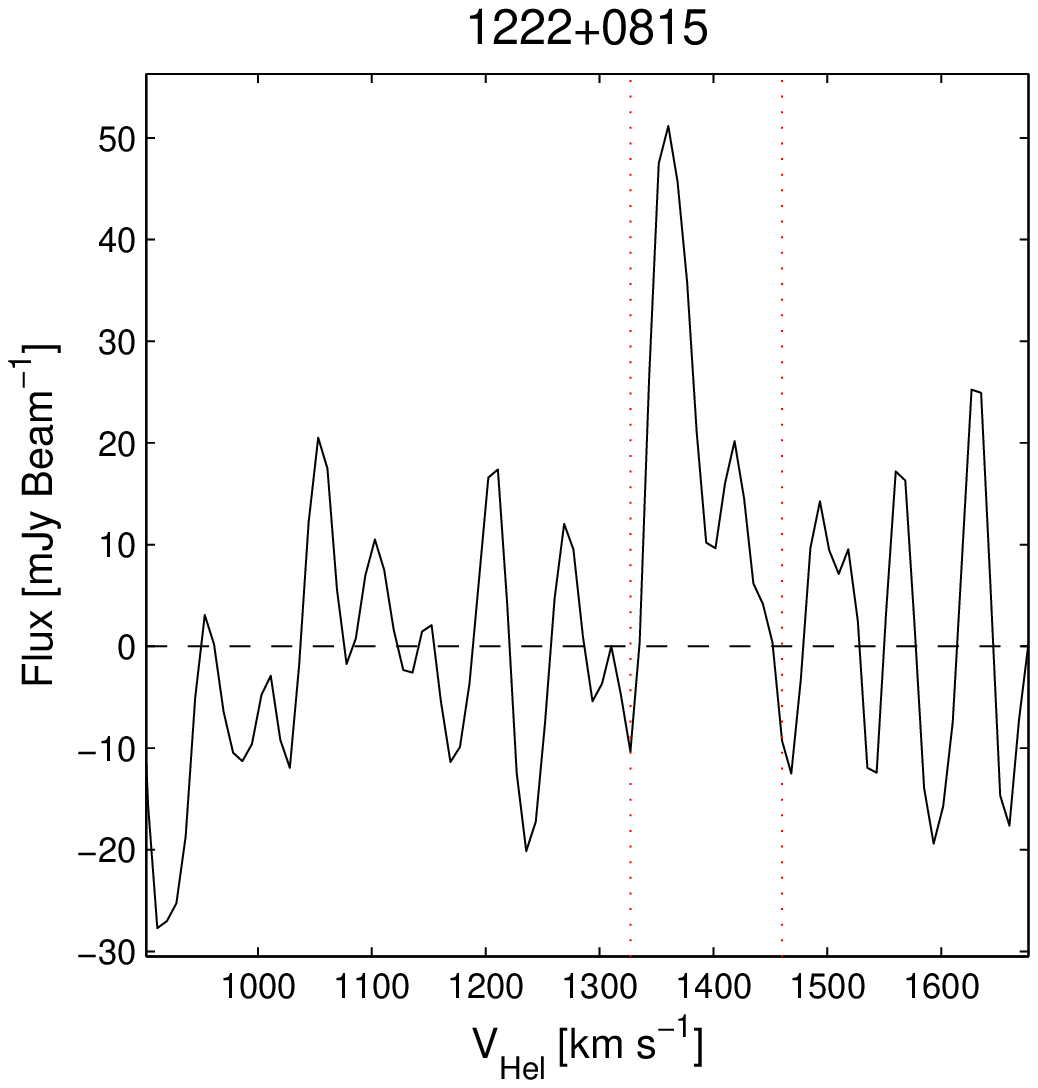}
 \includegraphics[width=0.22\textwidth]{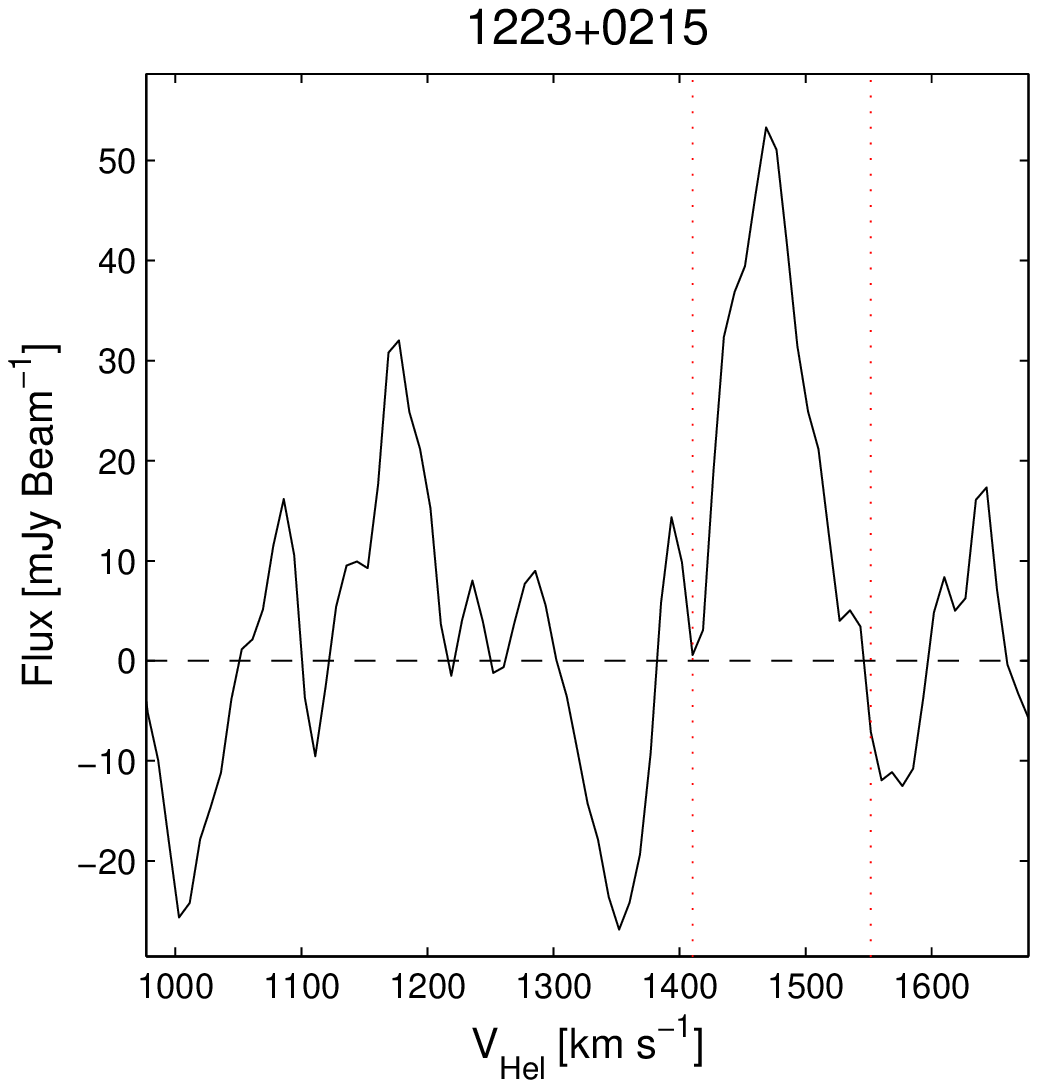}
 \includegraphics[width=0.22\textwidth]{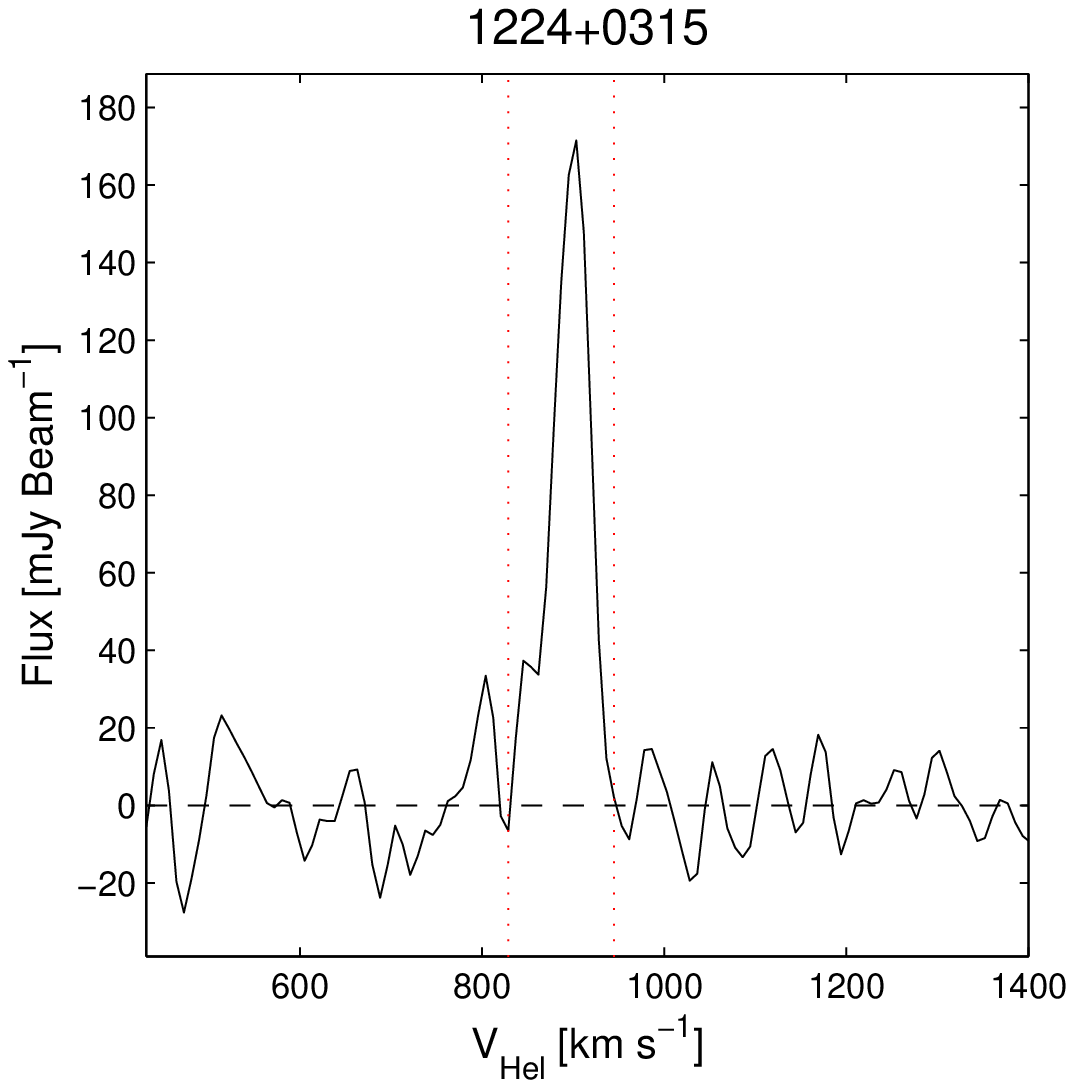}
 \includegraphics[width=0.22\textwidth]{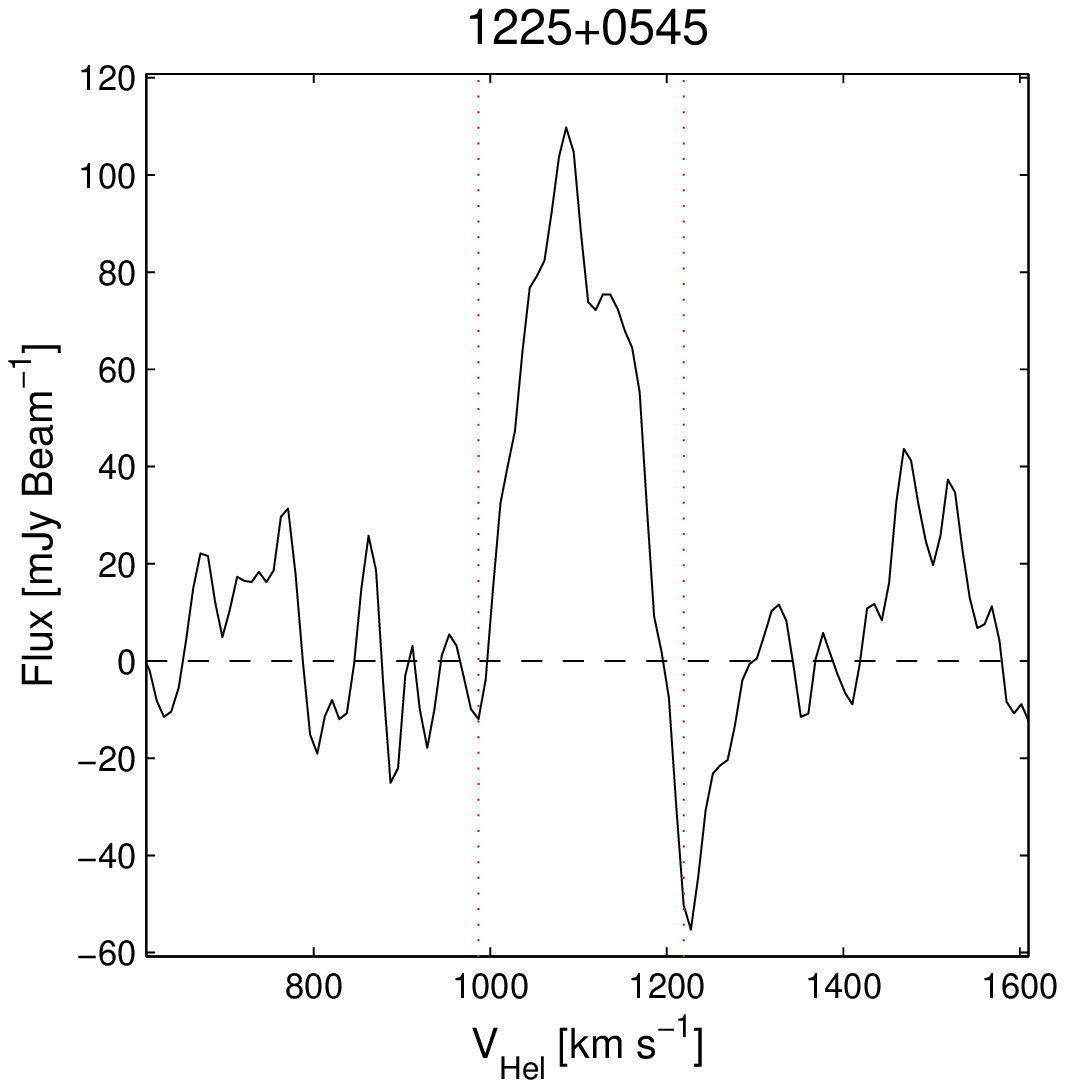}
 \includegraphics[width=0.22\textwidth]{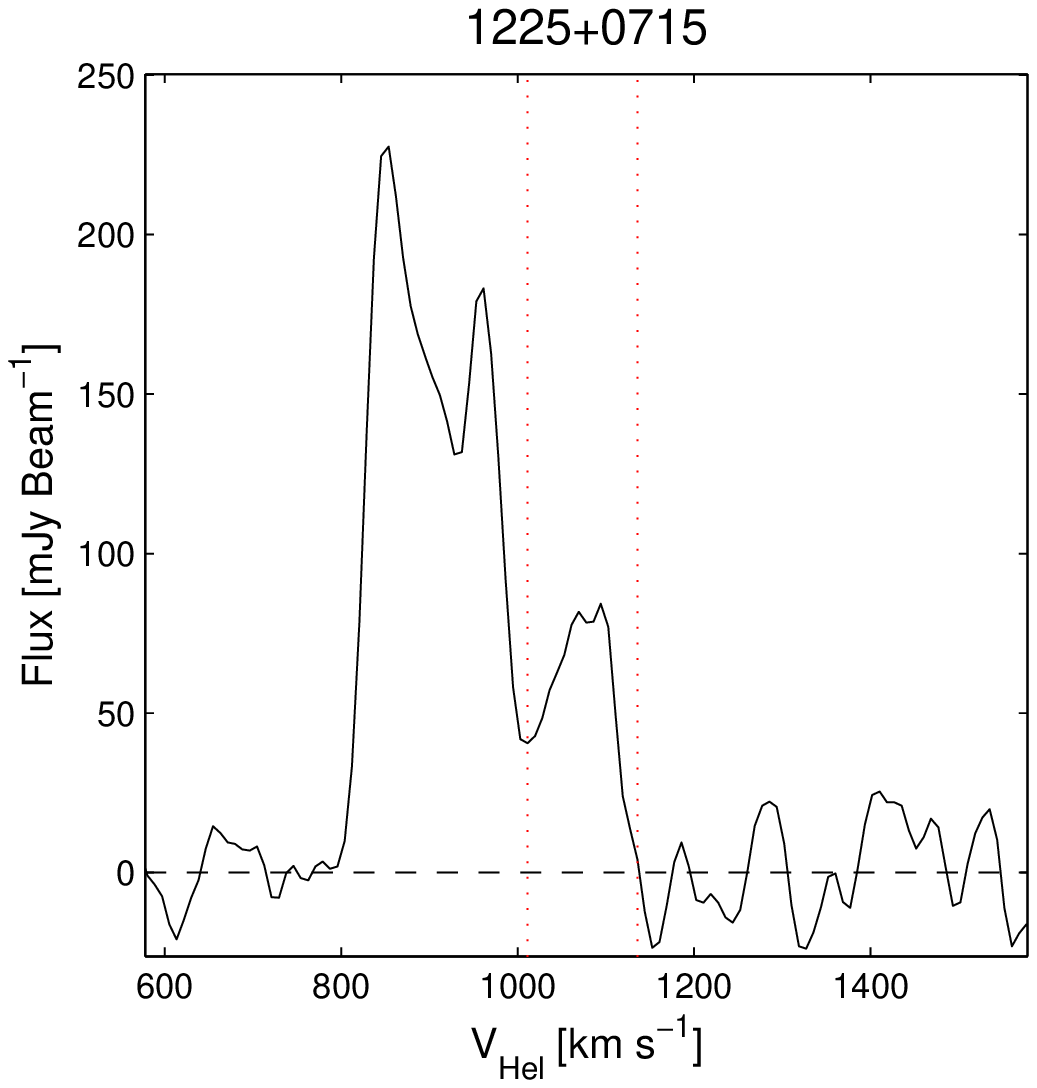}
 \includegraphics[width=0.22\textwidth]{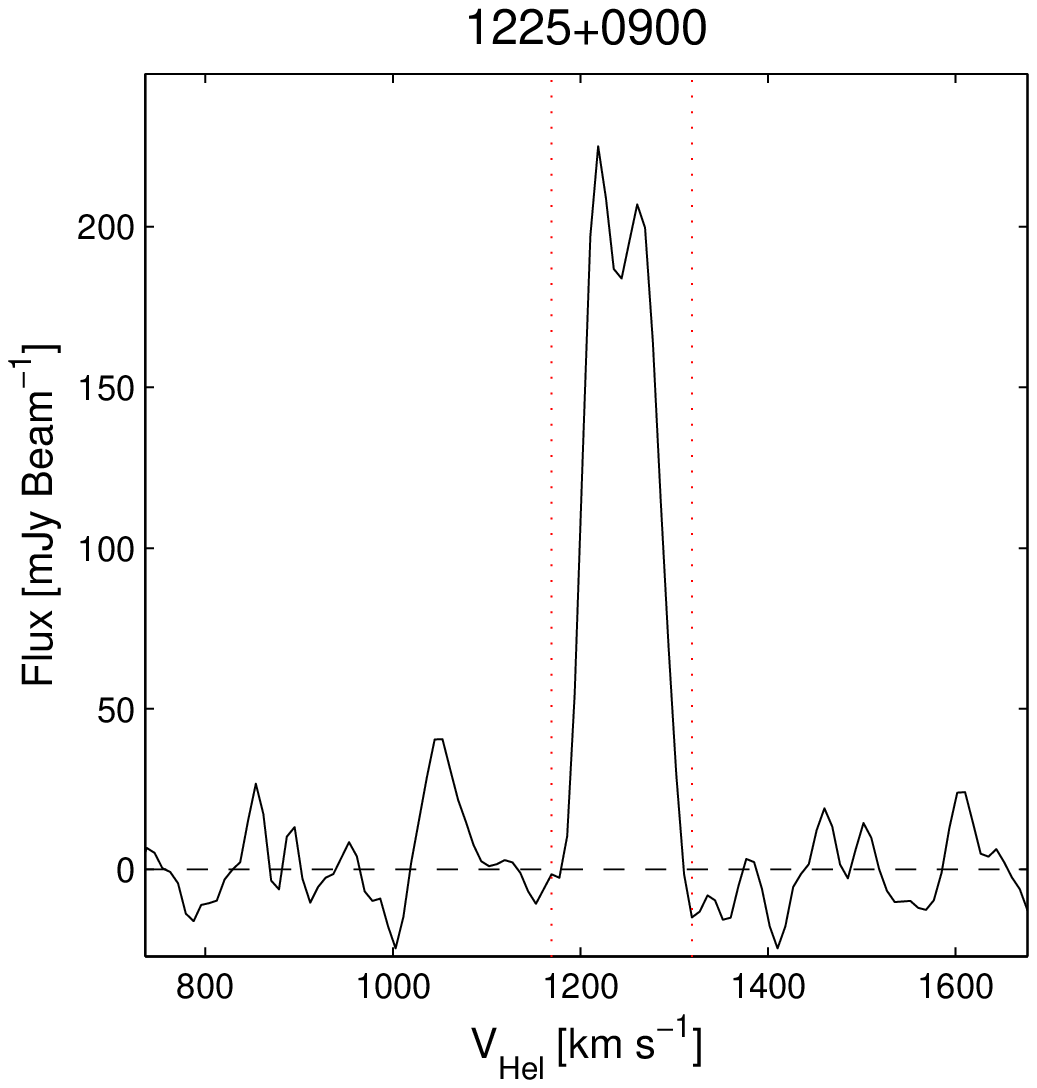}
 \includegraphics[width=0.22\textwidth]{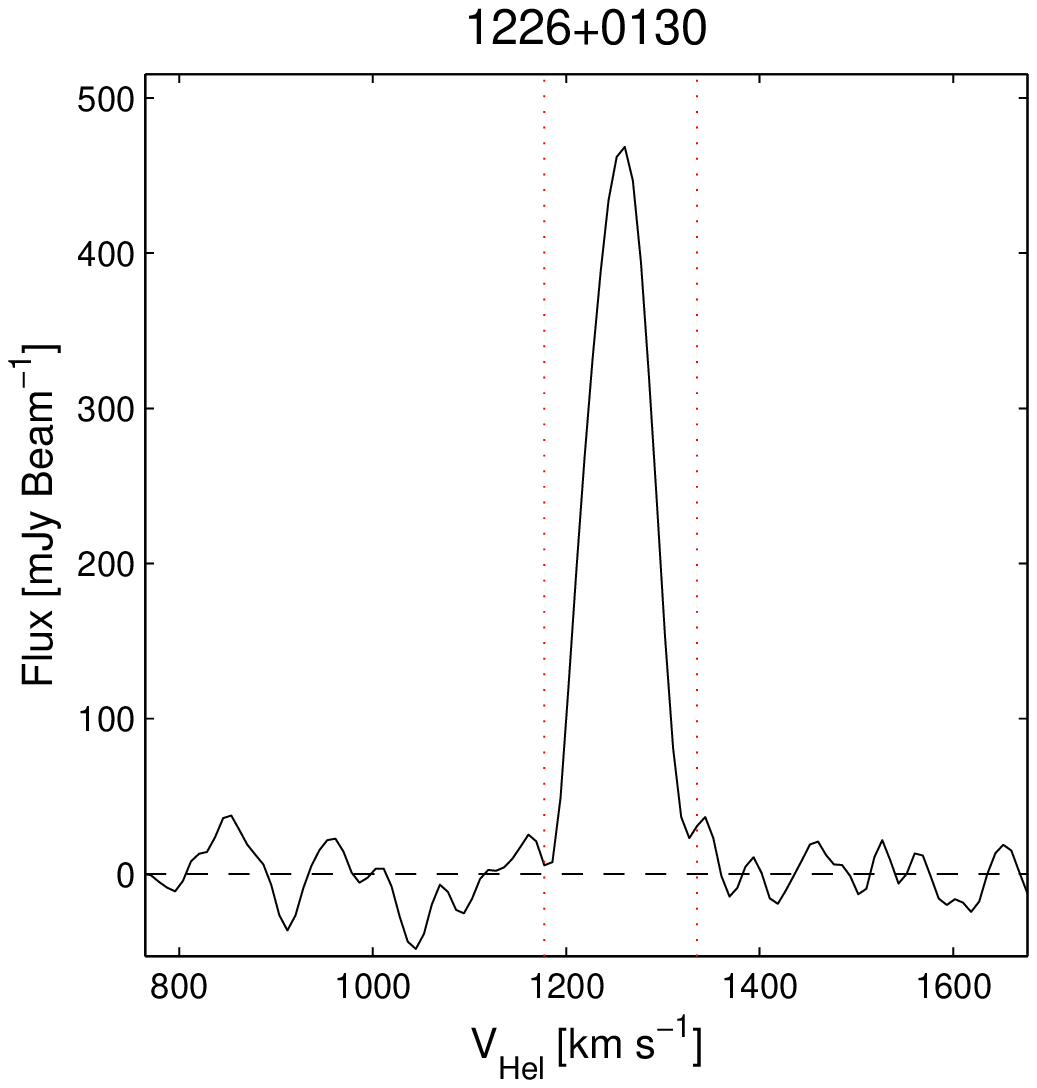}
 \includegraphics[width=0.22\textwidth]{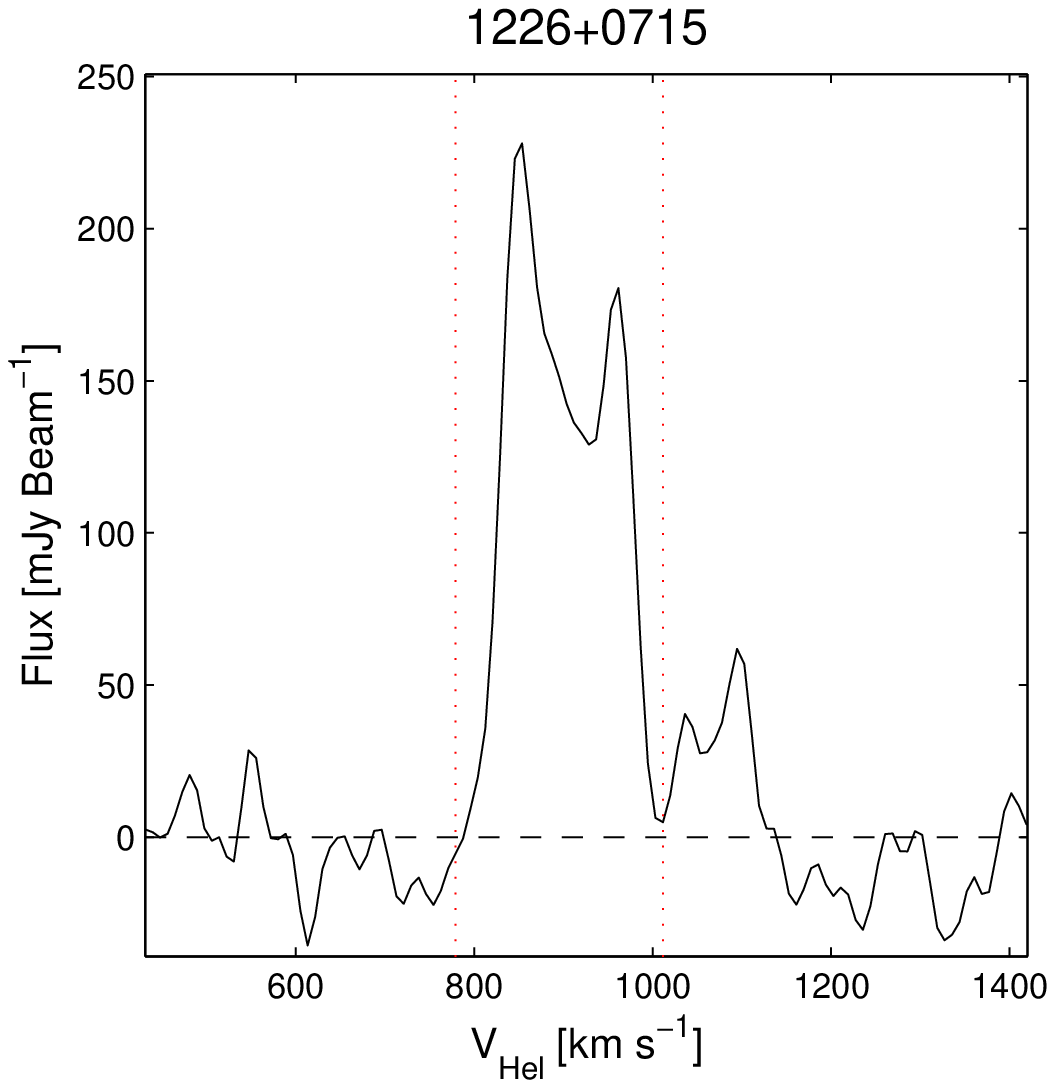}
 \includegraphics[width=0.22\textwidth]{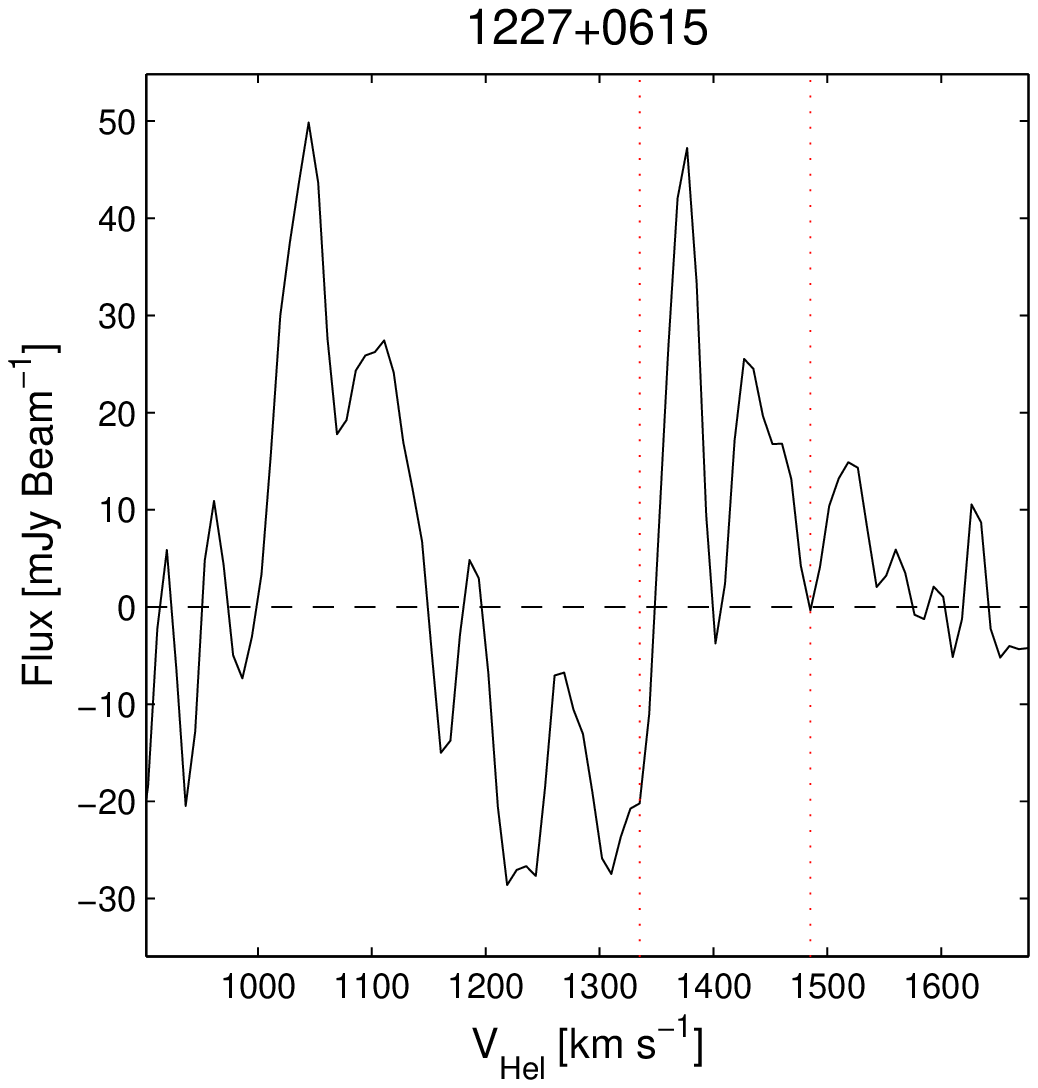}
 \includegraphics[width=0.22\textwidth]{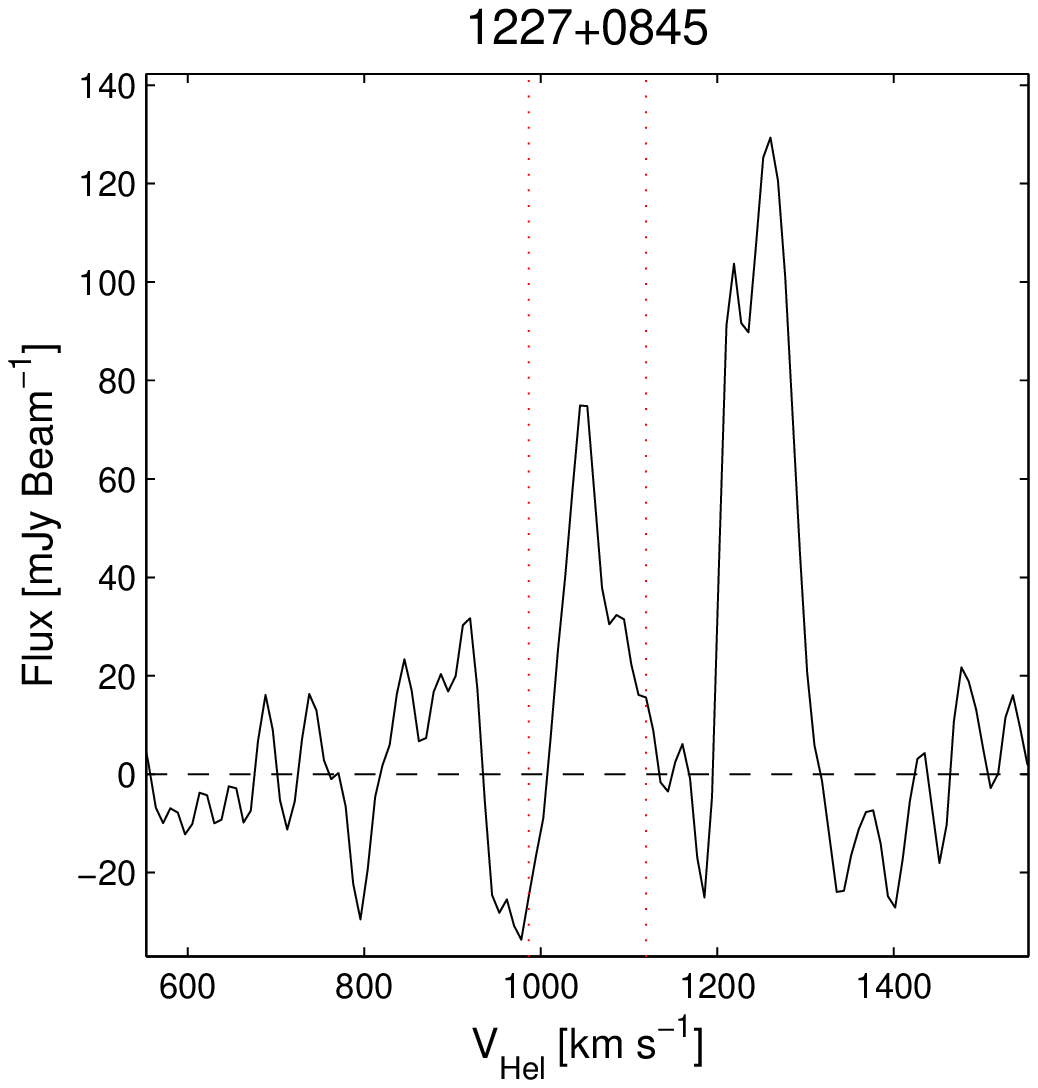}
 \includegraphics[width=0.22\textwidth]{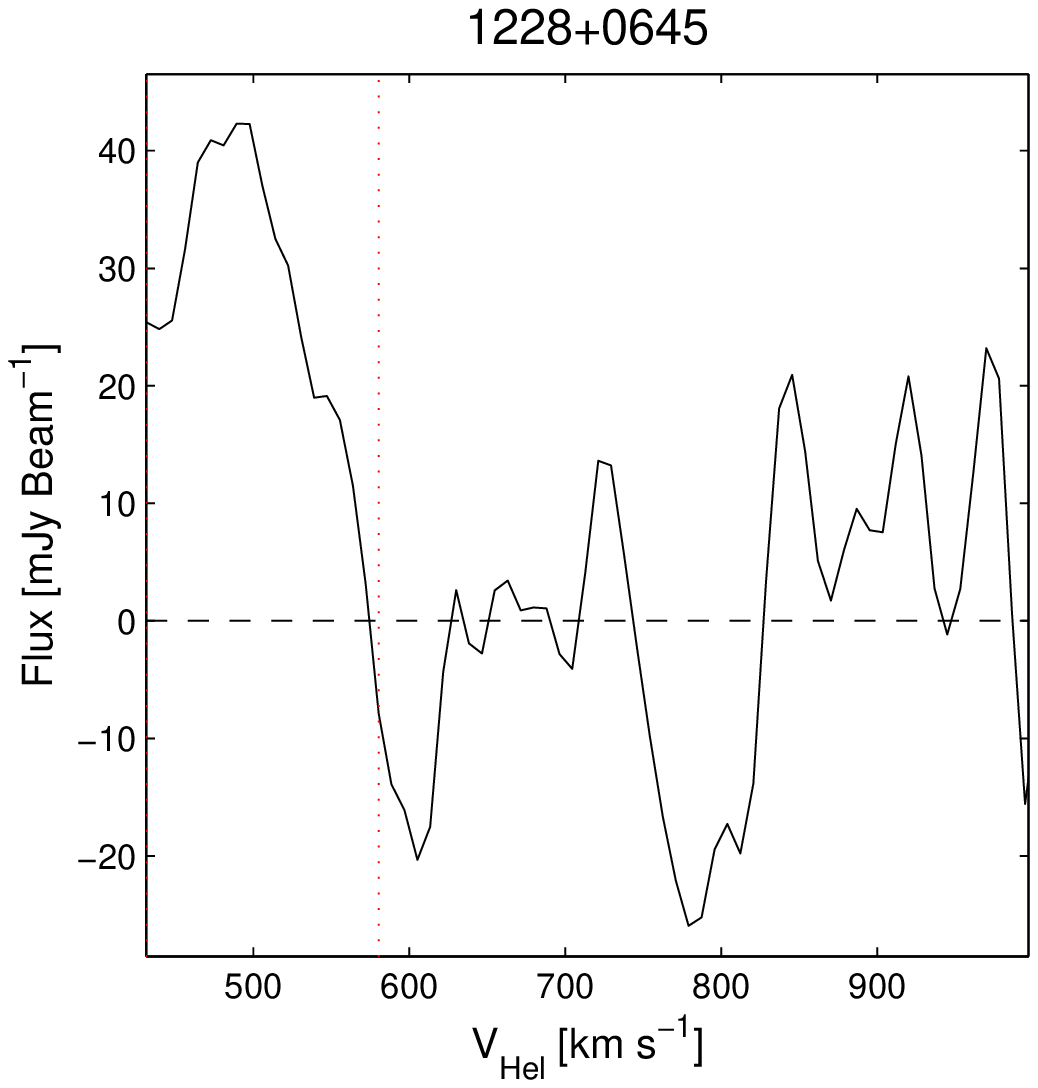}                                                        
                                                         
 \end{center}                                            
{\bf Fig~\ref{all_spectra}.} (continued)                                        
 
\end{figure*}

\begin{figure*}
  \begin{center}

 \includegraphics[width=0.22\textwidth]{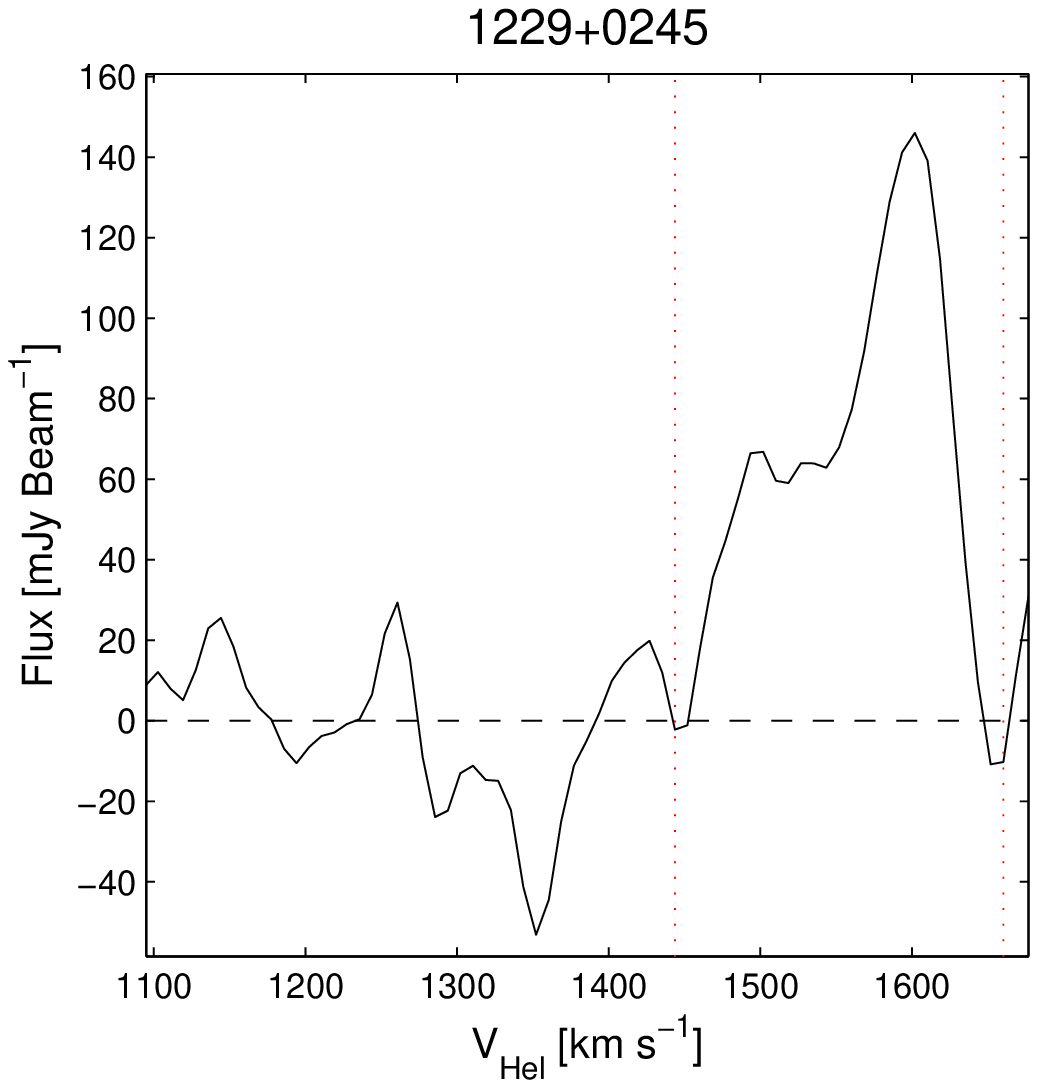}
 \includegraphics[width=0.22\textwidth]{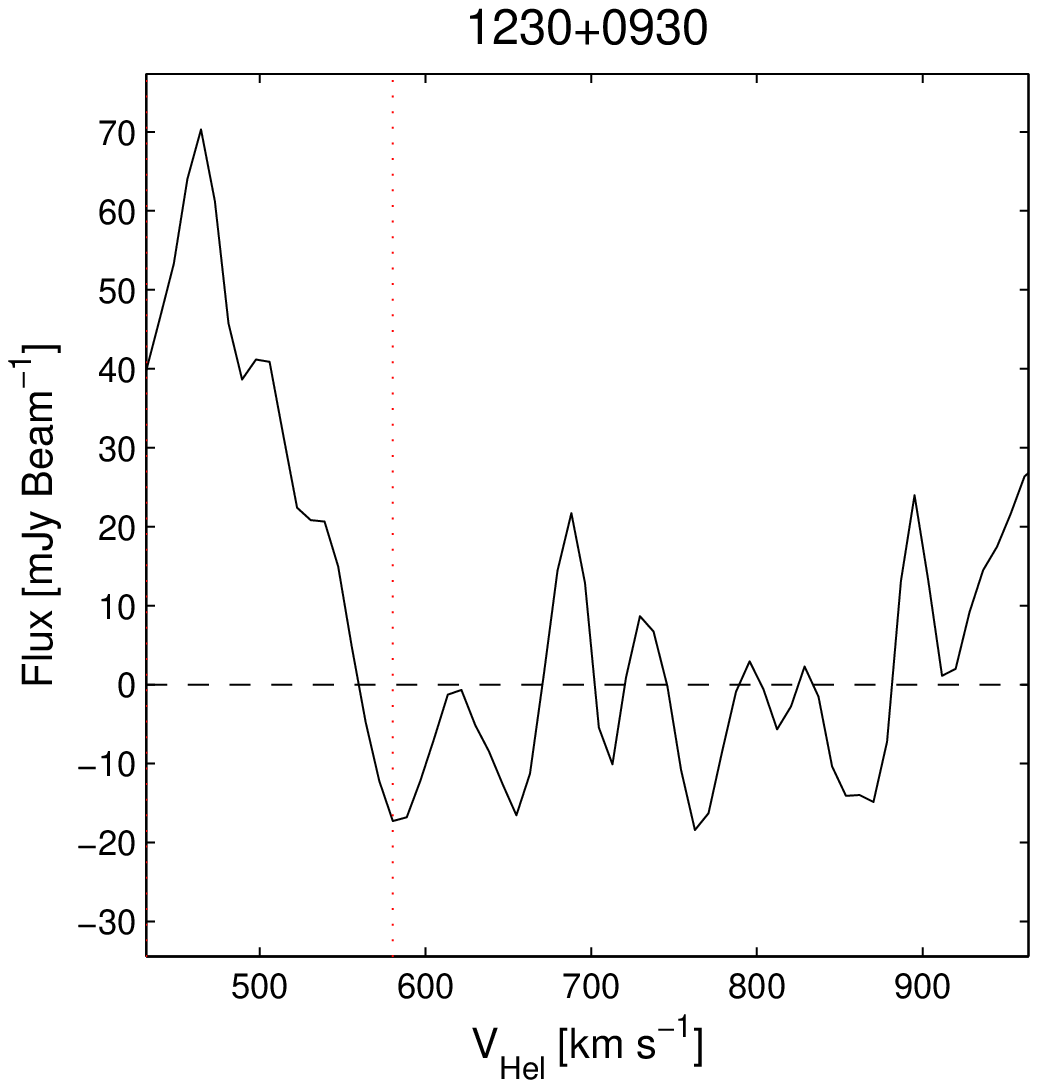}
 \includegraphics[width=0.22\textwidth]{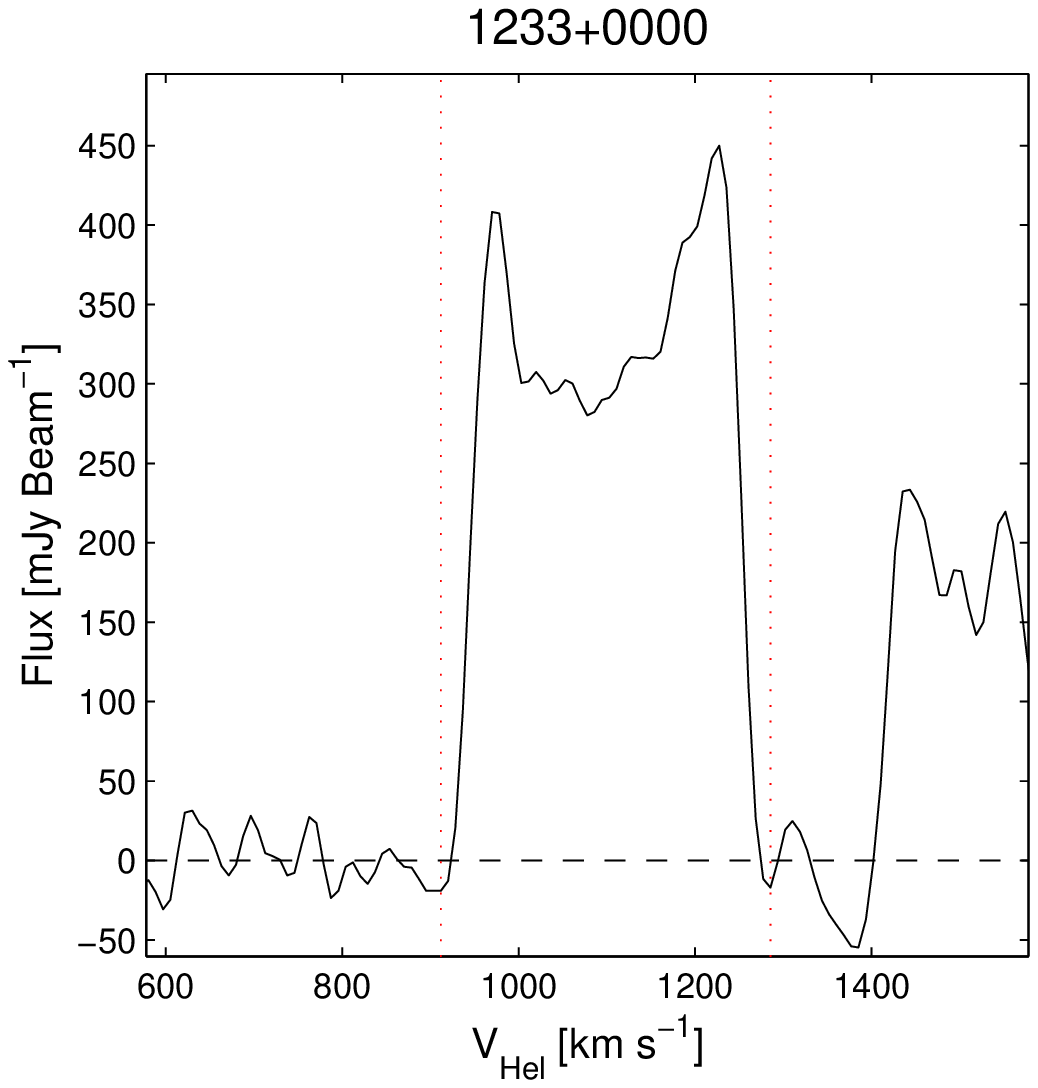}
 \includegraphics[width=0.22\textwidth]{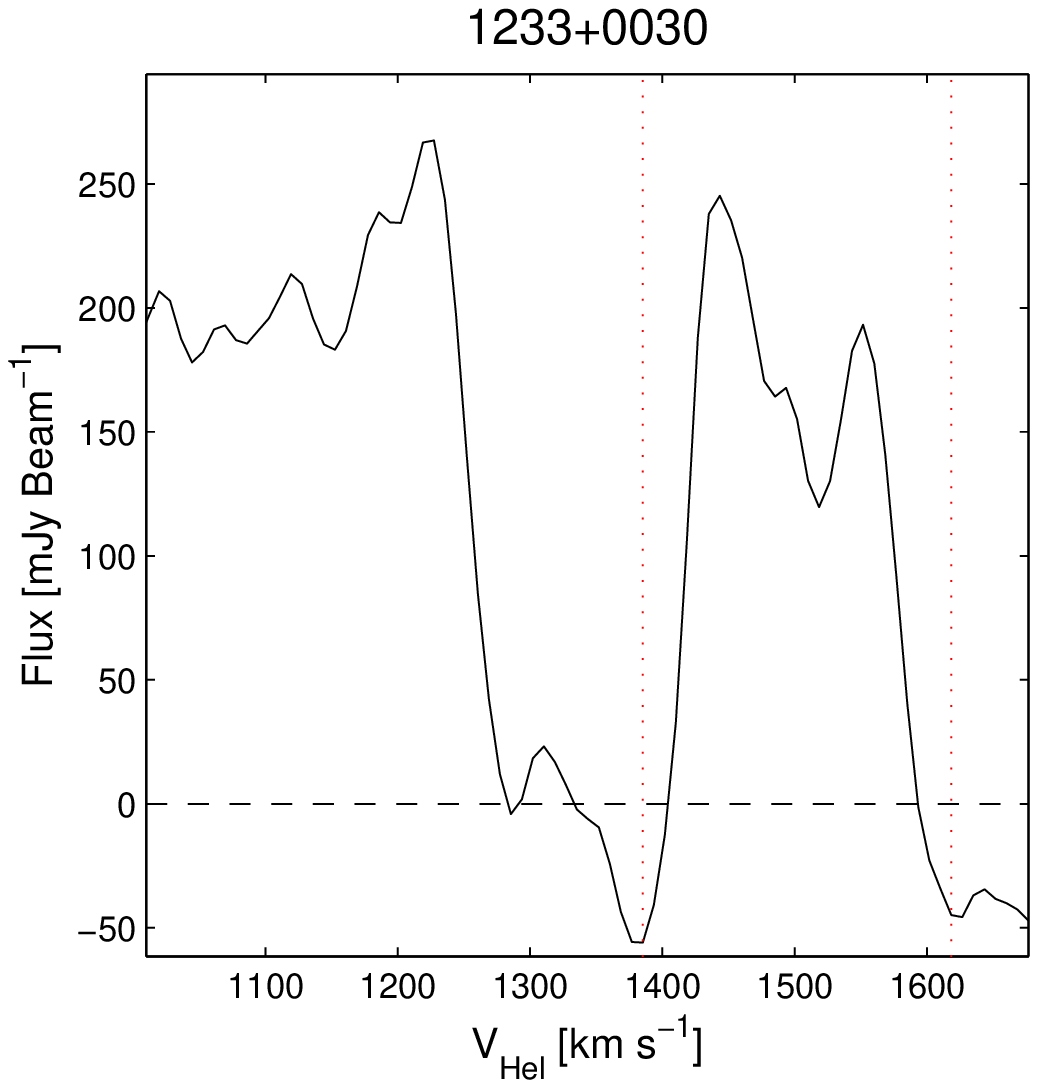}
 \includegraphics[width=0.22\textwidth]{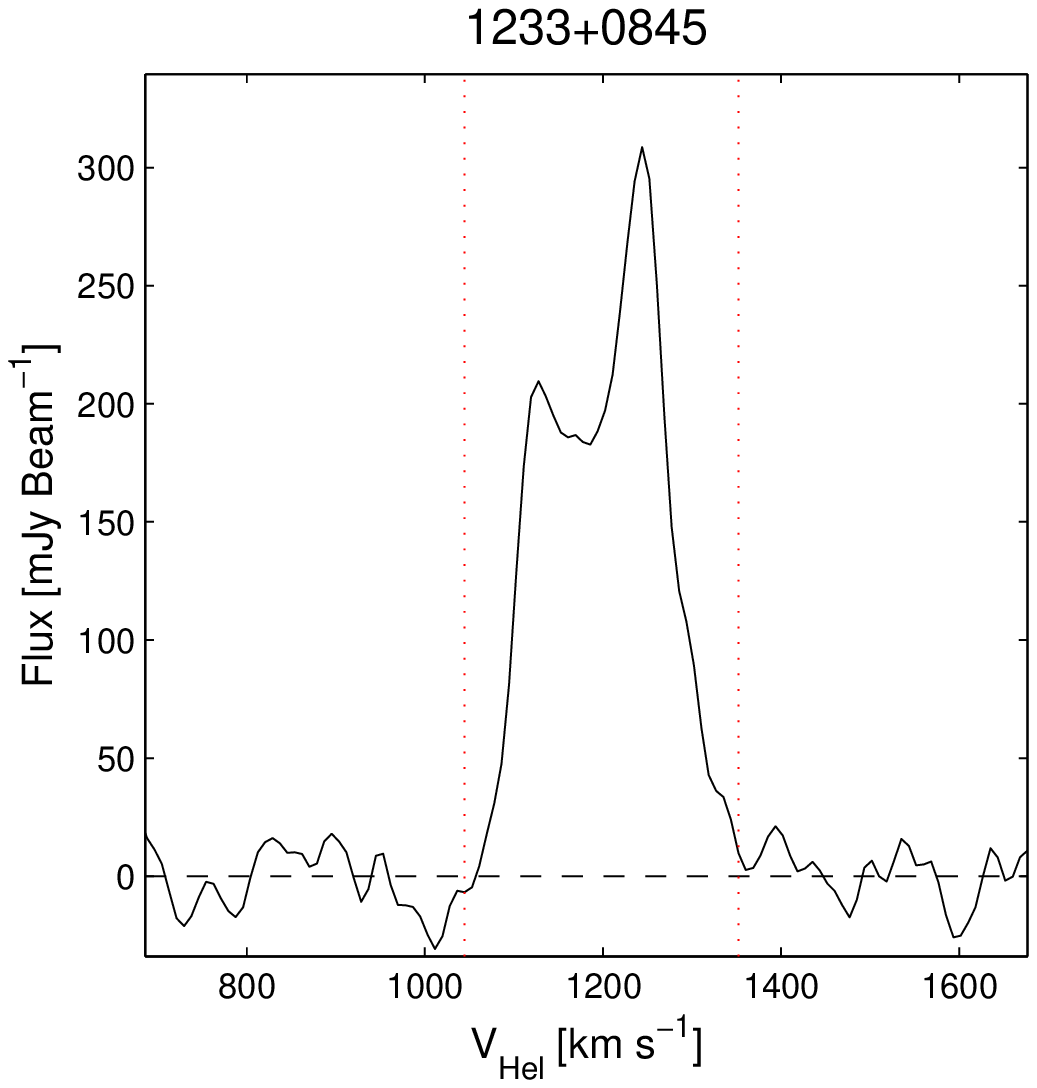}
 \includegraphics[width=0.22\textwidth]{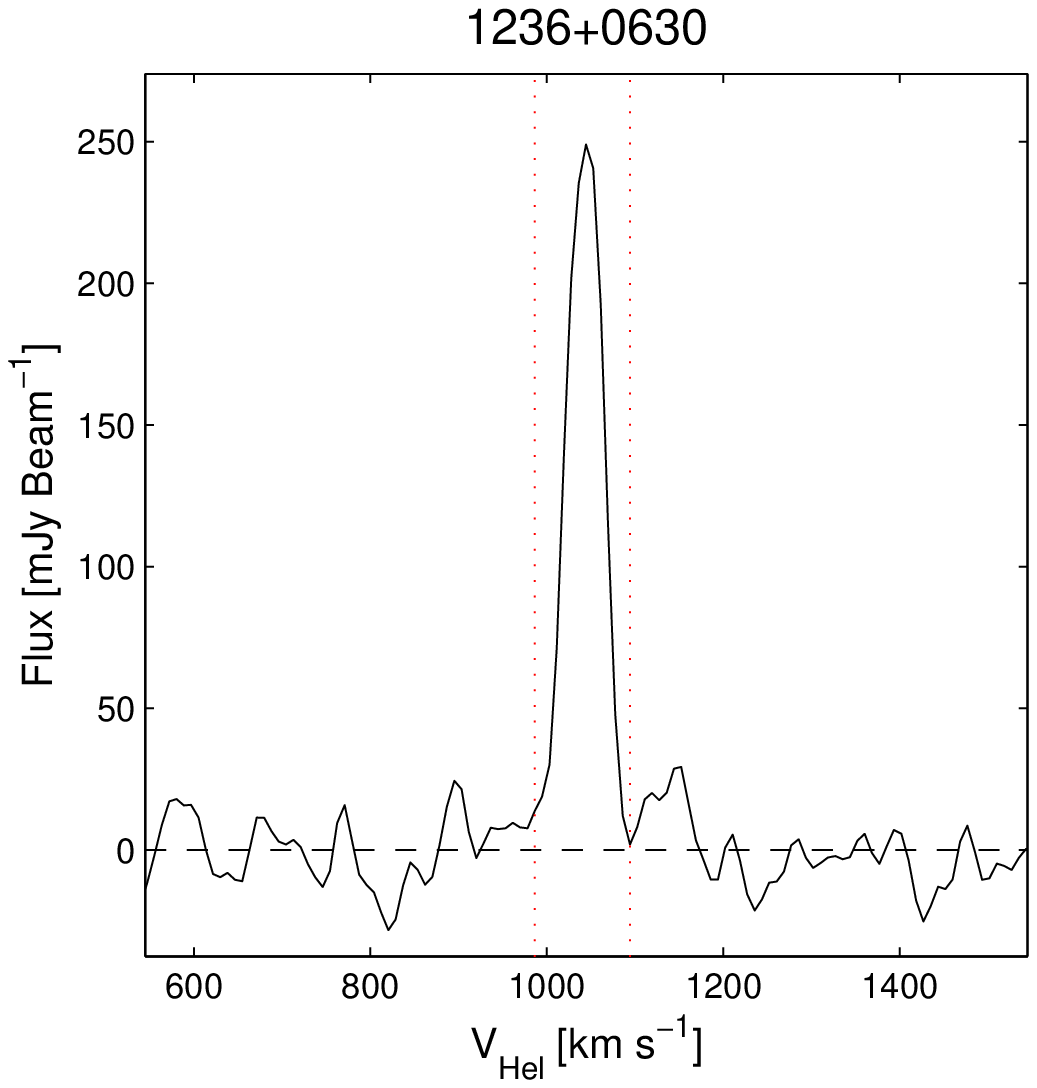}
 \includegraphics[width=0.22\textwidth]{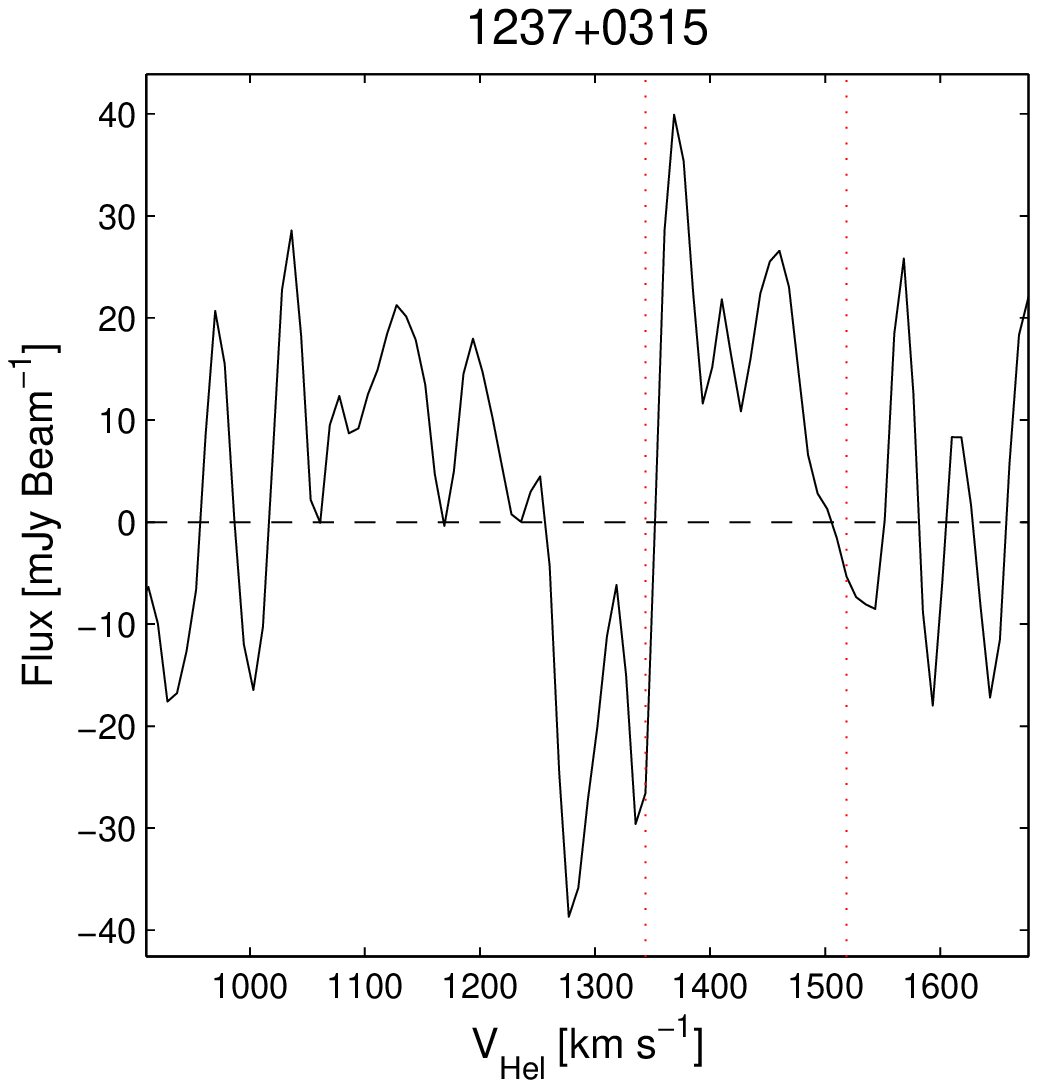}
 \includegraphics[width=0.22\textwidth]{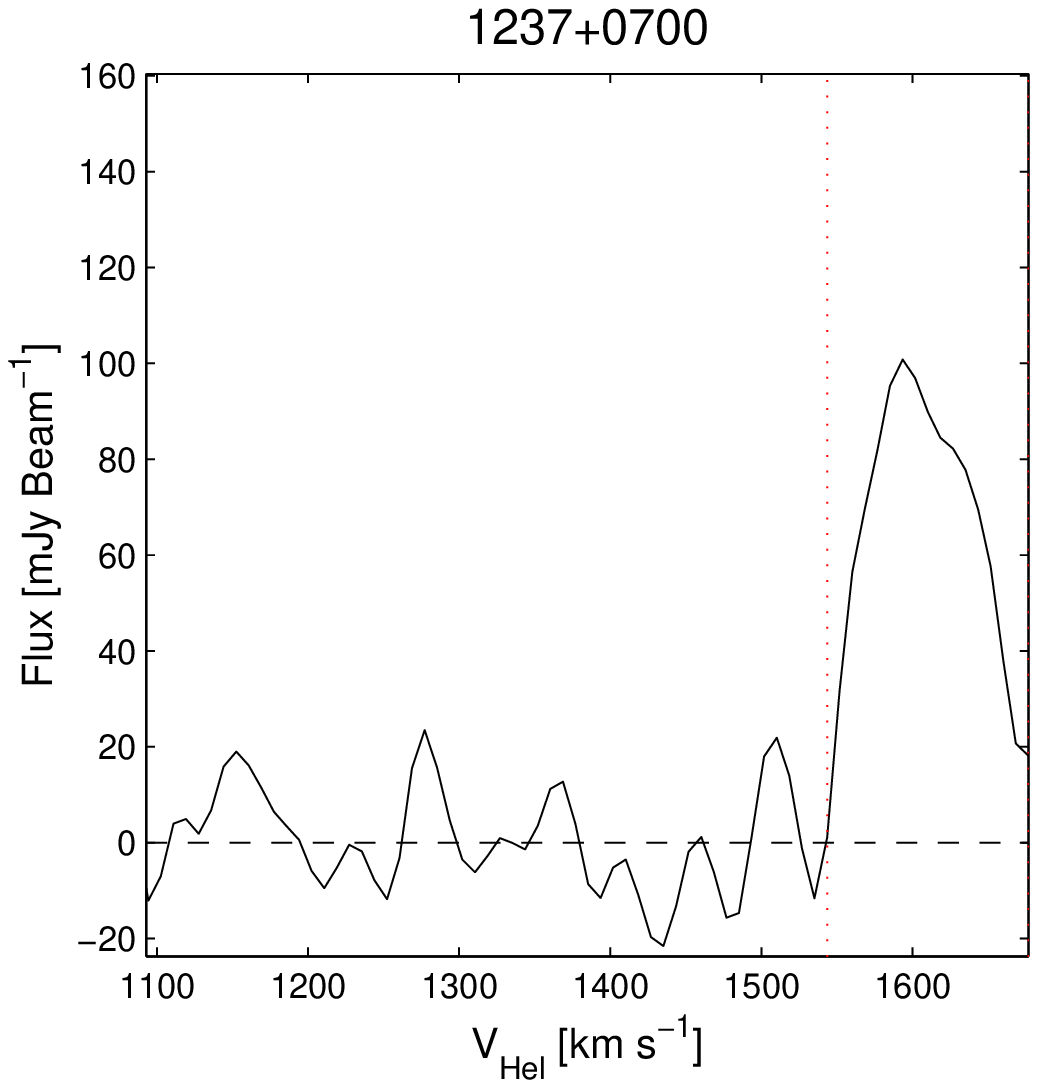}
 \includegraphics[width=0.22\textwidth]{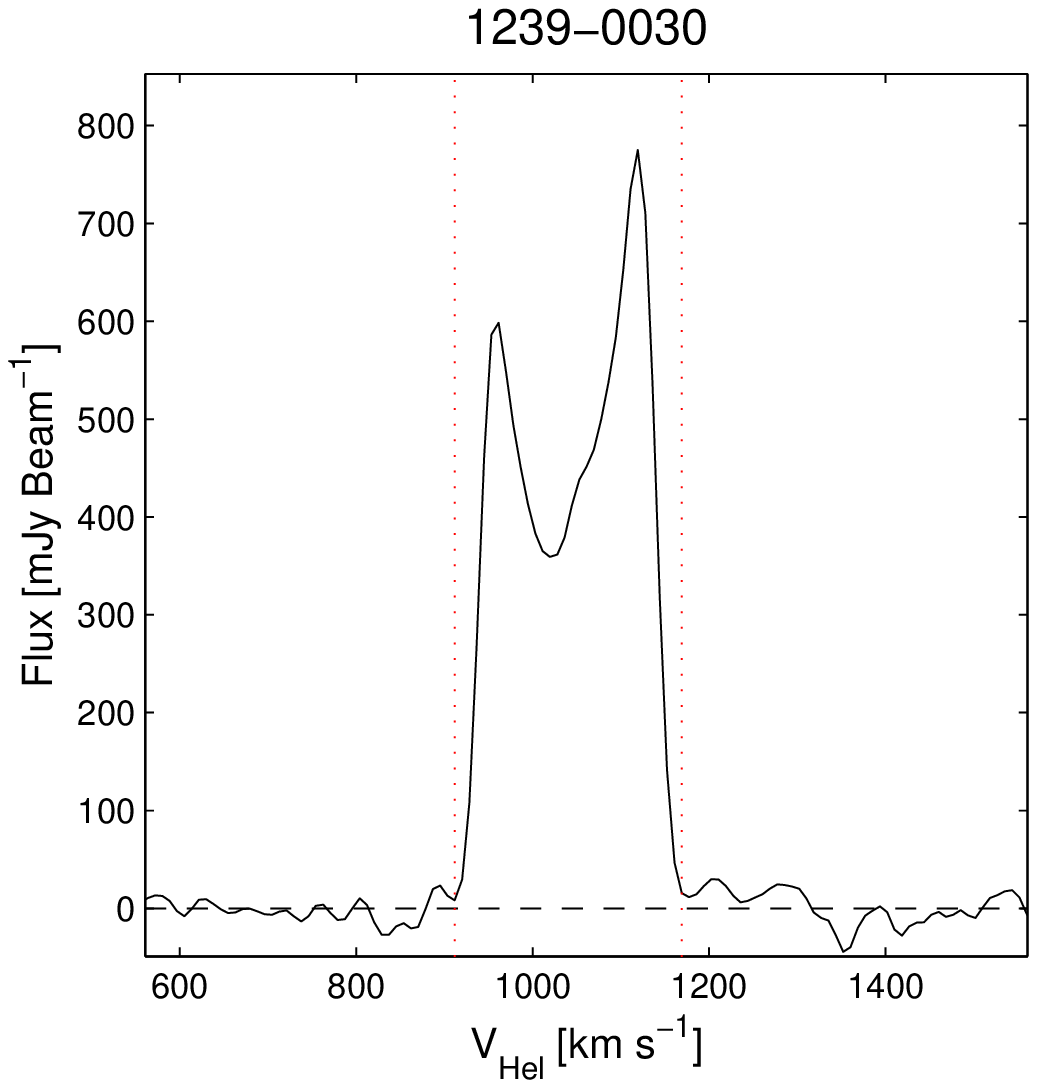}
 \includegraphics[width=0.22\textwidth]{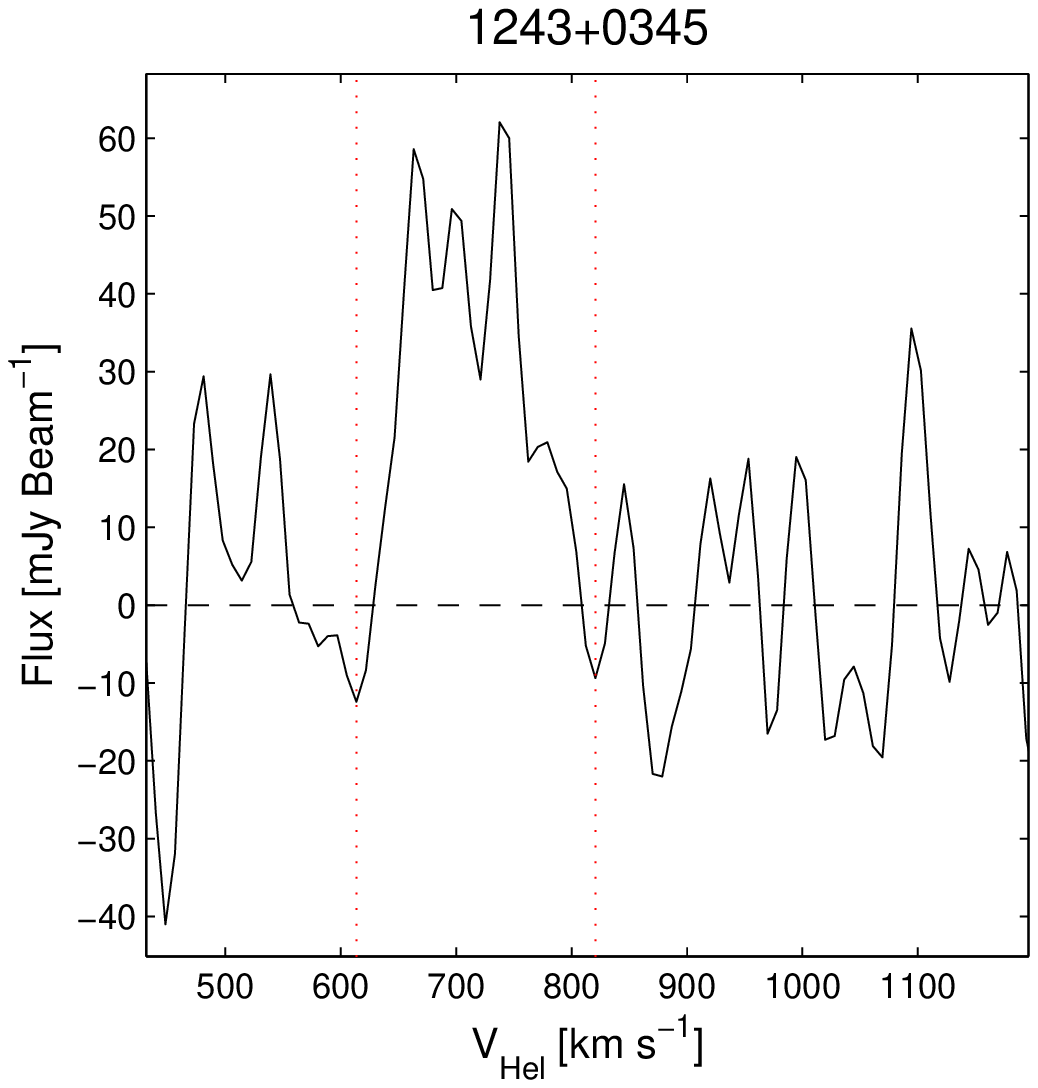}
 \includegraphics[width=0.22\textwidth]{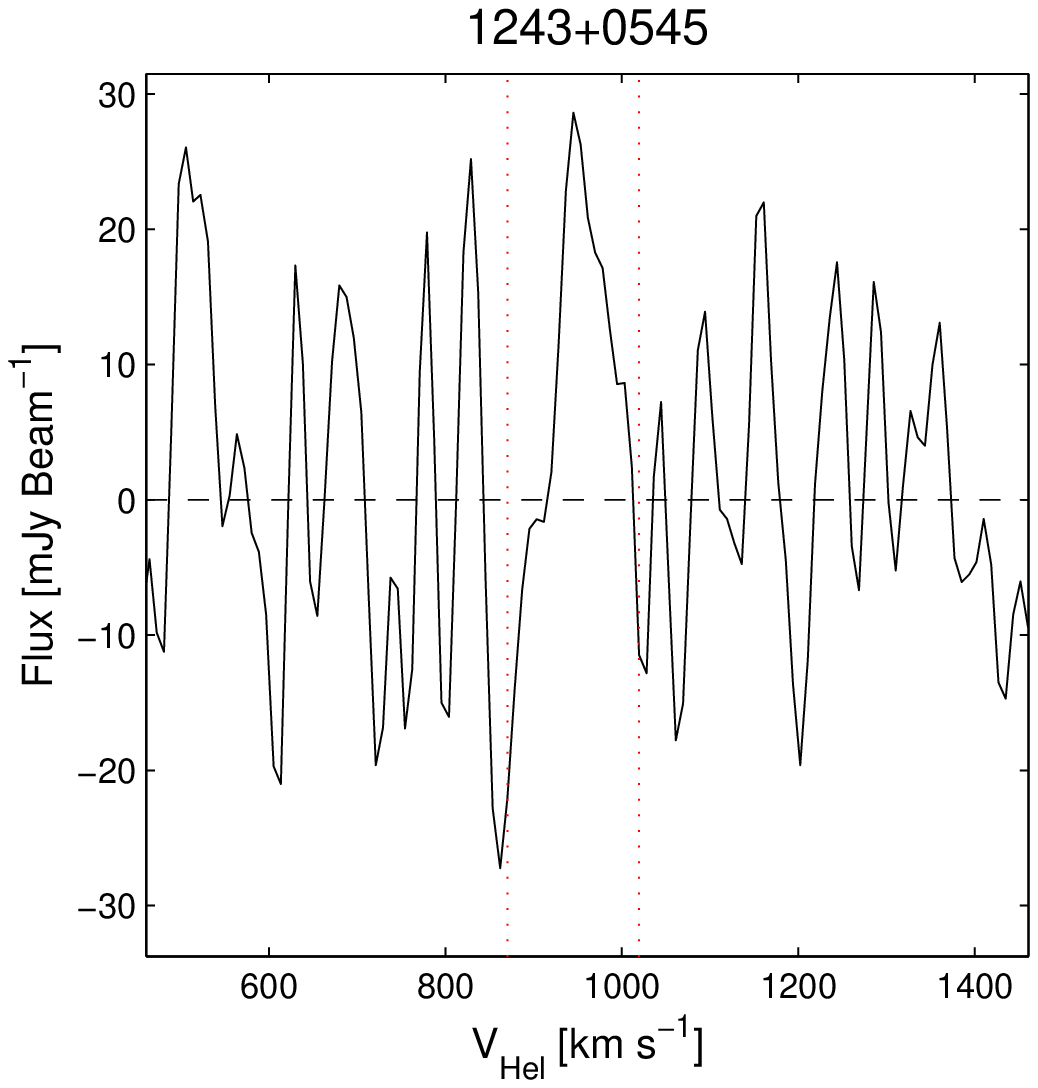}
 \includegraphics[width=0.22\textwidth]{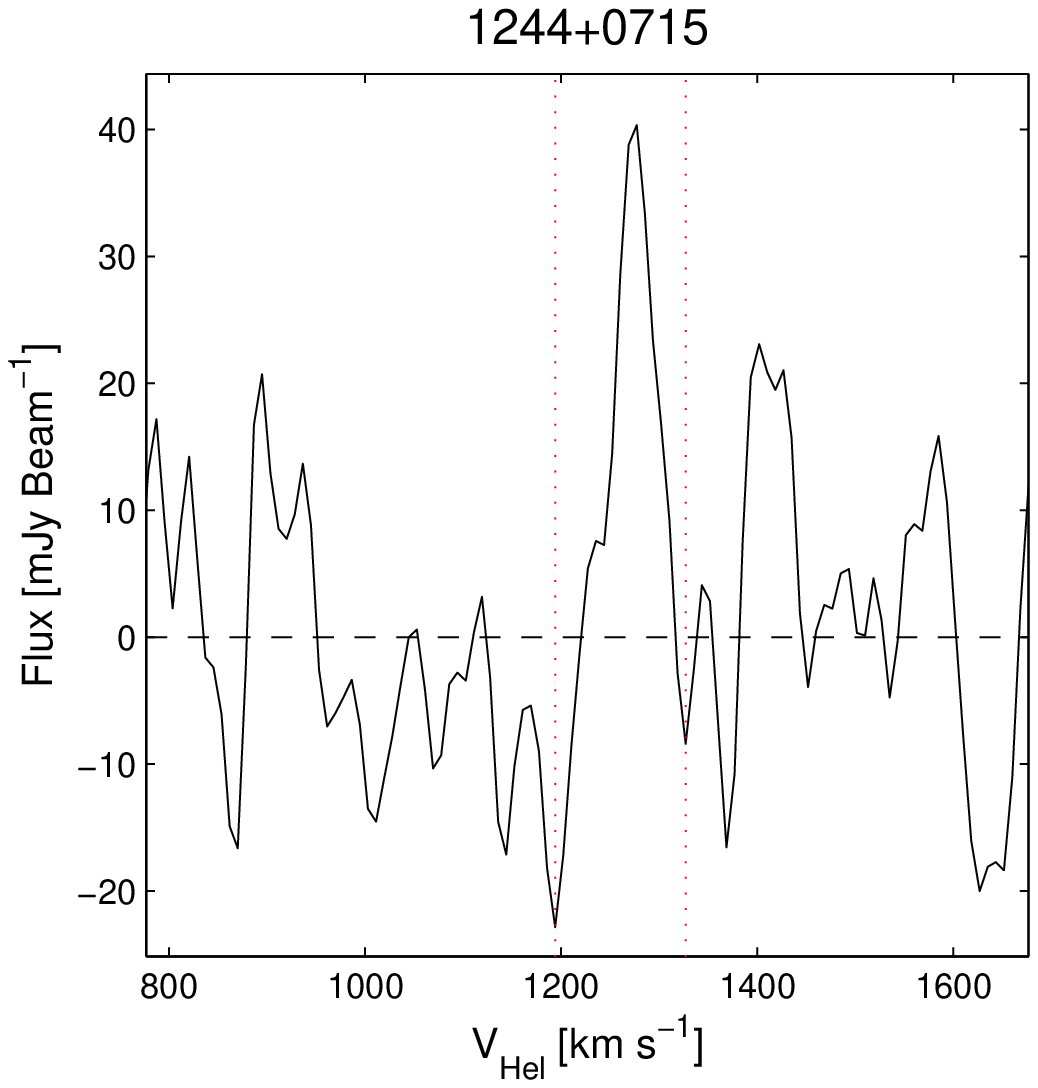}
 \includegraphics[width=0.22\textwidth]{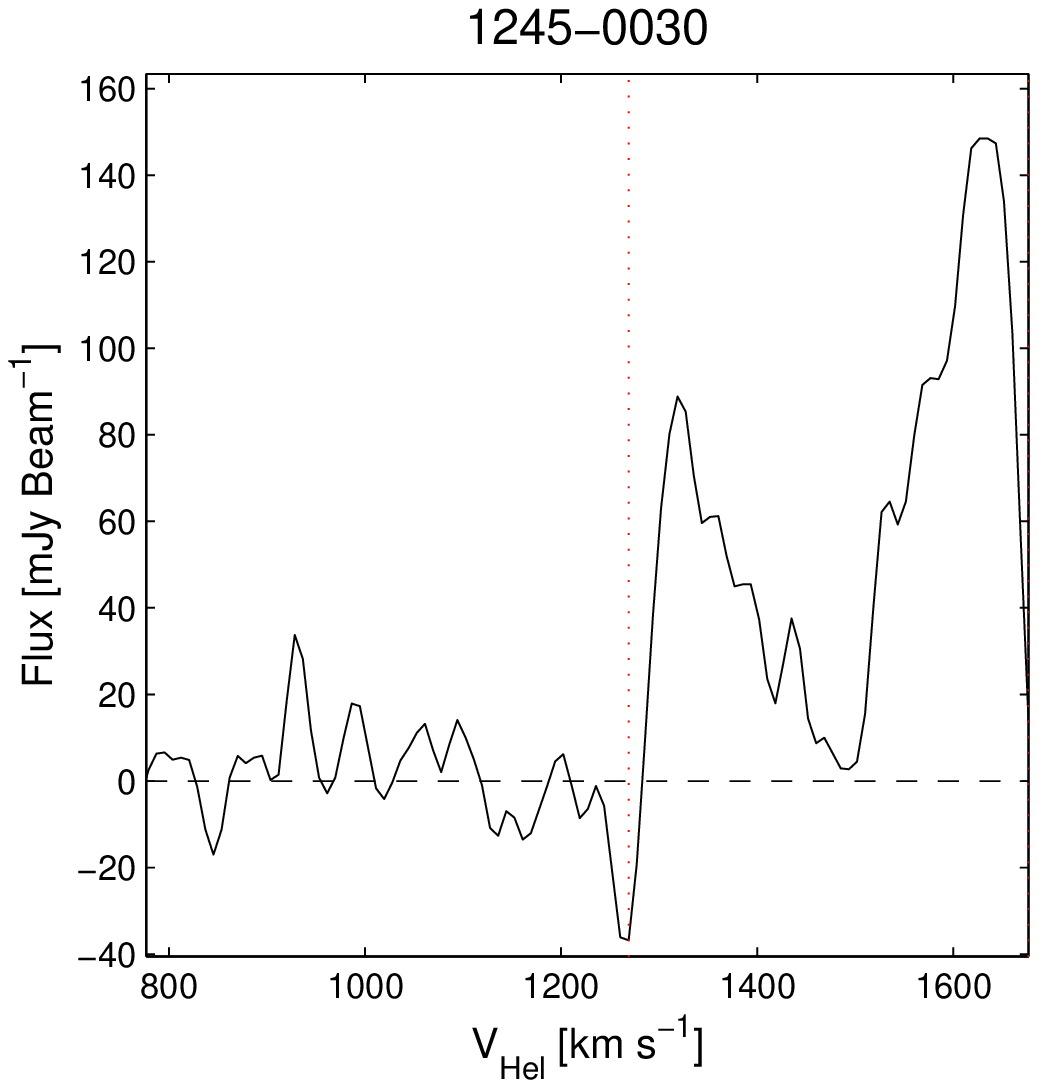}
 \includegraphics[width=0.22\textwidth]{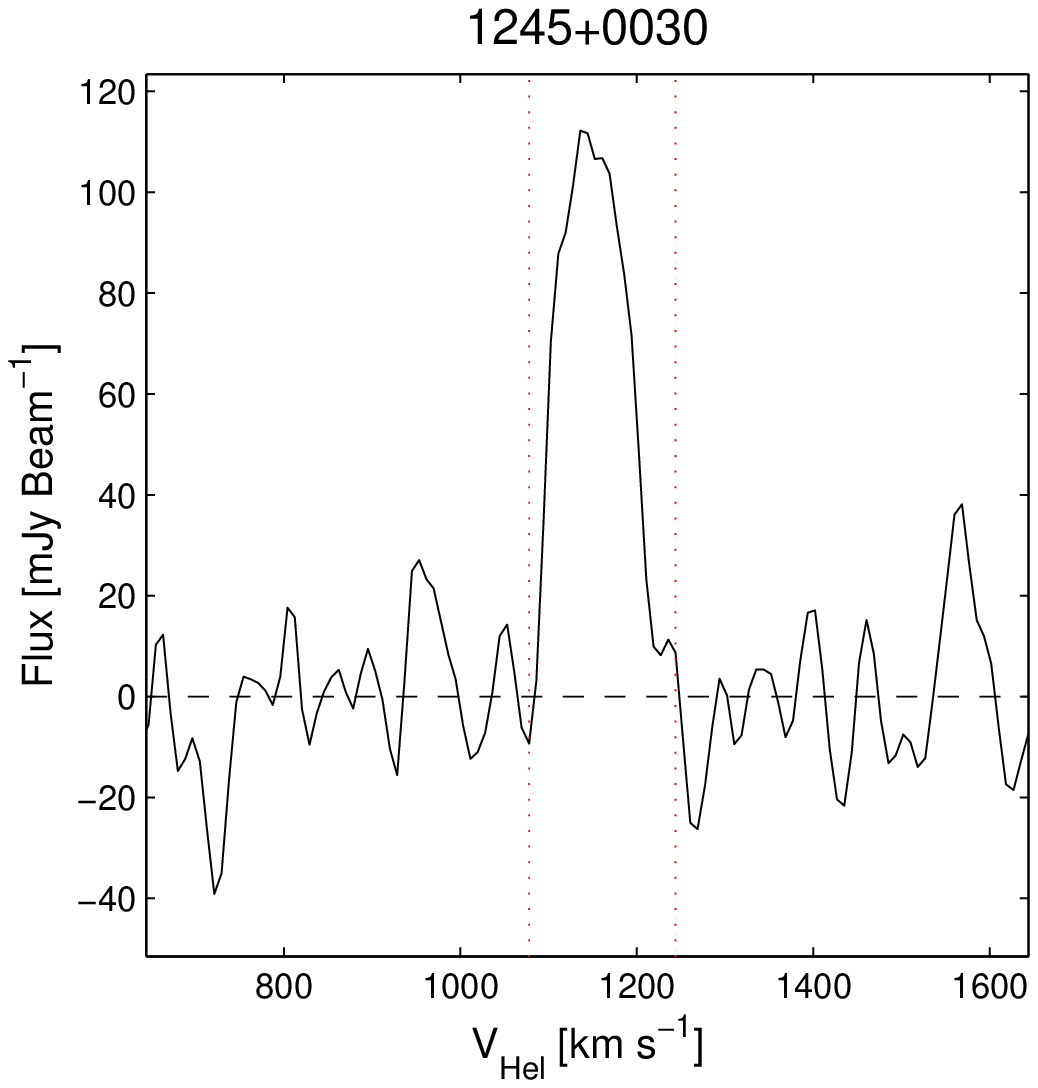}
 \includegraphics[width=0.22\textwidth]{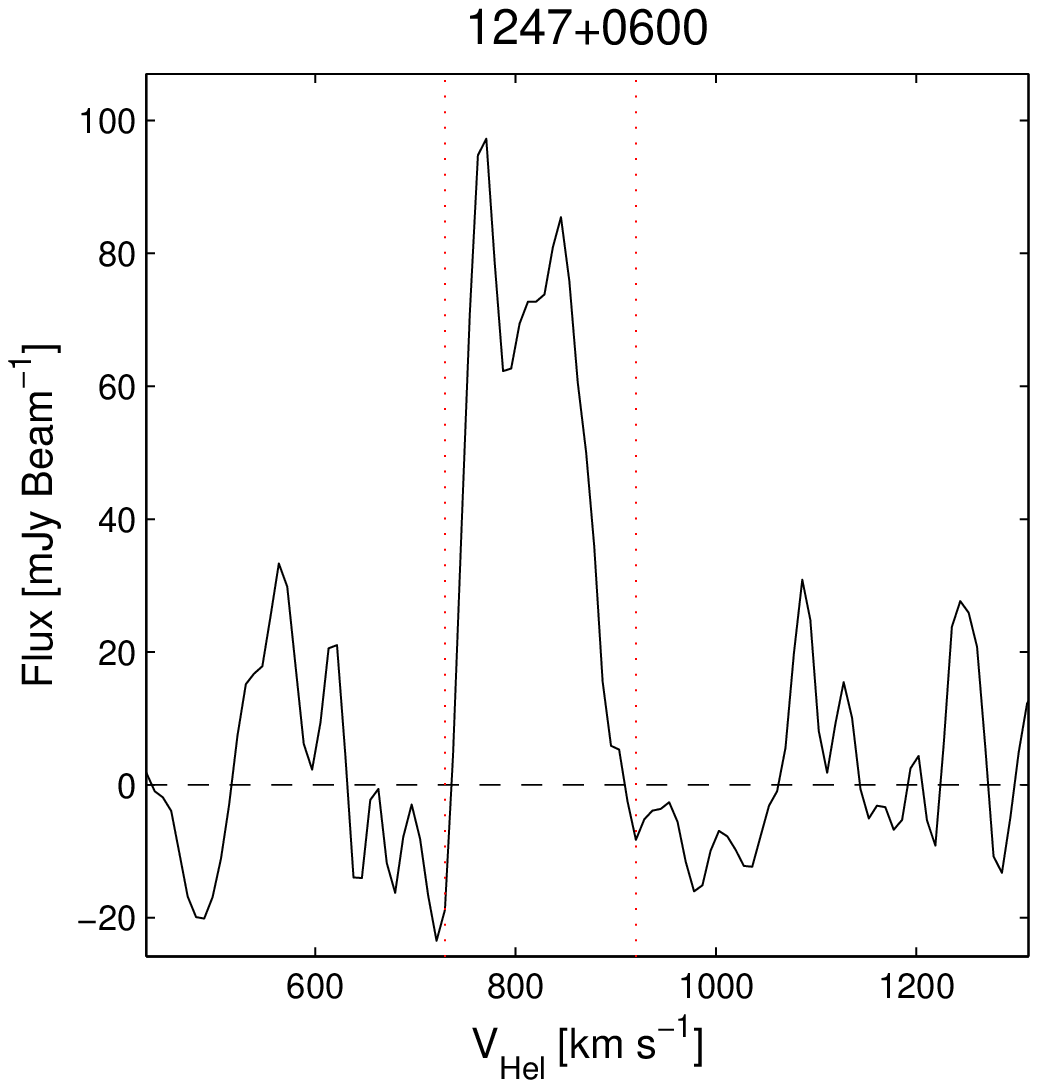}
 \includegraphics[width=0.22\textwidth]{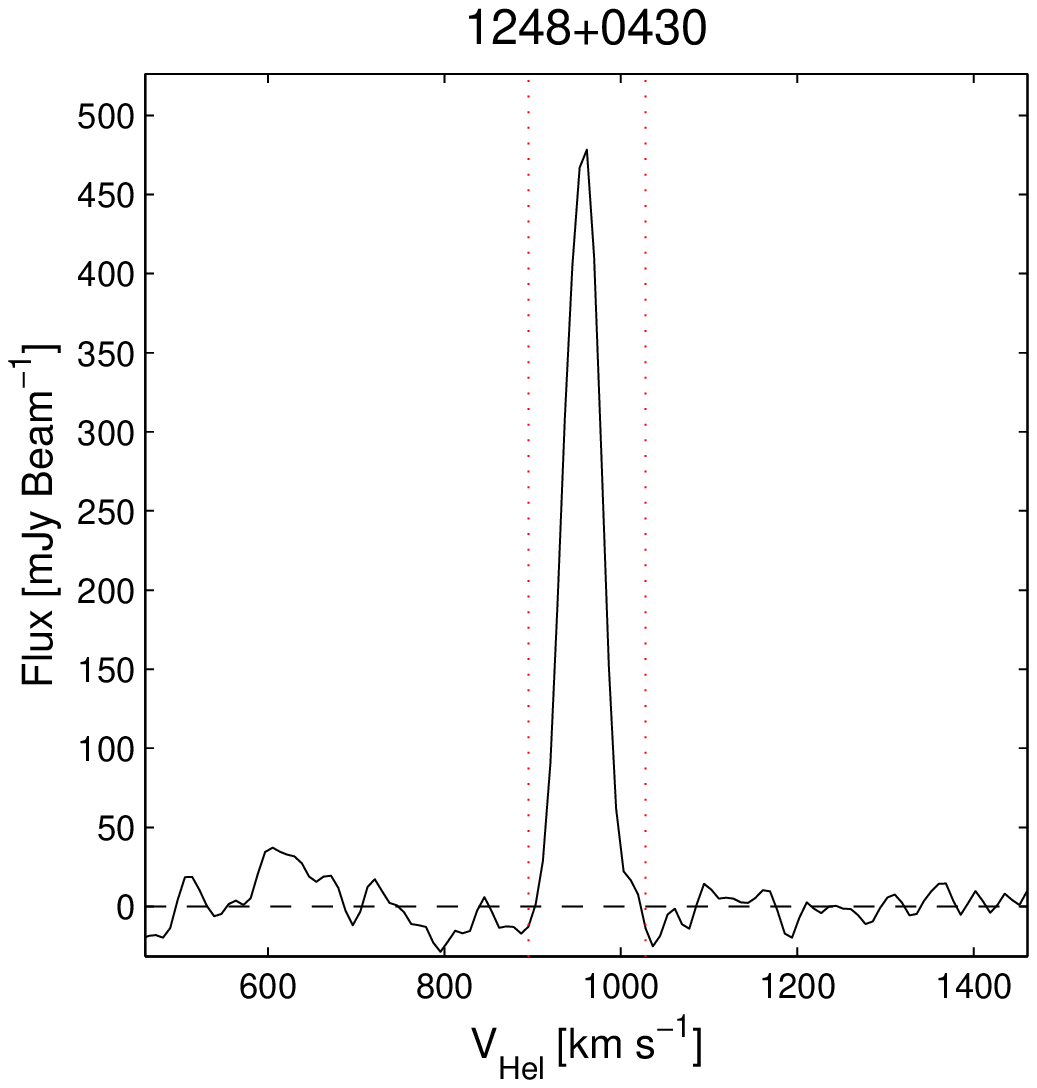}
 \includegraphics[width=0.22\textwidth]{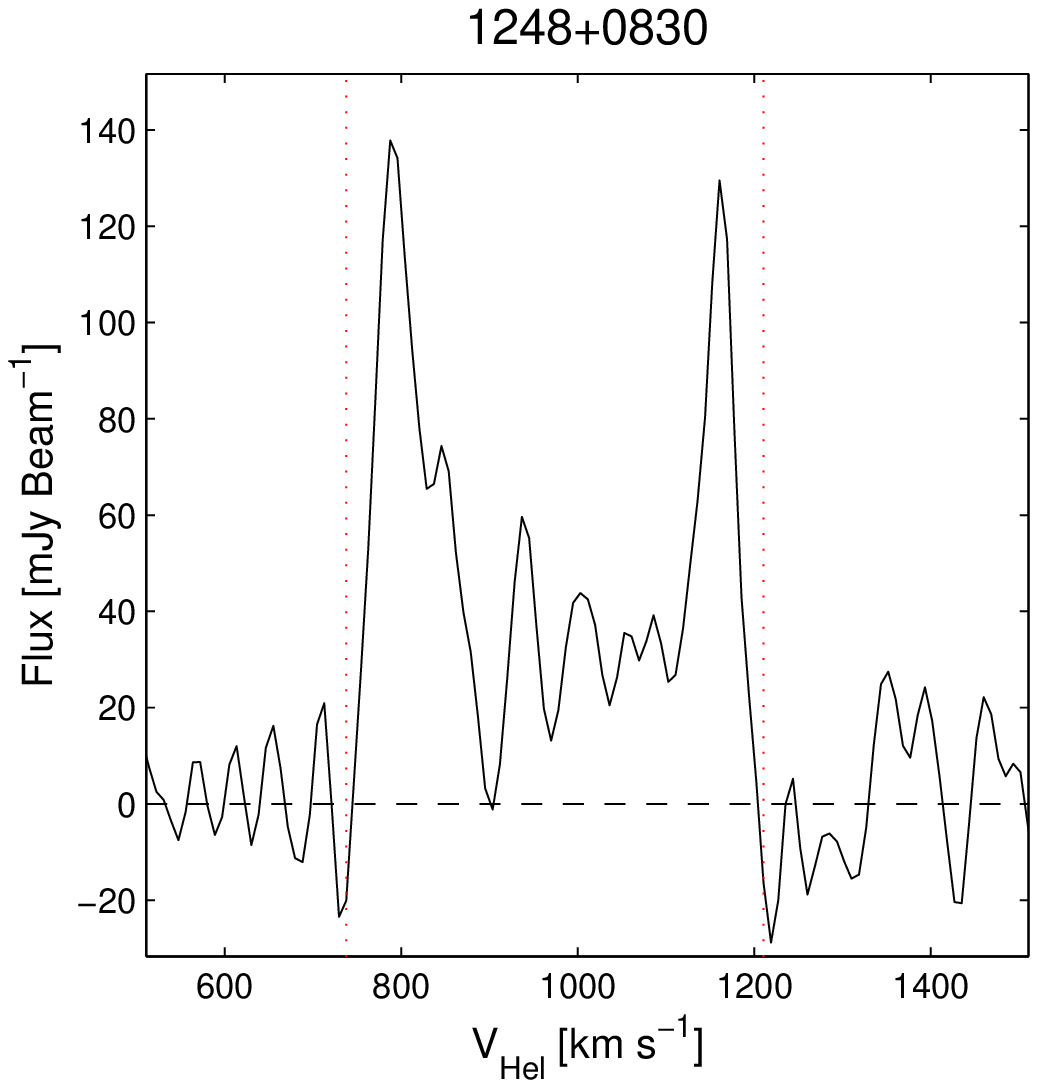}
 \includegraphics[width=0.22\textwidth]{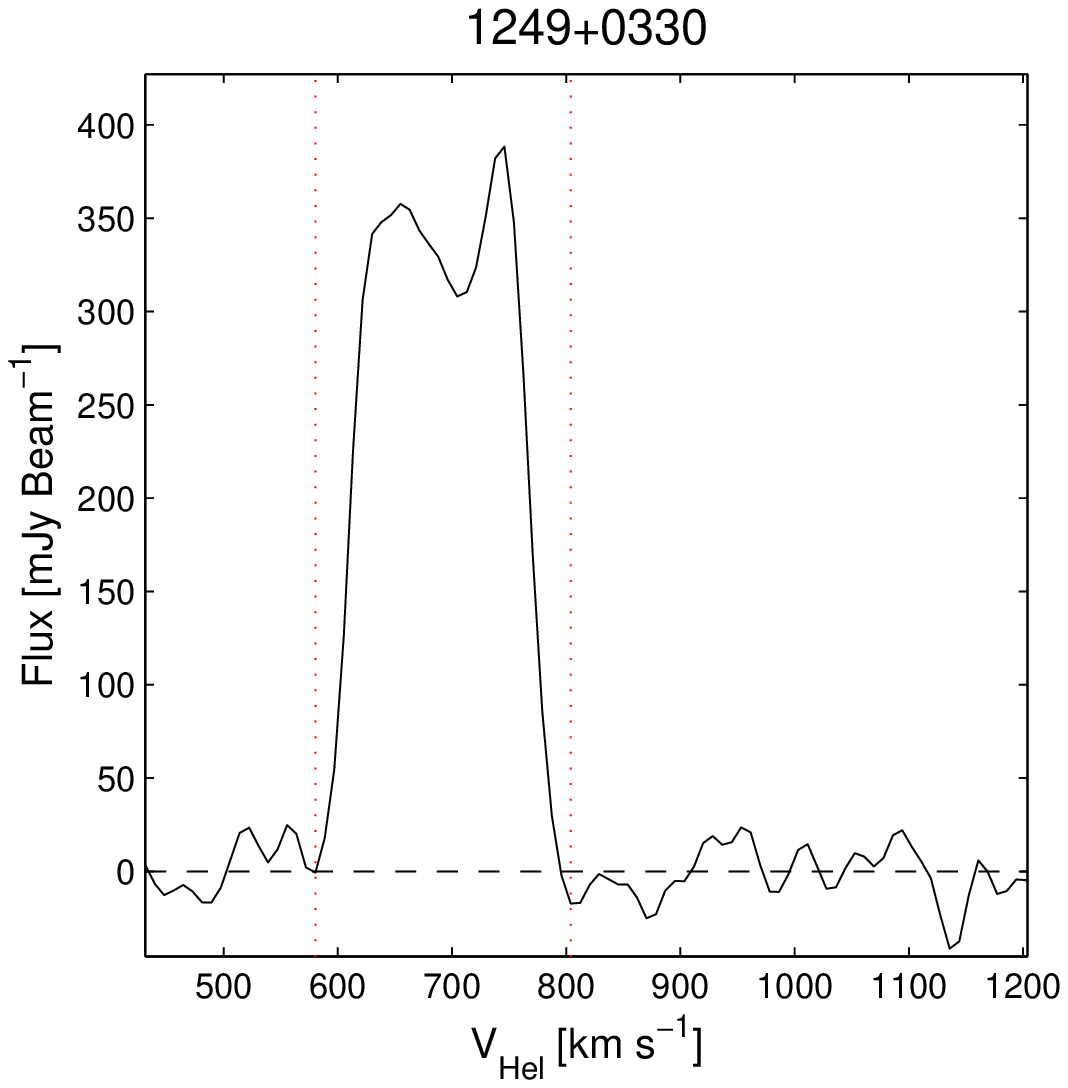}
 \includegraphics[width=0.22\textwidth]{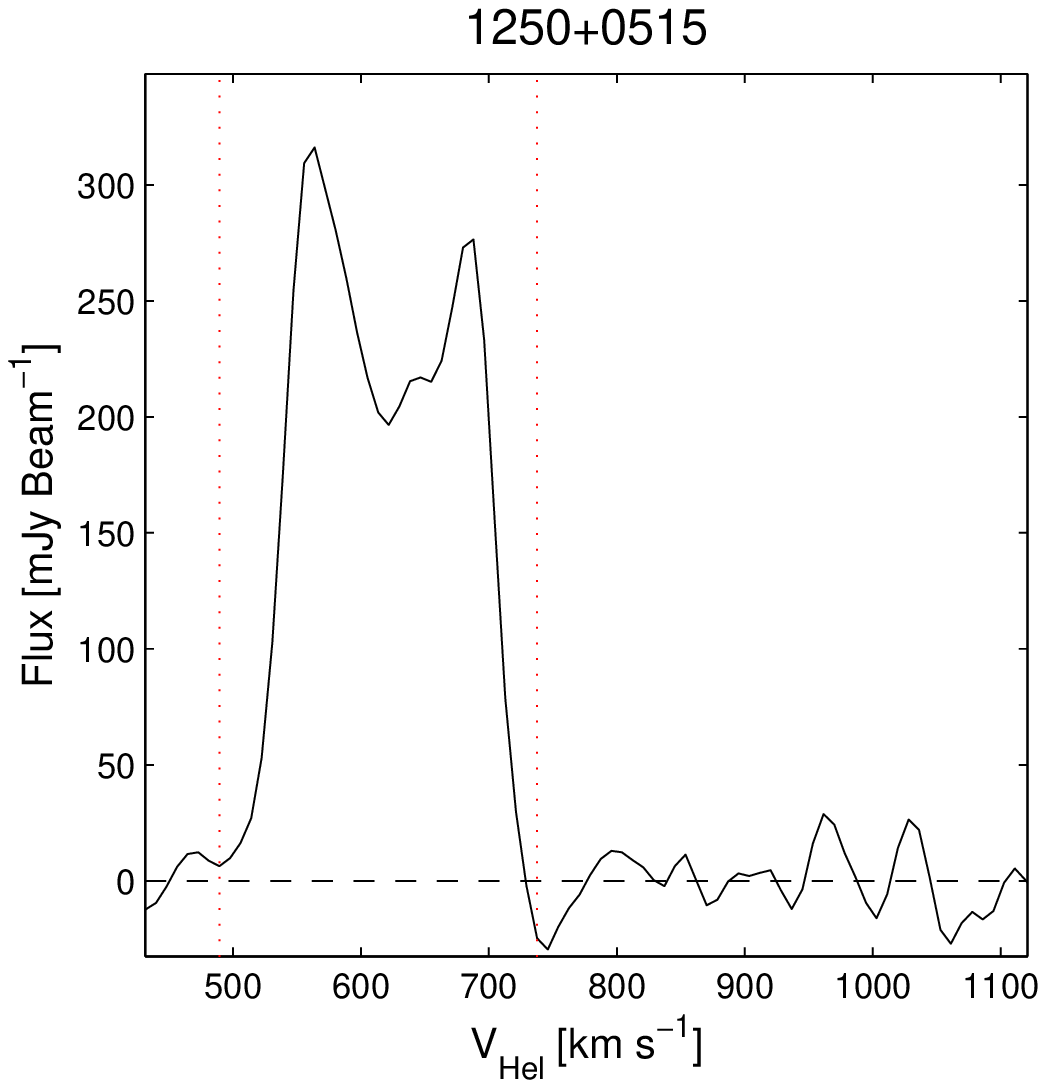}
 \includegraphics[width=0.22\textwidth]{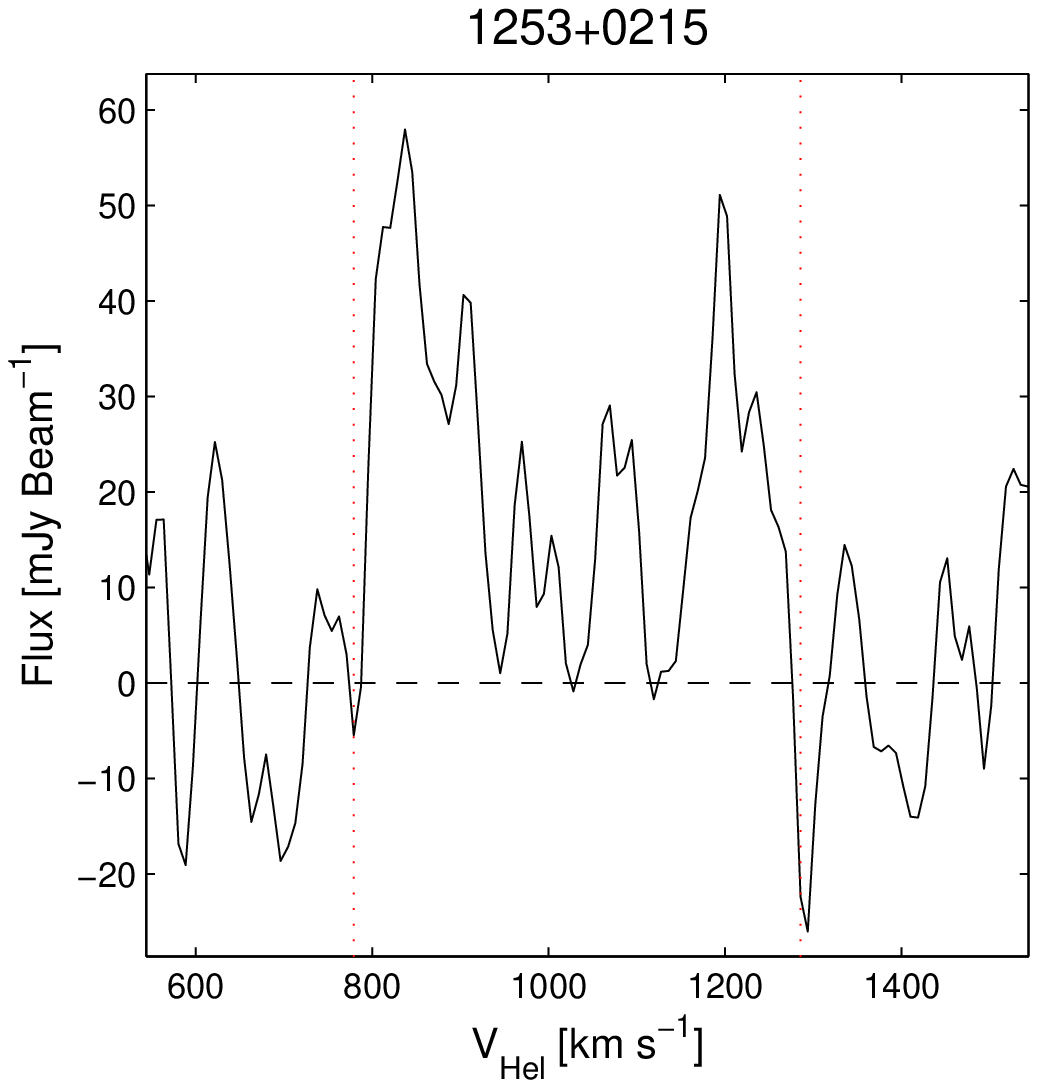}                                                        
                                                         
 \end{center}                                            
{\bf Fig~\ref{all_spectra}.} (continued)                                        
 
\end{figure*}

\begin{figure*}
  \begin{center}

 \includegraphics[width=0.22\textwidth]{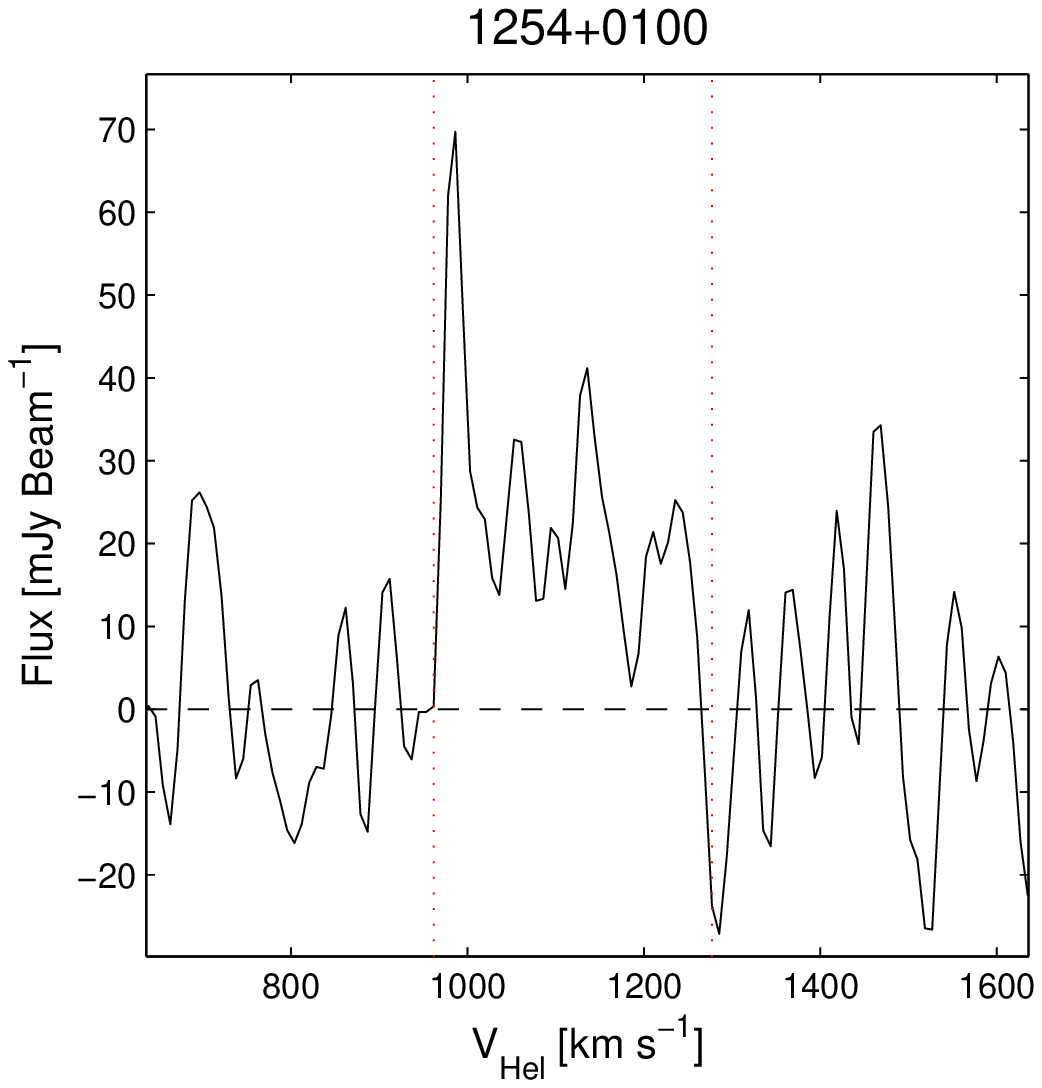}
 \includegraphics[width=0.22\textwidth]{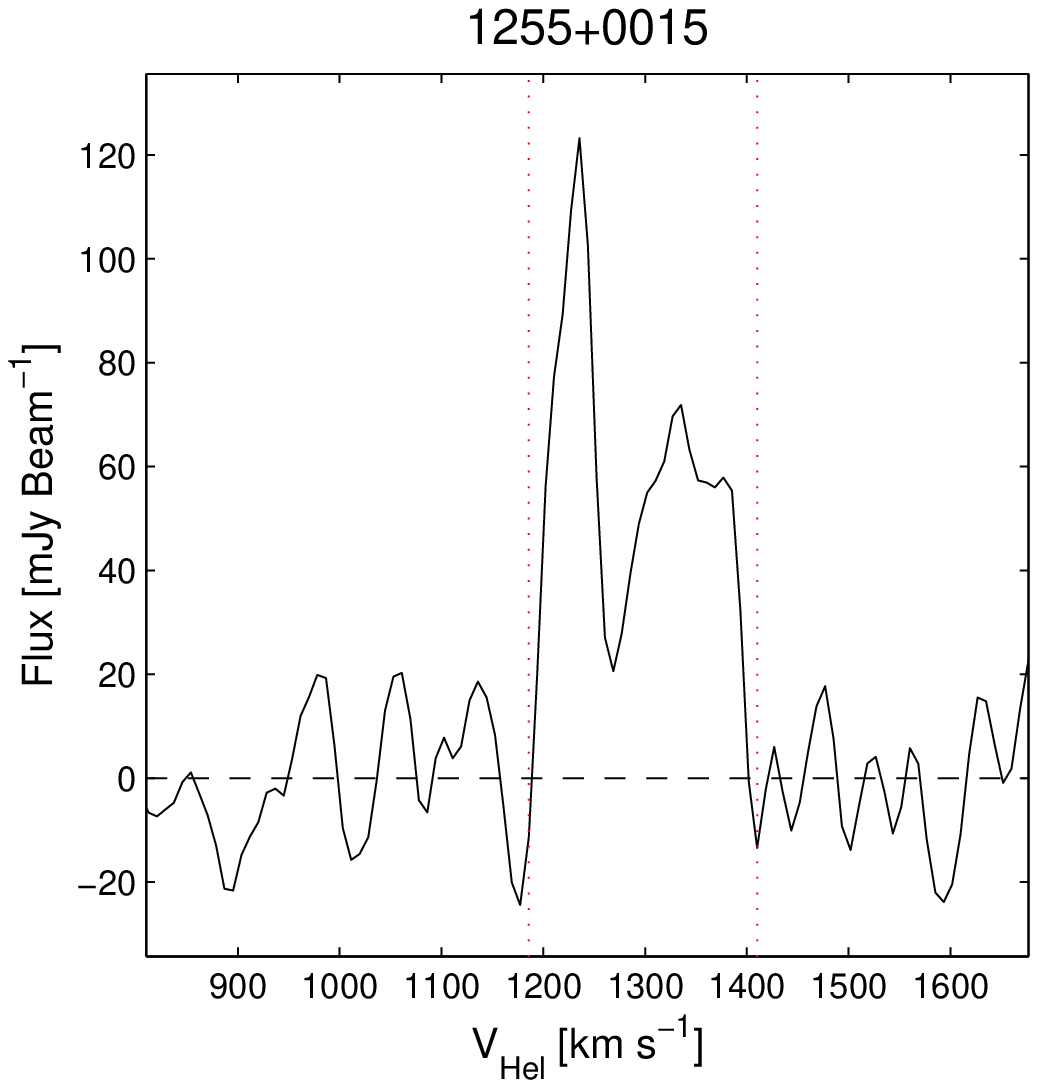}
 \includegraphics[width=0.22\textwidth]{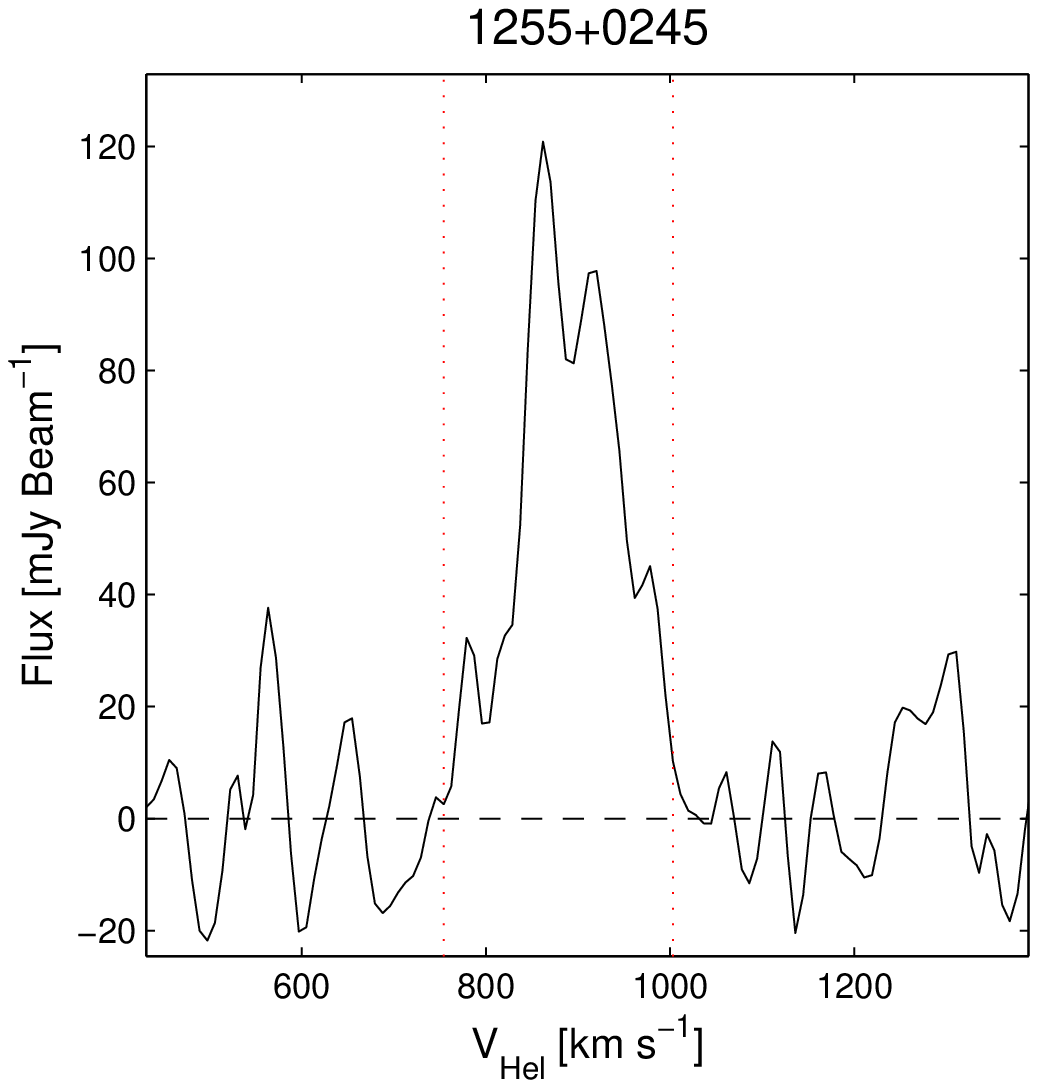}
 \includegraphics[width=0.22\textwidth]{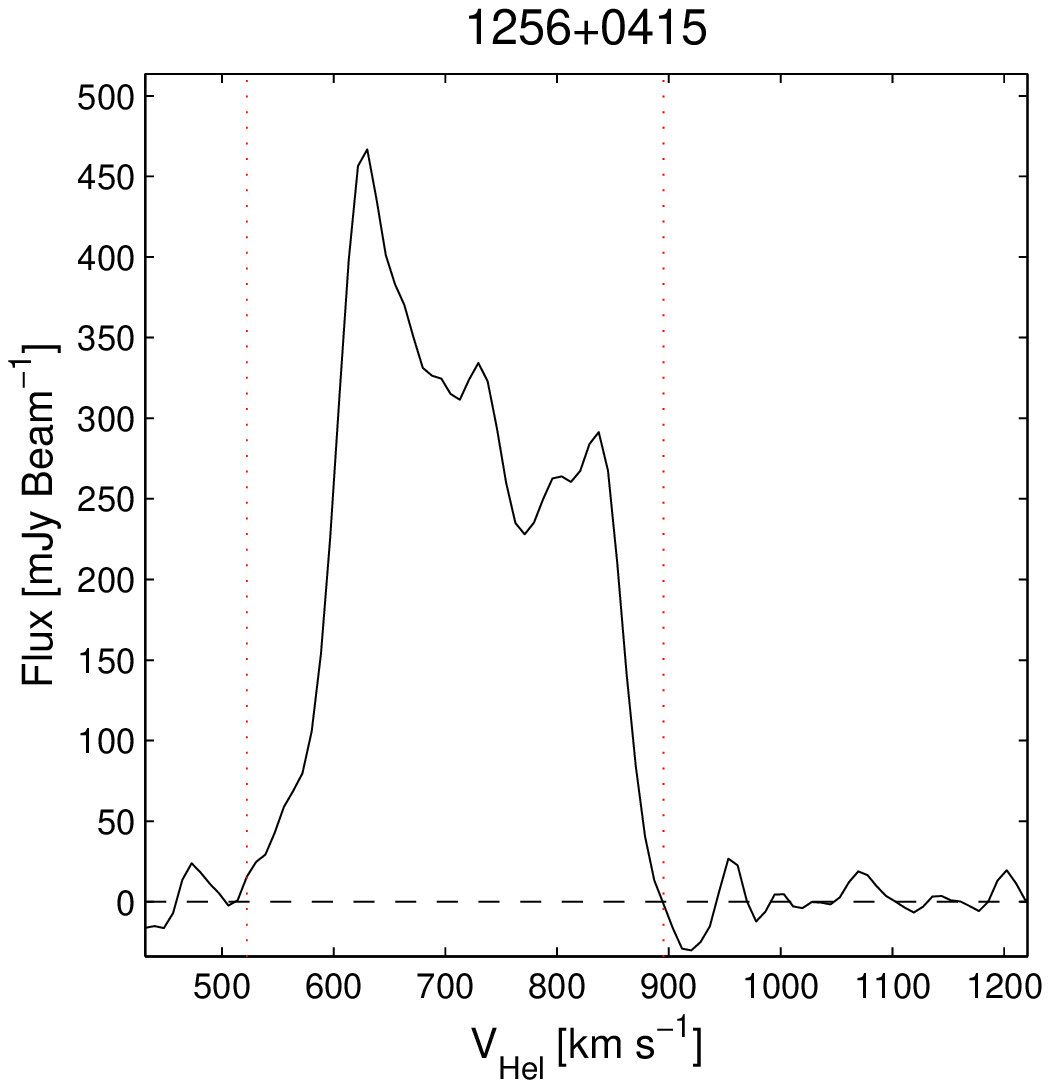}
 \includegraphics[width=0.22\textwidth]{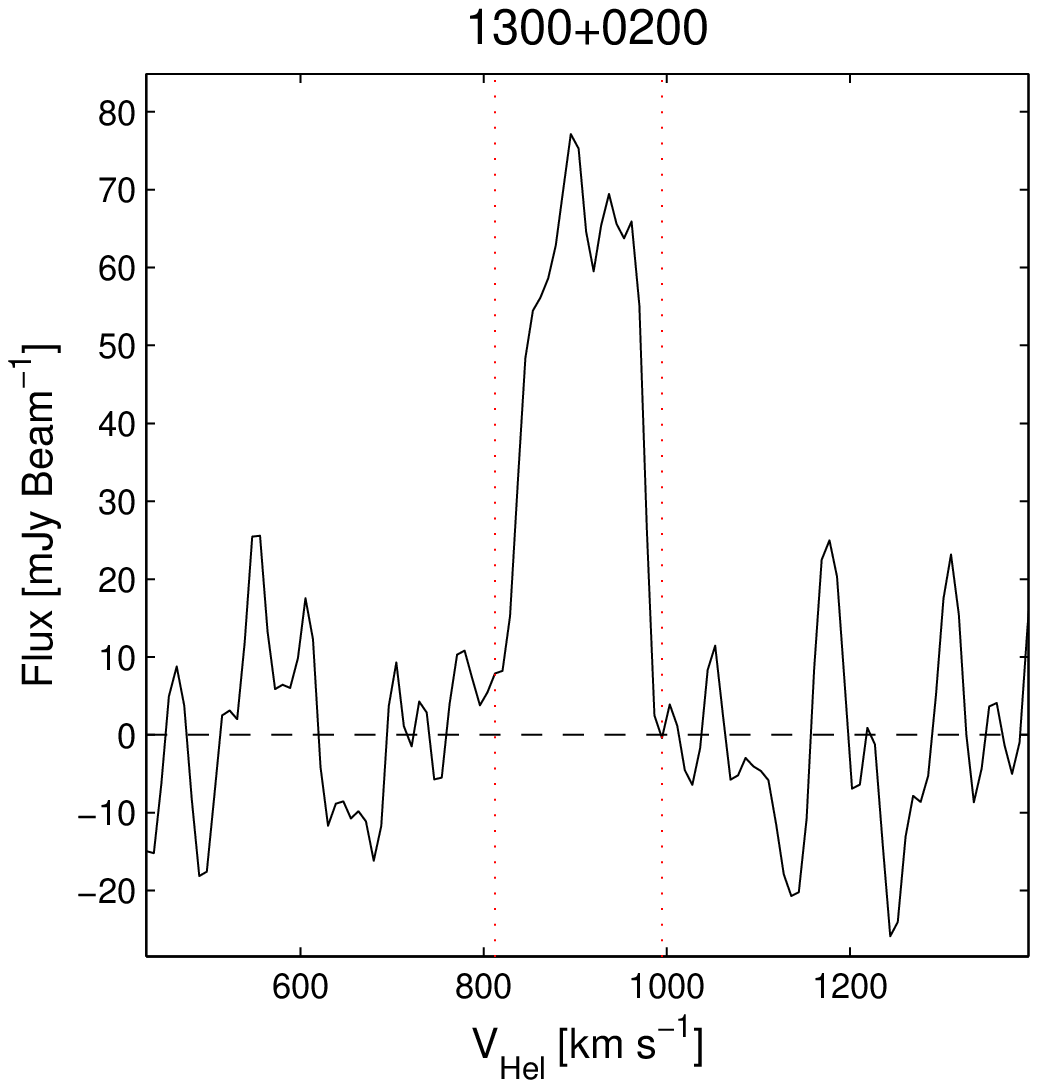}
 \includegraphics[width=0.22\textwidth]{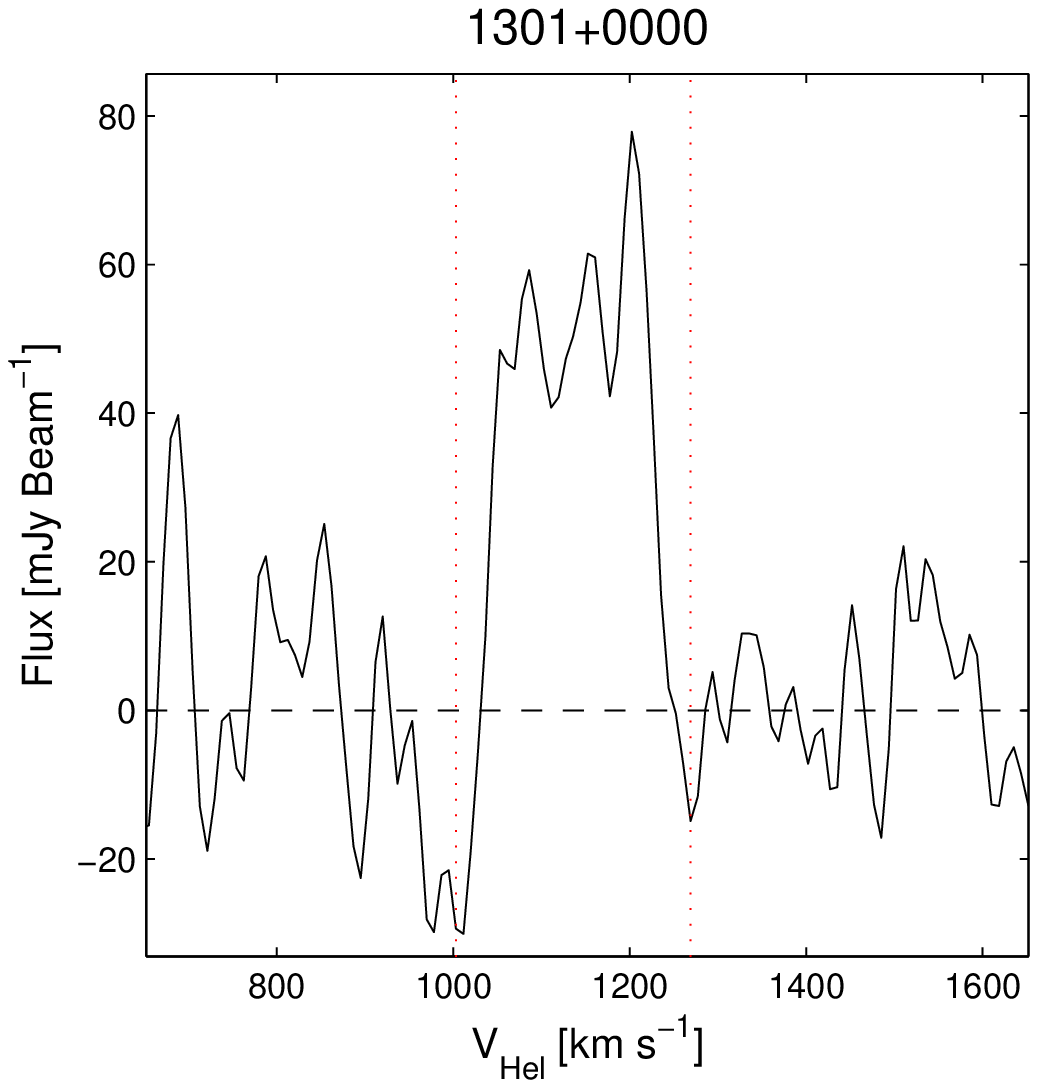}
 \includegraphics[width=0.22\textwidth]{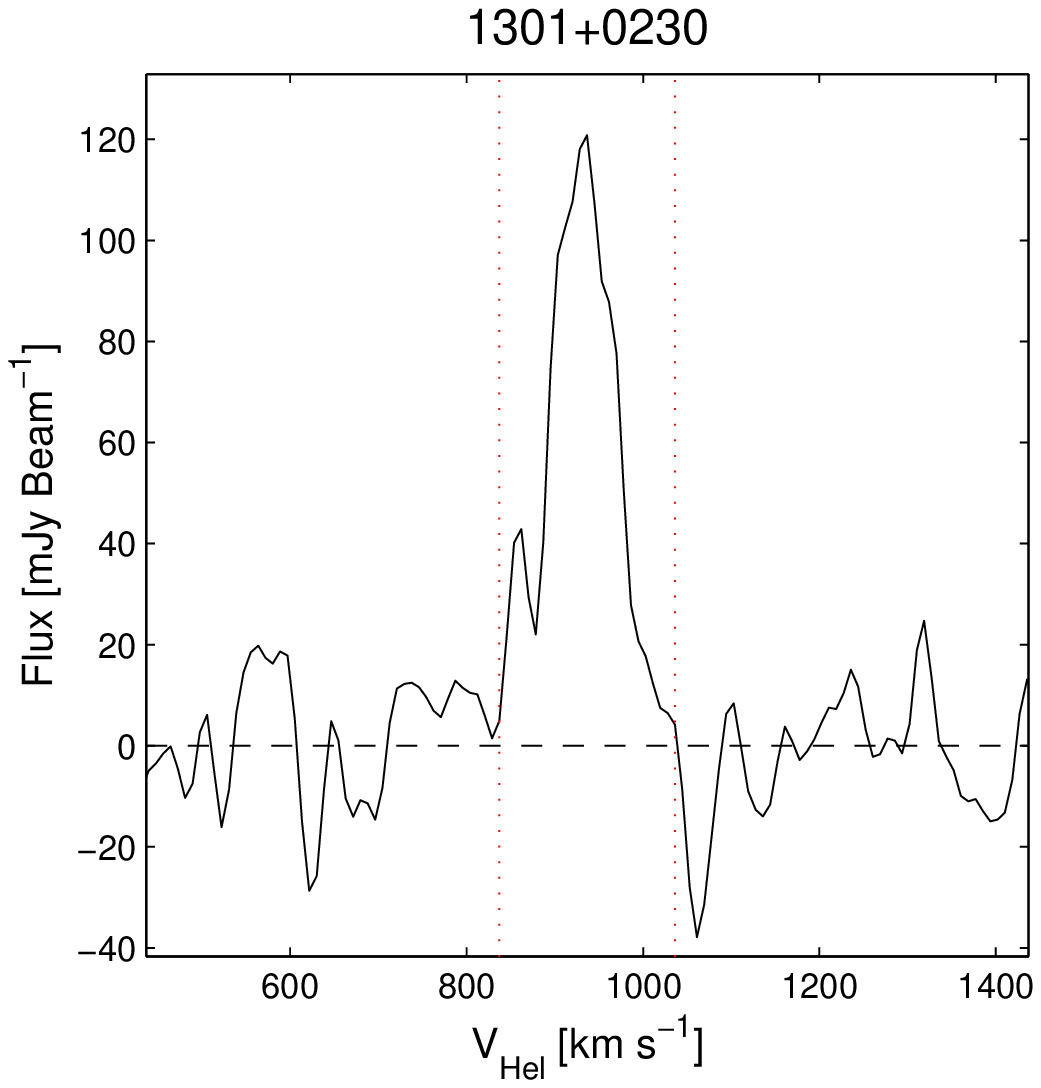}
 \includegraphics[width=0.22\textwidth]{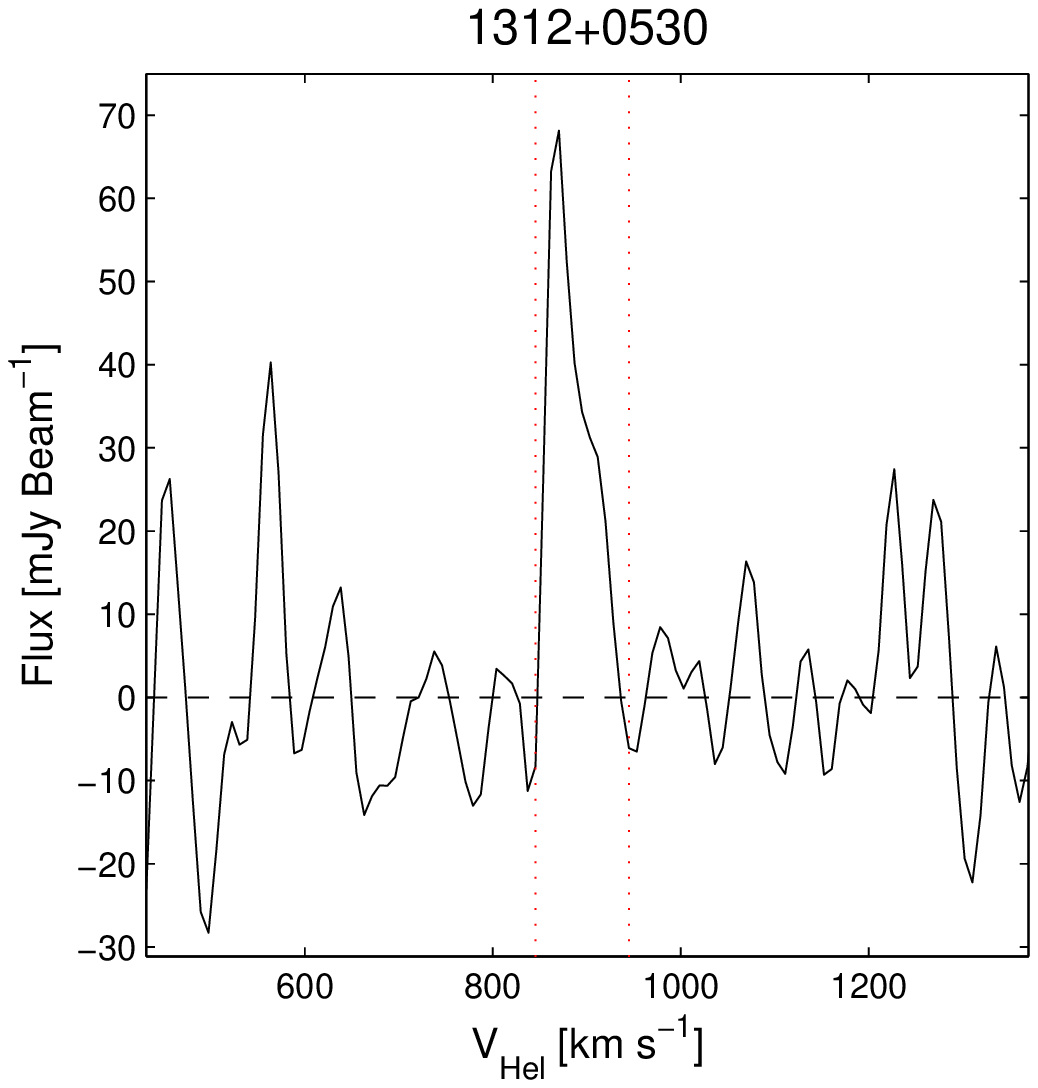}
 \includegraphics[width=0.22\textwidth]{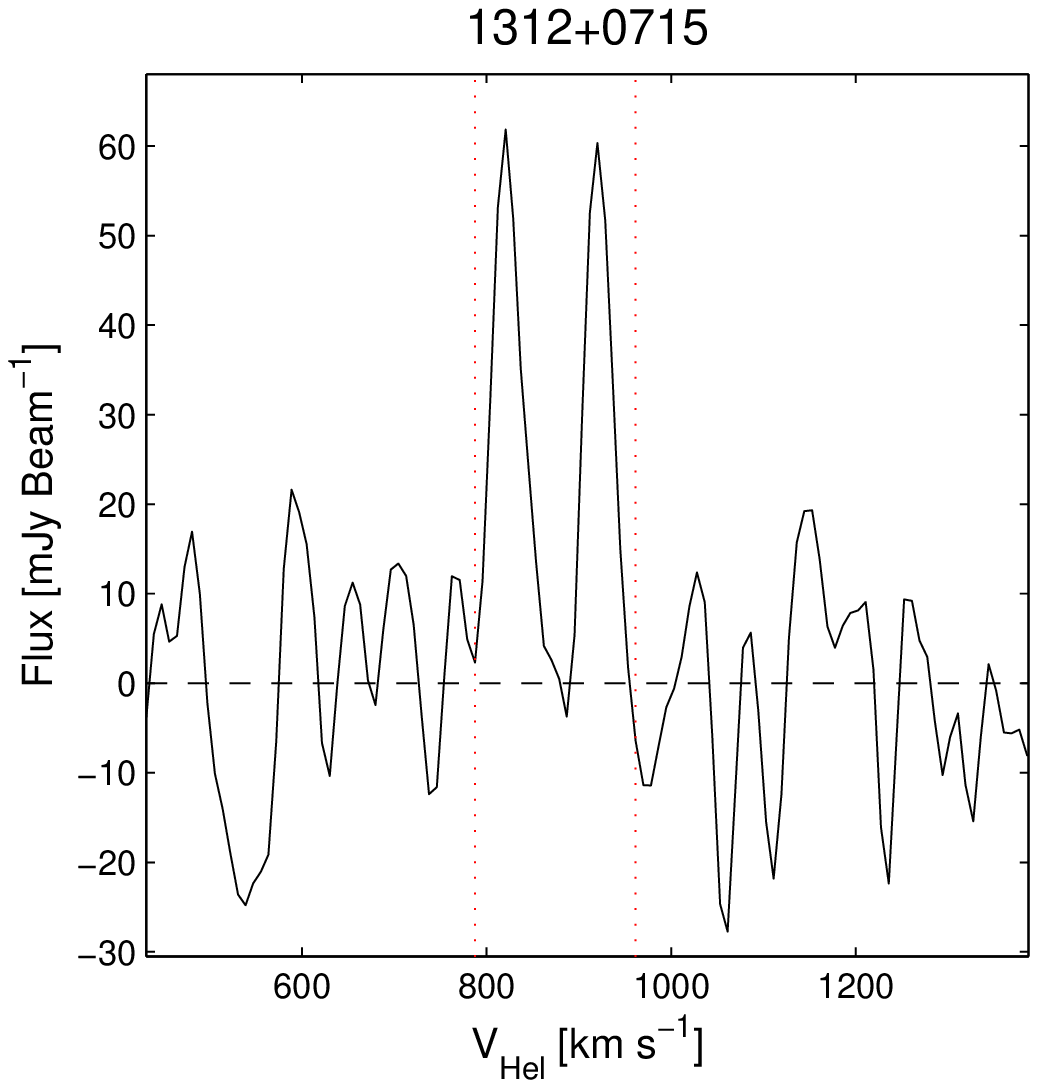}
 \includegraphics[width=0.22\textwidth]{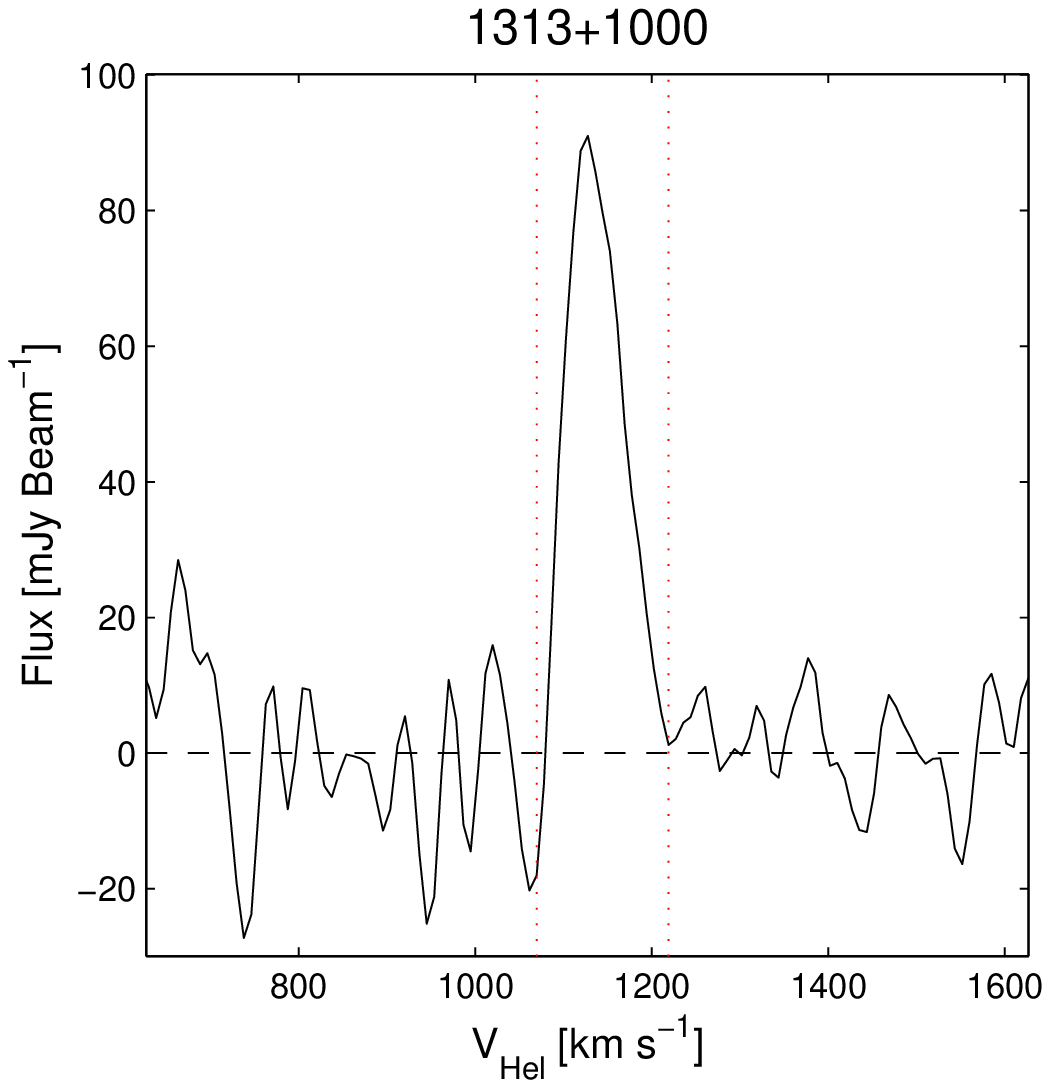}
 \includegraphics[width=0.22\textwidth]{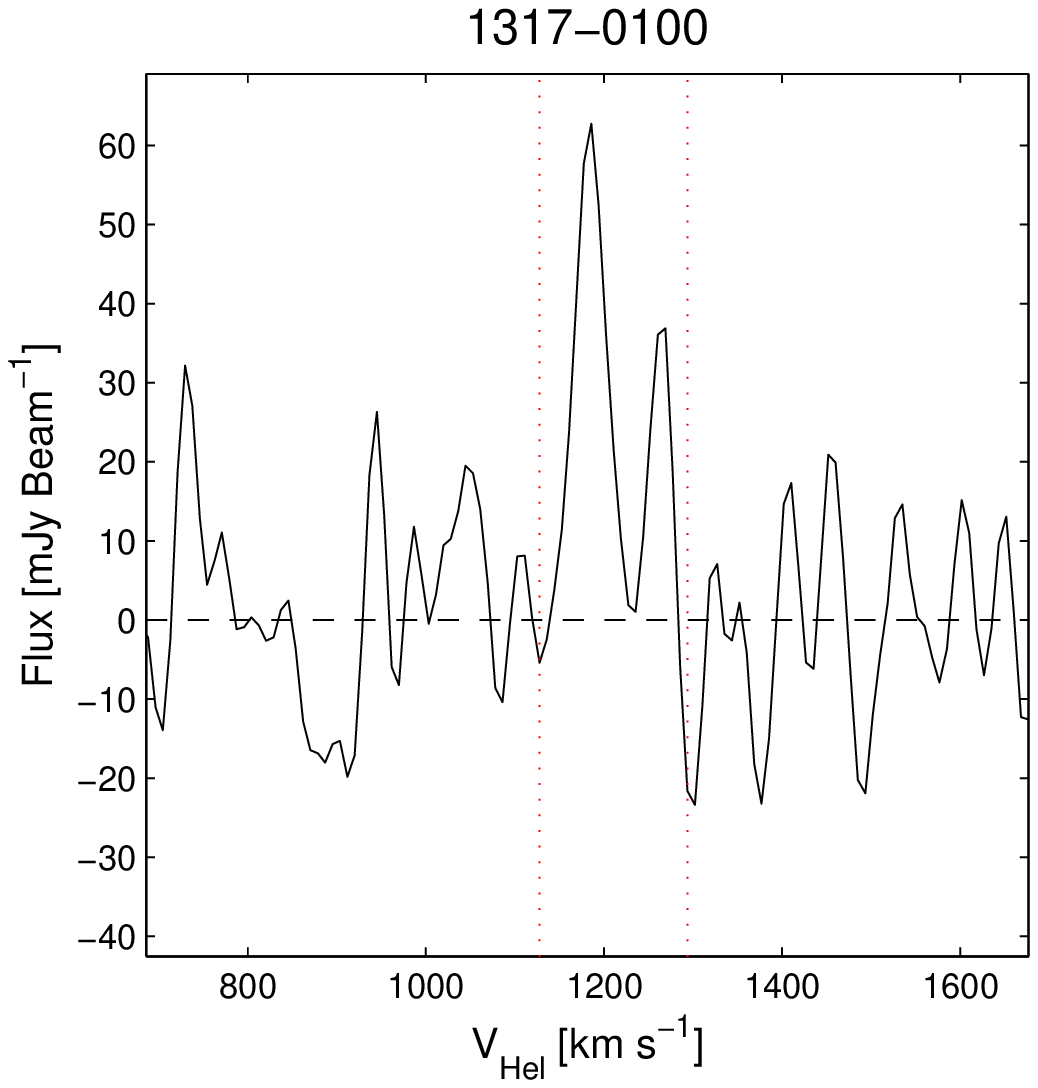}
 \includegraphics[width=0.22\textwidth]{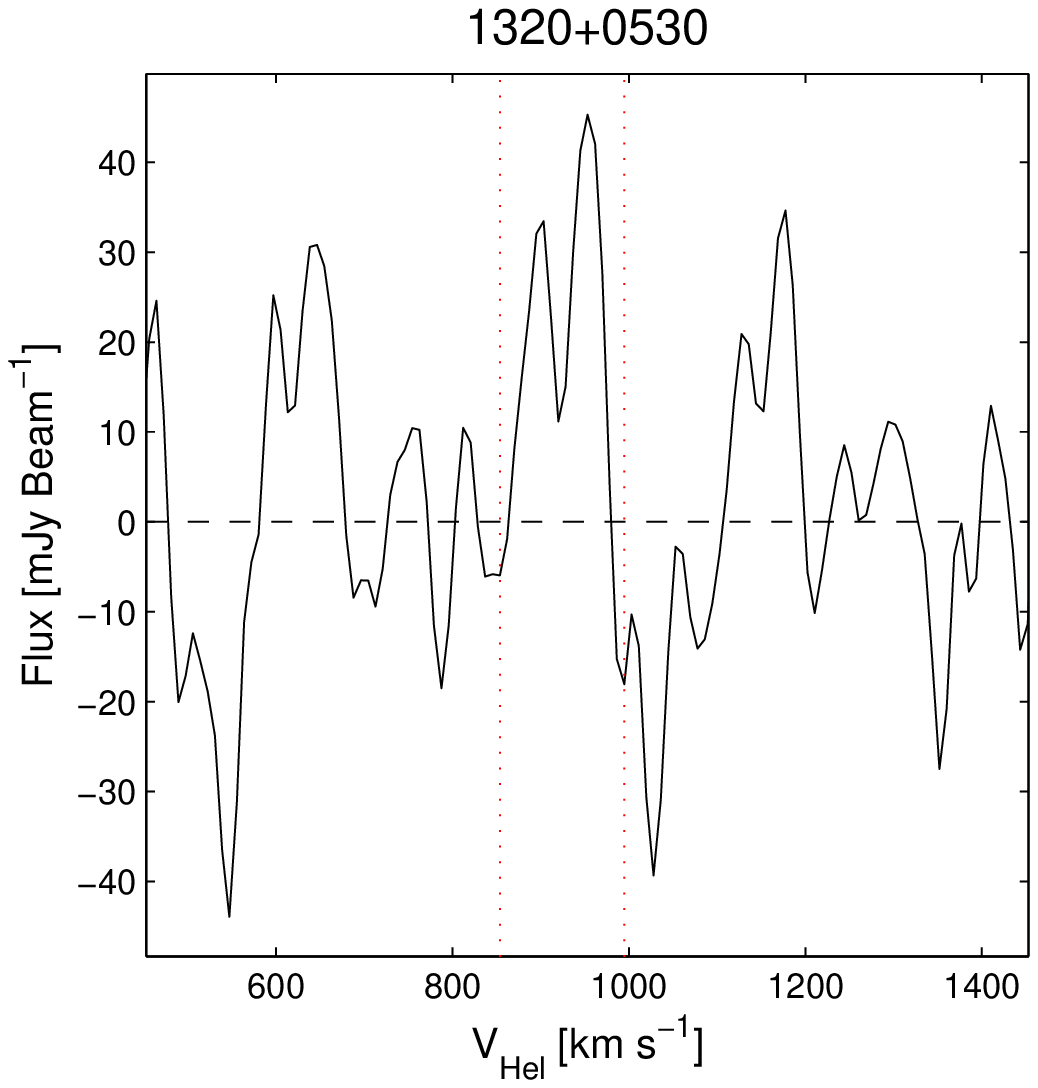}
 \includegraphics[width=0.22\textwidth]{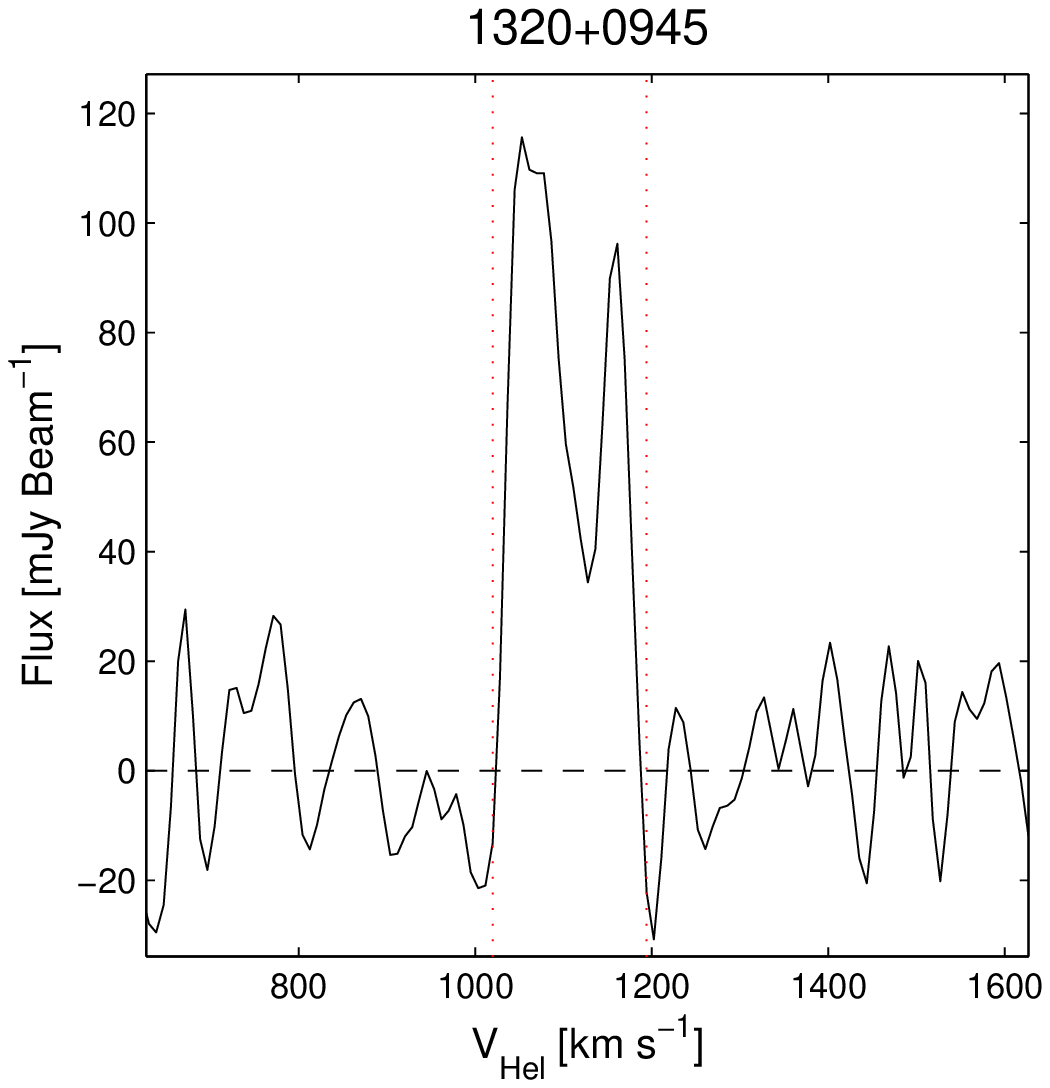}
 \includegraphics[width=0.22\textwidth]{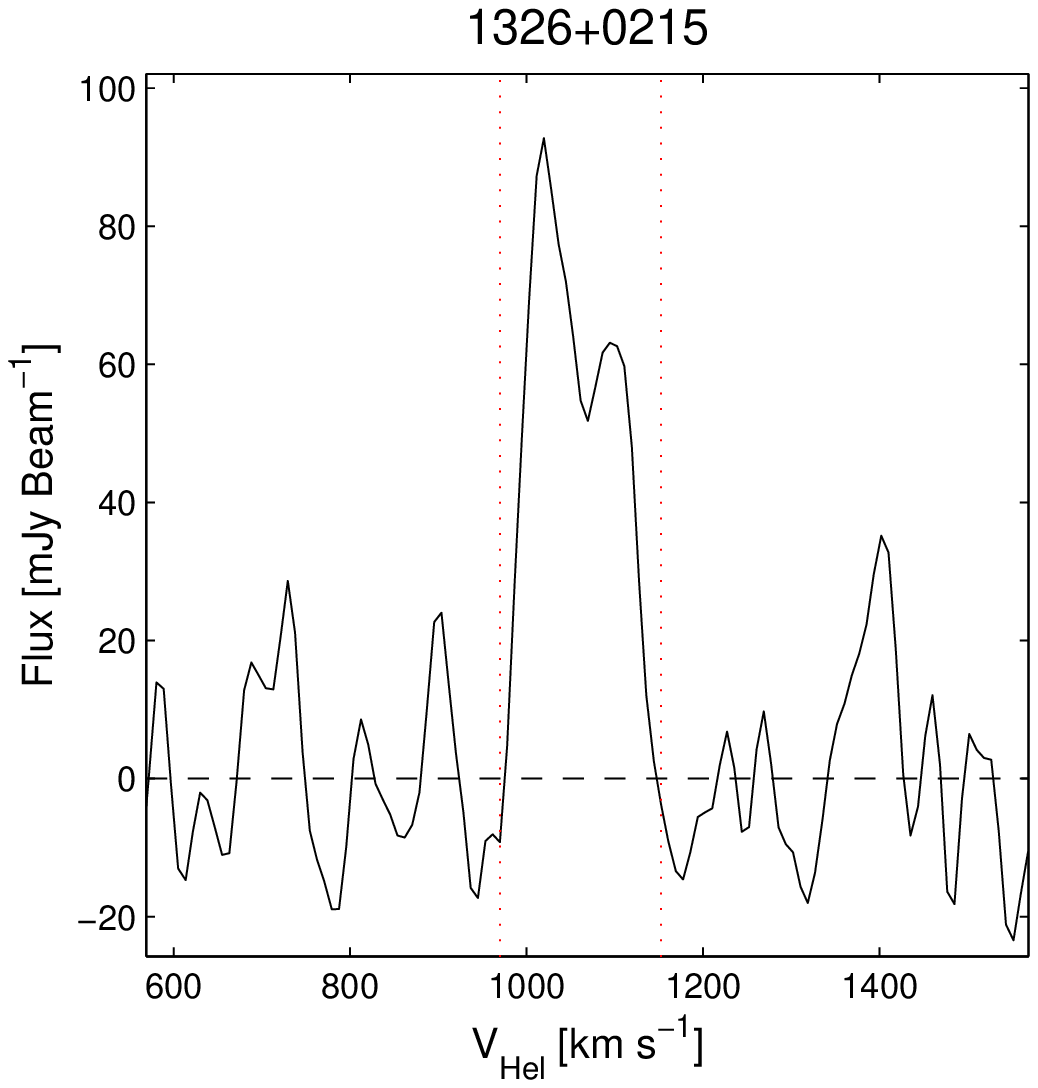}
 \includegraphics[width=0.22\textwidth]{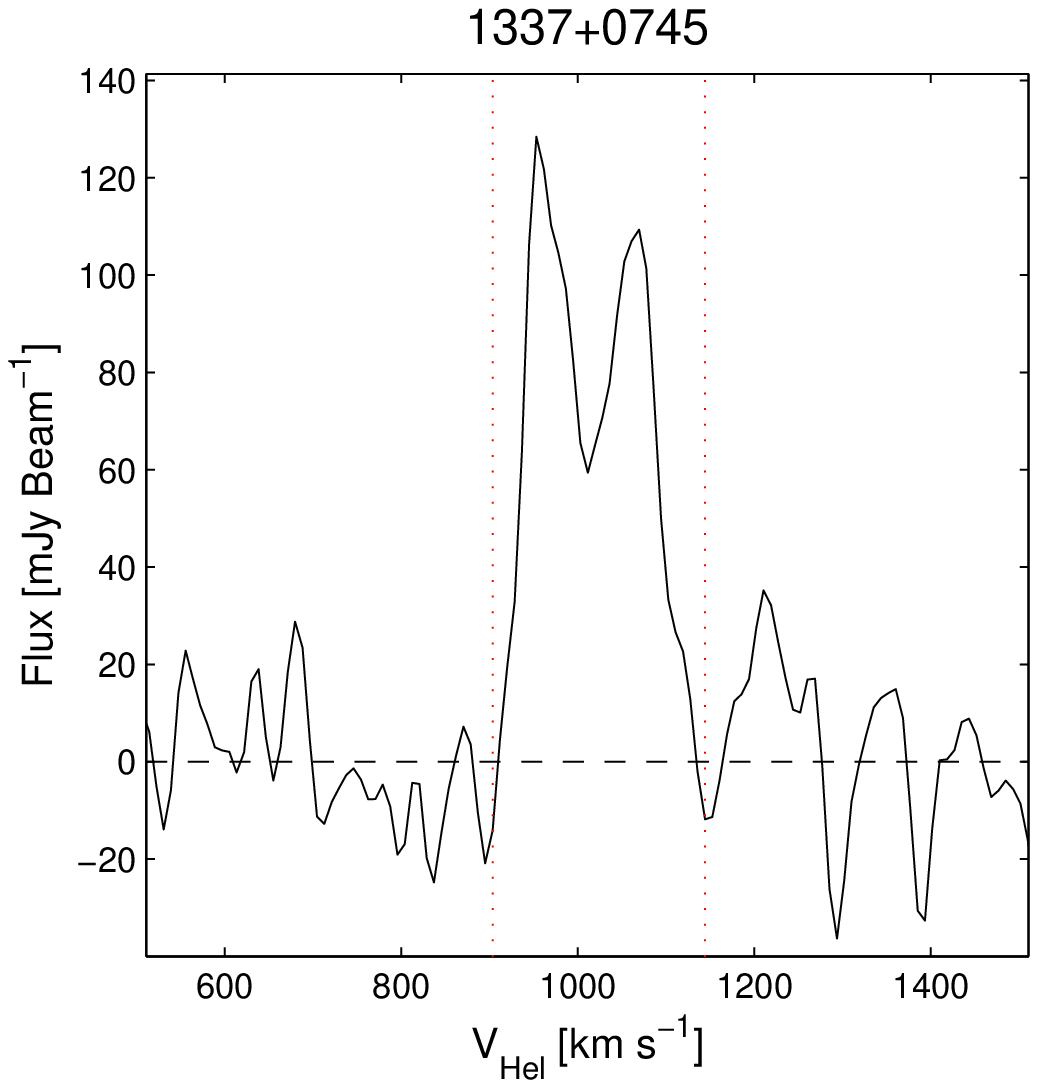}
 \includegraphics[width=0.22\textwidth]{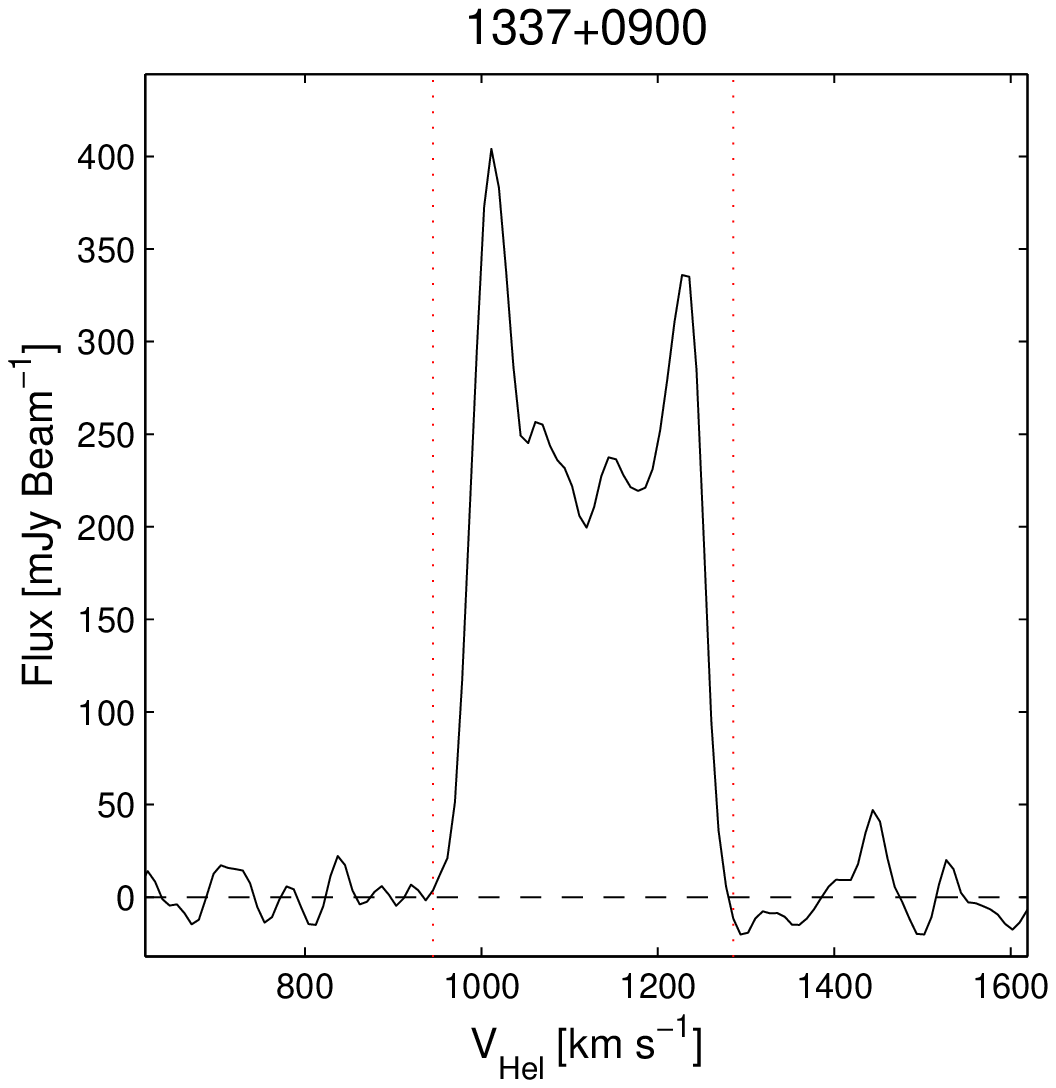}
 \includegraphics[width=0.22\textwidth]{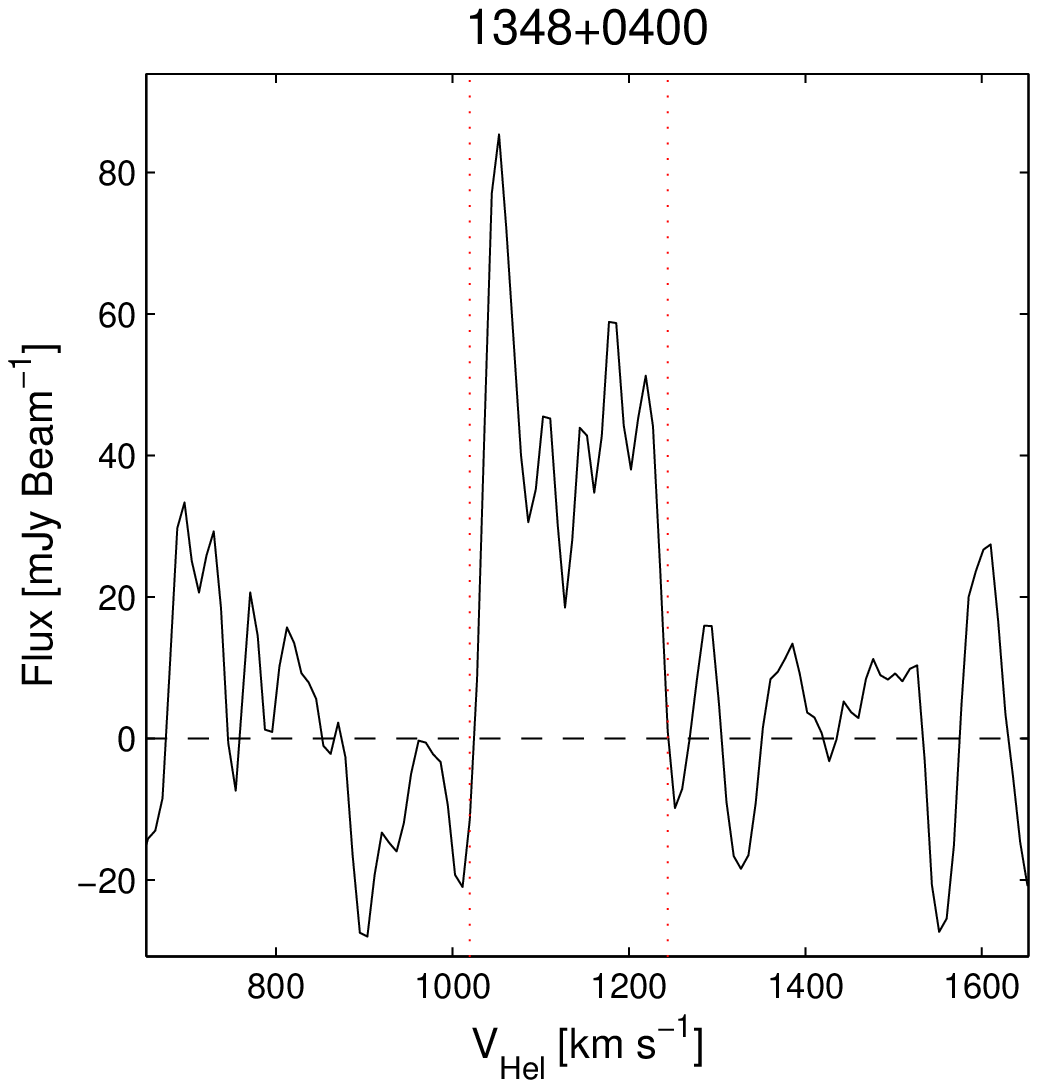}
 \includegraphics[width=0.22\textwidth]{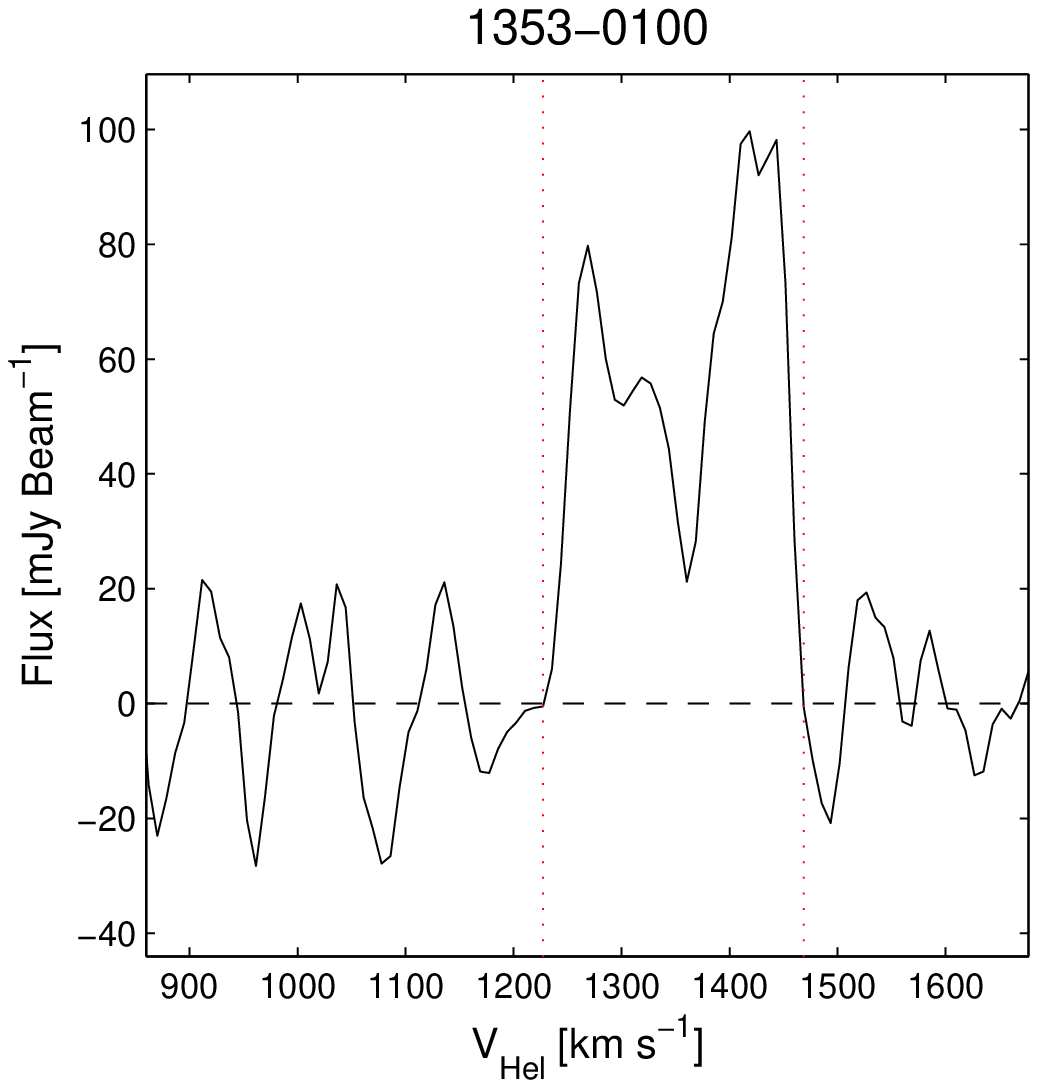}
 \includegraphics[width=0.22\textwidth]{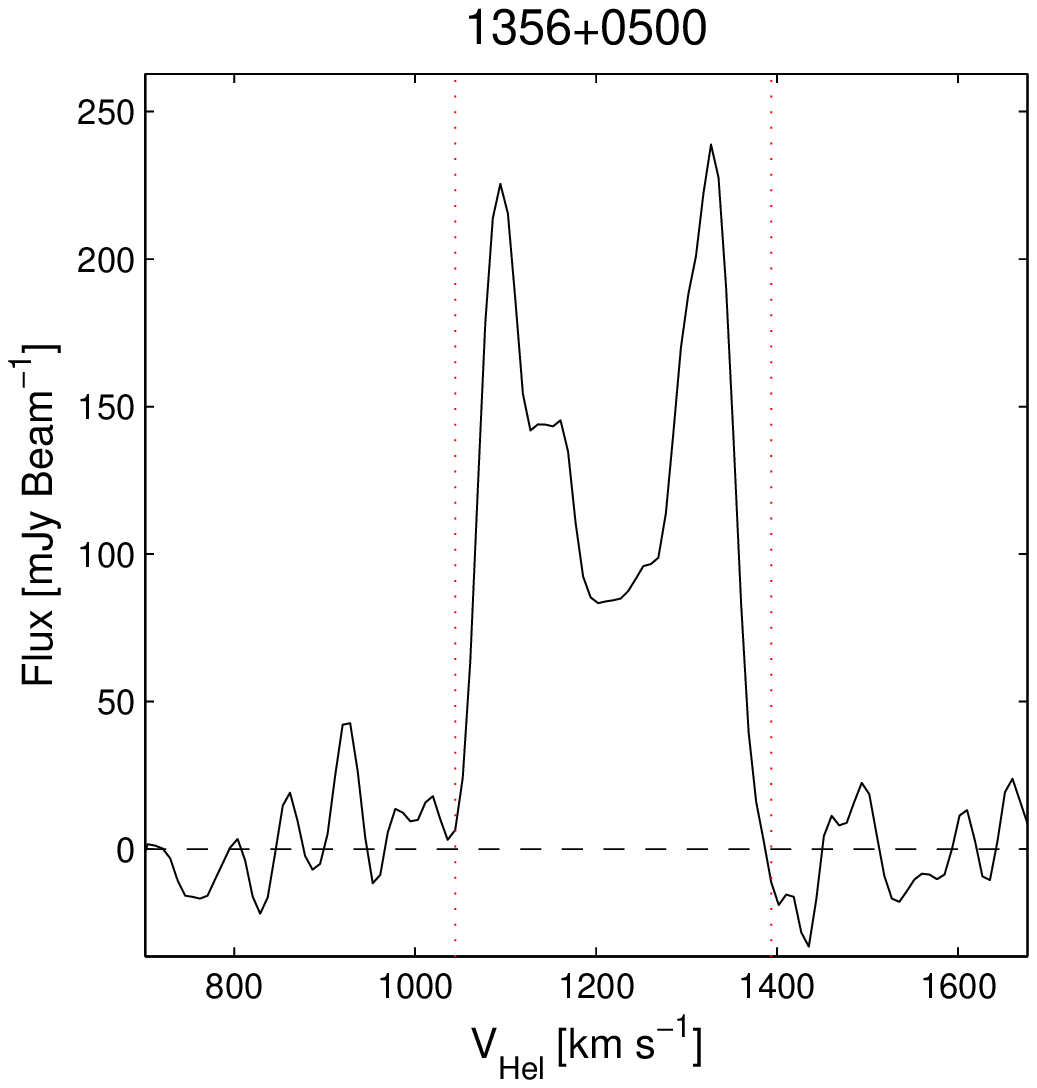}
 \includegraphics[width=0.22\textwidth]{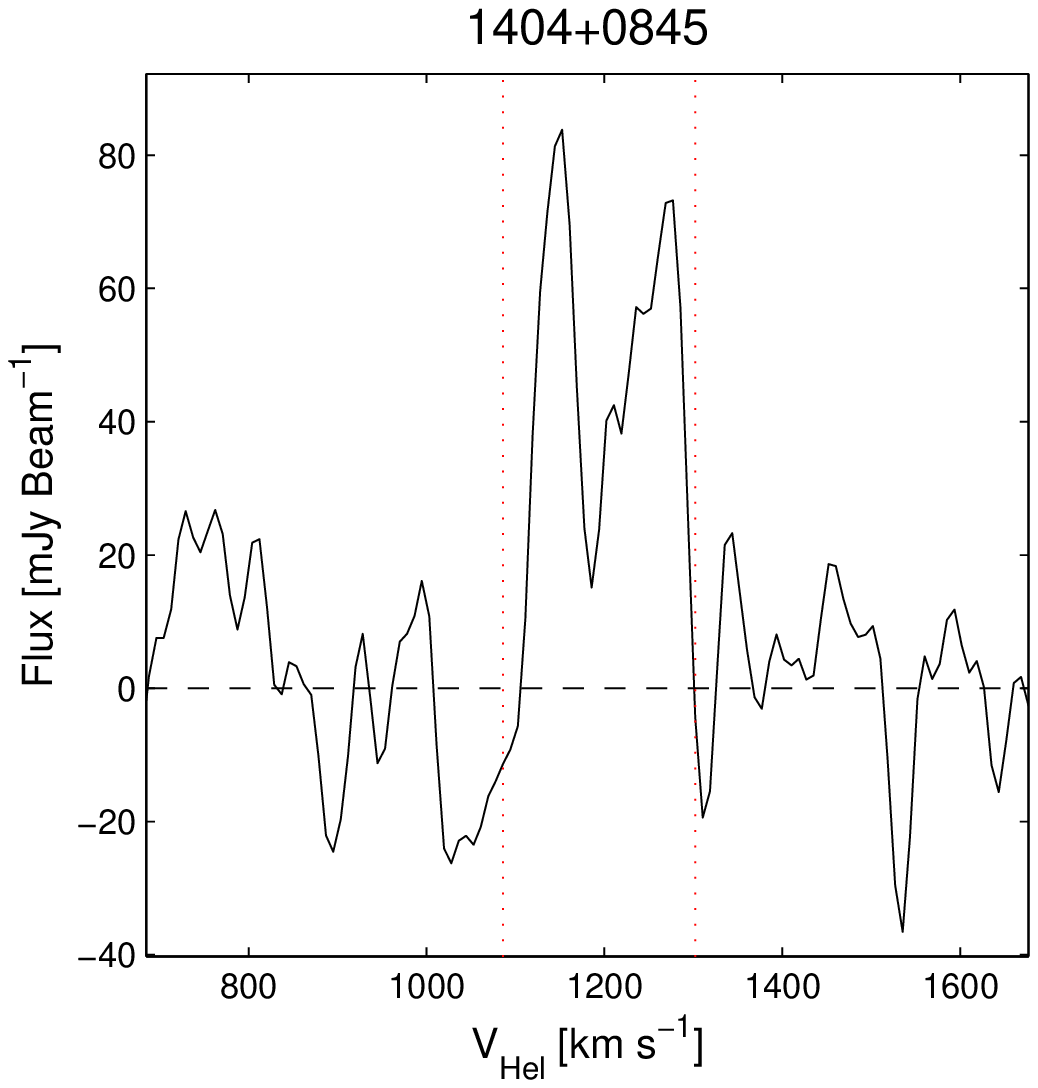}                                                        
                                                         
 \end{center}                                            
{\bf Fig~\ref{all_spectra}.} (continued)                                        
 
\end{figure*}

\begin{figure*}
  \begin{center}

 \includegraphics[width=0.22\textwidth]{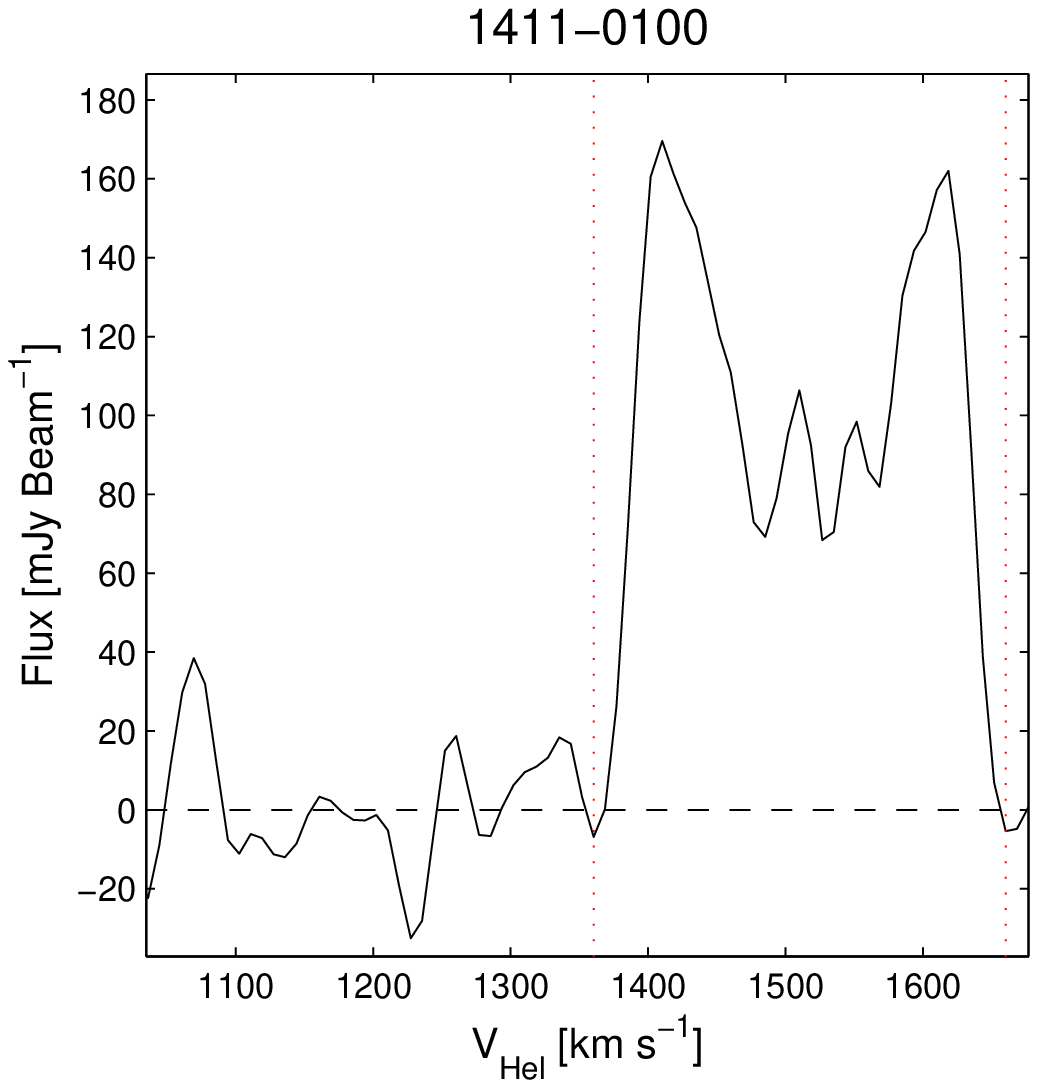}
 \includegraphics[width=0.22\textwidth]{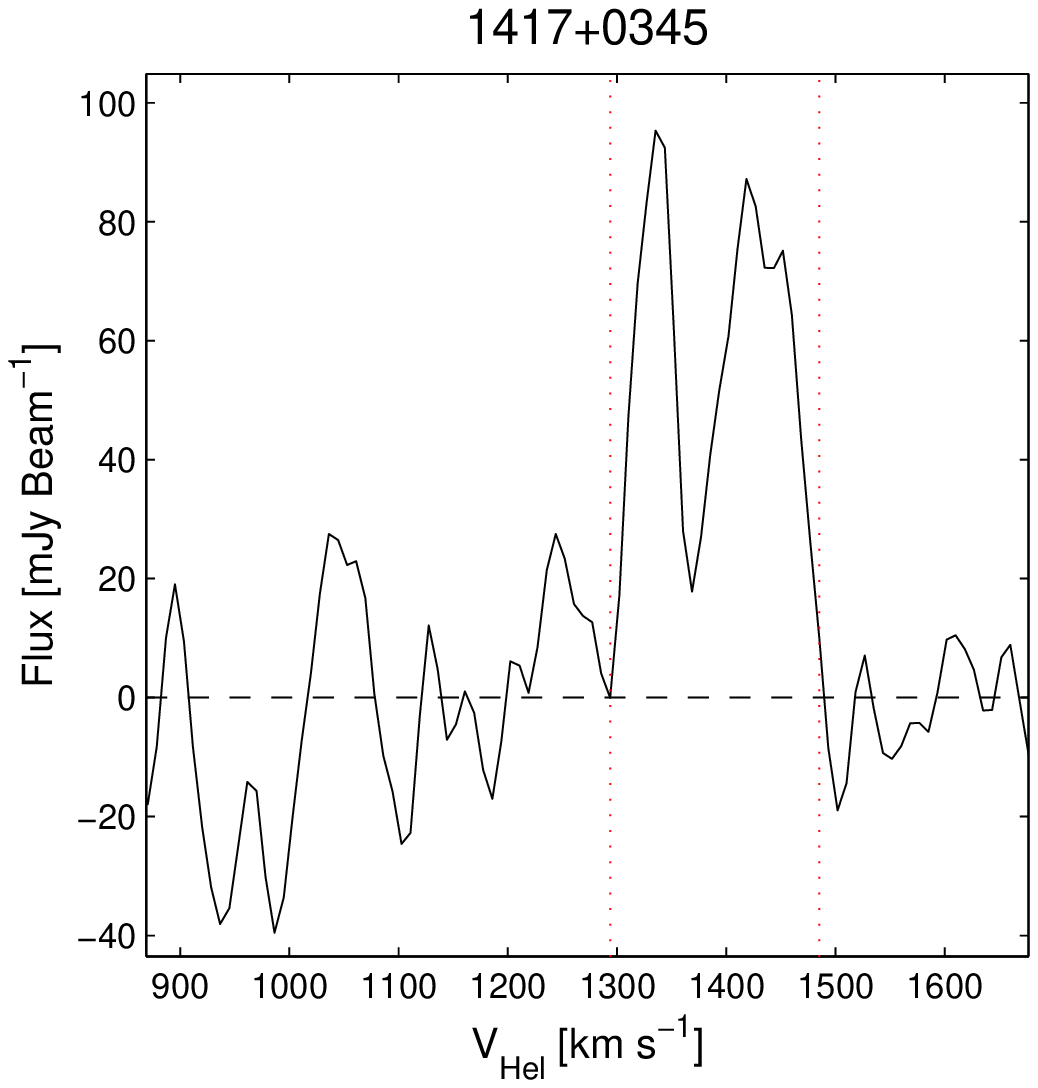}
 \includegraphics[width=0.22\textwidth]{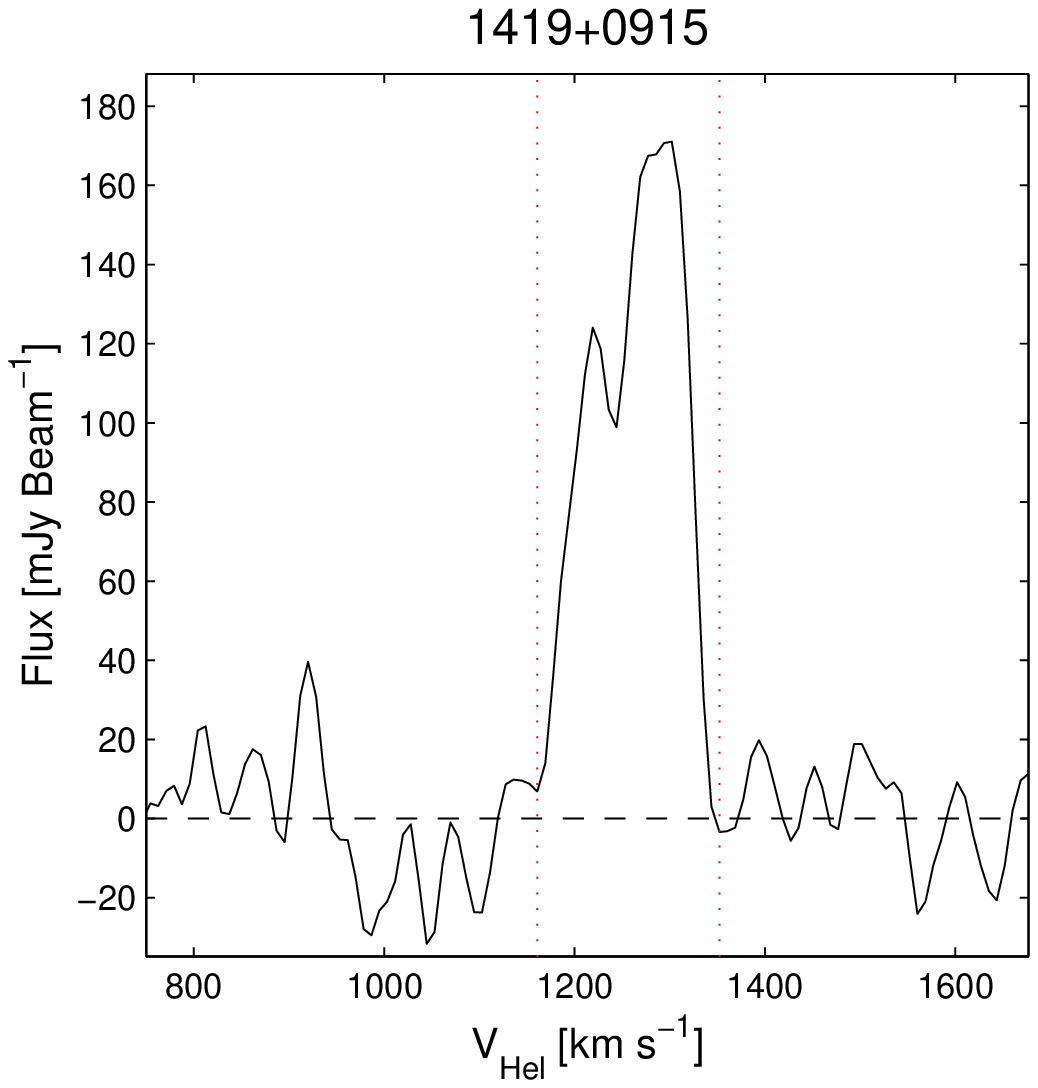}
 \includegraphics[width=0.22\textwidth]{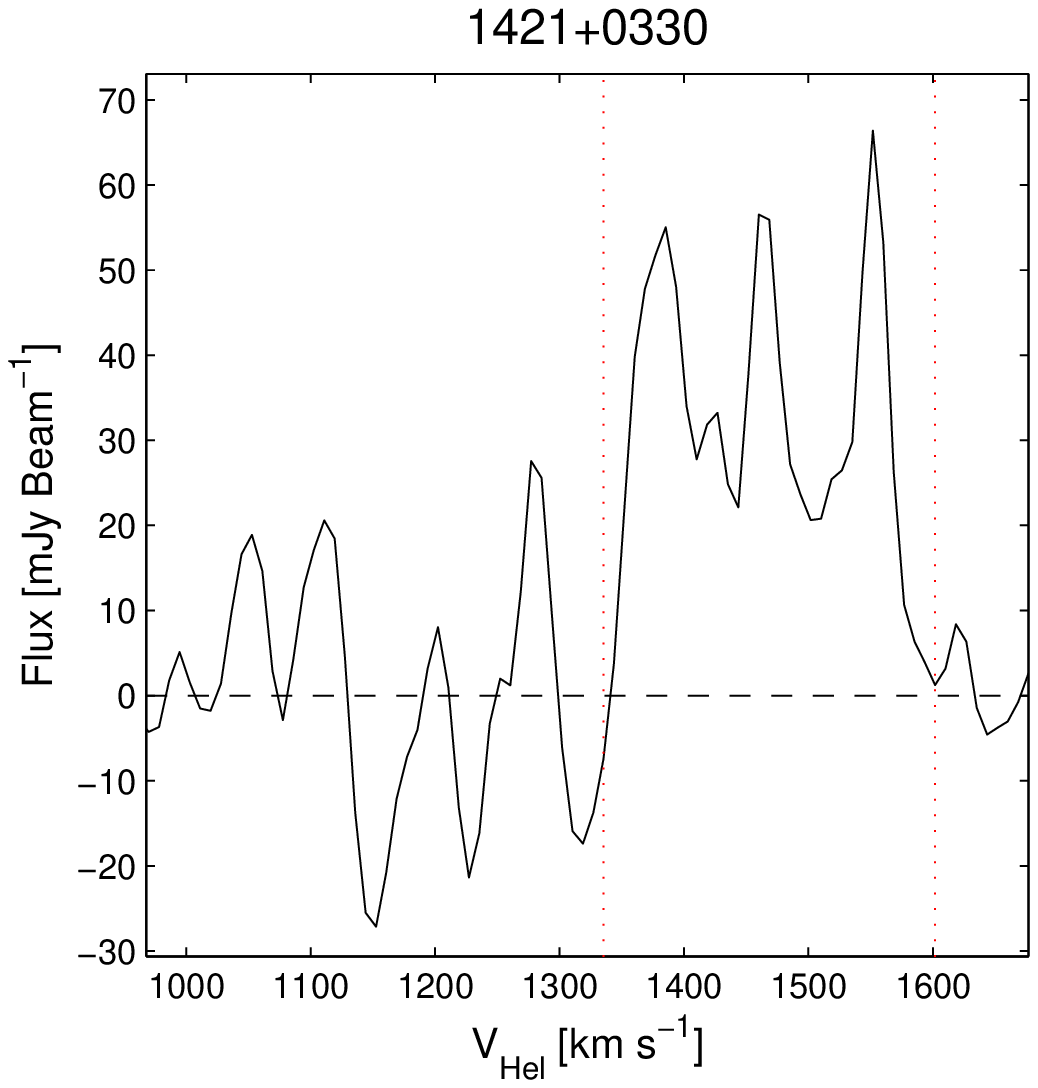}
 \includegraphics[width=0.22\textwidth]{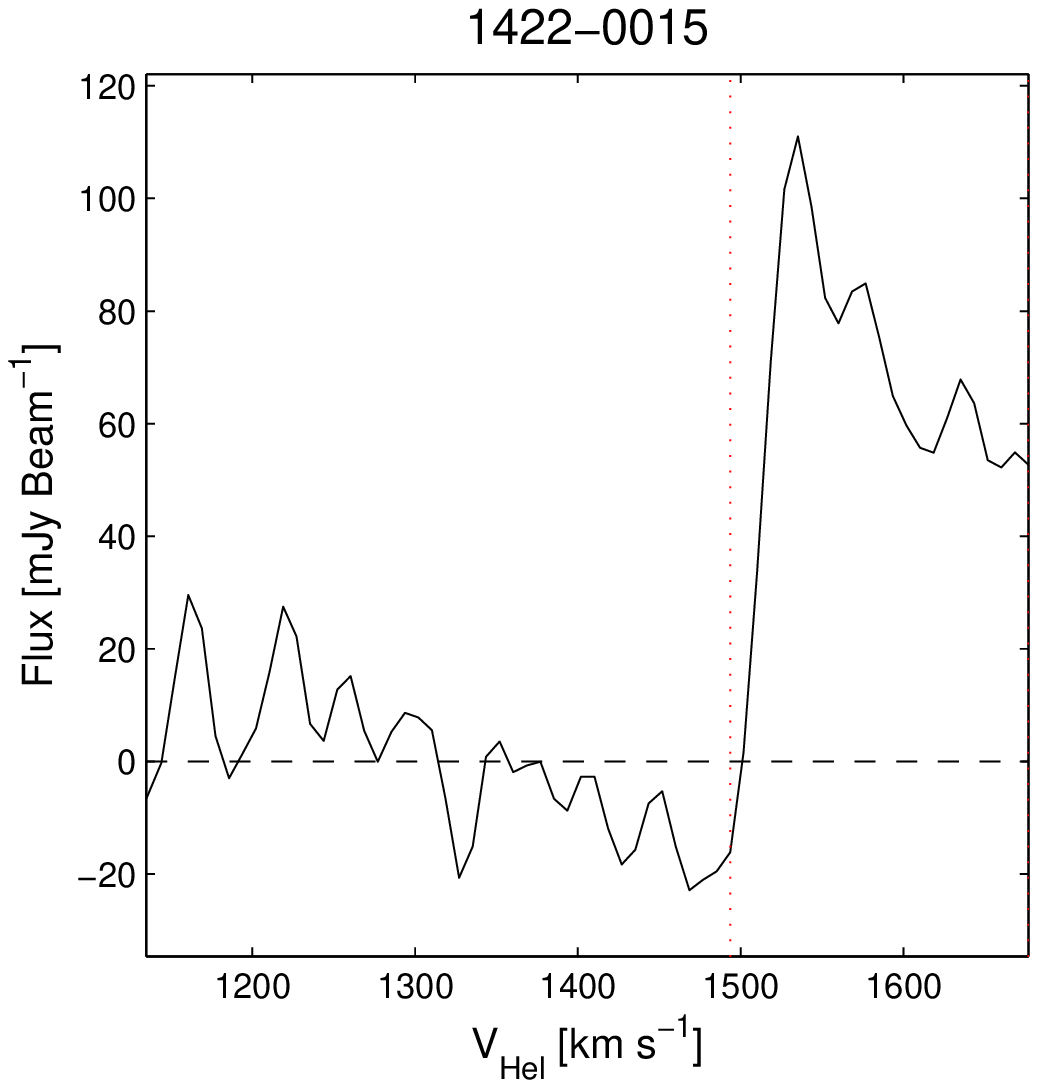}
 \includegraphics[width=0.22\textwidth]{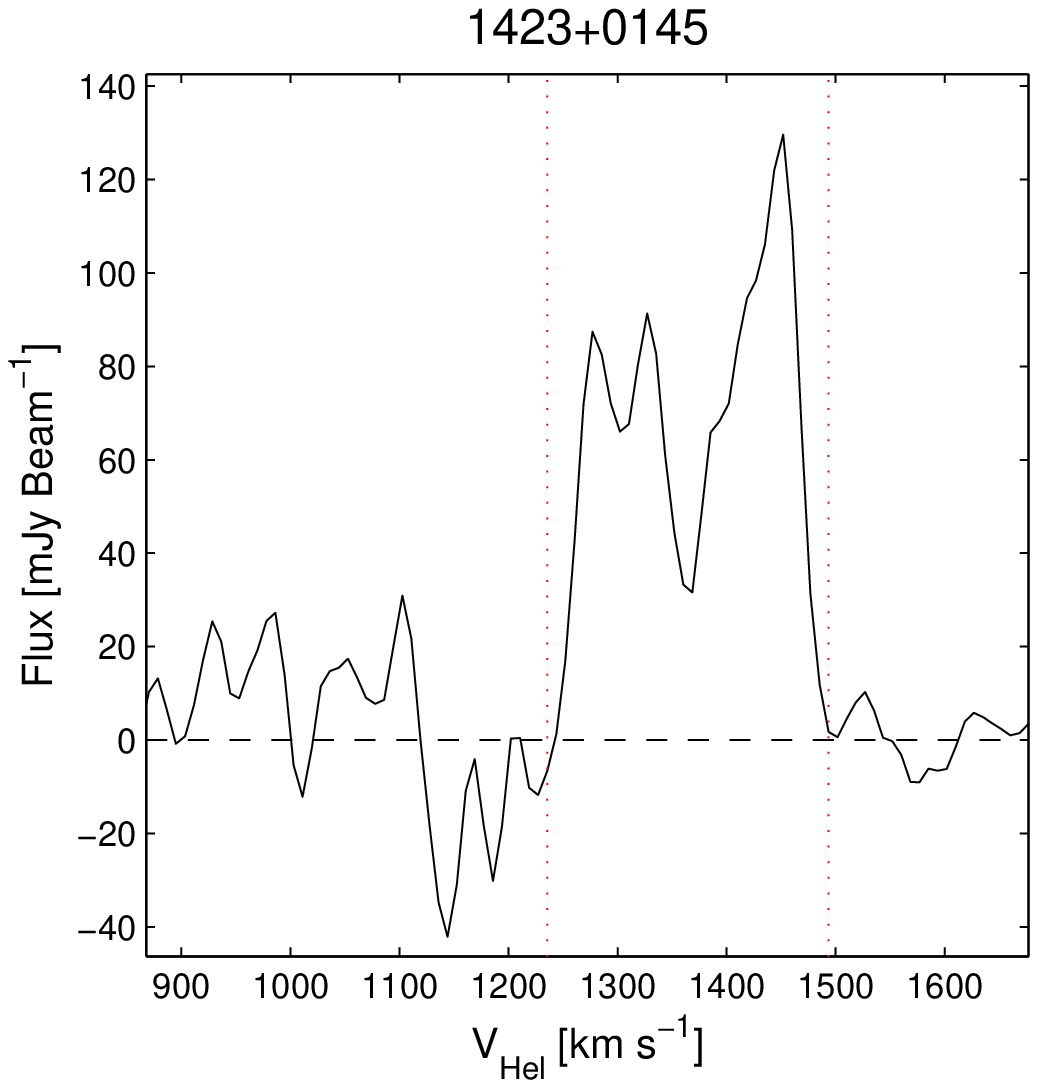}
 \includegraphics[width=0.22\textwidth]{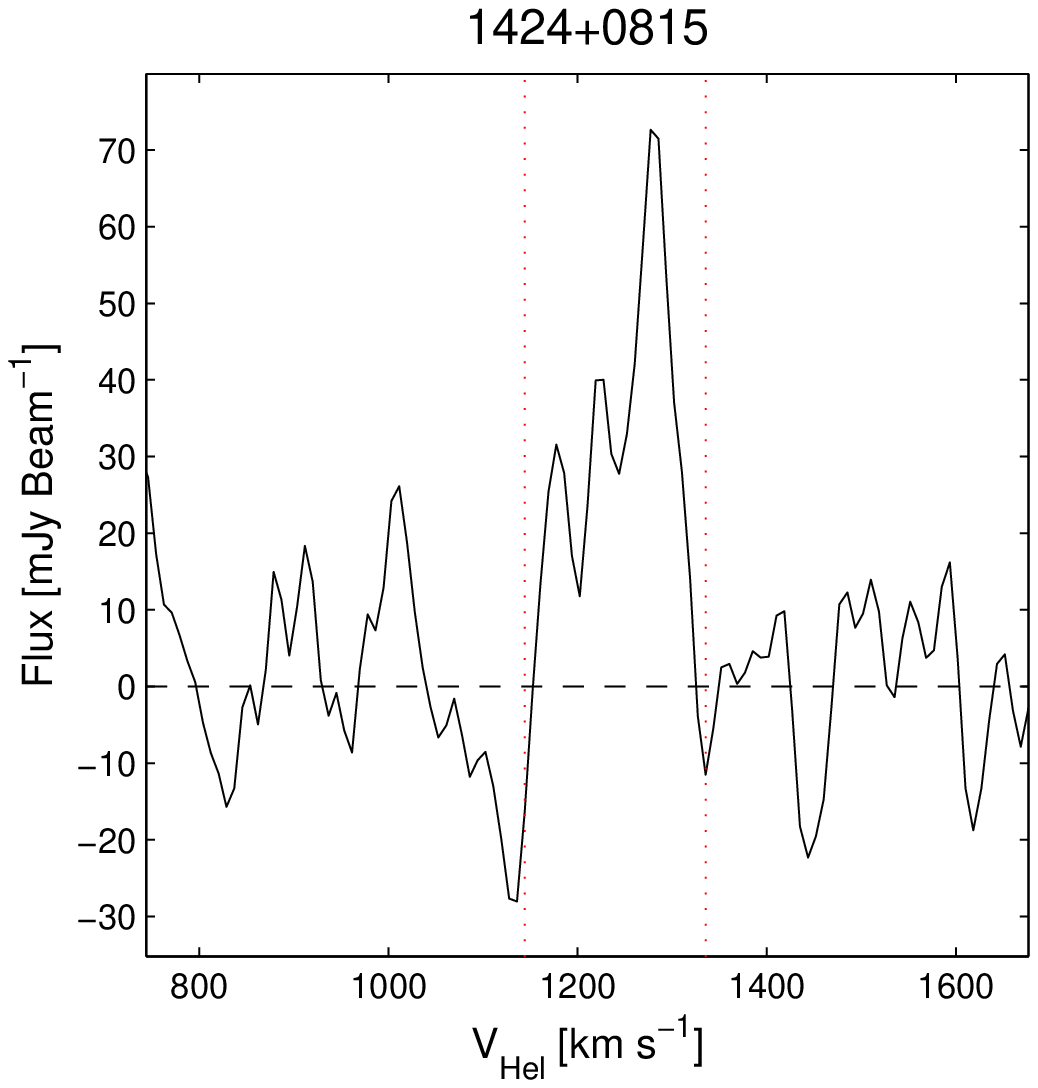}
 \includegraphics[width=0.22\textwidth]{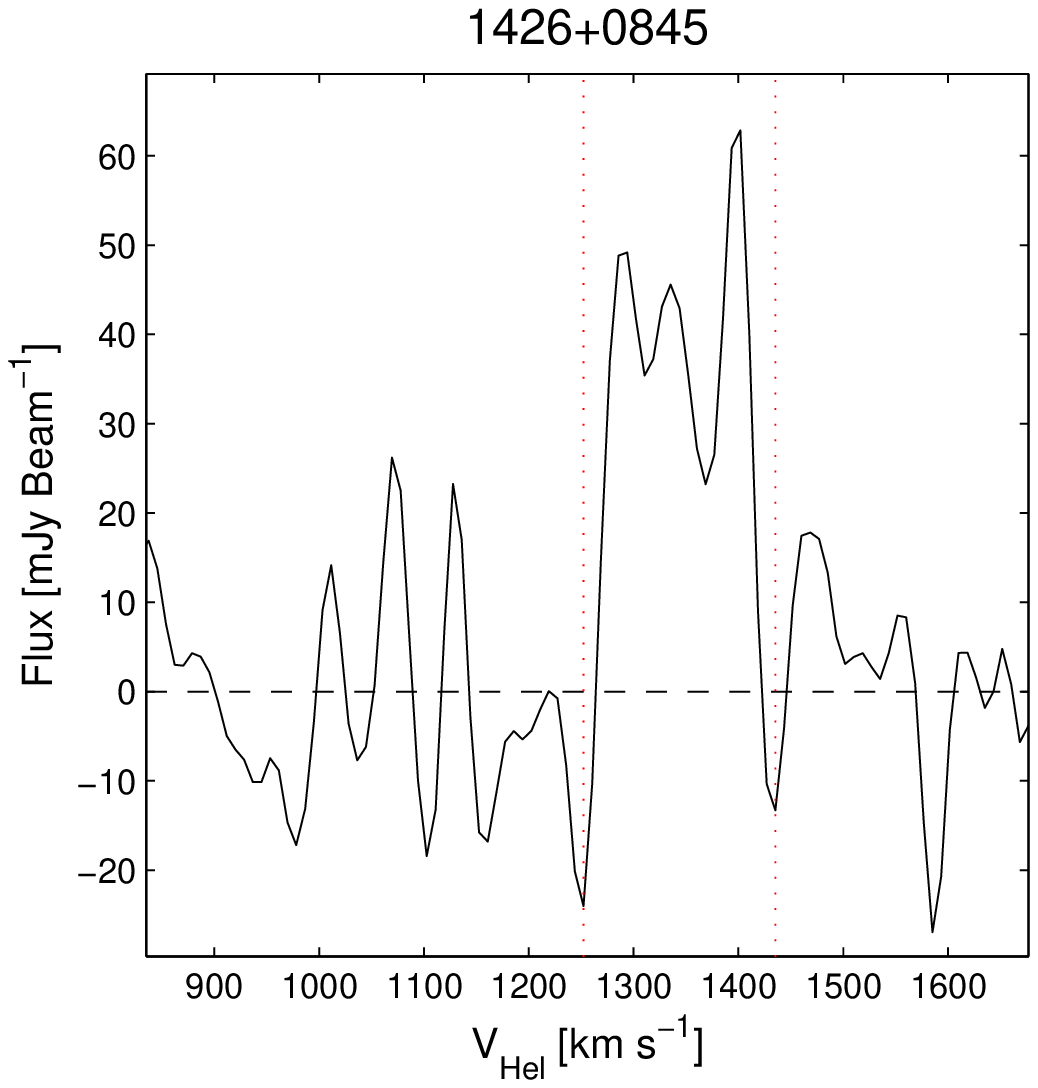}
 \includegraphics[width=0.22\textwidth]{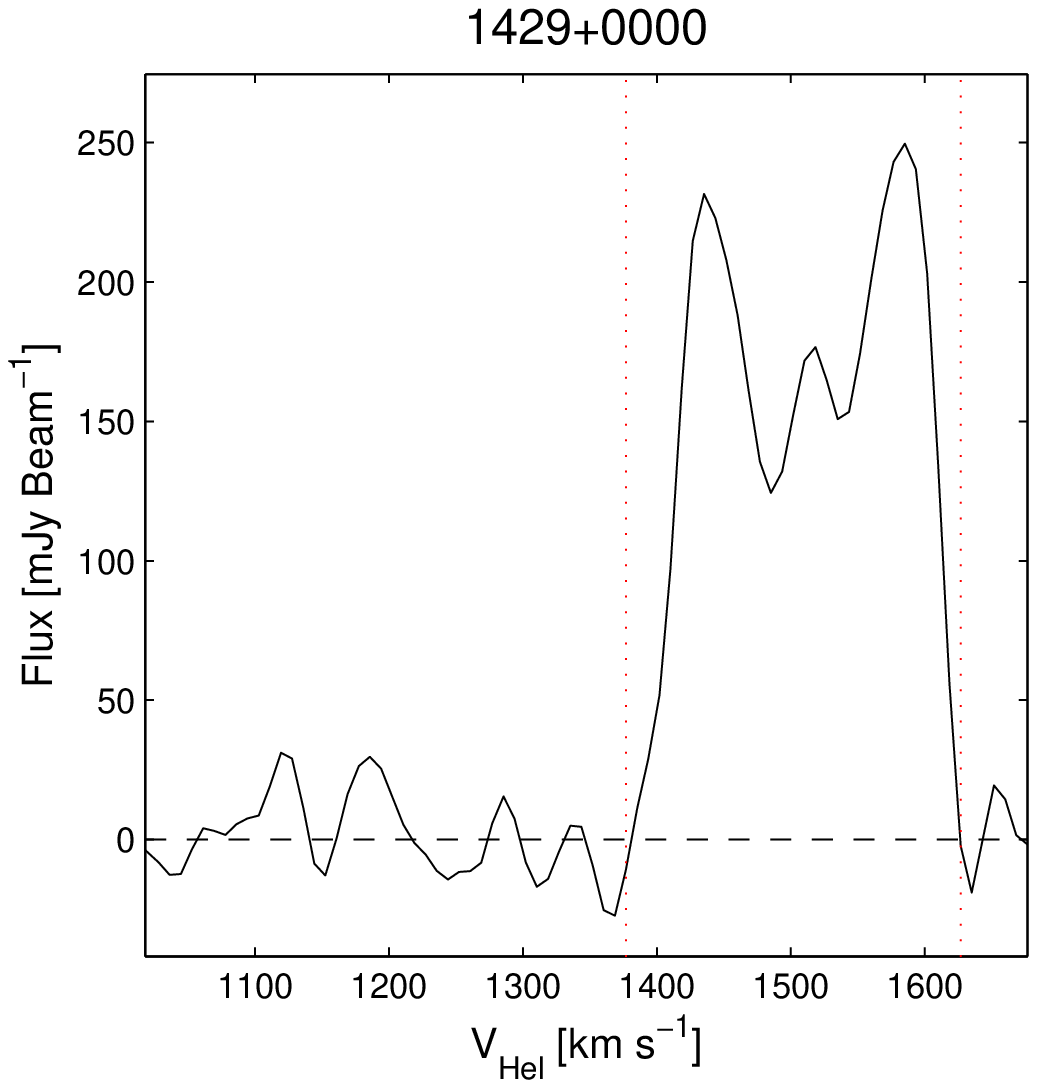}
 \includegraphics[width=0.22\textwidth]{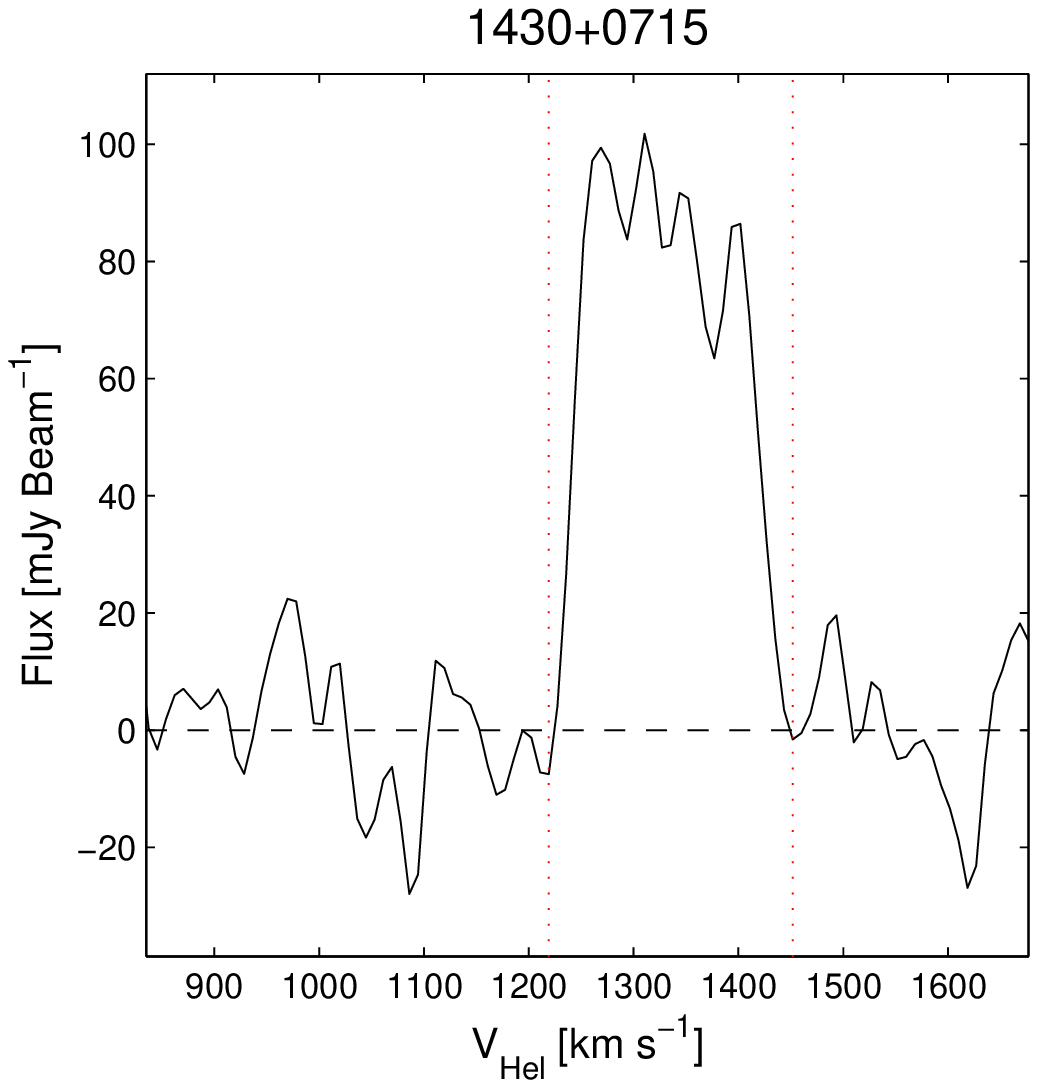}
 \includegraphics[width=0.22\textwidth]{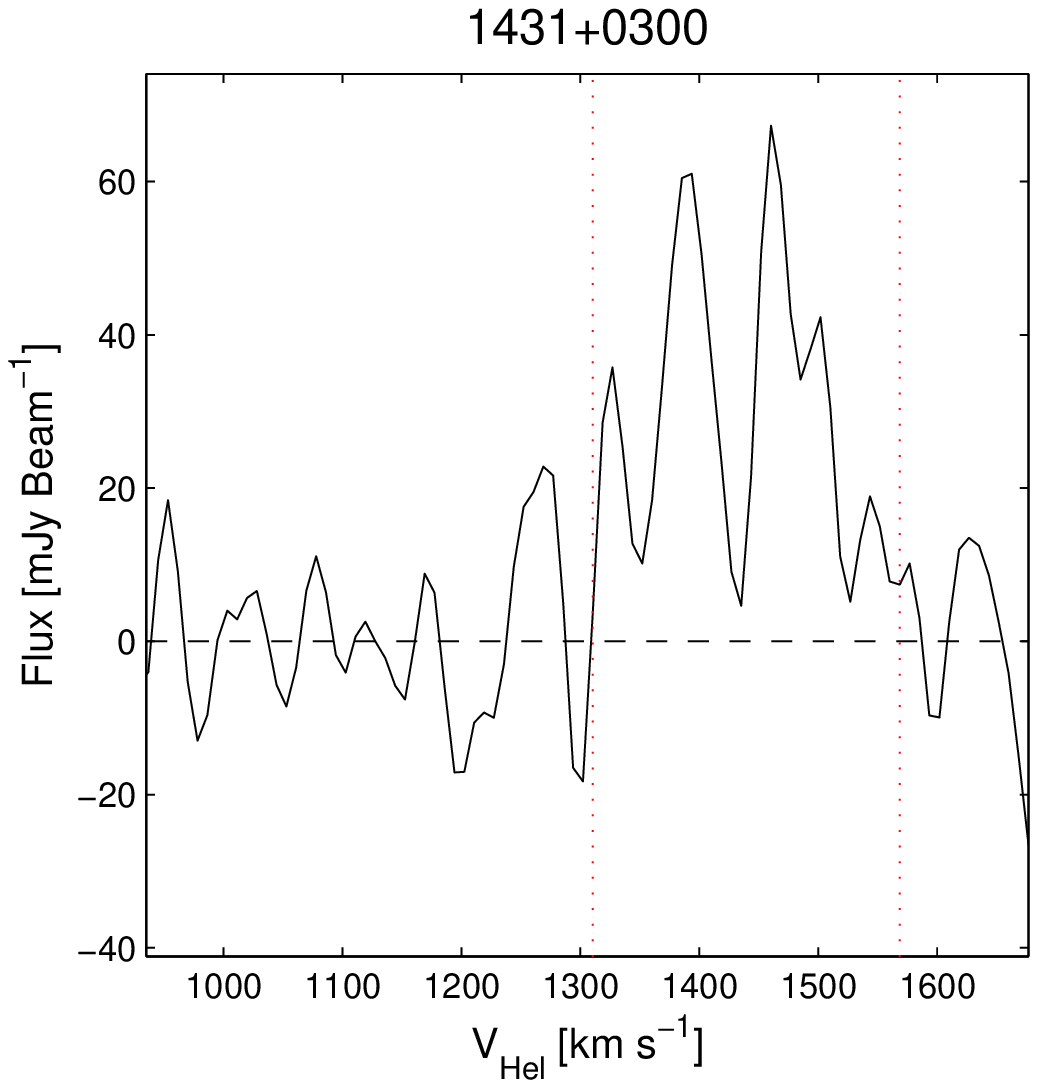}
 \includegraphics[width=0.22\textwidth]{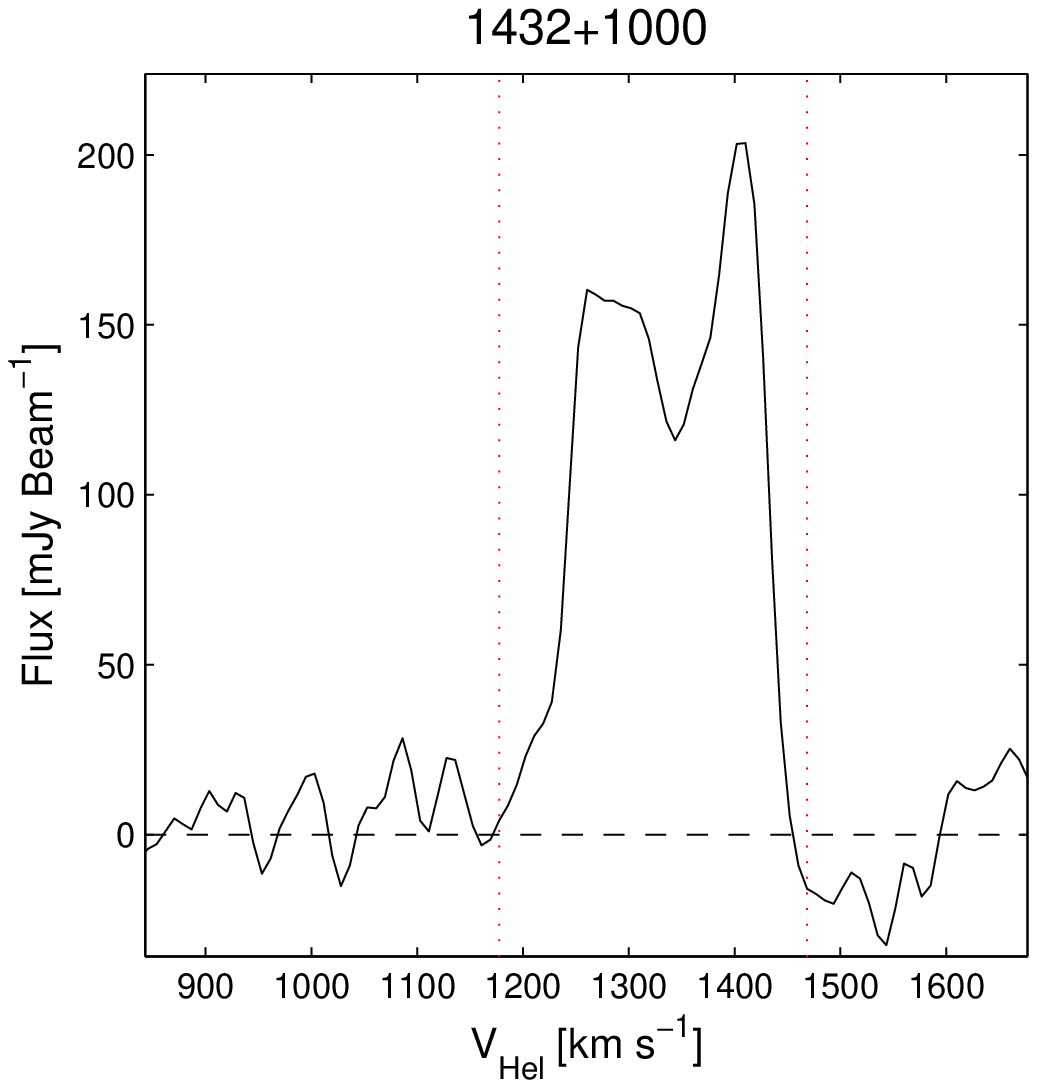}
 \includegraphics[width=0.22\textwidth]{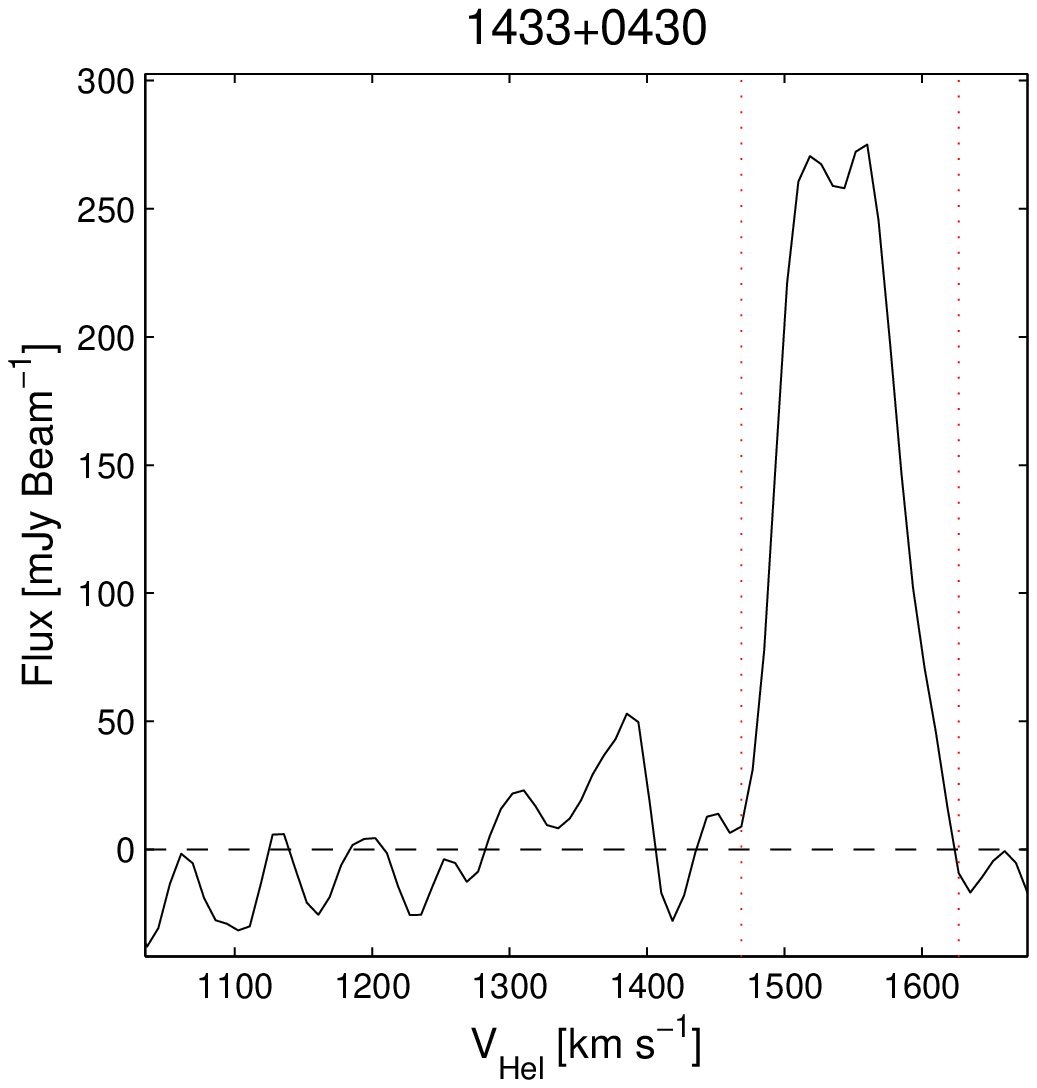}
 \includegraphics[width=0.22\textwidth]{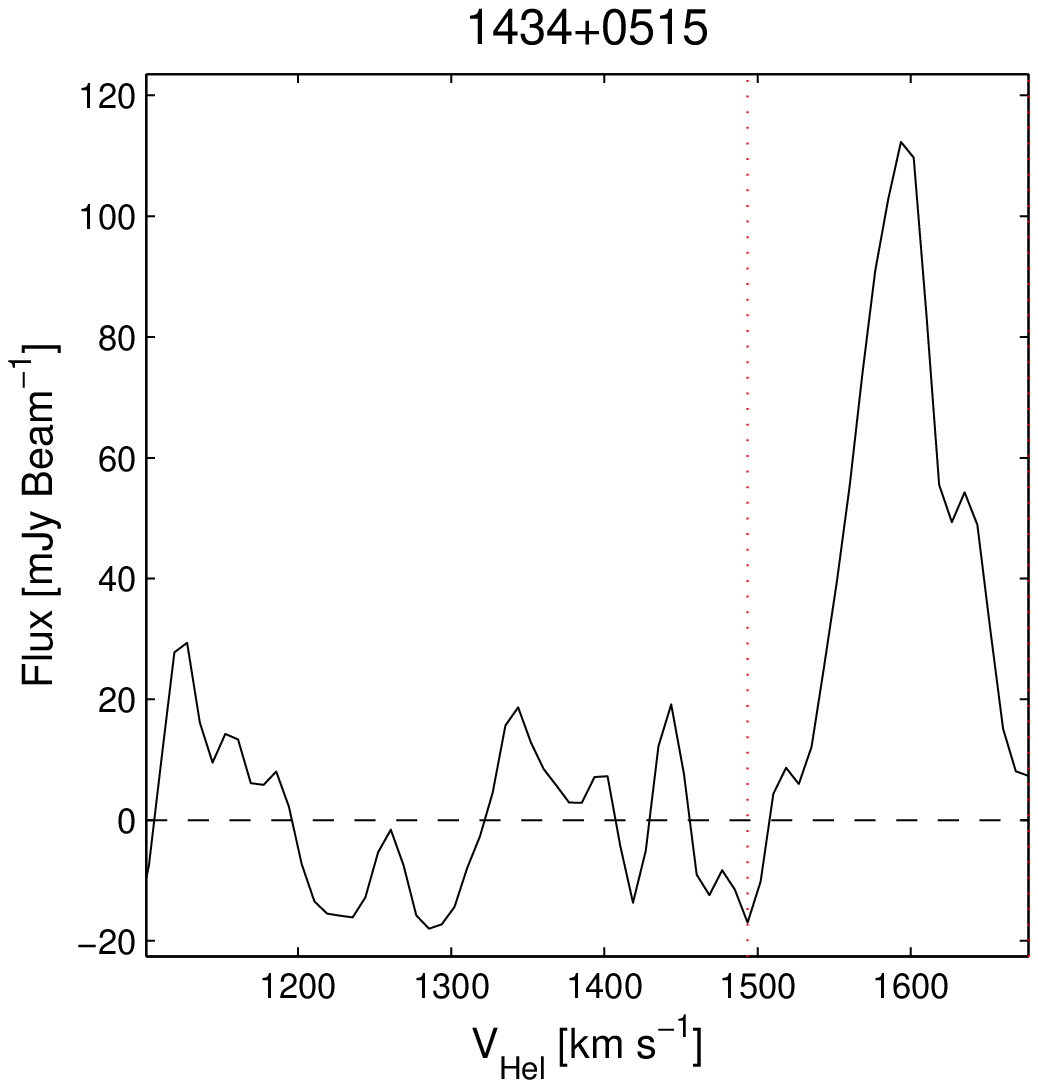}
 \includegraphics[width=0.22\textwidth]{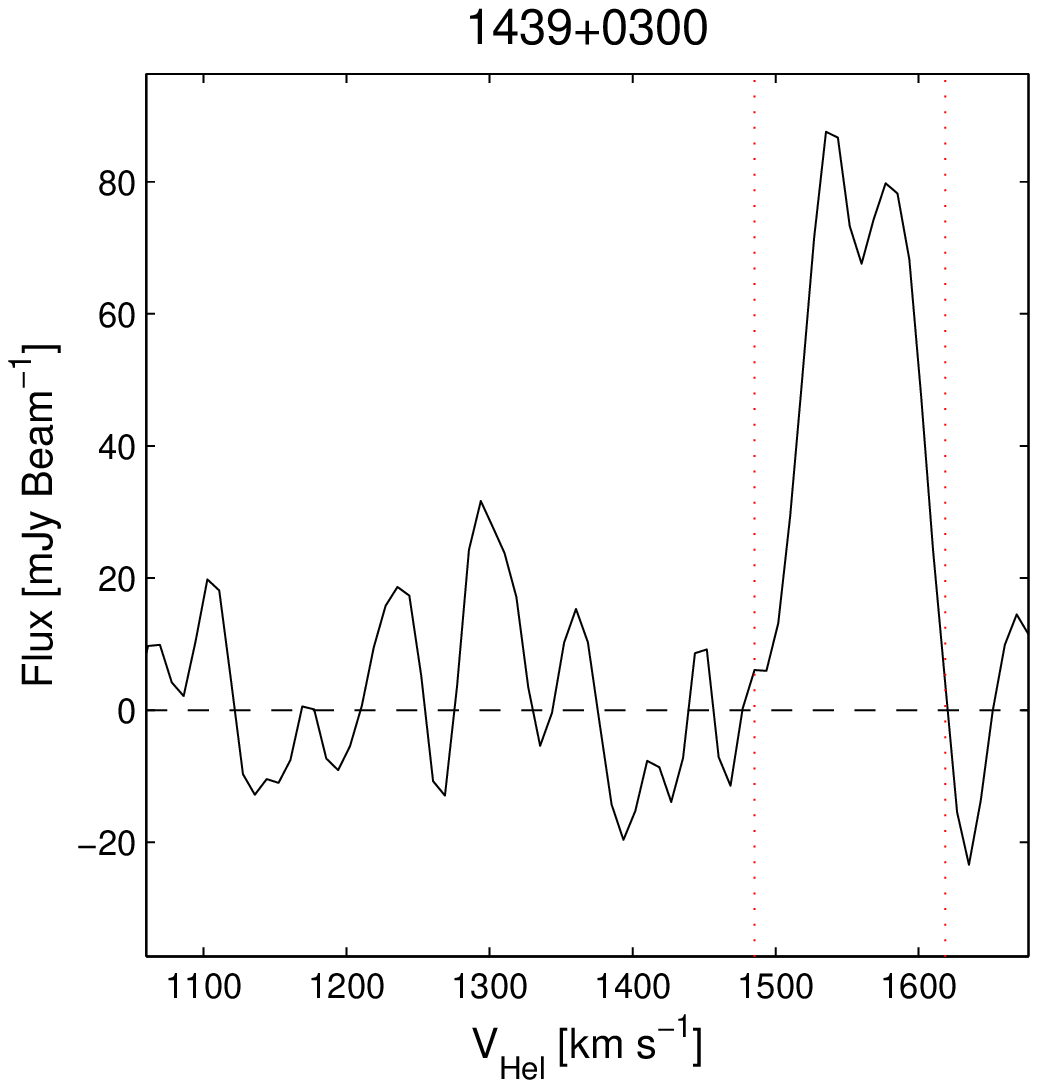}
 \includegraphics[width=0.22\textwidth]{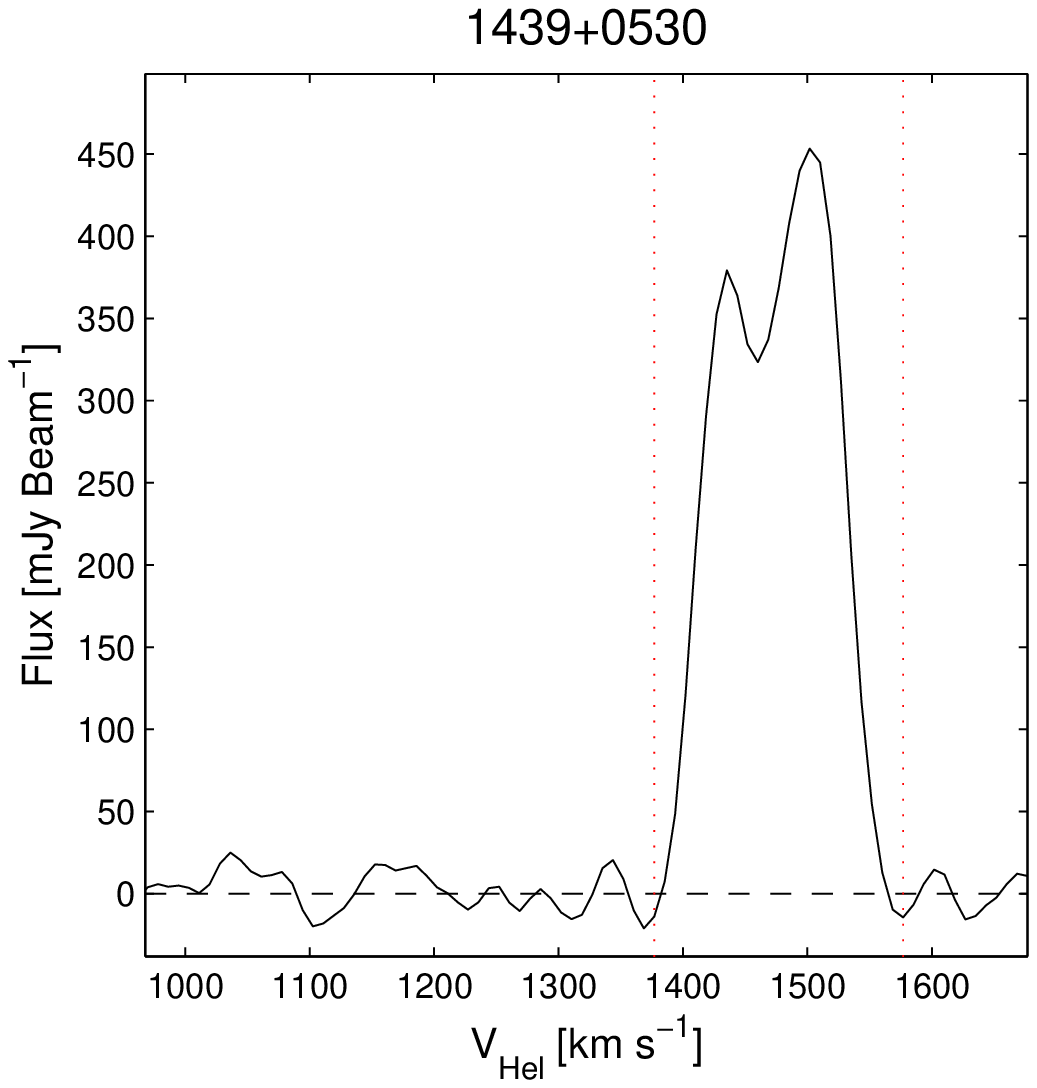}
 \includegraphics[width=0.22\textwidth]{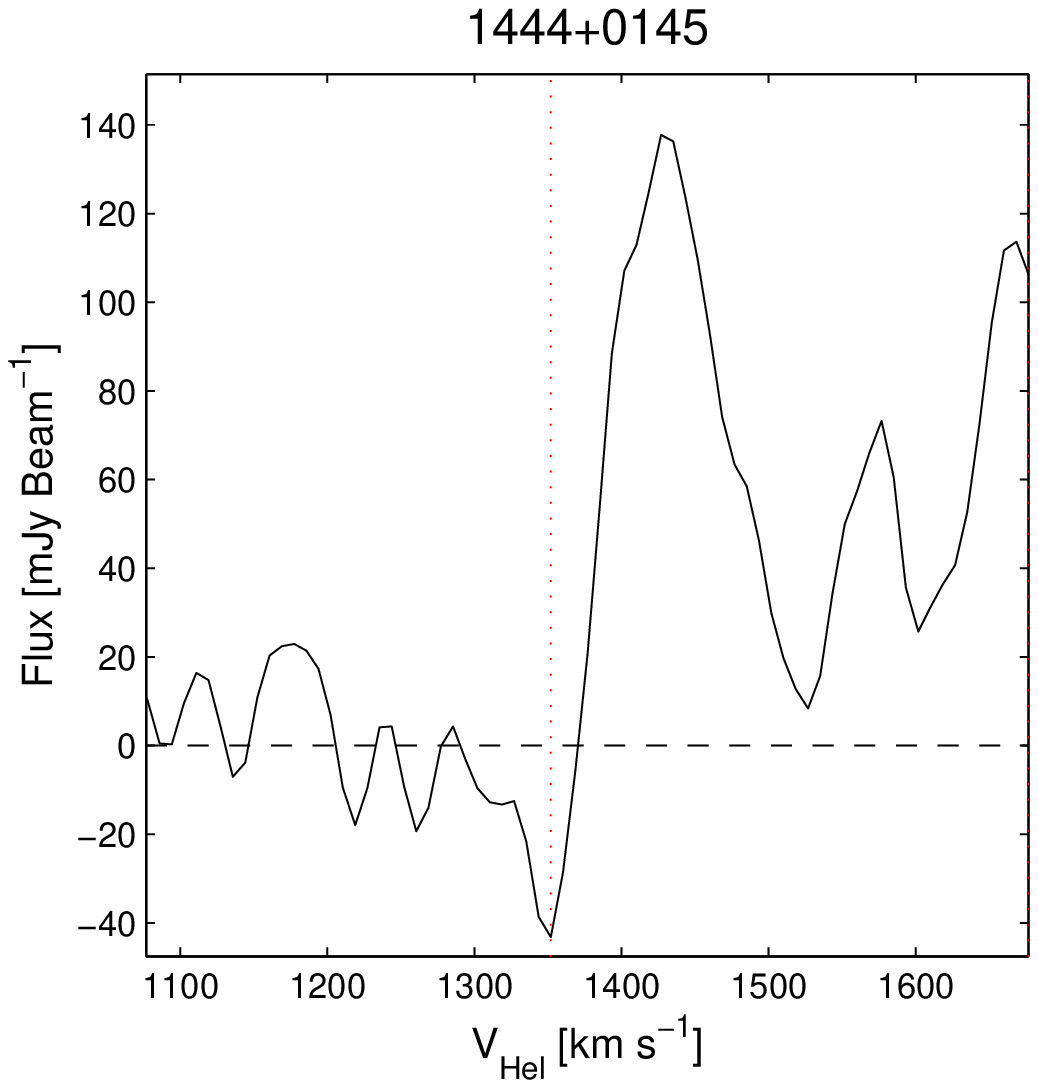}
 \includegraphics[width=0.22\textwidth]{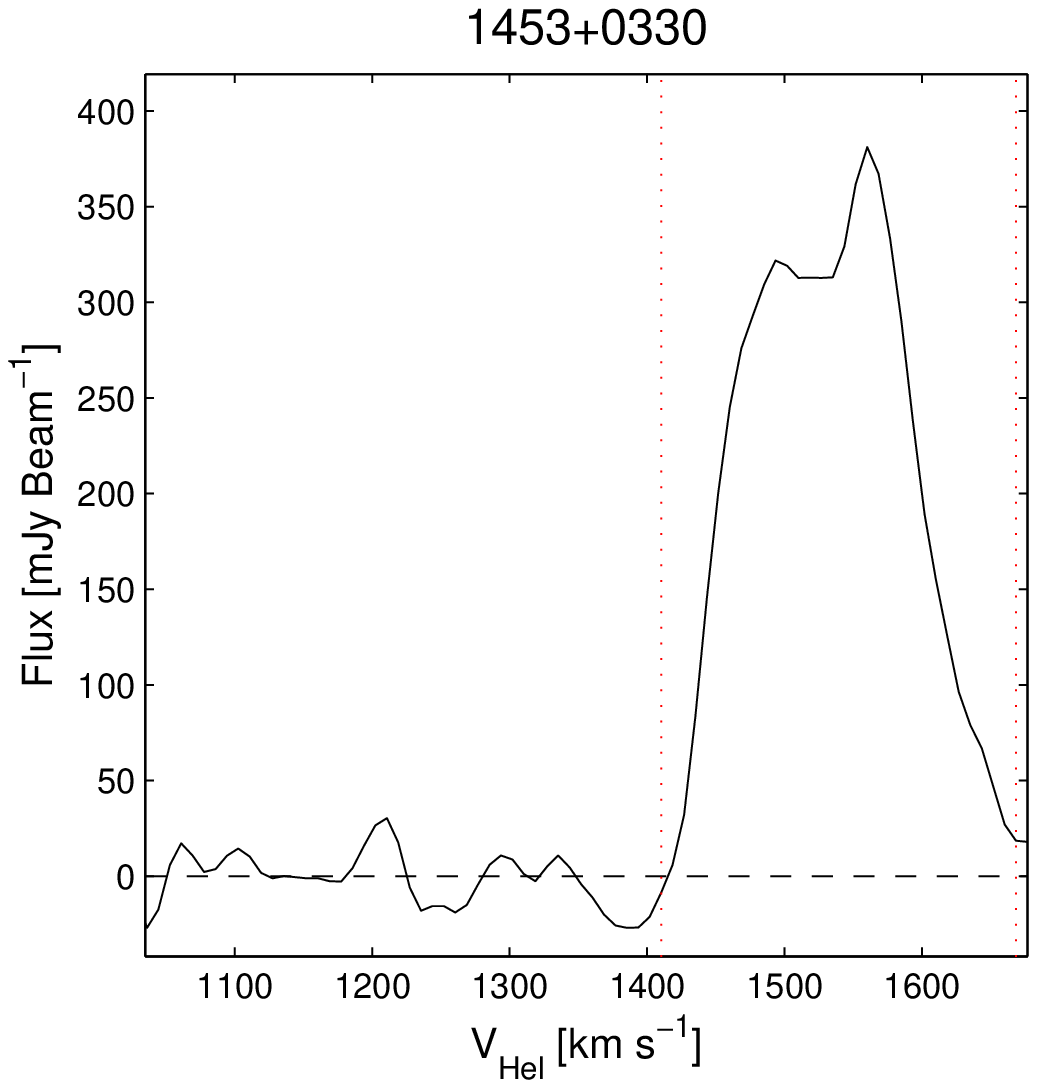}
 \includegraphics[width=0.22\textwidth]{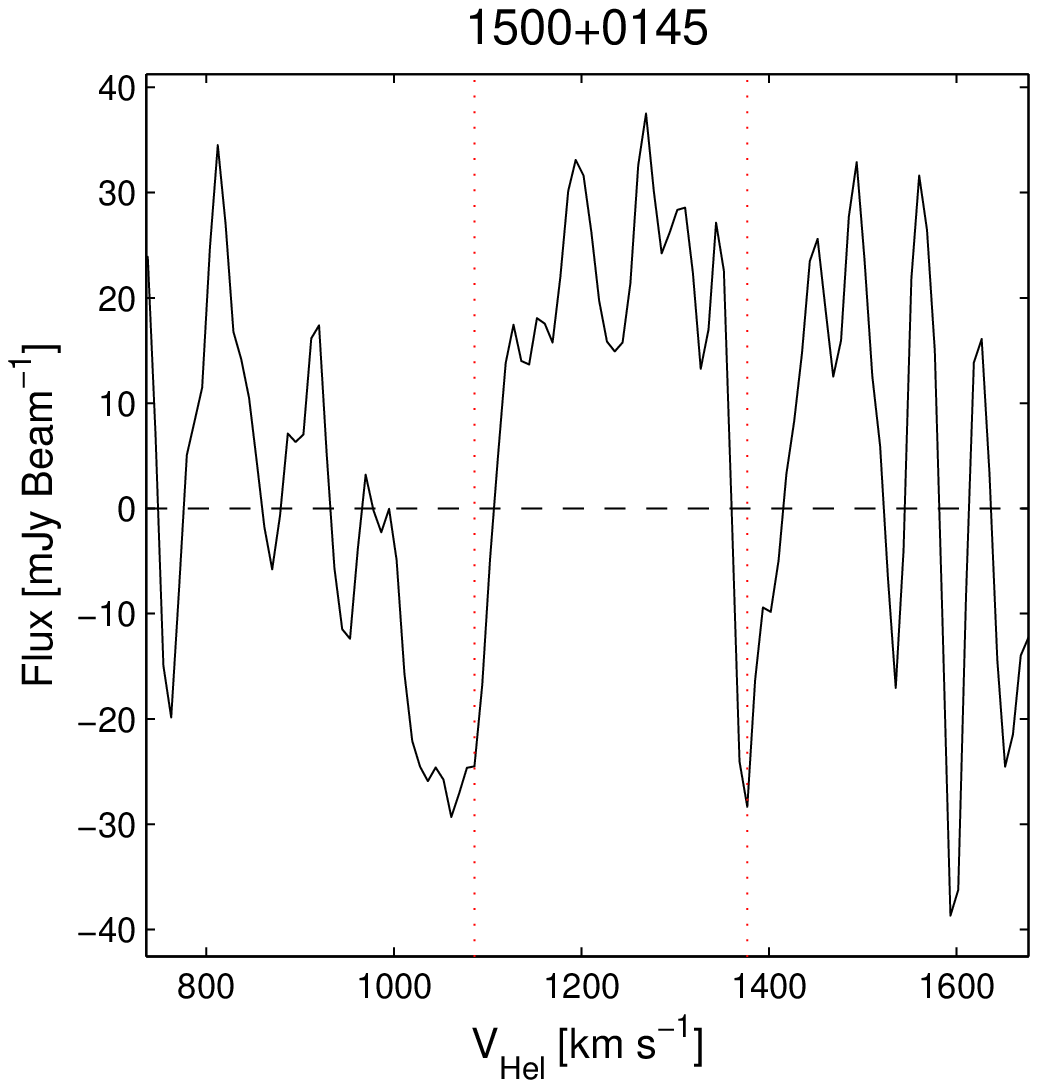}
 \includegraphics[width=0.22\textwidth]{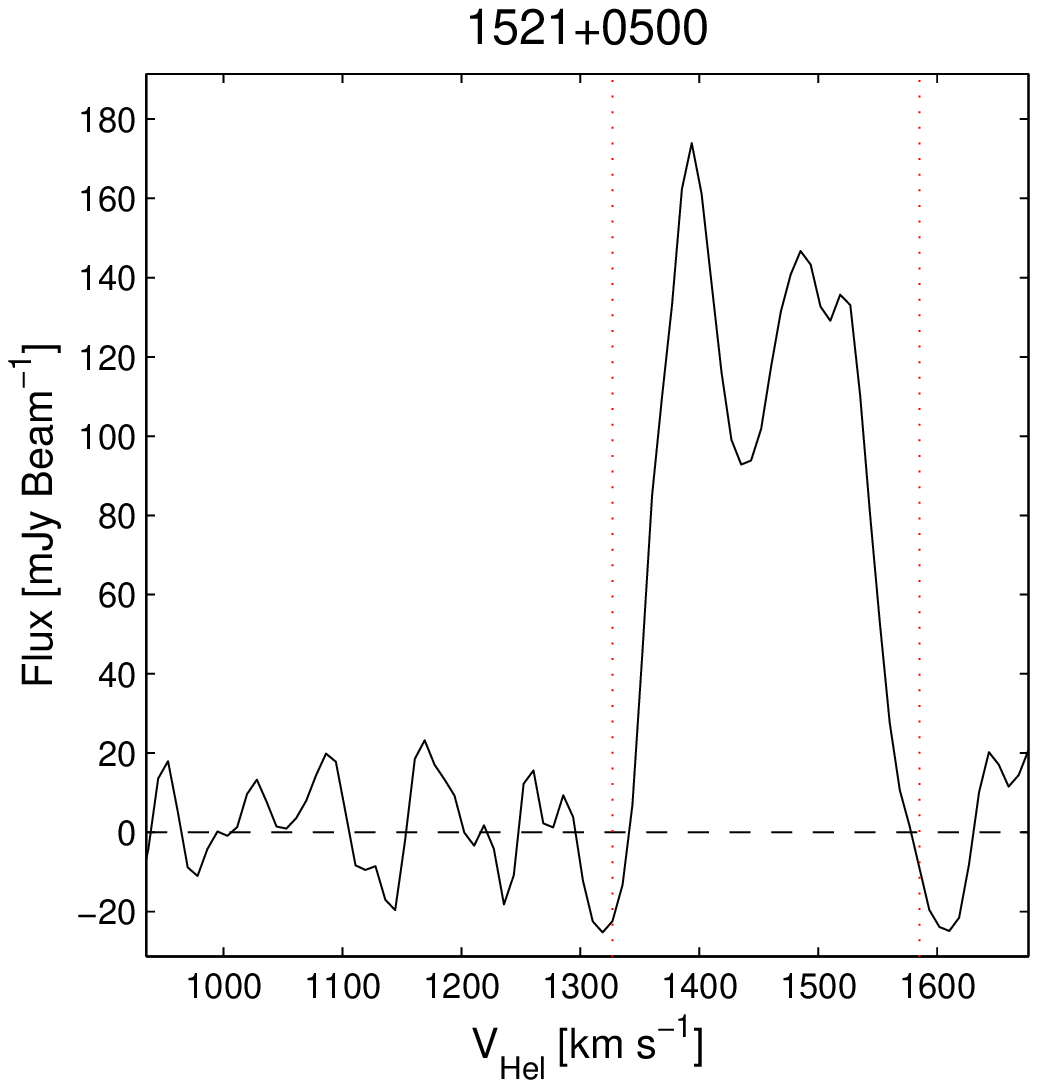}                                                        
                                                         
 \end{center}                                            
{\bf Fig~\ref{all_spectra}.} (continued)                                        
 
\end{figure*}

\begin{figure*}
  \begin{center}

 \includegraphics[width=0.22\textwidth]{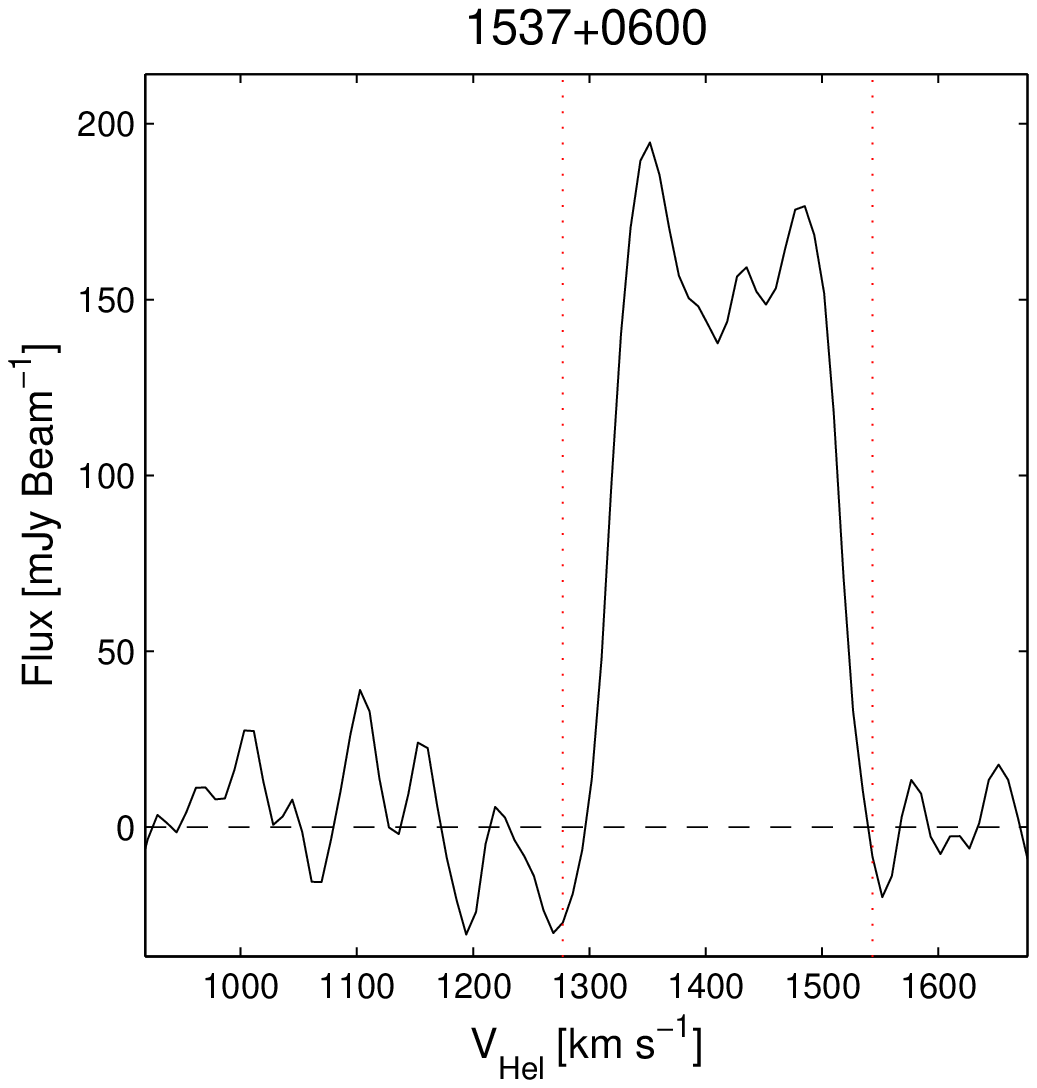}
 \includegraphics[width=0.22\textwidth]{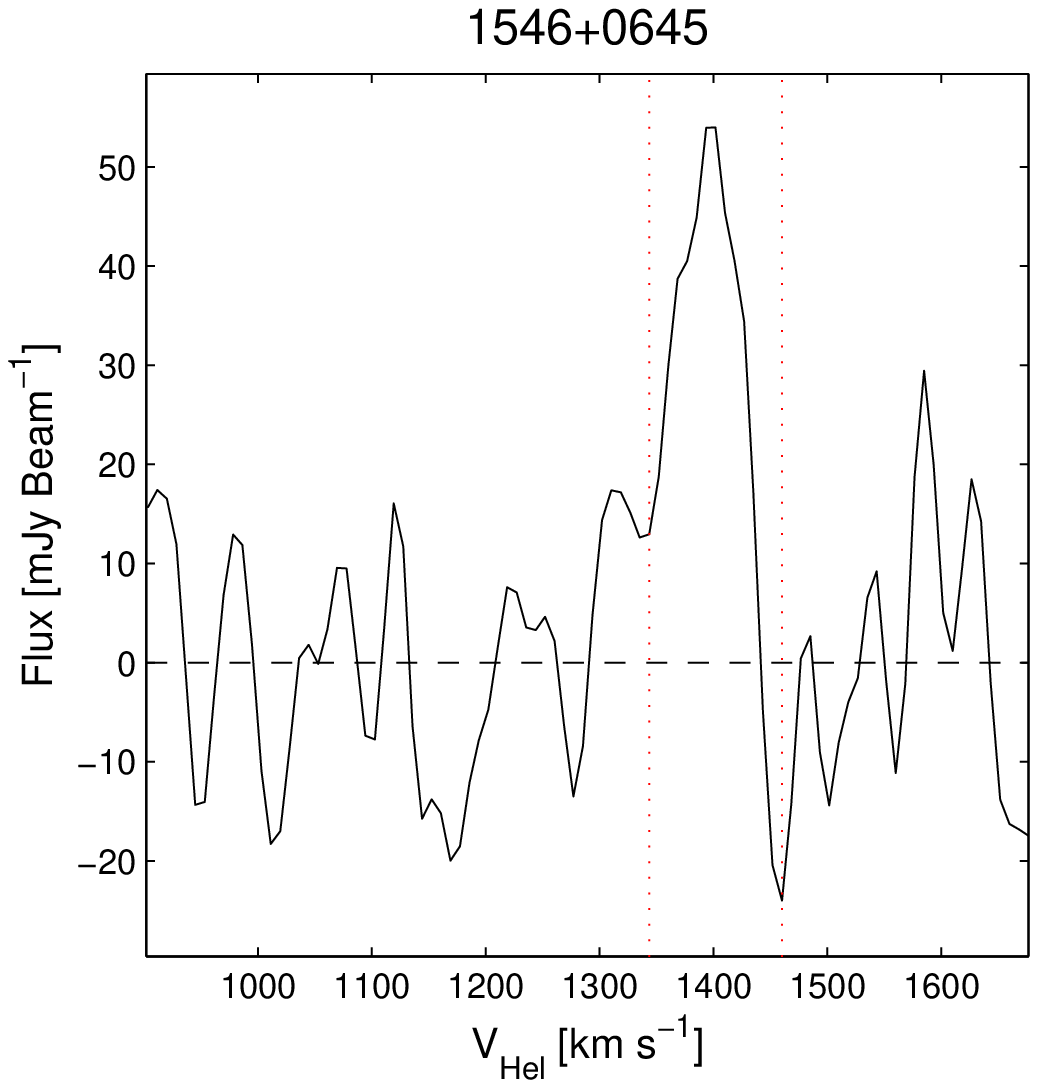}
 \includegraphics[width=0.22\textwidth]{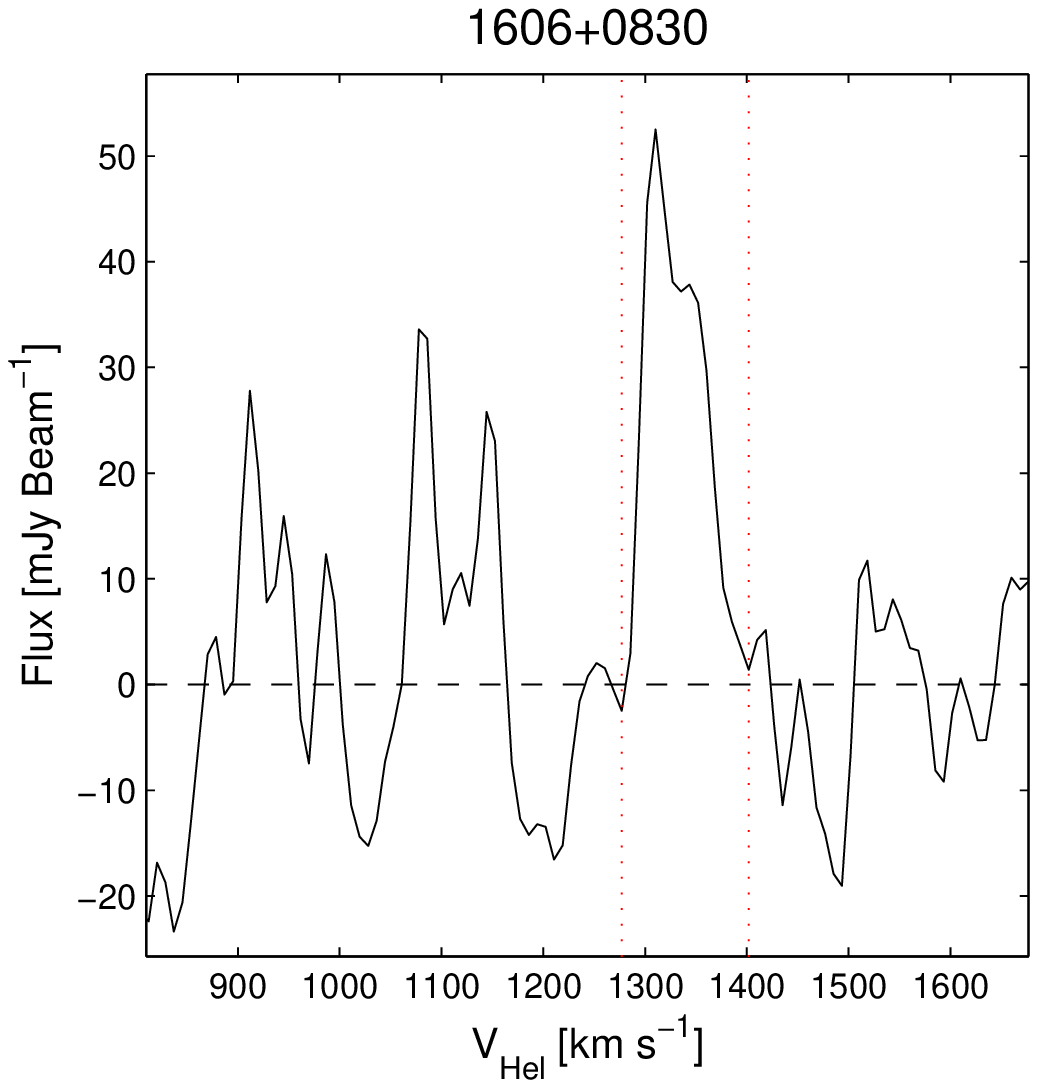}
 \includegraphics[width=0.22\textwidth]{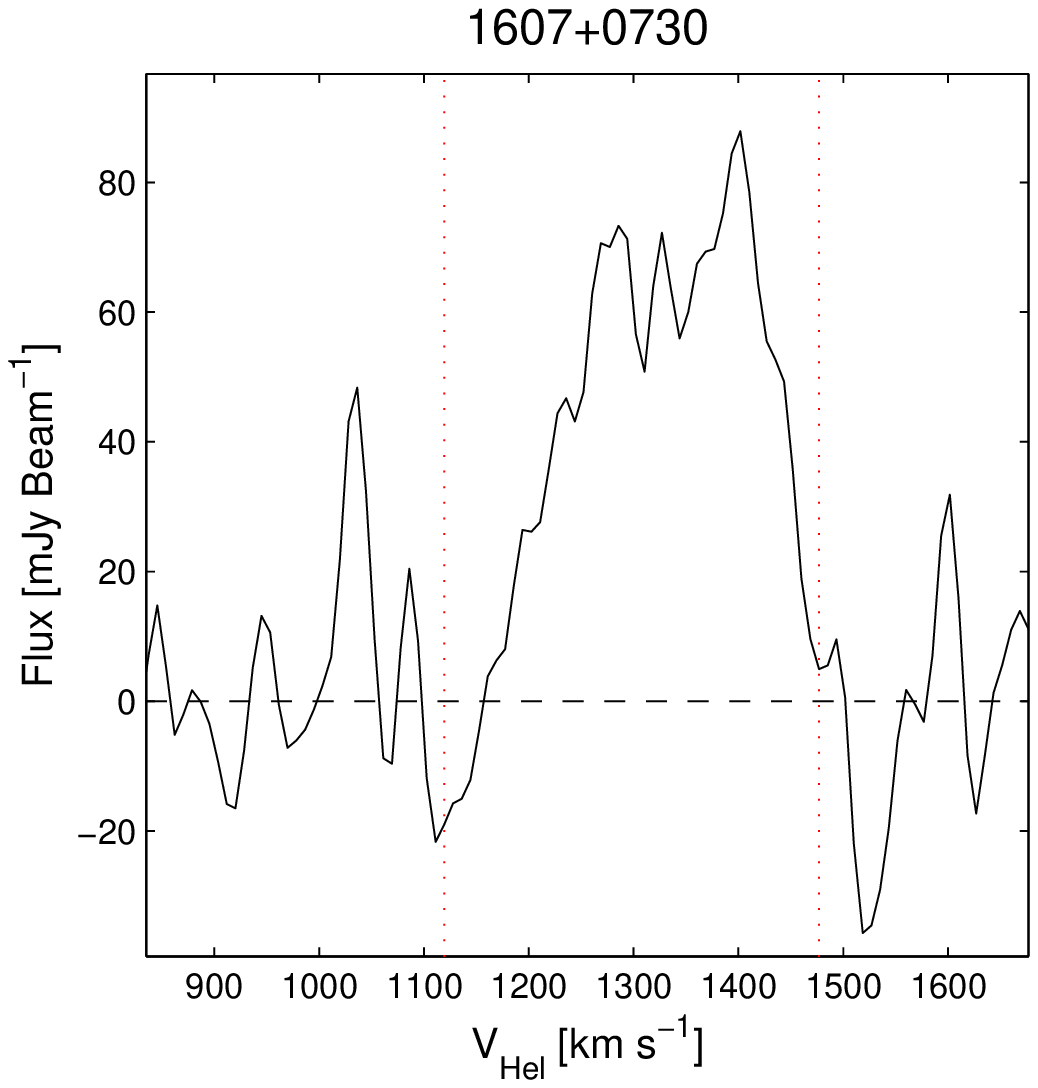}
 \includegraphics[width=0.22\textwidth]{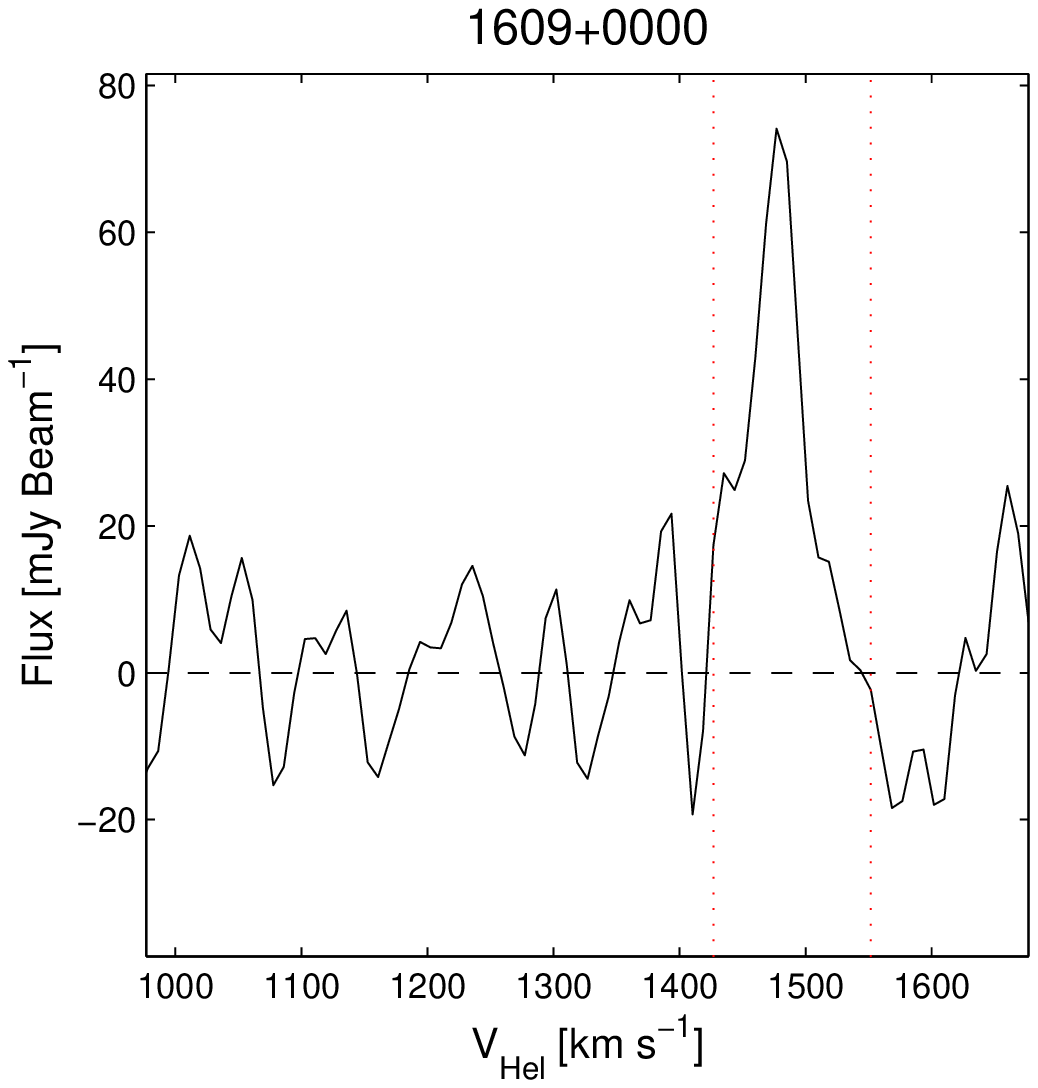}
 \includegraphics[width=0.22\textwidth]{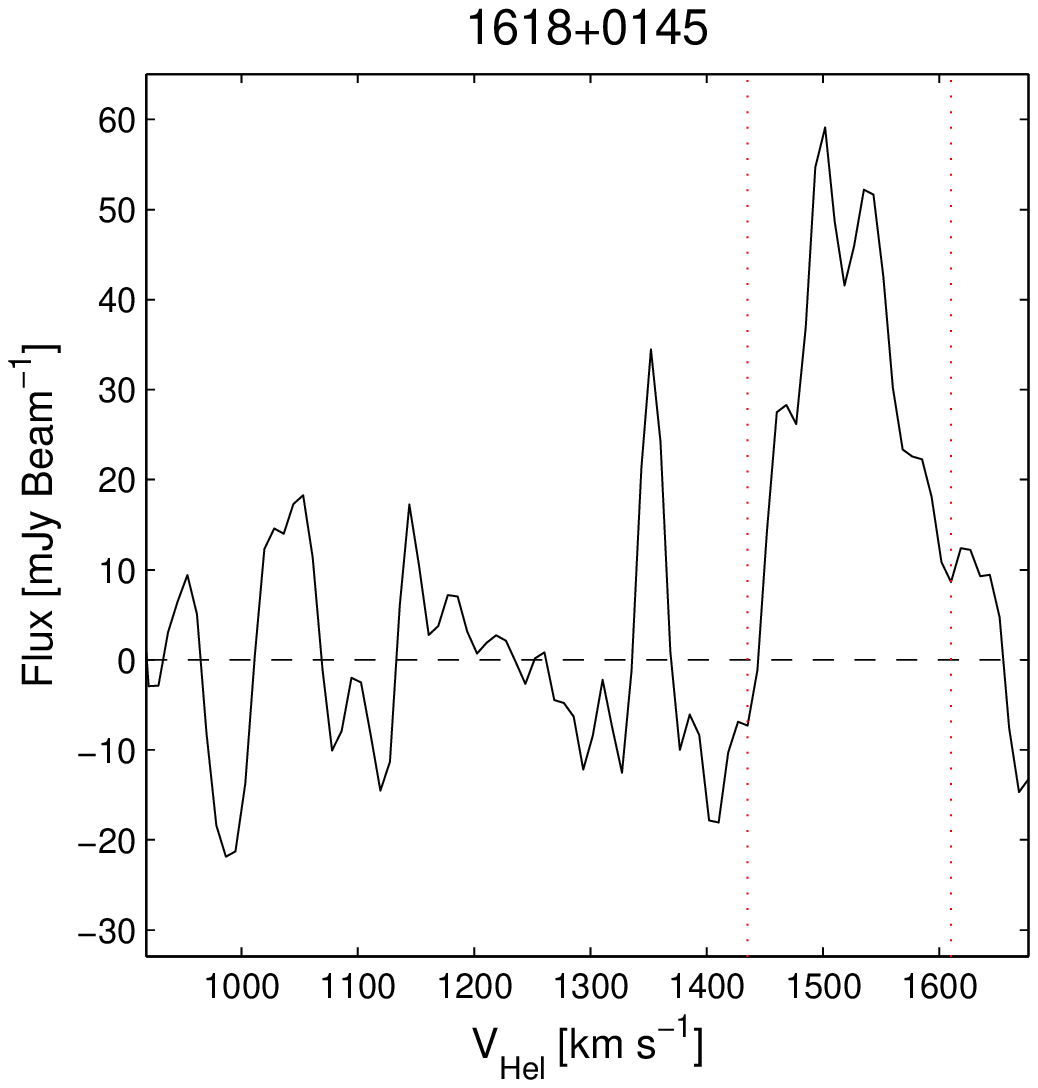}
 \includegraphics[width=0.22\textwidth]{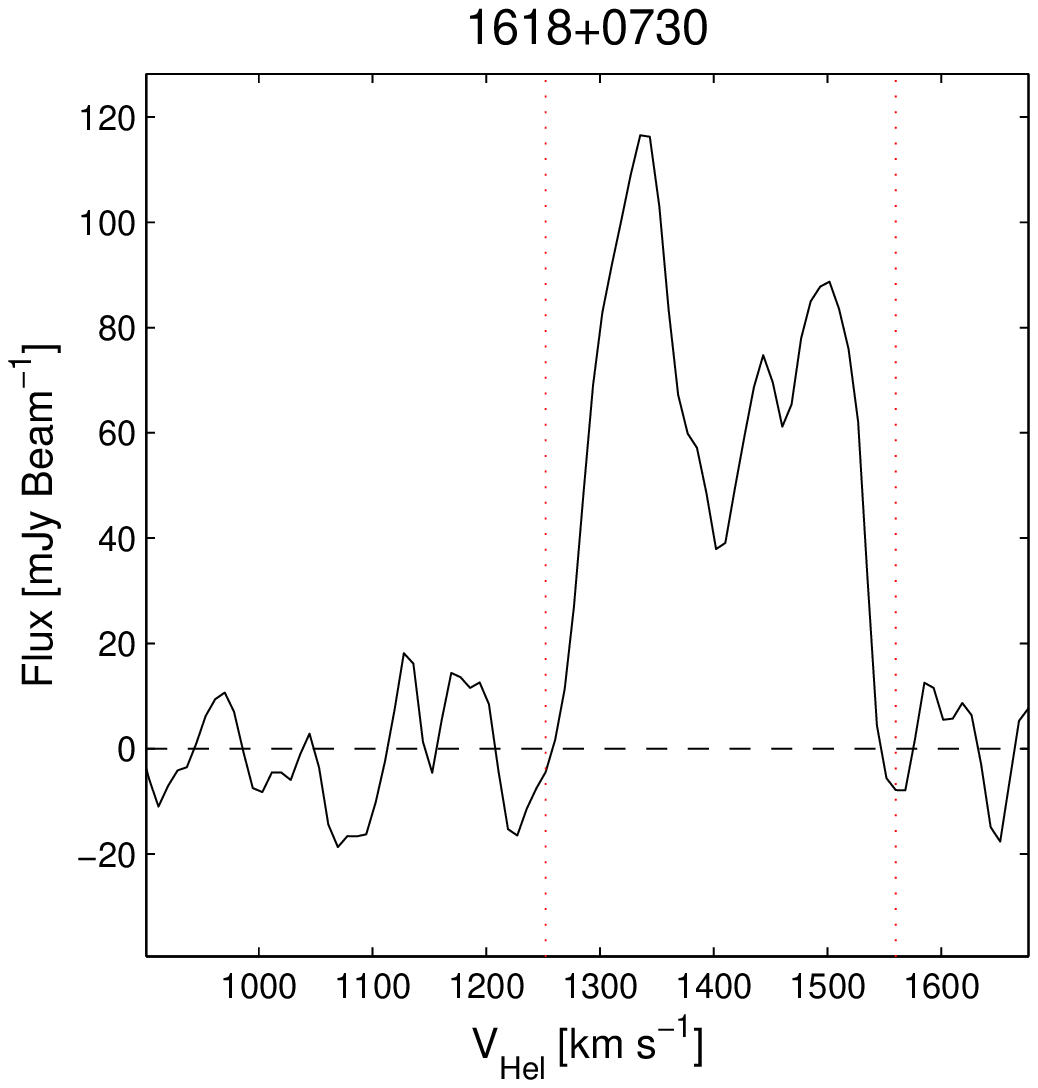}
 \includegraphics[width=0.22\textwidth]{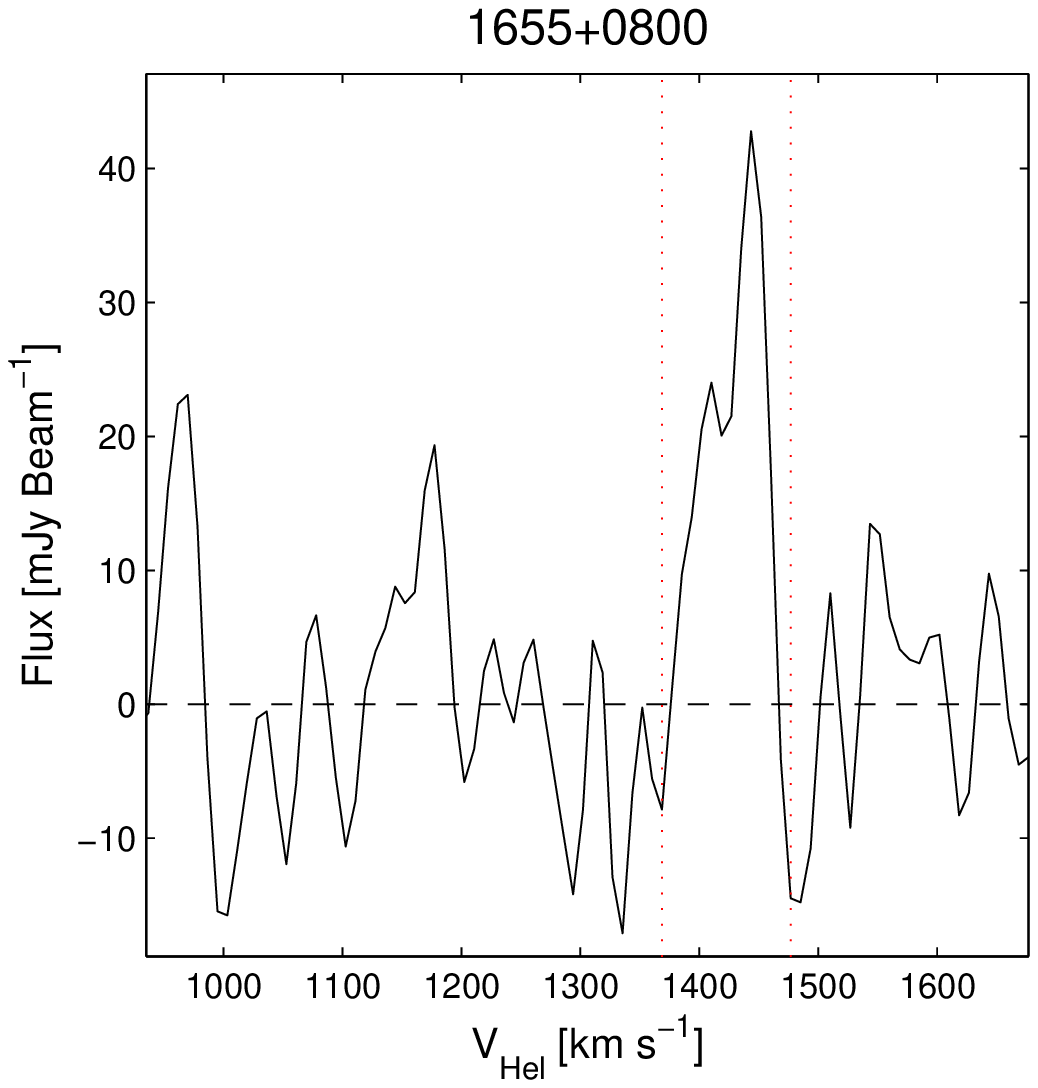}

 \end{center}                                            
{\bf Fig~\ref{all_spectra}.} (continued)                                        
 
\end{figure*}

\end{appendix}

\end{document}